\documentclass{book}
\usepackage{authblk}
\usepackage{roboto}

\usepackage[english]{babel}
\usepackage[letterpaper,top=2cm,bottom=2cm,left=3cm,right=3cm,marginparwidth=1.75cm]{geometry}

\usepackage{amsfonts}
\UseRawInputEncoding

\usepackage{pdflscape}
\usepackage{rotating}
\usepackage{longtable}
\usepackage{array}
\usepackage{float}
\usepackage{amsmath}
\usepackage{graphicx}
\usepackage{inconsolata}
\usepackage[colorlinks=true, allcolors=blue]{hyperref}
\usepackage{enumitem}
\usepackage{tikz}  
\usetikzlibrary{trees, shapes, positioning}

\usepackage{listings}
\usepackage{xcolor}
\usepackage[T1]{fontenc}
\usepackage{IEEEtrantools}  
\usepackage{epigraph}  
\usepackage[hang,flushmargin]{footmisc} 

\DeclareFontFamily{T1}{zi4}{}
\DeclareFontFamily{TS1}{zi4}{}

\DeclareFontShape{T1}{zi4}{m}{n}{<-> ssub * cmss/m/n}{}
\DeclareFontShape{TS1}{zi4}{m}{n}{<-> ssub * cmss/m/n}{}
\DeclareFontShape{T1}{zi4}{m}{it}{<-> ssub * cmss/m/it}{}
\DeclareFontShape{TS1}{zi4}{m}{it}{<-> ssub * cmss/m/it}{}

\DeclareFontShape{T1}{zi4}{bx}{n}{<-> ssub * cmss/bx/n}{}
\DeclareFontShape{T1}{zi4}{bx}{it}{<-> ssub * cmss/bx/it}{}

\DeclareFontShape{T1}{zi4}{b}{n}{<-> ssub * zi4/bx/n}{}
\DeclareFontShape{T1}{zi4}{b}{it}{<-> ssub * zi4/bx/it}{}

\DeclareFontShape{T1}{zi4}{m}{sl}{<-> ssub * zi4/m/it}{}
\DeclareFontShape{T1}{zi4}{bx}{sl}{<-> ssub * zi4/bx/it}{}
\DeclareFontShape{TS1}{zi4}{m}{sl}{<->ssub*zi4/m/it}{}

\definecolor{bgcolor}{rgb}{0.97,0.97,0.97}
\definecolor{codeblue}{rgb}{0.1,0.1,0.8}
\definecolor{codegreen}{rgb}{0,0.4,0}
\definecolor{codegray}{rgb}{0.4,0.4,0.4}
\definecolor{codepurple}{rgb}{0.5,0,0.5}
\definecolor{codered}{rgb}{0.6,0.2,0.2}
\definecolor{lightgray}{rgb}{0.9,0.9,0.9}
\definecolor{darkgray}{rgb}{0.6,0.6,0.6}  

\makeatletter
\renewcommand{\paragraph}{
  \@startsection{paragraph}{4}{\z@}{1ex}{-1em}{\normalfont\normalsize\bfseries\color{gray}}}
\makeatother

\lstdefinestyle{python}{
    language=Python,
    basicstyle=\ttfamily\small\color{black}\usefont{T1}{zi4}{m}{n},  
    keywordstyle=\bfseries\color{codeblue},  
    stringstyle=\color{codegreen},  
    commentstyle=\slshape\color{codegray},  
    showstringspaces=false,
    numbers=left,
    numberstyle=\tiny\color{codegray},  
    stepnumber=1,
    numbersep=8pt,
    frame=single,
    rulecolor=\color{darkgray},  
    breaklines=true,
    backgroundcolor=\color{bgcolor},
    tabsize=4,
    captionpos=b,
    morekeywords={self},  
}

\lstdefinestyle{cmd}{
    language=bash,
    basicstyle=\ttfamily\small\color{black}\usefont{T1}{zi4}{m}{n},  
    keywordstyle=\bfseries\color{blue},
    stringstyle=\color{codegreen},
    commentstyle=\itshape\color{gray},
    showstringspaces=false,
    numbers=none,
    frame=single,
    rulecolor=\color{darkgray},  
    breaklines=true,
    backgroundcolor=\color{bgcolor},
    tabsize=4,
    captionpos=b,
}

\title{Advanced Deep Learning Methods for Protein Structure Prediction and Design}
\footnotesize

\author{
    Yichao Zhang\textsuperscript{*}\thanks{The University of Texas at Dallas, yichao.zhang.us@gmail.com},
    Ningyuan Deng\textsuperscript{*}\thanks{Renmin University, dnylily@gmail.com}, 
    Xinyuan Song\textsuperscript{*}\thanks{Emory University, xsong30@emory.edu},
    Ziqian Bi\textsuperscript{*}\thanks{Indiana University, bizi@iu.edu}, 
    Tianyang Wang\textsuperscript{*}\thanks{University of Liverpool, tianyangwang0305@gmail.com}, 
    Zheyu Yao\thanks{University of Liverpool, pszyao2@liverpool.ac.uk},
    Keyu Chen\thanks{Georgia Institute of Technology, kchen637@gatech.edu}, 
    Ming Li\thanks{Georgia Institute of Technology, mli694@gatech.edu}, 
    Qian Niu\thanks{Kyoto University, niu.qian.f44@kyoto-u.ac.jp}, 
    Junyu Liu\thanks{Kyoto University, liu.junyu.82w@kyoto-u.ac.jp}, 
    Benji Peng\thanks{AppCubic, benji@appcubic.com}, 
    Sen Zhang\thanks{Rutgers University, sen.z@rutgers.edu}, 
    Ming Liu\thanks{Purdue University, liu3183@purdue.edu}, 
    Li Zhang\thanks{University of science and technology, zanly20@mail.ustc.edu.cn}, 
    Xuanhe Pan\thanks{University of Wisconsin-Madison, xpan73@wisc.edu}, 
    Jinlang Wang\thanks{University of Wisconsin-Madison, jinlang.wang@wisc.edu}, 
    Pohsun Feng\thanks{National Taiwan Normal University, 41075018h@ntnu.edu.tw}, 
    Yizhu Wen\thanks{University of Hawaii, yizhuw@hawaii.edu}, 
    Lawrence K.Q. Yan\thanks{Hong Kong University of Science and Technology, kqyan@connect.ust.hk}, 
    Hong-Ming Tseng\thanks{School of Visual Arts, htseng@sva.edu}, 
    Yan Zhong\thanks{Nanyang technology University, n2409733d@e.ntu.edu.sg}, 
    Yunze Wang\thanks{University of Edinburgh, y.wang-861@sms.ed.ac.uk}, 
    Ziyuan Qin\thanks{Case Western Reserve University, ziyuan.qin@case.edu}, 
    Bowen Jing\thanks{University of Manchester, bowen.jing@postgrad.manchester.ac.uk}, 
    Junjie Yang\thanks{Pingtan Research Institute of Xiamen University, youngboy@xmu.edu.cn}, 
    Jun Zhou\thanks{The University of Texas at Dallas, jun.zhou.tx@gmail.com},
    Chia Xin Liang\thanks{JTB Technology Corp, cxldun@gmail.com},
    Junhao Song\textsuperscript{$\dagger$}\thanks{Imperial College London, junhao.song23@imperial.ac.uk}
}
\date{}  

\begin{document}

\maketitle

\begingroup
\renewcommand\thefootnote{}\footnote{
    \textsuperscript{*} Equal contribution \\
    \textsuperscript{$\dagger$} Corresponding author
}
\addtocounter{footnote}{0}
\endgroup

\epigraph{"When visual and audio events occur together, it suggests that there might be a common, underlying event that produced both signals."}{\textit{Andrew Owens}}

\tableofcontents  

\chapter*{Introduction}

In the last few decades, protein structure prediction and design have emerged as critical areas in computational biology and bioinformatics~\cite{Jnes2024DeepLF, Wicky2022HallucinatingSP, Ma2024BeyondCB, Hansen2024CarvingOA}. Understanding protein structures is essential because the three-dimensional (3D) conformation of a protein directly influences its function, interactions, and involvement in various biological processes. Despite remarkable advances in experimental methods, the complexity of protein folding and the limitations of traditional techniques have posed significant challenges to researchers.

With the advent of deep learning, the field has experienced an unprecedented leap forward. Notably, the development of \textbf{AlphaFold} by DeepMind represents a landmark achievement \cite{jumper2021highly, jumper2021applying}. AlphaFold's success in the 2020 CASP14 competition has provided an extraordinary leap in our ability to predict protein structures, solving many aspects of the "protein folding problem." This breakthrough has catalyzed the design of new proteins, aiding in drug development, molecular engineering, and understanding disease mechanisms.

\section*{Historical Approaches to Protein Structure Prediction}

Protein structure prediction has a rich history, rooted in a combination of experimental techniques and computational methods. Before diving into the modern advances brought by deep learning, it is crucial to understand the traditional approaches that laid the foundation for today's achievements.

\subsection*{Experimental Techniques}
For decades, determining the structure of proteins relied heavily on experimental methods. The most prominent techniques include:

\subsubsection*{X-ray Crystallography}
X-ray crystallography has long been considered the gold standard for determining protein structures at atomic resolution \cite{drenth2007principles}. This technique involves crystallizing the protein and bombarding it with X-rays. The resulting diffraction pattern allows researchers to infer the protein's structure by solving complex mathematical equations. Although highly accurate, X-ray crystallography is time-consuming, requires pure protein crystals, and is not applicable to all protein types, especially membrane proteins and intrinsically disordered proteins.

\subsubsection*{Nuclear Magnetic Resonance (NMR) Spectroscopy}
NMR spectroscopy is an alternative method, particularly useful for studying proteins in solution \cite{cavanagh1996protein}. It measures the magnetic properties of atomic nuclei to infer distances and angles between atoms. NMR can provide information about protein dynamics and is valuable for proteins that cannot be crystallized. However, its resolution is generally lower than that of X-ray crystallography, and it is limited to smaller proteins (typically under 30 kDa).

\subsubsection*{Cryo-Electron Microscopy (Cryo-EM)}
Cryo-EM has recently gained popularity as a powerful technique for studying the structures of large macromolecular complexes and membrane proteins \cite{cheng2018single}. The technique involves flash-freezing proteins and imaging them with an electron microscope. Cryo-EM has seen a remarkable increase in resolution in recent years, reaching near-atomic resolution for certain structures. It has the advantage of not requiring crystallization, making it more versatile for challenging targets. However, it requires specialized equipment and computational resources.

\subsection*{Computational Methods Prior to Deep Learning}
While experimental methods are invaluable, they are expensive and labor-intensive. To complement these techniques, various computational approaches have been developed over the years, allowing researchers to predict protein structures in silico.

\subsubsection*{Homology Modeling (Comparative Modeling)}
One of the earliest and most widely used computational methods is homology modeling \cite{webb2016comparative}. This approach leverages the evolutionary conservation of protein structures, using known structures of homologous proteins (proteins with similar sequences) to predict the structure of an unknown protein. Homology modeling is particularly effective when a closely related structure exists in the protein database. However, its accuracy decreases significantly when applied to proteins with low sequence similarity to known structures.

\subsubsection*{Threading (Template-Based Modeling)}
Threading, also known as fold recognition, is another traditional method \cite{yang2015tasser}. It aligns the target protein sequence with a set of structural templates to identify the most probable fold for the protein. This method is particularly useful for proteins that lack a clear homolog in the sequence database but may still share a common fold with known proteins. While threading can predict structures for proteins without high sequence homology, its accuracy is often limited by the availability of suitable structural templates.

\subsubsection*{Molecular Dynamics (MD) Simulations}
Molecular dynamics simulations are a physics-based approach that predicts protein structures by simulating the atomic-level movements of proteins over time \cite{abraham2015gromacs}. MD simulations provide detailed information about protein dynamics, folding pathways, and conformational changes. However, they are computationally expensive and are typically only feasible for small proteins or short timescales. Moreover, MD simulations depend heavily on the accuracy of the force fields used to model the protein's atomic interactions.

\section*{The Rise of Deep Learning in Protein Structure Prediction}

The advent of deep learning marked a paradigm shift in the field of protein structure prediction~\cite{Mardikoraem2024EvoSeqMLAD, Jnes2024DeepLF, Saharkhiz2024TheSO, Goudy2023InSE}. Traditional methods, while useful, were limited by the complexity of protein folding and the lack of comprehensive datasets. Deep learning models, with their ability to learn complex patterns from large datasets, have dramatically improved the accuracy of structure prediction, transforming the field.

\subsection*{AlphaFold: A Breakthrough in Protein Folding}
\textbf{AlphaFold}, developed by DeepMind, represents the most significant breakthrough in protein structure prediction to date. AlphaFold's success at CASP14 was a major turning point. The model achieved near-experimental accuracy in predicting protein structures, outperforming all previous methods by a wide margin.

AlphaFold utilizes deep neural networks to learn the relationships between amino acid sequences and their corresponding three-dimensional structures. By incorporating evolutionary information and geometric constraints, AlphaFold can accurately predict protein folds even for proteins with no known homologous structures. The model also includes a confidence score for each predicted structure, allowing researchers to assess the reliability of the predictions.

The impact of AlphaFold on the scientific community has been profound. It has been applied to a wide range of biological problems, including drug discovery, understanding protein-protein interactions, and studying disease-related mutations. Its predictions have been made available through the AlphaFold Protein Structure Database, which contains the structures of over 200 million proteins. This resource is revolutionizing our understanding of the proteome and enabling new discoveries across multiple fields.

\subsection*{Other Advances in Deep Learning}
Following AlphaFold's success, several other deep learning models have emerged, contributing to the ongoing revolution in protein structure prediction and design:

\subsubsection*{RoseTTAFold}
Developed by the Baker Lab at the University of Washington, \textbf{RoseTTAFold} is another deep learning model designed for protein structure prediction \cite{baek2021accurate}. Like AlphaFold, RoseTTAFold uses a neural network to predict protein structures, but it employs a different architecture, making it faster and more accessible for certain applications. RoseTTAFold has been successfully used in de novo protein design, enabling the creation of novel proteins with specific functions.

\subsubsection*{ProteinMPNN}
\textbf{ProteinMPNN} is a deep learning model tailored for protein design rather than prediction \cite{dauparas2022robust}. It generates protein sequences that are likely to fold into desired structures, which is invaluable for designing proteins with specific functions, such as enzymes or therapeutic agents. ProteinMPNN is part of the growing field of protein design, which holds the potential to revolutionize biotechnology and synthetic biology.

\subsubsection*{OmegaFold}
\textbf{OmegaFold} is a more recent entrant in the field of protein structure prediction \cite{wu2022high}. Building on the success of AlphaFold, OmegaFold uses deep learning to tackle even more complex protein structures, including those with significant disorder or multi-domain architecture. OmegaFold seeks to address some of the remaining challenges in the field, pushing the boundaries of what can be achieved through AI-driven protein prediction.

\section*{The Nobel Prize Discussion}
In 2024, Demis Hassabis and John Jumper, the scientists behind AlphaFold, were awarded the Nobel Prize in Chemistry for their groundbreaking work in protein structure prediction using artificial intelligence. AlphaFold's ability to predict protein structures with high accuracy has been hailed as a major scientific achievement. The AI-driven system has revolutionized biology and medicine, providing crucial insights for drug development, disease understanding, and synthetic biology.

The AlphaFold Protein Structure Database, which offers open access to more than 200 million protein structures, has significantly accelerated research in various fields, enabling scientists around the world to explore previously inaccessible proteins. Its impact spans from pharmaceuticals to agriculture, contributing to vaccine development and efforts to combat antibiotic resistance. AlphaFold's success highlights the transformative potential of AI in addressing global scientific challenges.

\section*{Future Directions}
The future of protein structure prediction and design is bright, with deep learning continuing to drive progress in the field. As models like AlphaFold, RoseTTAFold, and OmegaFold evolve, we can expect even greater accuracy and efficiency in predicting protein structures. Moreover, advances in protein design tools such as ProteinMPNN will open new doors for creating novel proteins with specific functions, offering exciting possibilities for biotechnology, medicine, and materials science.

The integration of deep learning with other computational techniques, such as molecular dynamics and quantum mechanics, could further enhance our ability to predict protein behavior in complex environments. Additionally, as experimental data grows and computational power increases, the synergy between experimental and computational methods will continue to refine our understanding of protein folding and dynamics.

\section*{Conclusion}
The intersection of deep learning and protein structure prediction represents one of the most exciting frontiers in modern biology. The progress made by AlphaFold and other deep learning models has not only solved long-standing challenges but also paved the way for future innovations in protein design and biomedical research. As we continue to explore this field, we are likely to witness even more transformative discoveries, bringing us closer to solving critical problems in health, disease, and biotechnology.

This book will guide you through the history, current state, and future directions of protein structure prediction and design, highlighting the profound impact of deep learning technologies on the biological sciences. From traditional methods to the cutting-edge AI models, this journey will illuminate the fascinating world of proteins, the building blocks of life.
\chapter{The History of Protein Structure Prediction and Protein Design}

\section{The Early Research and Methods}

Protein structure prediction and design have a rich history, dating back to the early days of molecular biology when scientists began to understand the relationship between a protein's amino acid sequence and its three-dimensional (3D) structure. In this chapter, we will explore the early research and methods that laid the foundation for modern computational approaches, including the role of Python and deep learning in protein structure prediction.

\subsection{Understanding Protein Structure}

Proteins are made up of long chains of amino acids, and their function is determined by their 3D structure. The structure of a protein can be broken down into four levels:

\begin{itemize}
    \item \textbf{Primary structure}: The linear sequence of amino acids in the polypeptide chain. This is the simplest representation of a protein, and understanding it is critical for predicting higher-level structures.
    \item \textbf{Secondary structure}: Local folding patterns of the polypeptide chain, primarily the formation of $\alpha$-helices and $\beta$-sheets. These structures are stabilized by hydrogen bonds between the backbone atoms of the amino acids.
    \item \textbf{Tertiary structure}: The overall 3D conformation of the entire polypeptide chain, formed by interactions between side chains (R groups) of amino acids.
    \item \textbf{Quaternary structure}: The arrangement of multiple polypeptide chains (subunits) into a functional protein complex.
\end{itemize}

Before we dive into modern deep learning approaches, it is important to understand the traditional methods used in protein structure prediction and design.

\subsection{Homology Modeling}

One of the earliest computational methods for predicting protein structure is homology modeling (also known as comparative modeling). Homology modeling is based on the observation that proteins with similar amino acid sequences tend to have similar structures. If the structure of a homologous protein is known, it can serve as a template to predict the structure of an unknown protein.

\subsubsection{Example of Homology Modeling}

Let's consider an example where we want to predict the structure of a protein whose sequence is similar to a protein with a known structure. Suppose we have the following sequences:

\textbf{Target protein sequence:}

\begin{lstlisting}[style=python]
target_sequence = "MQSLKSTALIVGLNQRLAALVDTRAYAEGFYGDPATLG"  
\end{lstlisting}

\textbf{Template protein sequence (known structure):}

\begin{lstlisting}[style=python]
template_sequence = "MQSLKATALVVGLNQRLSALVDTRAFAGGFLGDPATLG"
\end{lstlisting}

Notice the high degree of similarity between the target and template sequences. Using homology modeling, we can align these two sequences and model the target protein's structure based on the template structure.

Tools like \textbf{Modeller} in Python allow us to automate this process \cite{webb2016comparative}. Here's an example of how we could perform homology modeling using Modeller in Python:

\begin{lstlisting}[style=python]
from modeller import *
from modeller.automodel import *

env = environ()
a = automodel(env, alnfile='alignment.ali',
              knowns='template', sequence='target')
a.starting_model = 1
a.ending_model = 5
a.make()
\end{lstlisting}

In this code, we create a homology model based on the known template structure, aligning the sequences and generating multiple models of the target protein to choose the best one.

\subsection{Threading (Template-Based Modeling)}

Threading, or fold recognition, is another early method used for protein structure prediction \cite{jones1999protein}. Unlike homology modeling, which relies on sequence similarity, threading can be used to predict the structure of proteins with no obvious sequence homology. This method works by "threading" the target protein sequence onto a library of known protein folds and evaluating which fold fits best.

Let's visualize this concept with a simple tree diagram to illustrate the different possible folds a protein could take.


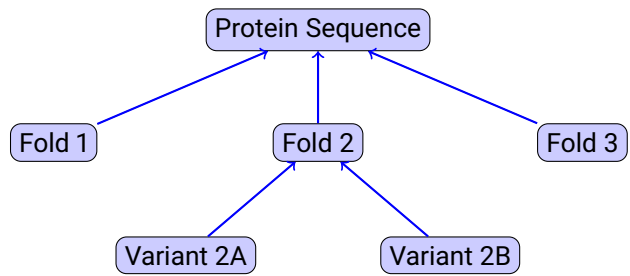
\begin{center}
\begin{tikzpicture}[sibling distance=10em, every node/.style = {shape=rectangle, rounded corners, draw, align=center, fill=blue!20}]
  \node (protein) {Protein Sequence}
    child { node (fold1) {Fold 1} }
    child { node (fold2) {Fold 2}
      child { node (variant2A) {Variant 2A} }
      child { node (variant2B) {Variant 2B} }
    }
    child { node (fold3) {Fold 3} };

  \tikzset{edge/.style={->, thick, blue}}
  \draw[edge] (fold1) -- (protein);  
  \draw[edge] (fold2) -- (protein);  
  \draw[edge] (fold3) -- (protein);  
  \draw[edge] (variant2A) -- (fold2);  
  \draw[edge] (variant2B) -- (fold2);  
\end{tikzpicture}
\end{center}

\subsubsection{Threading Example in Python}

Although threading is less commonly used today due to the success of deep learning methods, we can still implement it in Python using existing libraries like \textbf{HHpred}, a tool commonly used for threading:

\begin{lstlisting}[style=python]
from hhpred import HHpred
hhpred = HHpred()
target_sequence = "MQSLKSTALIVGLNQRLAALVDTRAYAEGFYGDPATLG"
hhpred.predict(target_sequence, database='pdb')
\end{lstlisting}

This simple example shows how you can use threading to predict a protein structure by comparing the target sequence to a database of known structures.

\subsection{Molecular Dynamics Simulations}

Molecular dynamics (MD) simulations provide a way to study the physical movements of atoms and molecules in proteins \cite{abraham2015gromacs}. Instead of predicting static structures, MD allows researchers to simulate the dynamics of proteins, including their folding pathways, interactions with ligands, and conformational changes.

In MD simulations, the forces acting on each atom in the protein are calculated, and the positions of the atoms are updated over time. This requires solving Newton's equations of motion for every atom in the system.

\subsubsection{Example of MD Simulation}

Let's consider an example using the popular MD simulation tool \textbf{GROMACS} \cite{abraham2015gromacs}. Below is an example of how to prepare a simulation system using GROMACS in a command-line environment.

\begin{lstlisting}[style=cmd]
# Prepare the protein structure for simulation
gmx pdb2gmx -f protein.pdb -o processed.gro -water spce

# Define the simulation box
gmx editconf -f processed.gro -o newbox.gro -c -d 1.0 -bt cubic

# Add water molecules
gmx solvate -cp newbox.gro -cs spc216.gro -o solvated.gro -p topol.top

# Add ions
gmx grompp -f ions.mdp -c solvated.gro -p topol.top -o ions.tpr
gmx genion -s ions.tpr -o solvated.gro -p topol.top -pname NA -nname CL -neutral
\end{lstlisting}

In this process:
\begin{itemize}
    \item We start by preparing the protein structure using the `pdb2gmx` command.
    \item We then define a box for the simulation and solvate the system with water molecules.
    \item Finally, we add ions to neutralize the system.
\end{itemize}

\subsection{Transition to Deep Learning in Protein Structure Prediction}

While traditional methods like homology modeling, threading, and MD simulations have been valuable in understanding protein structures, they are limited by factors such as the availability of homologous structures, computational cost, and the difficulty in modeling large, complex proteins. This led to the development of new approaches involving machine learning, and later, deep learning.

In recent years, deep learning has revolutionized protein structure prediction. The introduction of \textbf{AlphaFold} by DeepMind, in particular, has solved many challenges that previously limited the field \cite{jumper2021highly, jumper2021applying}. In the following sections, we will introduce how Python and deep learning frameworks such as \textbf{TensorFlow} \cite{abadi2016tensorflow} and \textbf{PyTorch} \cite{paszke2019pytorch} are used to predict and design protein structures.

\subsection{Python Basics for Protein Structure Prediction}

Before diving into deep learning, it's important for beginners to understand the basics of Python programming, as it is widely used in the computational biology community. Here's a simple Python example that introduces basic data structures used in protein sequence analysis.

\begin{lstlisting}[style=python]
# Define a protein sequence
protein_sequence = "MVLSPADKTNVKAAW"

# Iterate over each amino acid in the sequence
for amino_acid in protein_sequence:
    print(amino_acid)
\end{lstlisting}

This simple code defines a protein sequence and prints each amino acid. In real-world applications, Python is used to manipulate protein sequences, analyze data from biological databases, and interact with protein prediction tools.

\subsection{Introduction to Deep Learning in Protein Structure Prediction}

Deep learning has brought a new era to protein structure prediction, allowing for the creation of models that can predict protein structures from amino acid sequences with unprecedented accuracy～\cite{senior2020improved, Enireddy2022OneHotEncodingAL}. In the next chapters, we will explore how these deep learning models are built, trained, and applied to real-world protein prediction problems.

For example, \textbf{AlphaFold} uses a deep learning model to predict the 3D structure of proteins directly from their sequence. Python is at the heart of these innovations, with frameworks like \textbf{TensorFlow} and \textbf{PyTorch} being used to build and train these models.

Here's a simple Python code snippet that shows how a deep learning model can be defined in TensorFlow:

\begin{lstlisting}[style=python]
import tensorflow as tf
from tensorflow.keras import layers

# Define a simple sequential model
model = tf.keras.Sequential([
    layers.Dense(128, activation='relu', input_shape=(20,)),
    layers.Dense(64, activation='relu'),
    layers.Dense(1, activation='sigmoid')
])

# Compile the model
model.compile(optimizer='adam',
              loss='binary_crossentropy',
              metrics=['accuracy'])

# Model summary
model.summary()
\end{lstlisting}

This example defines a simple neural network model that could be adapted to predict protein features such as secondary structures. As we continue, we will build on this foundation to explore how deep learning is specifically applied to protein structure prediction and design.

\section{Summary}

In this section, we introduced the early methods used in protein structure prediction, including homology modeling, threading, and molecular dynamics simulations. These traditional approaches provided critical insights into protein structure, but they also had limitations. With the advent of deep learning~\cite{Liu2024ExploringPB, Dai2025ASO} and tools like AlphaFold, protein structure prediction has entered a new era, offering more accurate and scalable solutions. In the next sections, we will delve deeper into Python and deep learning techniques, providing practical examples for beginners to explore protein structure prediction and design.

\section{Milestones in Protein Structure Prediction}

The history of protein structure prediction is a story of gradual breakthroughs and technological advancements. Understanding the key milestones in this field is essential for appreciating the role that Python and deep learning now play in modern approaches. In this section, we will explore some of the most important milestones in protein structure prediction, providing detailed explanations and practical examples to help beginners and newcomers to the field.

\subsection{Early Experimental Methods}

The earliest protein structure predictions were heavily reliant on experimental methods. These methods provided accurate insights into protein structure but were time-consuming and costly.

\subsubsection{X-ray Crystallography}

X-ray crystallography was the first major experimental technique used to determine protein structures \cite{kendrew1958three}. This technique involves crystallizing the protein, directing X-rays at the crystal, and analyzing the diffraction patterns to reconstruct the protein's 3D structure.

\begin{itemize}
    \item \textbf{Example}: The structure of myoglobin, the first protein structure to be determined using X-ray crystallography, was solved in 1958 by John Kendrew.
\end{itemize}

Although X-ray crystallography provides highly accurate structural information, it requires well-formed crystals, which can be difficult to produce for many proteins, especially membrane proteins.

\subsubsection{NMR Spectroscopy}

Nuclear Magnetic Resonance (NMR) spectroscopy emerged as an alternative to X-ray crystallography for studying proteins in solution \cite{wuthrich1986nmr}. NMR exploits the magnetic properties of atomic nuclei to infer the distances and angles between atoms, allowing researchers to construct a 3D model of the protein.

\begin{itemize}
    \item \textbf{Example}: The structure of the first protein, a small peptide, was determined using NMR in 1985. This marked a significant step toward studying proteins in their native environment.
\end{itemize}

While useful for smaller proteins, NMR is limited in its ability to handle larger proteins and complexes due to signal overlap and complexity in interpreting the data.

\subsubsection{Cryo-Electron Microscopy (Cryo-EM)}

Cryo-electron microscopy (Cryo-EM) is a more recent technique that allows scientists to visualize protein structures at near-atomic resolution without the need for crystallization \cite{dubochet2017nobel}. Proteins are flash-frozen and observed under an electron microscope, making it possible to study larger and more complex structures, such as protein complexes and viruses.

\begin{itemize}
    \item \textbf{Example}: The structure of the ribosome, a massive molecular machine responsible for protein synthesis, was determined using Cryo-EM, for which the 2017 Nobel Prize in Chemistry was awarded.
\end{itemize}

\subsection{The Computational Era: Homology Modeling and Threading}

As computing technology advanced, researchers began developing computational methods for predicting protein structures. These methods allowed scientists to predict structures more quickly and without relying solely on experimental data.

\subsubsection{Homology Modeling}

Homology modeling (also known as comparative modeling) was one of the earliest computational methods developed \cite{krieger2003homology}. It is based on the assumption that proteins with similar sequences will have similar structures. By aligning a target sequence with a known template, researchers can build a model of the unknown structure.

\textbf{Example of Homology Modeling Using Python}:

In Python, we can use the Modeller library to perform homology modeling. Here is a simple example:

\begin{lstlisting}[style=python]
from modeller import *
from modeller.automodel import *

# Set up the environment
env = environ()

# Define the alignment file and the known template structure
a = automodel(env, alnfile='alignment.ali', knowns='template', sequence='target')

# Generate multiple models
a.starting_model = 1
a.ending_model = 5

# Build the models
a.make()
\end{lstlisting}

In this example, we use the Modeller library to generate multiple models of the target protein, based on its similarity to a known structure. Homology modeling relies heavily on the availability of high-quality template structures in databases like the Protein Data Bank (PDB).

\subsubsection{Threading (Fold Recognition)}

Threading, or fold recognition, is another computational method that predicts the structure of a protein by threading its sequence onto known protein folds \cite{jones1999protein}. Unlike homology modeling, threading does not require a high sequence similarity between the target and template proteins. It compares the target sequence against a library of known folds and identifies the best fit.

\textbf{Threading Example in Python}:

Using the HHpred tool for threading in Python:

\begin{lstlisting}[style=python]
from hhpred import HHpred

# Initialize HHpred
hhpred = HHpred()

# Define the target sequence
target_sequence = "MQSLKSTALIVGLNQRLAALVDTRAYAEGFYGDPATLG"

# Predict the structure using threading
hhpred.predict(target_sequence, database='pdb')
\end{lstlisting}

Threading allows researchers to predict protein structures even when no close homologs are available, making it a valuable tool for hard-to-predict proteins.

\subsection{Molecular Dynamics Simulations}

While homology modeling and threading predict static structures, molecular dynamics (MD) simulations allow researchers to study the dynamic behavior of proteins over time. MD simulations model the physical movements of atoms in a protein, revealing insights into its flexibility, folding, and interactions.

\textbf{Example of Molecular Dynamics Simulation Using GROMACS}:

Here's an example of how to set up and run an MD simulation using GROMACS, a widely used molecular simulation tool:

\begin{lstlisting}[style=cmd]
# Step 1: Prepare the protein structure
gmx pdb2gmx -f protein.pdb -o processed.gro -water spce

# Step 2: Define the simulation box
gmx editconf -f processed.gro -o newbox.gro -c -d 1.0 -bt cubic

# Step 3: Solvate the system with water
gmx solvate -cp newbox.gro -cs spc216.gro -o solvated.gro -p topol.top

# Step 4: Add ions to neutralize the system
gmx grompp -f ions.mdp -c solvated.gro -p topol.top -o ions.tpr
gmx genion -s ions.tpr -o solvated.gro -p topol.top -pname NA -nname CL -neutral

# Step 5: Perform energy minimization
gmx grompp -f minim.mdp -c solvated.gro -p topol.top -o em.tpr
gmx mdrun -v -deffnm em

# Step 6: Run the molecular dynamics simulation
gmx grompp -f md.mdp -c em.gro -p topol.top -o md.tpr
gmx mdrun -v -deffnm md
\end{lstlisting}

In this process, the protein is placed in a simulation box, solvated with water, and ions are added to neutralize the system. Once energy minimization is complete, the MD simulation is run, revealing how the protein behaves in a dynamic environment.

\subsection{The Advent of Machine Learning and Deep Learning}

The introduction of machine learning and, more recently, deep learning has revolutionized protein structure prediction. Traditional methods such as homology modeling and threading are limited by the availability of template structures. Deep learning, however, allows researchers to predict protein structures directly from sequences, often with remarkable accuracy.

\subsubsection{AlphaFold: A Breakthrough in Protein Structure Prediction}

The development of AlphaFold by DeepMind represents one of the most significant milestones in protein structure prediction. AlphaFold uses deep learning to predict protein structures with near-experimental accuracy, even for proteins with no known homologs. AlphaFold's success at the 2020 CASP14 competition (Critical Assessment of Structure Prediction) demonstrated the power of deep learning in this field.

\textbf{AlphaFold and Python}:

Python is at the heart of many deep learning frameworks, including TensorFlow and PyTorch, which are used to build models like AlphaFold. Let's look at a simple example of a deep learning model using TensorFlow:

\begin{lstlisting}[style=python]
import tensorflow as tf
from tensorflow.keras import layers

# Define a basic sequential model
model = tf.keras.Sequential([
    layers.Dense(128, activation='relu', input_shape=(20,)),
    layers.Dense(64, activation='relu'),
    layers.Dense(1, activation='sigmoid')
])

# Compile the model
model.compile(optimizer='adam',
              loss='binary_crossentropy',
              metrics=['accuracy'])

# Summary of the model
model.summary()
\end{lstlisting}

This code defines a simple neural network that could be adapted for a variety of tasks, including protein feature prediction. AlphaFold and similar models use much more sophisticated architectures, but the fundamental principles of neural networks remain the same.

\subsubsection{RoseTTAFold: An Alternative Approach}

Another deep learning-based protein structure prediction model is RoseTTAFold, developed by the Baker Lab at the University of Washington \cite{baek2021accurate}. Like AlphaFold, RoseTTAFold uses deep learning to predict protein structures, but it offers a more accessible framework for certain types of predictions.

Both AlphaFold and RoseTTAFold have opened new doors for researchers, allowing for more accurate and rapid predictions of protein structures. These tools have also had a profound impact on related fields such as drug discovery, enzyme engineering, and synthetic biology.

\subsection{Summary of Key Milestones in Protein Structure Prediction}

To summarize, the milestones in protein structure prediction can be categorized into several key eras:

\begin{itemize}
    \item \textbf{Early Experimental Methods}: Techniques such as X-ray crystallography, NMR spectroscopy, and Cryo-EM laid the foundation for understanding protein structures.
    \item \textbf{Computational Methods}: Homology modeling and threading introduced faster, more scalable ways to predict protein structures based on known templates.
    \item \textbf{Molecular Dynamics Simulations}: These simulations allowed researchers to explore the dynamic behavior of proteins in silico, providing insights into protein folding and flexibility.
    \item \textbf{Deep Learning}: Tools like AlphaFold and RoseTTAFold have revolutionized the field by enabling accurate predictions directly from sequences, even for previously unknown structures.
\end{itemize}

\begin{itemize}
    \item Protein Structure Prediction
    \begin{itemize}
        \item Early Experimental Methods
        \begin{itemize}
            \item X-ray Crystallography
            \item NMR Spectroscopy
            \item Cryo-EM
        \end{itemize}
        \item Computational Methods
        \begin{itemize}
            \item Homology Modeling
            \item Threading
        \end{itemize}
        \item Deep Learning Methods
        \begin{itemize}
            \item AlphaFold
            \item RoseTTAFold
        \end{itemize}
    \end{itemize}
\end{itemize}

These milestones demonstrate the rapid progress made in the field of protein structure prediction, from early experimental techniques to the cutting-edge use of deep learning. In the next sections, we will explore how these tools are used in practice and provide hands-on examples for readers to try on their own.

\section{The Development of Protein Design and Its Applications}

Protein design has emerged as one of the most exciting and impactful areas in modern biology. While traditional protein structure prediction focuses on determining the structure of natural proteins, protein design aims to create new proteins with specific, desirable functions. This chapter will introduce the development of protein design, explain the fundamental concepts behind it, and provide real-world examples of its applications. We will explore the role of Python and deep learning in protein design and give step-by-step instructions to help beginners understand the field.

\subsection{What is Protein Design?}

Protein design is the process of creating novel proteins that do not exist in nature~\cite{Pearce2021DeepLT}. These proteins can be designed to have specific structural features or to perform certain biological functions, such as catalyzing chemical reactions, binding to target molecules, or serving as scaffolds in drug development. The basic goal of protein design is to manipulate amino acid sequences in such a way that the designed sequence will fold into a stable and functional 3D structure.

Protein design has broad applications in areas such as biotechnology, medicine, and synthetic biology. With the advent of computational methods, particularly machine learning and deep learning, the ability to design functional proteins has become more efficient and accurate.

\subsection{Key Concepts in Protein Design}

Before diving into practical examples, it is essential to understand a few key concepts in protein design:

\begin{itemize}
    \item \textbf{Amino Acid Sequence}: A protein's primary structure is the sequence of amino acids. The design process starts by determining the sequence of amino acids that will result in the desired structure and function.
    \item \textbf{Energy Minimization}: Proteins adopt their functional conformations by folding into the lowest energy state. During protein design, computational methods are used to ensure that the designed sequence will fold into a stable, low-energy structure.
    \item \textbf{Backbone and Side Chain Conformation}: The backbone of the protein refers to the main chain of the polypeptide, while side chains are the functional groups attached to the backbone. Designing proteins often involves optimizing the positions of both backbone and side chains.
\end{itemize}

The design of a protein often involves iterating through multiple sequences and evaluating their ability to fold into the desired structure. Tools such as deep learning and evolutionary algorithms have greatly improved the speed and accuracy of this process.

\subsection{Early Approaches to Protein Design}

Before deep learning, early approaches to protein design relied on more traditional computational techniques. One of the most well-known early approaches is \textbf{Rosetta}, a software suite for predicting protein structures and designing new proteins \cite{leaver2011rosetta3}. Rosetta uses physics-based methods to predict how a given sequence of amino acids will fold and evaluates the stability of different designs based on energy calculations.

\subsubsection{Example of Protein Design Using Rosetta}

Rosetta can be used to design new protein sequences. Below is an example command sequence showing how Rosetta can be used to design a protein:

\begin{lstlisting}[style=cmd]
# Run a protein design simulation with Rosetta
rosetta_scripts.default.linuxgccrelease \
    -s input_protein.pdb \
    -parser:protocol design_script.xml \
    -out:file:scorefile design_score.sc \
    -out:path:pdb designed_protein.pdb
\end{lstlisting}

This example uses Rosetta's design protocol to generate a new protein structure based on an input structure file (`input\_protein.pdb`). The output will include the newly designed protein structure (`designed\_protein.pdb`) and a score file (`design\_score.sc`) that evaluates the quality of the design.

While Rosetta is still widely used, more recent methods involving deep learning have transformed protein design by speeding up the design process and improving the accuracy of predictions.

\subsection{Deep Learning in Protein Design}

The advent of deep learning has dramatically changed the landscape of protein design~\cite{Hiranuma2022ProteinSA, Gao2022UnderstandingBD, Manshour2024IntegratingPS}. By leveraging large datasets of protein structures and sequences, deep learning models can learn the relationships between sequence and structure, making it possible to design proteins more efficiently~\cite{Feng2024MHTAPredSSAH, Wang2024ProteinDU, Sato2023RecentTI, Limbu2022ANH}.

One of the most notable deep learning models for protein design is \textbf{ProteinMPNN}, a model developed to design proteins based on a desired backbone structure \cite{dauparas2022robust}. ProteinMPNN generates amino acid sequences that are predicted to fold into the desired structure with a high degree of stability.

\subsubsection{Protein Design Using ProteinMPNN in Python}

ProteinMPNN is a neural network model specifically tailored for protein design. Below is a simplified Python example showing how ProteinMPNN can be used for protein design:

\begin{lstlisting}[style=python]
import torch
from protein_mpnn import ProteinMPNN

# Load the pre-trained ProteinMPNN model
model = ProteinMPNN()

# Define a target protein backbone (simplified for illustration)
backbone_structure = "target_backbone.pdb"

# Generate protein sequences that are likely to fold into the target structure
designed_sequences = model.design(backbone_structure)

# Display the designed sequences
for sequence in designed_sequences:
    print(sequence)
\end{lstlisting}

In this Python example, we use ProteinMPNN to generate amino acid sequences that are likely to fold into the target backbone structure (`target\_backbone.pdb`). The model returns a set of designed sequences that can be tested for stability and function.

ProteinMPNN is an example of how deep learning models can be trained on large datasets of protein structures and sequences to optimize the design process. These models are capable of generating new proteins that may have never existed in nature, opening up exciting possibilities for new biological functions and applications.

\subsection{Applications of Protein Design}

Protein design has a wide range of applications across several fields, from medicine to materials science. Below, we outline some of the key applications where protein design is making a significant impact.

\subsubsection{Drug Development}

One of the most exciting applications of protein design is in drug development. By designing proteins with specific binding sites, researchers can create therapeutic proteins that target disease-related molecules with high precision. For example, proteins can be engineered to bind specifically to cancer cells, marking them for destruction by the immune system.

\textbf{Example: Designing a Protein Drug for Cancer Therapy}

Imagine we are designing a protein that can specifically bind to a receptor on cancer cells. We can use Python and ProteinMPNN to design a sequence with high affinity for the cancer cell receptor:

\begin{lstlisting}[style=python]
# Define the target cancer receptor structure
receptor_structure = "cancer_receptor.pdb"

# Use ProteinMPNN to design a protein that binds to the receptor
designed_drug_sequences = model.design(receptor_structure)

# Display the designed drug sequences
for sequence in designed_drug_sequences:
    print(sequence)
\end{lstlisting}

This type of protein design allows for the creation of highly specific therapeutic proteins, which can be used in targeted cancer therapies with fewer side effects than traditional treatments.

\subsubsection{Enzyme Engineering}

Enzymes are proteins that catalyze biochemical reactions, and their design has many industrial and medical applications. By engineering new enzymes, we can create catalysts that are more efficient or that work under extreme conditions, such as high temperatures or acidic environments.

\textbf{Example: Designing a New Enzyme for Industrial Catalysis}

Let's design a protein that can act as an enzyme to speed up a chemical reaction in an industrial process:

\begin{lstlisting}[style=python]
# Define the backbone structure of the desired enzyme
enzyme_backbone = "enzyme_backbone.pdb"

# Use ProteinMPNN to generate enzyme sequences
designed_enzyme_sequences = model.design(enzyme_backbone)

# Display the designed enzyme sequences
for sequence in designed_enzyme_sequences:
    print(sequence)
\end{lstlisting}

By designing novel enzymes, industries such as biofuels and pharmaceuticals can improve the efficiency of their production processes, reducing costs and environmental impact.

\subsubsection{Materials Science}

Protein design is also making an impact in materials science. By designing proteins that self-assemble into specific structures, researchers can create new materials with unique properties, such as strength, flexibility, or conductivity.

\begin{itemize}
    \item \textbf{Example}: Scientists have designed proteins that can self-assemble into nanoscale fibers, which can be used in tissue engineering or as scaffolds for drug delivery.
\end{itemize}

\subsection{Challenges in Protein Design}

Despite the tremendous advances in protein design, several challenges remain. Designing proteins that fold correctly and maintain stability in different environments is still difficult, and there are often trade-offs between stability and function.

\subsubsection{Folding and Stability}

One of the primary challenges in protein design is ensuring that the designed sequence will fold into the desired 3D structure. Proteins are highly sensitive to environmental factors such as temperature, pH, and solvent conditions. Small changes in the amino acid sequence can lead to misfolding or aggregation, which can render the protein non-functional or even harmful.

\subsubsection{Trade-Offs Between Stability and Function}

Sometimes, increasing the stability of a protein can come at the cost of its functionality. For example, rigidifying certain regions of a protein might increase its stability but reduce its flexibility, making it less effective at binding to other molecules.

\textbf{Example: Exploring Stability and Function Trade-Offs Using Deep Learning}

Using deep learning models, we can explore how different mutations affect both the stability and function of a protein. Here's how we might simulate this trade-off:

\begin{lstlisting}[style=python]
# Define the target structure for a functional protein
functional_protein_backbone = "functional_protein.pdb"

# Design sequences with a focus on stability
stable_sequences = model.design(functional_protein_backbone, focus_on='stability')

# Design sequences with a focus on function
functional_sequences = model.design(functional_protein_backbone, focus_on='function')

# Compare stability and function of designed sequences
for stability_seq, function_seq in zip(stable_sequences, functional_sequences):
    print(f"Stable sequence: {stability_seq}")
    print(f"Functional sequence: {function_seq}")
\end{lstlisting}

In this example, we use the same protein backbone but design two sets of sequences: one optimized for stability and the other for function. Comparing these designs allows us to understand the trade-offs involved.

\subsection{Summary}

In this section, we explored the development of protein design, from early approaches like Rosetta to modern deep learning models such as ProteinMPNN. We covered key concepts such as energy minimization and the challenges associated with protein folding and stability. We also highlighted the diverse applications of protein design, including drug development, enzyme engineering, and materials science.

Protein design is a rapidly evolving field with exciting opportunities for new discoveries. The integration of Python and deep learning has made it more accessible to researchers, enabling faster and more accurate designs. As we continue to explore the frontiers of protein design, we will see even more impactful applications that push the boundaries of what proteins can achieve.
\chapter{Fundamentals of Protein Structure and Function}
\section{Primary, Secondary, Tertiary, and Quaternary Structures}

Proteins are essential molecules in all living organisms. They are responsible for a wide range of biological functions, from catalyzing chemical reactions to providing structural support to cells \cite{scheeff2003fundamentals,buxbaum2007fundamentals,kessel2018introduction}. Understanding protein structure is crucial because the function of a protein is directly linked to its shape. Proteins have a hierarchical structure that can be classified into four levels: primary, secondary, tertiary, and quaternary structures. In this section, we will explore each of these levels in detail, breaking them down into simple, easy-to-understand concepts, with practical Python examples to help reinforce the ideas.

\subsection{Primary Structure}

The \textbf{primary structure} \cite{lehrman2017protein} of a protein refers to the linear sequence of amino acids in its polypeptide chain. Amino acids are the building blocks of proteins, and there are 20 standard amino acids. These amino acids are linked together by covalent bonds, known as peptide bonds, to form a long chain.

The sequence of amino acids is critical because it determines how the protein will fold into its three-dimensional structure and ultimately how it will function. Even a single change in the sequence, known as a mutation, can have drastic effects on the protein's function.

\subsubsection{Example of Primary Structure Representation}

Let's take a simple example of a protein primary structure, which is a sequence of amino acids:

\begin{lstlisting}[style=python]
# Example of a protein primary structure (amino acid sequence)
protein_sequence = "MVLSPADKTNVKAAW"
print(f"Primary structure: {protein_sequence}")
\end{lstlisting}

This Python code snippet represents the primary structure of a protein using a string of amino acid symbols. Each letter corresponds to an amino acid, where 'M' stands for methionine, 'V' stands for valine, 'L' for leucine, and so on. The order of these amino acids determines how the protein will fold.

\subsection{Secondary Structure}

The \textbf{secondary structure} \cite{pauling1951configuration, pelton2000spectroscopic} of a protein refers to local folding patterns within the polypeptide chain. These patterns are primarily stabilized by hydrogen bonds between the backbone atoms of the polypeptide. The two most common types of secondary structures are:

\begin{itemize}
    \item \textbf{$\alpha$-helix}: A right-handed coil where the backbone of the polypeptide forms a helix \cite{pauling1951configuration}.
    \item \textbf{$\beta$-sheet}: A sheet-like structure formed by multiple polypeptide strands lying side by side \cite{pauling1951configuration}.
\end{itemize}

These secondary structures play an important role in stabilizing the overall structure of the protein.

\subsubsection{Visualization of Secondary Structure}

To better understand the secondary structure, we can use a simple tree diagram to illustrate the relationship between the primary sequence and its local secondary structure elements.


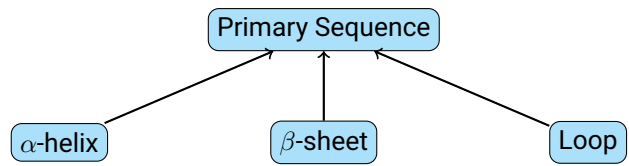
\begin{center}
\begin{tikzpicture}[sibling distance=10em, every node/.style={draw, align=center}]
  \node[rectangle, rounded corners, fill=cyan!30] (root) {Primary Sequence}
    child { node[rectangle, rounded corners, fill=cyan!30] (alpha) {$\alpha$-helix} }
    child { node[rectangle,  rounded corners,fill=cyan!30] (beta) {$\beta$-sheet} }
    child { node[rectangle,  rounded corners, fill=cyan!30] (loop) {Loop} };

  \tikzset{edge/.style={->, thick}}
  \draw[edge] (alpha) -- (root);
  \draw[edge] (beta) -- (root);
  \draw[edge] (loop) -- (root);
\end{tikzpicture}
\end{center}

In this diagram, we see how different parts of the primary sequence fold into distinct secondary structures, such as $\alpha$-helices and $\beta$-sheets.

\subsubsection{Python Example: Identifying Secondary Structure}

We can use Python to analyze a protein sequence and predict whether it will form an $\alpha$-helix or a $\beta$-sheet based on known patterns. Here's a simple example:

\begin{lstlisting}[style=python]
# Simplified function to predict secondary structure based on sequence patterns
def predict_secondary_structure(sequence):
    if "A" in sequence or "L" in sequence:  # Simplified rules
        return "Alpha helix"
    elif "V" in sequence or "I" in sequence:
        return "Beta sheet"
    else:
        return "Loop"

# Example protein sequence
protein_sequence = "MVLSPADKTNVKAAW"

# Predict the secondary structure
structure = predict_secondary_structure(protein_sequence)
print(f"Predicted secondary structure: {structure}")
\end{lstlisting}

This code is a highly simplified way to predict secondary structure based on the presence of certain amino acids. In practice, more sophisticated algorithms such as \textbf{DSSP} \cite{kabsch1983dictionary} or \textbf{PSIPRED} \cite{jones1999protein} are used for secondary structure prediction.

\subsection{Tertiary Structure}

The \textbf{tertiary structure} \cite{rees1977secondary} of a protein refers to its overall three-dimensional shape, which is formed when the secondary structures fold further due to interactions between the side chains (R groups) of the amino acids. These interactions include:

\begin{itemize}
    \item Hydrophobic interactions
    \item Hydrogen bonds
    \item Disulfide bridges (covalent bonds between cysteine residues)
    \item Ionic bonds
\end{itemize}

The tertiary structure is crucial for the protein's function. Many proteins only become functional when they adopt their proper tertiary structure.

\subsubsection{Example: Tertiary Structure Visualization}

Let's visualize the process by which a protein's secondary structures (e.g., $\alpha$-helices and $\beta$-sheets) come together to form the overall tertiary structure.

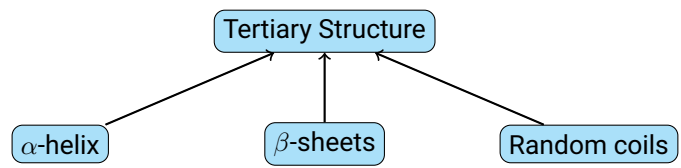
\begin{center}
\begin{tikzpicture}[sibling distance=10em, every node/.style={draw, align=center}]
  \node[rectangle, rounded corners, fill=cyan!30] (root) {Tertiary Structure}
    child { node[rectangle, rounded corners, fill=cyan!30] (alpha) {$\alpha$-helix} }
    child { node[rectangle,  rounded corners,fill=cyan!30] (beta) {$\beta$-sheets} }
    child { node[rectangle,  rounded corners, fill=cyan!30] (loop) {Random coils} };

  \tikzset{edge/.style={->, thick}}
  \draw[edge] (alpha) -- (root);
  \draw[edge] (beta) -- (root);
  \draw[edge] (loop) -- (root);
\end{tikzpicture}
\end{center}


In this illustration, secondary structures (such as helices and sheets) are organized in space to form the final tertiary structure.

\subsubsection{Python Example: Analyzing Tertiary Structure}

In practice, analyzing the tertiary structure requires knowledge of the three-dimensional coordinates of the protein's atoms. Here is a simplified Python example that shows how you might load and analyze the tertiary structure of a protein using a library like \textbf{Biopython} \cite{cock2009biopython}:

\begin{lstlisting}[style=python]
from Bio.PDB import PDBParser

# Parse a PDB file containing the tertiary structure of a protein
parser = PDBParser()
structure = parser.get_structure('protein', 'protein_structure.pdb')

# Extract and print the details of the tertiary structure
for model in structure:
    for chain in model:
        for residue in chain:
            print(residue)
\end{lstlisting}

This code uses Biopython to load a protein structure from a PDB (Protein Data Bank) file \cite{berman2000protein}. The structure of the protein can then be analyzed in detail, including its folding patterns and interactions between residues.

\subsection{Quaternary Structure}

The \textbf{quaternary structure} of a protein is the arrangement of multiple polypeptide chains (subunits) into a functional protein complex. Not all proteins have quaternary structures; only proteins that consist of more than one polypeptide chain (e.g., hemoglobin) have this level of organization.

In proteins with quaternary structures, the interaction between different subunits is crucial for their biological function. For example, hemoglobin, which carries oxygen in the blood, consists of four subunits, and the quaternary arrangement of these subunits is essential for its oxygen-binding ability.

\subsubsection{Example of Quaternary Structure}

Consider hemoglobin as an example of a protein with quaternary structure. Hemoglobin has four subunits: two alpha chains and two beta chains. The arrangement of these subunits allows hemoglobin to function properly.


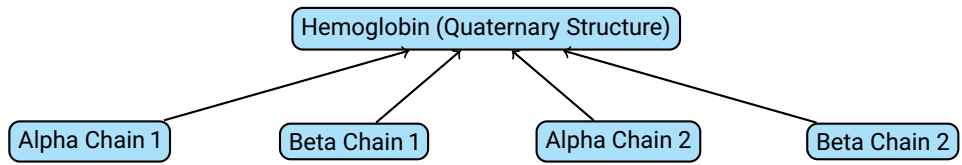
\begin{center}
\begin{tikzpicture}[sibling distance=10em, every node/.style={draw, align=center, font=\small, rounded corners}]
  \node[rectangle, rounded corners, fill=cyan!30, thick] (root) {Hemoglobin (Quaternary Structure)}
    child { node[rectangle, rounded corners, fill=cyan!30, thick] (alpha1) {Alpha Chain 1} }
    child { node[rectangle, rounded corners, fill=cyan!30, thick] (beta1) {Beta Chain 1} }
    child { node[rectangle, rounded corners, fill=cyan!30, thick] (alpha2) {Alpha Chain 2} }
    child { node[rectangle, rounded corners, fill=cyan!30, thick] (beta2) {Beta Chain 2} };

  \tikzset{edge/.style={->, thick}}
  \draw[edge] (alpha1) -- (root);
  \draw[edge] (beta1) -- (root);
  \draw[edge] (alpha2) -- (root);
  \draw[edge] (beta2) -- (root);
\end{tikzpicture}
\end{center}

This diagram illustrates how the four subunits of hemoglobin (two alpha and two beta chains) come together to form the functional protein.

\subsubsection{Python Example: Quaternary Structure Analysis}

Quaternary structures can also be analyzed using Python and Biopython. Here's an example of how to analyze the quaternary structure of a protein:

\begin{lstlisting}[style=python]
from Bio.PDB import PDBParser

# Parse the PDB file for a protein with quaternary structure (e.g., hemoglobin)
parser = PDBParser()
structure = parser.get_structure('hemoglobin', 'hemoglobin.pdb')

# Extract and print details about the subunits (chains)
for model in structure:
    for chain in model:
        print(f"Chain: {chain.id}")
\end{lstlisting}

This code uses Biopython to analyze the quaternary structure of a protein, such as hemoglobin, by extracting the individual chains (subunits) and examining their interactions.

\subsection{Summary of Protein Structures}

Understanding the different levels of protein structure is fundamental to studying their function:

\begin{itemize}
    \item \textbf{Primary Structure}: The sequence of amino acids in the polypeptide chain.
    \item \textbf{Secondary Structure}: Local folding patterns like $\alpha$-helices and $\beta$-sheets.
    \item \textbf{Tertiary Structure}: The overall three-dimensional shape of a single polypeptide chain.
    \item \textbf{Quaternary Structure}: The arrangement of multiple polypeptide chains into a functional complex.
\end{itemize}

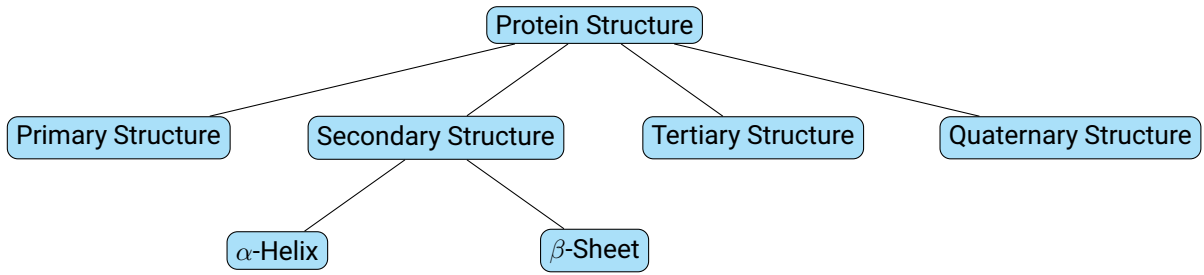
\begin{center}
\begin{tikzpicture}[sibling distance=12em, every node/.style = {shape=rectangle, rounded corners, draw, align=center,fill=cyan!30}]
  \node {Protein Structure}
    child { node {Primary Structure} }
    child { node {Secondary Structure}
      child { node {$\alpha$-Helix} }
      child { node {$\beta$-Sheet} }
    }
    child { node {Tertiary Structure} }
    child { node {Quaternary Structure} };
\end{tikzpicture}
\end{center}

Each level of protein structure plays a crucial role in determining how the protein will function in a biological system. As we continue in this book, we will explore how Python and deep learning can be used to model and predict these structures, helping researchers design new proteins and understand the mechanisms of existing ones.

\section{The Physics and Chemistry of Protein Folding}

Protein folding is one of the most fundamental processes in biology. The proper folding of a protein is essential for its function, and misfolding can lead to diseases such as Alzheimer's, Parkinson's, and cystic fibrosis. Protein folding is driven by a complex interplay of physical and chemical forces that determine how a linear chain of amino acids adopts its three-dimensional (3D) structure. In this section, we will explore the physics and chemistry behind protein folding, explain how these forces work, and provide practical Python examples to help you understand these concepts step by step.

\subsection{The Energy Landscape of Protein Folding}

Proteins fold because they seek to minimize their free energy. The concept of the \textbf{energy landscape} is a powerful way to understand how proteins fold \cite{onuchic1997theory}. Imagine the energy landscape as a funnel, where the unfolded protein is at the top, and the folded protein is at the bottom. As the protein folds, it moves through different conformational states, gradually lowering its free energy until it reaches the most stable, folded state.

\begin{itemize}
    \item \textbf{High Energy}: Unfolded or misfolded states are at higher energy levels.
    \item \textbf{Low Energy}: The native, correctly folded state is at the lowest energy level.
\end{itemize}

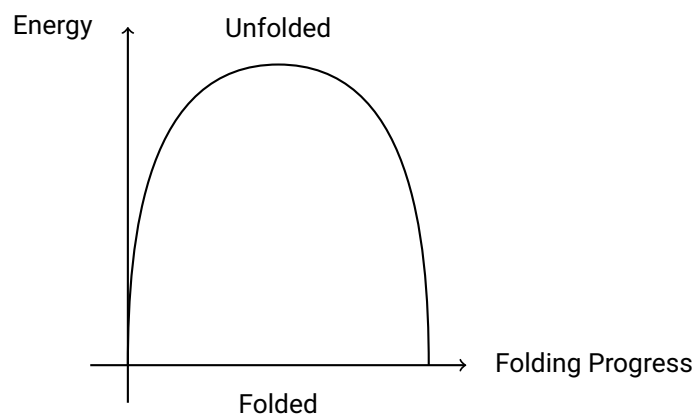
\begin{center}
\begin{tikzpicture}
    \draw[thick] (0,0) to[out=90,in=180] (2,4) to[out=0,in=90] (4,0);
    \node at (2,4.5) {Unfolded};
    \node at (2,-0.5) {Folded};
    \draw[thick,->] (-0.5,0) -- (4.5,0);
    \node at (6.2,0) {Folding Progress};
    \draw[thick,->] (0,-0.5) -- (0,4.5);
    \node at (-1.0,4.5) {Energy};
\end{tikzpicture}
\end{center}

In this diagram, the energy landscape is depicted as a funnel, where the protein moves toward lower energy as it folds. The goal is to find the most stable, low-energy state, which corresponds to the protein's native structure.

\subsubsection{Python Example: Simulating an Energy Landscape}

To understand this concept better, we can simulate a simple energy landscape using Python. Below is an example that simulates the movement of a protein down an energy landscape:

\begin{lstlisting}[style=python]
import numpy as np
import matplotlib.pyplot as plt

# Define the energy landscape as a quadratic function (simplified)
def energy_landscape(x):
    return 0.5 * x**2

# Simulate protein folding by generating random positions along the landscape
x_values = np.linspace(-2, 2, 100)
energy_values = energy_landscape(x_values)

# Plot the energy landscape
plt.plot(x_values, energy_values)
plt.xlabel('Folding Progress')
plt.ylabel('Energy')
plt.title('Simplified Energy Landscape of Protein Folding')
plt.show()
\end{lstlisting}

In this code, we define a simplified energy landscape as a quadratic function and plot it to represent the folding process. As the protein folds, it moves toward the minimum energy point on the graph, which represents the folded state.

\subsection{Forces Involved in Protein Folding}

Several physical and chemical forces drive the process of protein folding. Understanding these forces is crucial for modeling and predicting protein structures. The major forces include:

\begin{itemize}
    \item \textbf{Hydrophobic Interactions}: Hydrophobic amino acids tend to cluster together to avoid water, driving the folding of the protein's core \cite{kauzmann1959some}.
    \item \textbf{Hydrogen Bonds}: These occur between the backbone atoms and sometimes between side chains, stabilizing the secondary structure (e.g., $\alpha$-helices and $\beta$-sheets) \cite{baker2000structural}.
    \item \textbf{Van der Waals Forces}: Weak interactions between atoms that help stabilize the folded structure \cite{israelachvili2011intermolecular}.
    \item \textbf{Electrostatic Interactions}: Attractions between positively and negatively charged side chains \cite{dill1990dominant}.
    \item \textbf{Disulfide Bonds}: Covalent bonds that form between cysteine residues, locking parts of the protein in place \cite{thornton1981disulphide}.
\end{itemize}

\subsubsection{Hydrophobic Interactions}

Hydrophobic interactions are a key force in protein folding \cite{kauzmann1959some}. Hydrophobic amino acids, such as leucine, isoleucine, and valine, are repelled by water and tend to cluster in the interior of the protein, away from the aqueous environment.

\textbf{Example: Hydrophobic Core Formation}

As proteins fold, hydrophobic amino acids \cite{kauzmann1959some} are buried inside the protein structure, while hydrophilic (water-attracting) amino acids remain on the outside. We can simulate this concept with a simple Python example:

\begin{lstlisting}[style=python]
# Define a simple protein sequence with hydrophobic (H) and hydrophilic (P) residues
protein_sequence = "HPPHHPPHHPHPPHPPH"

# Function to count hydrophobic and hydrophilic residues
def count_residues(sequence):
    hydrophobic = sequence.count("H")
    hydrophilic = sequence.count("P")
    return hydrophobic, hydrophilic

# Count the residues
hydrophobic, hydrophilic = count_residues(protein_sequence)
print(f"Hydrophobic residues: {hydrophobic}, Hydrophilic residues: {hydrophilic}")
\end{lstlisting}

In this example, we categorize the amino acids in a simple sequence as either hydrophobic ('H') or hydrophilic ('P'). Understanding the balance of hydrophobic and hydrophilic residues is important for predicting how a protein will fold.

\subsubsection{Hydrogen Bonds}

Hydrogen bonds form between the hydrogen atom of one molecule and an electronegative atom (such as oxygen or nitrogen) of another \cite{pauling1951configuration}. In proteins, hydrogen bonds stabilize $\alpha$-helices and $\beta$-sheets by connecting the backbone atoms.

\textbf{Example: Hydrogen Bonding in a $\beta$-Sheet}

In a $\beta$-sheet, hydrogen bonds occur between strands of the protein chain. Below is a simplified Python example to count potential hydrogen bonds based on a sequence:

\begin{lstlisting}[style=python]
# Define a simple beta sheet pattern
beta_sheet_sequence = "HHPPHHPPHH"

# Function to predict potential hydrogen bonds
def predict_hydrogen_bonds(sequence):
    return sequence.count("H") // 2  # Simplified rule

# Predict the number of hydrogen bonds
hydrogen_bonds = predict_hydrogen_bonds(beta_sheet_sequence)
print(f"Potential hydrogen bonds: {hydrogen_bonds}")
\end{lstlisting}

Here, we assume that each pair of 'H' (hydrophobic) residues forms a hydrogen bond. In reality, the formation of hydrogen bonds is more complex, but this example provides a simple introduction to the concept.

\subsubsection{Van der Waals Forces}

Van der Waals forces are weak interactions that occur between atoms when they are in close proximity \cite{israelachvili2011intermolecular}. These forces are crucial for the fine-tuning of protein folding, helping to stabilize the compact structure. Although these forces are weak individually, they become significant when many atoms are involved.

\subsubsection{Electrostatic Interactions}

Electrostatic interactions, or salt bridges, occur between charged amino acid side chains, such as lysine (positive) and aspartic acid (negative). These interactions can stabilize the folded structure by bringing together oppositely charged residues \cite{zhou2018electrostatic}.

\subsubsection{Disulfide Bonds}

Disulfide bonds are strong covalent bonds that form between two cysteine residues in a protein \cite{thornton1981disulphide}. These bonds act like molecular "staples," holding different parts of the protein together.

\textbf{Example: Disulfide Bond Formation}

In Python, we can simulate the formation of disulfide bonds between cysteine residues in a protein:

\begin{lstlisting}[style=python]
# Define a protein sequence with cysteine (C) residues
protein_sequence = "MCCVPVCCK"

# Function to predict potential disulfide bonds
def predict_disulfide_bonds(sequence):
    cysteines = sequence.count("C")
    return cysteines // 2  # Each disulfide bond requires two cysteines

# Predict the number of disulfide bonds
disulfide_bonds = predict_disulfide_bonds(protein_sequence)
print(f"Potential disulfide bonds: {disulfide_bonds}")
\end{lstlisting}

This simple example calculates the number of potential disulfide bonds based on the number of cysteine residues in the sequence. Disulfide bonds are particularly important for proteins that need to maintain a stable structure in harsh environments, such as those outside the cell.

\subsection{Kinetics and Thermodynamics of Protein Folding}

Protein folding is governed by both kinetic and thermodynamic principles \cite{dill1995principles}:

\begin{itemize}
    \item \textbf{Kinetics}: Refers to the speed at which a protein folds. Some proteins fold very quickly, while others require more time or assistance from molecular chaperones.
    \item \textbf{Thermodynamics}: Refers to the stability of the folded protein. The folded state must be thermodynamically favorable, meaning it has the lowest free energy.
\end{itemize}

Proteins fold spontaneously because their native structure is thermodynamically stable, but the pathway to reach that structure can involve multiple intermediates.

\subsubsection{Python Simulation: Protein Folding Kinetics}

Let's simulate a simple model of protein folding kinetics in Python, where we track the progress of folding over time:

\begin{lstlisting}[style=python]
import numpy as np
import matplotlib.pyplot as plt

# Define the folding kinetics (simplified as a linear progress)
time = np.linspace(0, 10, 100)
folding_progress = 1 - np.exp(-0.5 * time)  # Exponential approach to folded state

# Plot the folding kinetics
plt.plot(time, folding_progress)
plt.xlabel('Time')
plt.ylabel('Folding Progress')
plt.title('Protein Folding Kinetics')
plt.show()
\end{lstlisting}

In this example, we use a simple exponential function to simulate the protein's progress toward its folded state over time. As time increases, the protein approaches a fully folded state.

\subsection{Chaperones and Assisted Folding}

Not all proteins fold spontaneously. Some proteins require the assistance of molecular chaperones-specialized proteins that help guide the folding process and prevent misfolding or aggregation.

\textbf{Example: Chaperone-Assisted Folding}

In large, complex proteins, chaperones can help by providing a protective environment where the protein can fold without interference from the cellular environment \cite{balchin2016chaperone}. Here's a simplified Python simulation that models the presence of a chaperone:

\begin{lstlisting}[style=python]
# Simulate the effect of a chaperone on folding speed
def folding_with_chaperone(time, chaperone_assistance=True):
    if chaperone_assistance:
        return 1 - np.exp(-1.0 * time)  # Faster folding with chaperone
    else:
        return 1 - np.exp(-0.5 * time)  # Slower folding without chaperone

# Generate time data
time = np.linspace(0, 10, 100)

# Plot folding progress with and without chaperone
plt.plot(time, folding_with_chaperone(time, chaperone_assistance=True), label="With Chaperone")
plt.plot(time, folding_with_chaperone(time, chaperone_assistance=False), label="Without Chaperone")
plt.xlabel('Time')
plt.ylabel('Folding Progress')
plt.title('Effect of Chaperones on Protein Folding')
plt.legend()
plt.show()
\end{lstlisting}

In this example, we simulate how the presence of a chaperone speeds up the protein folding process. Proteins that need help folding rely on chaperones to reach their native state efficiently.

\subsection{Summary of Protein Folding Forces}

Protein folding is driven by several key physical and chemical forces:

\begin{itemize}
    \item Hydrophobic interactions guide non-polar residues toward the interior of the protein.
    \item Hydrogen bonds stabilize secondary structures like $\alpha$-helices and $\beta$-sheets.
    \item Van der Waals forces help fine-tune the structure by stabilizing close contacts between atoms.
    \item Electrostatic interactions and salt bridges bring charged residues together.
    \item Disulfide bonds provide strong covalent links between parts of the protein.
\end{itemize}

The folding process is spontaneous because it leads to a thermodynamically favorable, low-energy state. However, proteins may require assistance from chaperones to avoid misfolding and aggregation, especially in crowded cellular environments.

By understanding the physics and chemistry of protein folding, researchers can use computational tools like Python and deep learning models to predict how proteins fold and design new proteins with desired properties.

\section{The Relationship Between Structure and Protein Design}

Protein structure is intimately connected to protein function, and this relationship forms the basis for protein design. Understanding how the structure of a protein dictates its function allows researchers to engineer new proteins with desired behaviors. In this section, we will explore the relationship between protein structure and protein design in detail, providing practical examples that illustrate how changes in structure can lead to functional modifications. We will also cover the role of Python and deep learning in facilitating the design process.

\subsection{The Central Role of Structure in Protein Function}

The function of a protein is determined by its three-dimensional (3D) structure, which is itself a consequence of the linear sequence of amino acids (the primary structure). This relationship can be summarized in the following way:

\begin{itemize}
    \item \textbf{Primary structure}: The sequence of amino acids, which determines how the protein will fold.
    \item \textbf{Secondary and tertiary structure}: The local folding patterns (e.g., $\alpha$-helices and $\beta$-sheets) and the overall 3D shape of the protein, which create the specific functional regions, such as active sites or binding pockets.
    \item \textbf{Quaternary structure}: In multi-subunit proteins, the arrangement of subunits can affect how the protein interacts with other molecules and performs its function.
\end{itemize}

The shape of a protein determines its interactions with other molecules, which is the foundation of its biological activity. For instance, enzymes have active sites that are shaped to bind specific substrates, while antibodies have regions that recognize specific antigens. Altering the structure, even slightly, can dramatically change a protein's function, either enhancing it, reducing it, or creating new functional capabilities.

\subsubsection{Example: Structure-Function Relationship in Enzymes}

Consider an enzyme that catalyzes a biochemical reaction. The active site of the enzyme is shaped to bind a specific substrate, which then undergoes a chemical transformation. The enzyme's structure is crucial for positioning the substrate in the correct orientation and stabilizing the transition state.

We can illustrate this using a simple tree diagram:

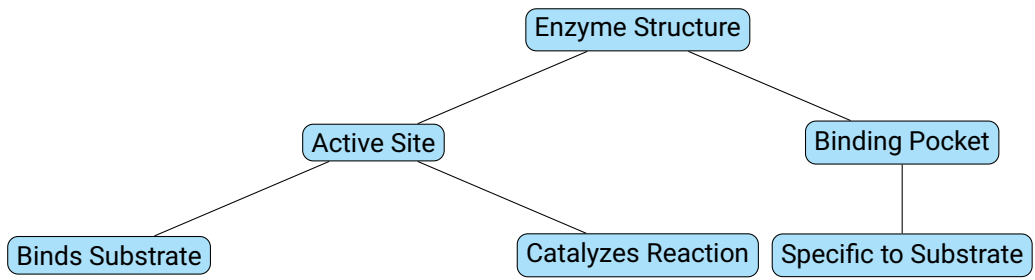
\begin{center}
\begin{tikzpicture}[sibling distance=20em, every node/.style = {shape=rectangle, rounded corners, draw, align=center,fill=cyan!30}]
  \node {Enzyme Structure}
    child { node {Active Site} 
      child { node {Binds Substrate} }
      child { node {Catalyzes Reaction} }
    }
    child { node {Binding Pocket} 
      child { node {Specific to Substrate} }
    };
\end{tikzpicture}
\end{center}

In this diagram, the enzyme's structure creates an active site that binds the substrate and facilitates the reaction.

\subsection{Protein Design: Modifying Structure to Alter Function}

Protein design is the process of modifying the structure of a protein to achieve a specific function. This can involve changing the amino acid sequence to create a new structure or improving an existing function. The design process relies on understanding how specific changes in the sequence will lead to changes in the protein's structure and function.

There are several strategies for protein design:

\begin{itemize}
    \item \textbf{Rational design}: This involves making targeted changes to a protein's sequence based on knowledge of its structure and function \cite{wilson2015rational}. For example, changing a single amino acid in an enzyme's active site might improve its catalytic efficiency.
    \item \textbf{Directed evolution}: This approach mimics natural evolution by introducing random mutations into a protein and selecting variants with improved functions \cite{arnold1998design}. It is a trial-and-error method that can lead to unexpected results.
    \item \textbf{De novo design}: This involves designing a completely new protein from scratch, often using computational methods to predict how a specific sequence will fold and function \cite{huang2016coming}.
\end{itemize}

\subsubsection{Rational Design Example: Improving Enzyme Catalysis}

Let's consider an enzyme whose function we want to improve. We know that changing a particular amino acid in the active site could enhance the enzyme's ability to bind its substrate. Using Python, we can model this change and predict its effects.

Here's a simple example using Python to mutate an amino acid in a sequence and analyze its potential effect:

\begin{lstlisting}[style=python]
# Original enzyme sequence
enzyme_sequence = "MVLSPADKTNVKAAWGK"

# Function to mutate an amino acid at a specific position
def mutate_sequence(sequence, position, new_amino_acid):
    sequence_list = list(sequence)
    sequence_list[position] = new_amino_acid
    return ''.join(sequence_list)

# Mutate the enzyme to improve substrate binding (e.g., changing 'A' to 'R' at position 10)
mutated_sequence = mutate_sequence(enzyme_sequence, 10, 'R')
print(f"Original sequence: {enzyme_sequence}")
print(f"Mutated sequence: {mutated_sequence}")
\end{lstlisting}

In this example, we mutate the enzyme sequence at position 10 to improve its substrate binding. By changing alanine ('A') to arginine ('R'), we hope to introduce a stronger electrostatic interaction between the enzyme and the substrate.

\subsection{The Role of Deep Learning in Protein Design}

Deep learning has revolutionized the field of protein design by enabling the prediction of protein structures and functions from amino acid sequences. By training neural networks on large datasets of known protein structures, deep learning models can predict how a new sequence will fold and what functional properties it might exhibit.

One of the most prominent tools in this space is \textbf{AlphaFold}, which uses deep learning to predict the 3D structure of a protein based on its amino acid sequence \cite{jumper2021highly}. Another tool, \textbf{ProteinMPNN}, allows researchers to design new sequences that are predicted to fold into a desired structure \cite{dauparas2022robust}.

\subsubsection{Example: Using AlphaFold to Predict Structure from Sequence}

AlphaFold uses deep learning to predict the 3D structure of a protein. Below is an example workflow that shows how you can use Python to interface with AlphaFold:

\begin{lstlisting}[style=cmd]
# Step 1: Download AlphaFold model
git clone https://github.com/deepmind/alphafold.git
cd alphafold

# Step 2: Install dependencies
pip install -r requirements.txt

# Step 3: Run AlphaFold on a protein sequence
python run_alphafold.py --fasta_path=my_protein.fasta --output_dir=./output
\end{lstlisting}

This example shows how to use AlphaFold to predict the structure of a protein from its sequence. Once the model is run, the output will include the predicted 3D structure, which can be visualized using molecular visualization tools like PyMOL \cite{delano2002pymol} or Chimera \cite{pettersen2004ucsf}.

\subsubsection{Example: Protein Design Using ProteinMPNN in Python}

ProteinMPNN is a deep learning model designed to generate protein sequences that are likely to fold into a desired 3D structure. Here's how you can use ProteinMPNN to design a new protein:

\begin{lstlisting}[style=python]
import torch
from protein_mpnn import ProteinMPNN

# Load the pre-trained ProteinMPNN model
model = ProteinMPNN()

# Define the desired protein backbone (simplified for illustration)
backbone_structure = "desired_backbone.pdb"

# Use ProteinMPNN to generate new sequences that fold into the desired structure
designed_sequences = model.design(backbone_structure)

# Display the designed sequences
for sequence in designed_sequences:
    print(sequence)
\end{lstlisting}

In this example, we use ProteinMPNN to design a new protein sequence that will fold into a specified backbone structure. This method is especially powerful for creating novel proteins with new functions, such as enzymes for industrial applications or therapeutic proteins for medicine.

\subsection{Applications of Structure-Based Protein Design}

Protein design has wide-ranging applications in many fields, including biotechnology, medicine, and synthetic biology. Below are some key examples:

\subsubsection{Drug Discovery}

One of the most important applications of protein design is in drug discovery. By designing proteins with specific binding sites, researchers can develop therapeutic proteins that interact with disease-related molecules, such as cancer cell receptors or viral proteins.

\textbf{Example: Designing a Drug to Bind to a Cancer Receptor}

Let's design a protein that specifically binds to a receptor found on the surface of cancer cells. We can use deep learning models to predict the structure of a protein that binds tightly to the cancer receptor, improving the efficacy of targeted therapies.

\begin{lstlisting}[style=python]
# Define the target cancer receptor structure
receptor_structure = "cancer_receptor.pdb"

# Use ProteinMPNN to design a protein that binds to the cancer receptor
designed_drug_sequences = model.design(receptor_structure)

# Display the designed drug sequences
for sequence in designed_drug_sequences:
    print(sequence)
\end{lstlisting}

In this example, we design a protein that binds to a specific cancer receptor. This type of targeted therapy can reduce side effects by delivering the drug directly to the cancer cells.

\subsubsection{Enzyme Engineering}

Enzymes are proteins that catalyze biochemical reactions, and their design has many industrial and environmental applications. Protein design can be used to improve the efficiency of enzymes or to create enzymes that work under extreme conditions, such as high temperatures or acidic environments.

\textbf{Example: Designing an Enzyme for Industrial Catalysis}

Let's design a new enzyme that is more efficient at catalyzing a chemical reaction in an industrial process:

\begin{lstlisting}[style=python]
# Define the backbone of the desired enzyme
enzyme_backbone = "enzyme_backbone.pdb"

# Use ProteinMPNN to generate enzyme sequences that fold into this backbone
designed_enzyme_sequences = model.design(enzyme_backbone)

# Display the designed enzyme sequences
for sequence in designed_enzyme_sequences:
    print(sequence)
\end{lstlisting}

This type of protein design is critical for industries such as biofuels, where more efficient enzymes can help reduce costs and improve sustainability.

\subsubsection{Synthetic Biology}

In synthetic biology, protein design is used to create entirely new biological systems, including artificial proteins that do not exist in nature. These proteins can be designed to perform specific tasks, such as assembling into nanostructures or acting as sensors.

\textbf{Example: Designing Proteins for Nanotechnology}

Proteins can be engineered to self-assemble into nanostructures with specific shapes and functions. These structures can be used in areas like drug delivery, tissue engineering, or as molecular machines.

\begin{itemize}
    \item \textbf{Example}: Scientists have designed proteins that self-assemble into nanoscale fibers, which can be used to deliver drugs to specific cells in the body.
\end{itemize}

\subsection{Challenges in Protein Design}

Despite the advancements in deep learning and computational tools, protein design remains a challenging task. Some of the key challenges include:

\begin{itemize}
    \item \textbf{Folding Prediction}: While models like AlphaFold have made great progress in predicting protein structures, accurately predicting how a novel sequence will fold remains difficult for very large or complex proteins.
    \item \textbf{Stability vs. Function}: Designing a protein that is both stable and highly functional can be challenging. Often, stabilizing a protein can reduce its flexibility, which may be required for its function.
    \item \textbf{Environmental Factors}: Proteins must function in specific environments, such as the high temperature or low pH of industrial processes. Designing proteins that maintain their structure and function under these conditions is still an area of active research.
\end{itemize}

\subsection{Summary}

The relationship between protein structure and design is fundamental to understanding how to create new proteins with specific functions. By manipulating the structure of a protein, researchers can alter its function, improving its efficiency or even creating entirely new functions. Python and deep learning tools, such as AlphaFold and ProteinMPNN, have significantly enhanced our ability to design proteins with precision, enabling applications in drug discovery, enzyme engineering, and synthetic biology.

Understanding the interplay between protein structure and function is key to successful protein design. As we continue to advance our computational tools and deepen our knowledge of protein folding, the potential for designing novel proteins that solve real-world problems is vast.
\chapter{Algorithms and Tools for Protein Design and Structure Prediction}

In this chapter, we explore the traditional algorithms used in protein design and structure prediction. These algorithms form the foundation for modern approaches, including deep learning methods. We will examine each method in detail, explaining the underlying principles and providing concrete examples that guide beginners through the concepts.

\section{Traditional Algorithms}

Traditional algorithms for protein structure prediction and design are based on well-established computational and physical principles \cite{pan2011introduction}. These methods include homology modeling, threading, molecular dynamics simulations, and energy function calculations. Each of these approaches plays an essential role in understanding protein structure and folding.

\subsection{Homology Modeling}

\textbf{Homology modeling} \cite{krieger2003homology} (also known as comparative modeling) is a method used to predict the 3D structure of a protein based on the structure of a homologous protein with a known structure. This approach is grounded in the assumption that proteins with similar sequences will adopt similar structures. Homology modeling is one of the most widely used techniques for structure prediction due to its accuracy when a homologous structure is available.

\subsubsection{Steps in Homology Modeling}

The homology modeling process generally follows these steps:

\begin{enumerate}
    \item \textbf{Identify a template}: Find a homologous protein with a known 3D structure.
    \item \textbf{Align sequences}: Align the sequence of the target protein with the sequence of the template protein.
    \item \textbf{Build a model}: Use the template structure as a guide to build the 3D structure of the target protein.
    \item \textbf{Refine the model}: Perform energy minimization and refinement to improve the accuracy of the model.
    \item \textbf{Evaluate the model}: Assess the quality of the model using structural validation tools.
\end{enumerate}

\subsubsection{Example: Homology Modeling Using Python and Modeller}

Let's look at an example of how to perform homology modeling using the Python library \textbf{Modeller} \cite{webb2016comparative}. Modeller is a popular tool for building protein models based on a known template.

\begin{lstlisting}[style=python]
from modeller import *
from modeller.automodel import *

# Create an environment for modeling
env = environ()

# Set up the alignment and template file
a = automodel(env, alnfile='alignment.ali', knowns='template', sequence='target')

# Generate five models of the target protein
a.starting_model = 1
a.ending_model = 5
a.make()

# Select the best model based on the lowest energy
\end{lstlisting}

In this code, we align the target sequence with a known template and generate five models of the target protein. Modeller will select the best model based on the energy of each structure, ensuring that the final model has a structure similar to the template.

\subsubsection{Limitations of Homology Modeling}

While homology modeling is powerful, it has limitations. The accuracy of the predicted model depends on the similarity between the target and the template. If no closely related template is available, the model may not be reliable. In such cases, other methods, such as threading or ab initio methods, are used.

\subsection{Threading}

\textbf{Threading} \cite{madej1995threading}, also known as \textbf{fold recognition}, is a method used to predict protein structure by aligning the target sequence to a library of known structural folds. Unlike homology modeling, threading does not require a high sequence similarity between the target and template proteins. Instead, it looks for structural similarities, even when the sequences are very different.

\subsubsection{How Threading Works}

The basic idea behind threading is to "thread" the target sequence through various known protein structures (templates) and evaluate which fold best accommodates the sequence. This evaluation is based on scoring functions that measure how well the sequence fits into the template's structure.

\subsubsection{Example: Using HHpred for Threading}

One commonly used tool for threading is \textbf{HHpred}, which is available in Python. Here is an example of how to use HHpred for threading:

\begin{lstlisting}[style=python]
from hhpred import HHpred

# Initialize HHpred
hhpred = HHpred()

# Provide the target sequence for threading
target_sequence = "MSTGILSPTQDLRSV... (truncated for clarity)"

# Perform threading using the PDB database
hhpred.predict(target_sequence, database='pdb')
\end{lstlisting}

In this example, we use HHpred to search for structural templates in the Protein Data Bank (PDB). The target sequence is threaded through the templates to identify the best match based on structural similarity.

\subsubsection{Advantages and Limitations of Threading}

Threading is advantageous when no close homologous structures are available, making it a valuable tool for predicting novel protein folds. However, threading requires accurate scoring functions, and the method's success depends on the availability of suitable structural templates in the database.

\subsection{Molecular Dynamics Simulation}

\textbf{Molecular dynamics (MD) simulation} \cite{hollingsworth2018molecular} is a computational method that models the physical movements of atoms and molecules in a protein over time. Unlike homology modeling and threading, which predict static structures, MD simulations provide insights into the dynamic behavior of proteins. These simulations are useful for studying protein folding, conformational changes, and interactions with other molecules, such as ligands or solvents.

\subsubsection{How Molecular Dynamics Simulation Works}

MD simulations involve solving Newton's equations of motion for the atoms in a protein. The forces acting on each atom are calculated based on a force field, which defines the potential energy of the system. The positions and velocities of the atoms are then updated over time to simulate the behavior of the protein.

\subsubsection{Example: Running a Molecular Dynamics Simulation with GROMACS}

\textbf{GROMACS} \cite{van2005gromacs} is one of the most popular tools for performing MD simulations. Below is an example workflow for setting up and running a simulation in GROMACS:

\begin{lstlisting}[style=cmd]
# Step 1: Prepare the protein structure for simulation
gmx pdb2gmx -f protein.pdb -o processed.gro -water spce

# Step 2: Define the simulation box
gmx editconf -f processed.gro -o newbox.gro -c -d 1.0 -bt cubic

# Step 3: Solvate the protein
gmx solvate -cp newbox.gro -cs spc216.gro -o solvated.gro -p topol.top

# Step 4: Add ions to neutralize the system
gmx grompp -f ions.mdp -c solvated.gro -p topol.top -o ions.tpr
gmx genion -s ions.tpr -o solvated.gro -p topol.top -pname NA -nname CL -neutral

# Step 5: Perform energy minimization
gmx grompp -f minim.mdp -c solvated.gro -p topol.top -o em.tpr
gmx mdrun -v -deffnm em

# Step 6: Run the molecular dynamics simulation
gmx grompp -f md.mdp -c em.gro -p topol.top -o md.tpr
gmx mdrun -v -deffnm md
\end{lstlisting}

This example illustrates the steps involved in preparing a protein for simulation, solvating the system, adding ions, performing energy minimization, and finally running the MD simulation.

\subsubsection{Applications of Molecular Dynamics Simulation}

MD simulations are particularly useful for studying the folding process of small proteins, conformational changes in response to environmental factors, and protein-ligand interactions. They are also used to calculate free energy changes during folding and binding processes, helping to understand the thermodynamic properties of proteins.

\subsection{Energy Function and Folding Free Energy Calculations}

The behavior of a protein during folding is governed by its \textbf{energy function} \cite{lazaridis1999effective}, which defines the potential energy of the system. The goal of protein folding is to find the structure that minimizes the protein's free energy. The \textbf{free energy landscape} \cite{benkovic2008free} provides insights into the stability of different conformations and the pathways the protein takes as it folds.

\subsubsection{Understanding Energy Functions}

Energy functions are mathematical expressions that describe the interactions between atoms in a protein. These interactions include:

\begin{itemize}
    \item \textbf{Bonded interactions} \cite{mundlapati2018noncovalent}: Forces between atoms connected by covalent bonds, such as bond stretching and angle bending.
    \item \textbf{Non-bonded interactions} \cite{knegtel1998binding}: Forces between atoms that are not directly connected, including van der Waals forces and electrostatic interactions.
\end{itemize}

Energy functions are used in both MD simulations and folding calculations to model the forces acting on each atom.

\subsubsection{Free Energy and Protein Folding}

The folding process is driven by the minimization of the protein's free energy. The \textbf{Gibbs free energy} \cite{rietman2016thermodynamic} ($\Delta G$) is a measure of the thermodynamic favorability of the folding process. Proteins fold spontaneously when the change in free energy is negative ($\Delta G < 0$), meaning that the folded state is more stable than the unfolded state.

The free energy of folding is influenced by several factors:

\begin{itemize}
    \item \textbf{Enthalpy ($\Delta H$)}: The energy associated with bonds forming and breaking during folding.
    \item \textbf{Entropy ($\Delta S$)}: The disorder or randomness of the system. As a protein folds, it becomes more ordered, which results in a decrease in entropy.
\end{itemize}

The free energy change is calculated using the equation:

\[
\Delta G = \Delta H - T\Delta S
\]

Where $T$ is the temperature.

\subsubsection{Example: Free Energy Calculation Using Python}

Below is an example of how to calculate the Gibbs free energy of a folding process using Python:

\begin{lstlisting}[style=python]
# Define the enthalpy change (in kJ/mol) and entropy change (in kJ/mol*K)
delta_H = -300  # Enthalpy change
delta_S = -0.8  # Entropy change
temperature = 298  # Temperature in Kelvin

# Calculate the Gibbs free energy (in kJ/mol)
delta_G = delta_H - temperature * delta_S
print(f"Gibbs free energy change (ΔG): {delta_G} kJ/mol")
\end{lstlisting}

In this example, we calculate the Gibbs free energy for a protein folding process. A negative value of $\Delta G$ indicates that the folding process is thermodynamically favorable.

\subsubsection{Energy Minimization in Protein Folding}

Energy minimization is a critical step in protein structure prediction. It involves finding the conformation of the protein that has the lowest possible energy. This is done by adjusting the positions of atoms to reduce unfavorable interactions and improve the overall stability of the structure.

Energy minimization is often performed using techniques such as \textbf{gradient descent} \cite{senior2019protein} or \textbf{conjugate gradient methods} \cite{watowich1988stable}, which iteratively move the system toward the minimum energy state.

\subsubsection{Example: Energy Minimization Using GROMACS}

We can use GROMACS to perform energy minimization on a protein structure before running molecular dynamics simulations:

\begin{lstlisting}[style=cmd]
# Perform energy minimization
gmx grompp -f minim.mdp -c solvated.gro -p topol.top -o em.tpr
gmx mdrun -v -deffnm em
\end{lstlisting}

In this step, GROMACS adjusts the atomic positions in the protein to reduce the energy of the system, ensuring that the structure is stable and ready for further simulation or analysis.

\subsection{Summary of Traditional Algorithms}

Traditional algorithms for protein structure prediction and design provide powerful tools for understanding and modeling protein folding. Each method has its strengths and limitations:

\begin{itemize}
    \item \textbf{Homology modeling} \cite{krieger2003homology}: Effective for predicting the structure of proteins with homologous templates.
    \item \textbf{Threading} \cite{madej1995threading}: Useful for identifying structural folds when sequence similarity is low.
    \item \textbf{Molecular dynamics simulation} \cite{hollingsworth2018molecular}: Offers insights into the dynamic behavior of proteins and their folding pathways.
    \item \textbf{Energy function and folding free energy calculations} \cite{ruiz2013global}: Essential for understanding the thermodynamics of protein folding and stability.
\end{itemize}

These traditional methods serve as the foundation for more advanced techniques, including deep learning approaches that we will explore in later chapters.

\section{Modern Machine Learning and Deep Learning Methods}

Machine learning and deep learning have revolutionized the field of protein structure prediction and design. These modern methods allow us to analyze large datasets of protein sequences and structures, identifying patterns and relationships that are beyond the reach of traditional algorithms. In this section, we will explore how neural networks, convolutional neural networks \cite{lecun2015deep,zhangrethinking}, graph neural networks \cite{battaglia2018relational}, generative models, and self-supervised learning are applied to protein prediction and design. We will also provide detailed Python examples to help beginners understand the concepts.

\subsection{Neural Networks in Protein Prediction}

Neural networks are the foundation of modern deep learning methods in protein prediction. At a high level, a neural network consists of layers of interconnected neurons, where each neuron performs a simple computation. The network learns to map input data (such as protein sequences) to output predictions (such as protein structures) through a process called training.

In protein prediction, neural networks can be used for tasks like secondary structure prediction, where the network predicts whether a given segment of a protein will form an $\alpha$-helix, $\beta$-sheet, or coil. Neural networks are also used in more complex tasks, such as predicting the entire 3D structure of a protein from its sequence.

\subsubsection{Example: Simple Neural Network for Secondary Structure Prediction}

Let's build a simple neural network in Python using \textbf{TensorFlow} \cite{pang2020deep} to predict the secondary structure of a protein based on its amino acid sequence. This example uses a toy dataset for illustration purposes.

\begin{lstlisting}[style=python]
import tensorflow as tf
from tensorflow.keras import layers

# Define a simple model for secondary structure prediction
model = tf.keras.Sequential([
    layers.InputLayer(input_shape=(100, 20)),  # Example input: sequence of length 100, one-hot encoded
    layers.Dense(128, activation='relu'),
    layers.Dense(64, activation='relu'),
    layers.Dense(3, activation='softmax')  # Output: 3 classes (alpha-helix, beta-sheet, coil)
])

# Compile the model
model.compile(optimizer='adam', loss='categorical_crossentropy', metrics=['accuracy'])

# Summary of the model
model.summary()

# Example data (randomized for simplicity)
import numpy as np
X_train = np.random.rand(1000, 100, 20)  # 1000 protein sequences, each length 100, one-hot encoded
y_train = np.random.randint(3, size=(1000, 100, 3))  # Corresponding secondary structure labels

# Train the model (this is a simplified example)
model.fit(X_train, y_train, epochs=10)
\end{lstlisting}

In this code:
\begin{itemize}
    \item We define a neural network with two dense (fully connected) layers and an output layer that predicts secondary structure categories (e.g., $\alpha$-helix, $\beta$-sheet, coil).
    \item The input is a one-hot encoded protein sequence, where each amino acid is represented by a binary vector.
    \item We train the model on synthetic data for illustration purposes. In a real-world application, the model would be trained on actual protein sequence data.
\end{itemize}

\subsection{Convolutional Neural Networks and Graph Neural Networks}

\textbf{Convolutional neural networks (CNNs)} and \textbf{graph neural networks (GNNs)} are specialized types of neural networks that are particularly useful for protein structure prediction.

\subsubsection{Convolutional Neural Networks (CNNs) for Protein Prediction}

CNNs are commonly used for image recognition tasks, but they are also effective in protein prediction, especially when dealing with contact maps or distance matrices. In protein structure prediction, CNNs can be used to predict pairwise distances between amino acids, which is critical for determining the 3D structure.

\textbf{Contact Maps:} A contact map is a 2D matrix where each element represents the distance between two amino acids in a protein. CNNs can process these maps to predict the spatial relationships between amino acids.

\subsubsection{Example: Using a CNN to Predict a Protein Contact Map}

Here's an example of how to use a CNN to predict a protein contact map:

\begin{lstlisting}[style=python]
import tensorflow as tf
from tensorflow.keras import layers

# Define a CNN model for contact map prediction
model = tf.keras.Sequential([
    layers.Conv2D(64, (3, 3), activation='relu', input_shape=(100, 100, 1)),  # Input: contact map (100x100)
    layers.MaxPooling2D((2, 2)),
    layers.Conv2D(128, (3, 3), activation='relu'),
    layers.MaxPooling2D((2, 2)),
    layers.Flatten(),
    layers.Dense(256, activation='relu'),
    layers.Dense(10000, activation='sigmoid'),  # Output: flattened contact map
])

# Compile the model
model.compile(optimizer='adam', loss='binary_crossentropy', metrics=['accuracy'])

# Example input (random contact maps)
X_train = np.random.rand(1000, 100, 100, 1)  # 1000 contact maps of size 100x100
y_train = np.random.rand(1000, 10000)  # Flattened contact maps as labels

# Train the model
model.fit(X_train, y_train, epochs=10)
\end{lstlisting}

This CNN model takes a contact map as input and predicts a flattened version of the contact map as output. The network uses convolutional layers to capture local spatial relationships between amino acids.

\subsubsection{Graph Neural Networks (GNNs) in Protein Prediction}

Graph neural networks (GNNs) are particularly well-suited for representing proteins, as proteins can be naturally modeled as graphs. In a GNN, nodes represent amino acids, and edges represent interactions between them. This allows GNNs to capture the complex interactions in protein structures.

\textbf{Graph Representation:} In a protein, each amino acid can be treated as a node, and an edge can be drawn between two nodes if the corresponding amino acids are spatially close or have some interaction (such as a hydrogen bond).

\subsubsection{Example: Using a GNN for Protein Structure Prediction}

Let's define a simple GNN using \textbf{PyTorch Geometric}, a popular library for working with graph data in deep learning:

\begin{lstlisting}[style=python]
import torch
from torch_geometric.nn import GCNConv
from torch_geometric.data import Data

# Define a simple GNN model for protein structure prediction
class ProteinGNN(torch.nn.Module):
    def __init__(self):
        super(ProteinGNN, self).__init__()
        self.conv1 = GCNConv(20, 64)  # Input: 20 features per node (amino acid properties)
        self.conv2 = GCNConv(64, 128)
        self.fc = torch.nn.Linear(128, 3)  # Output: 3D coordinates

    def forward(self, data):
        x, edge_index = data.x, data.edge_index
        x = self.conv1(x, edge_index).relu()
        x = self.conv2(x, edge_index).relu()
        x = torch.nn.functional.global_mean_pool(x, data.batch)  # Pooling over nodes
        return self.fc(x)

# Example data (random for illustration)
node_features = torch.rand(100, 20)  # 100 nodes (amino acids), each with 20 features
edge_index = torch.randint(0, 100, (2, 200))  # 200 edges representing interactions
data = Data(x=node_features, edge_index=edge_index)

# Initialize and apply the GNN
model = ProteinGNN()
output = model(data)
print(output)
\end{lstlisting}

In this example, we define a simple GNN that takes a graph representation of a protein as input (nodes = amino acids, edges = interactions) and predicts 3D coordinates for each amino acid. GNNs are powerful because they can capture complex relationships between amino acids and provide insights into how proteins fold.

\subsection{Generative Models and Diffusion Models}

Generative models are deep learning models that can generate new data samples. In protein design, generative models can be used to create new protein sequences that are likely to fold into specific structures or have desired functions.

\subsubsection{Variational Autoencoders (VAEs) for Protein Design}

\textbf{Variational autoencoders (VAEs)} are generative models that learn to encode protein sequences into a lower-dimensional latent space and then decode them back into protein sequences \cite{kingma2013auto}. By sampling from this latent space, we can generate new protein sequences that have never been seen before.

\subsubsection{Example: Using a VAE for Protein Sequence Generation}

Here's a simple example of a VAE in Python for generating new protein sequences:

\begin{lstlisting}[style=python]
from tensorflow.keras import layers
from tensorflow.keras.models import Model
import tensorflow as tf

# Define the encoder
input_seq = layers.Input(shape=(100, 20))  # Input: one-hot encoded protein sequence
x = layers.Dense(64, activation='relu')(input_seq)
z_mean = layers.Dense(32)(x)
z_log_var = layers.Dense(32)(x)

# Define the sampling function
def sampling(args):
    z_mean, z_log_var = args
    epsilon = tf.random.normal(shape=(tf.shape(z_mean)[0], 32))
    return z_mean + tf.exp(0.5 * z_log_var) * epsilon

z = layers.Lambda(sampling)([z_mean, z_log_var])

# Define the decoder
decoder_input = layers.Input(shape=(32,))
decoder_output = layers.Dense(100 * 20, activation='softmax')(decoder_input)  # Reconstruct protein sequence
decoder_output = layers.Reshape((100, 20))(decoder_output)

# Build the VAE model
encoder = Model(input_seq, [z_mean, z_log_var, z])
decoder = Model(decoder_input, decoder_output)
vae = Model(input_seq, decoder(encoder(input_seq)[2]))

# Compile the VAE
vae.compile(optimizer='adam', loss='categorical_crossentropy')

# Example data (random for illustration)
X_train = np.random.rand(1000, 100, 20)  # 1000 protein sequences, one-hot encoded

# Train the VAE
vae.fit(X_train, X_train, epochs=10)
\end{lstlisting}

In this VAE model:
\begin{itemize}
    \item The encoder maps a protein sequence to a lower-dimensional latent space.
    \item The decoder reconstructs the protein sequence from the latent space.
    \item By sampling from the latent space, we can generate new protein sequences.
\end{itemize}

\subsubsection{Diffusion Models for Protein Design}

\textbf{Diffusion models} are another type of generative model that have been used to generate novel proteins \cite{ho2020denoising}. These models simulate a diffusion process, where random noise is gradually added to data, and then the noise is removed to generate new samples. This process can be applied to generate new protein sequences or structures.

\subsection{Self-supervised Learning and Masked Language Models}

Self-supervised learning is a paradigm where models learn to predict parts of the data that are masked or missing. In protein prediction, masked language models, similar to those used in natural language processing (NLP), have been adapted to predict amino acids in protein sequences \cite{devlin2018bert}. These models learn representations of protein sequences by predicting masked or missing residues.

\subsubsection{Example: Using a Masked Language Model for Protein Sequence Prediction}

\textbf{Transformers}, a type of model used in NLP, are commonly used in self-supervised learning for protein sequences. Here's an example of how a transformer-based masked language model can be implemented:

\begin{lstlisting}[style=python]
from transformers import BertTokenizer, TFBertForMaskedLM

# Load a pre-trained BERT model for masked language modeling
model = TFBertForMaskedLM.from_pretrained('Rostlab/prot_bert')

# Tokenize a protein sequence with masked positions
tokenizer = BertTokenizer.from_pretrained('Rostlab/prot_bert')
sequence = "MSTGILSPTQDLRSV"  # Example protein sequence
inputs = tokenizer(sequence, return_tensors="tf", padding=True)
masked_input = inputs.input_ids  # Simulate a masked input

# Predict the masked tokens (amino acids)
outputs = model(masked_input)
predicted_ids = tf.argmax(outputs.logits, axis=-1)

# Decode the predicted sequence
predicted_sequence = tokenizer.decode(predicted_ids[0])
print(f"Predicted sequence: {predicted_sequence}")
\end{lstlisting}

In this example, we use a pre-trained transformer model (BERT) to predict masked amino acids in a protein sequence. The model learns to fill in the missing parts of the sequence based on the surrounding context, which can be useful for tasks like predicting mutations or designing new sequences.

\subsection{Summary of Modern Machine Learning Methods}

Modern machine learning and deep learning methods have transformed protein design and structure prediction. Key approaches include:

\begin{itemize}
    \item \textbf{Neural networks}: Used for tasks such as secondary structure prediction and protein function classification.
    \item \textbf{CNNs and GNNs}: Capture spatial relationships and graph-like structures in proteins.
    \item \textbf{Generative models (VAEs and diffusion models)}: Generate new protein sequences and structures.
    \item \textbf{Self-supervised learning (masked language models)}: Learn representations of protein sequences by predicting masked amino acids.
\end{itemize}

These tools, combined with Python libraries like TensorFlow and PyTorch, have made it possible to predict, design, and generate proteins with unprecedented accuracy. In the following chapters, we will dive deeper into specific applications of these techniques in protein design.

\section{Tools for Protein Design}

Protein design is a rapidly evolving field that combines biology, chemistry, and computational tools to create new proteins with specific functions. There are several categories of tools used in protein design, ranging from sequence-based tools that focus on manipulating the amino acid sequence, to structure-based tools that focus on modifying or creating 3D structures, and finally, computer-aided engineering methods that allow researchers to predict and refine the behavior of designed proteins. In this section, we will explore these tools in detail, providing examples and explanations to help beginners understand the underlying principles and applications.

\subsection{Sequence-based Design Tools}

\textbf{Sequence-based design tools} focus on generating or modifying protein sequences to achieve specific functional or structural characteristics. These tools often work by optimizing amino acid sequences to ensure that they fold into stable and functional proteins. They are particularly useful when the overall structure of the protein is already known, but modifications are needed to improve function or stability.

\subsubsection{Rosetta Design for Sequence Optimization}

One of the most widely used sequence-based protein design tools is \textbf{Rosetta} \cite{leaver2011rosetta3}, a powerful software suite for protein structure prediction and design. Rosetta can be used to optimize protein sequences by modifying amino acids while maintaining the overall structure. The tool evaluates various mutations and selects the sequence that minimizes the energy of the folded protein.

\textbf{Example: Rosetta Sequence Optimization}

Below is a simple example of how Rosetta can be used to design a new protein sequence. In this example, we will modify an existing protein structure and generate optimized sequences.

\begin{lstlisting}[style=cmd]
# Run Rosetta to optimize the sequence of a protein
rosetta_scripts.default.linuxgccrelease \
    -s input_protein.pdb \
    -parser:protocol design.xml \
    -out:file:scorefile design_score.sc \
    -out:path:pdb designed_protein.pdb
\end{lstlisting}

In this example, Rosetta takes an input protein structure (input\_protein.pdb) and applies a sequence optimization protocol defined in the XML file (design.xml). The result is a new, optimized sequence that will fold into a low-energy structure. Rosetta evaluates the energy of different sequences and outputs the final designed protein structure (designed\_protein.pdb).

\subsubsection{ProteinMPNN: Neural Network for Sequence Design}

Another tool for sequence-based protein design is \textbf{ProteinMPNN} \cite{dauparas2022robust}, a neural network-based model that designs amino acid sequences to fold into a given protein backbone. ProteinMPNN generates sequences by minimizing the energy of the designed protein, ensuring that it will fold into the desired structure with high stability.

\textbf{Example: Sequence Design Using ProteinMPNN}

Here is a Python example using ProteinMPNN to design a sequence that matches a given protein backbone:

\begin{lstlisting}[style=python]
import torch
from protein_mpnn import ProteinMPNN

# Load the pre-trained ProteinMPNN model
model = ProteinMPNN()

# Define the target backbone structure
backbone_structure = "target_backbone.pdb"

# Use ProteinMPNN to generate amino acid sequences
designed_sequences = model.design(backbone_structure)

# Print the designed sequences
for sequence in designed_sequences:
    print(sequence)
\end{lstlisting}

In this example, we use ProteinMPNN to design new protein sequences based on a target protein backbone. The model generates sequences that are expected to fold into the desired structure with high stability, making it a powerful tool for protein sequence design.

\subsection{Structure-based Design Tools}

\textbf{Structure-based design tools} \cite{ghosh2014structure} focus on manipulating or designing the 3D structure of a protein. These tools are used when researchers want to change the protein's shape, optimize its folding, or create entirely new protein structures. Structure-based tools take advantage of knowledge about the physical and chemical interactions that govern protein folding and stability.

\subsubsection{PyMOL for Structure Visualization and Design}

\textbf{PyMOL} \cite{delano2002pymol} is a widely used tool for visualizing and editing protein structures. Although primarily used for visualization, PyMOL also allows researchers to make specific structural changes, such as mutating amino acids, creating disulfide bonds, or deleting specific residues. These modifications can be visualized in real-time, helping researchers assess their potential effects on protein stability and function.

\textbf{Example: Mutating a Residue in PyMOL}

Here is a simple example of how to use PyMOL to mutate a specific residue in a protein structure:

\begin{lstlisting}[style=cmd]
# Load a protein structure in PyMOL
pymol protein.pdb

# Select the residue you want to mutate (e.g., position 50)
select res50, resi 50

# Mutate the residue to alanine (A)
alter res50, resn="ALA"

# Save the mutated structure
save mutated_protein.pdb
\end{lstlisting}

In this example, PyMOL is used to mutate a specific residue (position 50) in a protein to alanine. The mutated protein structure can then be saved and used for further analysis or experiments.

\subsubsection{Foldit: Gamified Protein Design}

\textbf{Foldit} \cite{cooper2010predicting} is a unique tool that gamifies protein design, allowing users to interactively manipulate protein structures in a 3D environment. Foldit simplifies the complex process of protein folding, making it accessible to non-experts. Users are tasked with manipulating protein structures to achieve the lowest energy conformation, effectively designing stable proteins through a game-like interface.

\textbf{Example: Folding a Protein Using Foldit}

In Foldit, players are given a protein structure and tools to rotate, pull, and twist it into its optimal folded state. Although this approach is less formal than traditional tools, it has proven effective in solving challenging protein folding problems.

\subsection{Computer-Aided Protein Engineering Methods}

\textbf{Computer-aided protein engineering (CAPE)} \cite{bogle2010role} methods combine various computational tools, including molecular modeling, docking simulations, and energy calculations, to design proteins with specific functions. These methods are essential for creating proteins that perform specific tasks, such as binding to a particular target or catalyzing a specific reaction.

\subsubsection{Molecular Docking for Protein Design}

Molecular docking is a technique used to predict how a protein interacts with other molecules, such as ligands, small molecules, or other proteins. Docking simulations allow researchers to design proteins with improved binding affinity, making it a valuable tool for drug discovery and protein engineering.

\textbf{Example: Molecular Docking Using AutoDock}

Below is an example of how to use \textbf{AutoDock} \cite{morris2001autodock} to perform molecular docking between a protein and a ligand:

\begin{lstlisting}[style=cmd]
# Prepare the receptor and ligand files
prepare_receptor4.py -r protein.pdb -o protein_receptor.pdbqt
prepare_ligand4.py -l ligand.pdb -o ligand.pdbqt

# Run docking simulation
autodock4 -p docking_parameters.dpf -l docking_results.dlg
\end{lstlisting}

In this example, AutoDock is used to perform a docking simulation between a protein receptor and a ligand. The docking results show the predicted binding orientation and affinity, providing insights into how the protein interacts with the ligand.

\subsubsection{Molecular Dynamics for Protein Engineering}

Molecular dynamics (MD) simulations are used in protein engineering to study the dynamic behavior of proteins and their interactions with other molecules. MD simulations provide detailed information about how a protein's structure changes over time and how it responds to environmental factors such as temperature and pH.

\textbf{Example: Running Molecular Dynamics with GROMACS}

Here is an example workflow for running a molecular dynamics simulation using \textbf{GROMACS} \cite{van2005gromacs} to study protein-ligand interactions:

\begin{lstlisting}[style=cmd]
# Step 1: Prepare the protein structure
gmx pdb2gmx -f protein.pdb -o processed.gro -water spce

# Step 2: Define the simulation box
gmx editconf -f processed.gro -o newbox.gro -c -d 1.0 -bt cubic

# Step 3: Solvate the protein
gmx solvate -cp newbox.gro -cs spc216.gro -o solvated.gro -p topol.top

# Step 4: Add ions
gmx grompp -f ions.mdp -c solvated.gro -p topol.top -o ions.tpr
gmx genion -s ions.tpr -o solvated.gro -p topol.top -pname NA -nname CL -neutral

# Step 5: Perform energy minimization
gmx grompp -f minim.mdp -c solvated.gro -p topol.top -o em.tpr
gmx mdrun -v -deffnm em

# Step 6: Run the molecular dynamics simulation
gmx grompp -f md.mdp -c em.gro -p topol.top -o md.tpr
gmx mdrun -v -deffnm md
\end{lstlisting}

This GROMACS workflow sets up and runs a molecular dynamics simulation for a protein, providing insights into the stability and flexibility of the protein in different environments.

\subsubsection{Computational Protein Engineering with Deep Learning}

Deep learning is increasingly used in computer-aided protein engineering. By training neural networks on large datasets of protein structures and sequences, deep learning models can predict protein functions, generate new sequences, and guide the design of proteins with specific properties.

\textbf{Example: Using AlphaFold for Protein Structure Prediction}

\textbf{AlphaFold} \cite{jumper2021highly}, a deep learning model developed by DeepMind, predicts the 3D structure of proteins from their amino acid sequences. Below is an example of how to run AlphaFold to predict the structure of a protein:

\begin{lstlisting}[style=cmd]
# Step 1: Clone AlphaFold repository
git clone https://github.com/deepmind/alphafold.git
cd alphafold

# Step 2: Install dependencies
pip install -r requirements.txt

# Step 3: Run AlphaFold on a protein sequence
python run_alphafold.py --fasta_path=protein_sequence.fasta --output_dir=./output
\end{lstlisting}

In this example, AlphaFold takes a protein sequence (in FASTA format) as input and predicts the corresponding 3D structure. The results can be used to design proteins with specific structural features or to modify existing proteins for improved function.

\subsection{Summary of Protein Design Tools}

In this section, we explored various tools used in protein design:

\begin{itemize}
    \item \textbf{Sequence-based design tools} such as Rosetta and ProteinMPNN optimize amino acid sequences to create stable and functional proteins.
    \item \textbf{Structure-based design tools} like PyMOL and Foldit allow researchers to modify and visualize protein structures interactively.
    \item \textbf{Computer-aided protein engineering methods} such as molecular docking, molecular dynamics simulations, and deep learning enable the design of proteins with specific binding properties, stability, and functions.
\end{itemize}

These tools are essential for designing proteins for applications in drug discovery, enzyme engineering, and synthetic biology. As technology advances, these methods continue to become more accessible, making protein design a key area of innovation in biological research.
\chapter{Deep Learning Revolution in Protein Structure Prediction}

Deep learning has brought about a paradigm shift in the field of protein structure prediction, enabling researchers to achieve unprecedented accuracy and efficiency. One of the most groundbreaking advancements in this area has been the development of \textbf{AlphaFold} \cite{jumper2021highly} and its successor, \textbf{AlphaFold2} \cite{bryant2022improved}, by DeepMind. These models have revolutionized our ability to predict the three-dimensional structures of proteins directly from their amino acid sequences, a task that had previously been a major challenge in biology. In this chapter, we will provide a comprehensive introduction to AlphaFold and AlphaFold2, explain how they work, and explore their implications for protein design and structure prediction. This section is written for beginners, so it introduces the concepts in a clear and structured way, with practical examples to make the ideas accessible.

\section{Introduction to AlphaFold and AlphaFold2}

AlphaFold and AlphaFold2 represent a significant leap forward in computational biology. They utilize deep learning techniques to predict the 3D structure of proteins based on their amino acid sequences with remarkable accuracy. Understanding how these models work can help beginners appreciate the scale of their impact on protein science.

\subsection{The Challenge of Protein Structure Prediction}

Before the development of AlphaFold, predicting a protein's 3D structure from its amino acid sequence was a difficult and often unreliable task. Traditional methods, such as homology modeling, threading, and molecular dynamics simulations, provided some success, but they required either a known homologous structure or massive computational resources. Experimental techniques like X-ray crystallography, NMR spectroscopy, and cryo-EM could determine protein structures with high accuracy, but these methods are expensive, time-consuming, and not always feasible for all proteins.

The challenge stems from the fact that proteins are made up of long chains of amino acids that fold into highly complex, three-dimensional shapes. These shapes, in turn, dictate the protein's function. Even small changes in the amino acid sequence can result in vastly different folds and, consequently, different functions. Predicting how a linear sequence of amino acids will fold into a functional 3D structure is known as the \textbf{protein folding problem} \cite{dill2008protein}.

\subsubsection{The Importance of Solving the Protein Folding Problem}

Understanding the structure of a protein is key to understanding its function, and this knowledge has wide-ranging applications in areas such as drug discovery, enzyme design, and the study of diseases like Alzheimer's and Parkinson's, which are linked to protein misfolding. Therefore, solving the protein folding problem has been a major goal for biologists and computational scientists for decades.

\subsection{What is AlphaFold?}

\textbf{AlphaFold} is a deep learning-based model developed by DeepMind that predicts protein structures with remarkable accuracy. It was introduced during the Critical Assessment of protein Structure Prediction (CASP) competition in 2018 (AlphaFold1) and achieved even greater success in CASP14 in 2020 (AlphaFold2), where it demonstrated near-experimental accuracy.

AlphaFold works by taking the amino acid sequence of a protein as input and predicting the 3D coordinates of its atoms, which represent the folded structure. The key innovation behind AlphaFold is its ability to combine multiple sources of information, including evolutionary information from protein sequence alignments and physical constraints that govern how proteins fold.

\subsubsection{AlphaFold1 Overview}

The first version of AlphaFold, introduced in 2018, used deep learning and optimization techniques to predict protein structures. It employed convolutional neural networks (CNNs) to process multiple sequence alignments (MSAs) of related proteins. These alignments provided information about which amino acids tend to co-evolve, giving clues about their spatial relationships in the protein structure.

AlphaFold1's architecture was designed to predict a distance map between pairs of amino acids, which could then be converted into 3D coordinates. While AlphaFold1 was groundbreaking, it still had limitations, especially in cases where sequence data was sparse.

\textbf{Example: AlphaFold1 Workflow}

Here's a simplified step-by-step explanation of the AlphaFold1 workflow:

\begin{enumerate}
    \item \textbf{Input}: The amino acid sequence of the target protein.
    \item \textbf{Multiple Sequence Alignment (MSA)} \cite{bawono2017multiple}: The sequence is aligned with similar sequences from related proteins to gather evolutionary information.
    \item \textbf{Distance Prediction}: A convolutional neural network predicts the distances between pairs of amino acids in the protein.
    \item \textbf{Structure Optimization}: The predicted distances are converted into 3D coordinates, representing the protein's folded structure.
\end{enumerate}

\subsection{AlphaFold2: A Game-Changer}

\textbf{AlphaFold2}, introduced in 2020, was a significant improvement over AlphaFold1. It marked a transformative leap in the field by reaching near-experimental accuracy in predicting protein structures. AlphaFold2 outperformed all other methods in CASP14, achieving a median global distance test (GDT) score of 92.4, which is comparable to the accuracy of experimental methods like X-ray crystallography.

AlphaFold2 introduced several key innovations that made it much more accurate and efficient than its predecessor. These innovations include a new architecture that combines a transformer model with 3D structure predictions, as well as an end-to-end training approach that allows the model to directly optimize for structure accuracy.

\subsubsection{Key Innovations in AlphaFold2}

The key innovations that set AlphaFold2 apart from previous models include:

\begin{itemize}
    \item \textbf{End-to-End Training}: AlphaFold2 is trained end-to-end, meaning that the model directly learns to predict 3D structures from the amino acid sequence, rather than predicting intermediate representations (such as distance maps). This improves both accuracy and computational efficiency.
    \item \textbf{Attention Mechanism}: AlphaFold2 uses a transformer-based architecture with attention mechanisms that allow the model to focus on specific parts of the input sequence. This helps the model capture long-range dependencies between amino acids, which are crucial for accurate folding.
    \item \textbf{Recycling Mechanism}: AlphaFold2 introduces a recycling mechanism that iteratively refines its predictions. The model predicts a structure, evaluates the accuracy of that prediction, and then feeds the structure back into the model for further refinement. This iterative process leads to highly accurate final structures.
    \item \textbf{Equivariant Neural Networks}: AlphaFold2 employs equivariant neural networks, which ensure that the model's predictions are invariant to rotation and translation. This is important because the 3D structure of a protein can be rotated or shifted in space without changing its biological function.
\end{itemize}

\subsubsection{AlphaFold2 Architecture Overview}

AlphaFold2's architecture is complex, but here is a simplified explanation of its components:

\begin{itemize}
    \item \textbf{Input Features}: The input includes the amino acid sequence of the target protein and evolutionary information from MSAs.
    \item \textbf{Evoformer}: This is a transformer-based module that processes the input sequence and MSAs to generate feature representations. It captures both local and long-range interactions between amino acids.
    \item \textbf{Structure Module}: The structure module takes the output from the Evoformer and predicts the 3D coordinates of the protein's atoms. It uses equivariant neural networks to ensure that the predictions are physically realistic.
    \item \textbf{Recycling Module}: The predicted structure is fed back into the model, allowing the structure to be refined iteratively until the final prediction is achieved.
\end{itemize}

\textbf{Visualization of AlphaFold2's Workflow}

Here is a visual representation of AlphaFold2's workflow, which shows how the model refines its predictions over several iterations.

\begin{itemize}
    \item Input: Protein Sequence
    \begin{itemize}
        \item Multiple Sequence Alignment
        \item Evoformer: Transformer Model
        \begin{itemize}
            \item Captures Long-Range Interactions
        \end{itemize}
        \item Structure Module: 3D Prediction
        \begin{itemize}
            \item Equivariant Neural Network
        \end{itemize}
        \item Recycling: Iterative Refinement
    \end{itemize}
\end{itemize}

\subsubsection{Example: Running AlphaFold2 to Predict a Protein Structure}

Running AlphaFold2 requires significant computational resources, but here's a simplified workflow that shows how AlphaFold2 is used in practice:

\begin{lstlisting}[style=cmd]
# Clone the AlphaFold2 repository
git clone https://github.com/deepmind/alphafold.git
cd alphafold

# Install dependencies
pip install -r requirements.txt

# Run AlphaFold2 on a protein sequence
python run_alphafold.py --fasta_path=protein_sequence.fasta --output_dir=./output
\end{lstlisting}

In this example, AlphaFold2 takes a protein sequence in FASTA format and predicts its 3D structure, which is saved in the output directory. The model uses the input sequence along with MSAs to generate an accurate prediction of the protein's folded structure.

\subsection{Applications of AlphaFold and AlphaFold2}

AlphaFold and AlphaFold2 have a wide range of applications in both basic and applied science. Some of the most significant applications include:

\begin{itemize}
    \item \textbf{Drug Discovery}: By predicting the structures of proteins that play key roles in diseases, AlphaFold can help researchers design drugs that specifically target these proteins. This is particularly important for diseases where the target protein's structure was previously unknown.
    \item \textbf{Enzyme Engineering}: AlphaFold can assist in designing enzymes with improved catalytic activity or stability by predicting how changes in the sequence will affect the 3D structure and function of the enzyme.
    \item \textbf{Synthetic Biology}: In synthetic biology, researchers design new proteins with specific functions. AlphaFold can predict how designed proteins will fold, helping to ensure that they adopt the desired structures.
    \item \textbf{Understanding Diseases}: Many diseases, such as Alzheimer's and cystic fibrosis, are linked to protein misfolding. AlphaFold can be used to study how certain mutations lead to misfolding, providing insights into the molecular mechanisms of these diseases.
\end{itemize}

\subsection{Limitations of AlphaFold and AlphaFold2}

While AlphaFold2 has revolutionized the field of protein structure prediction, it is important to recognize its limitations:

\begin{itemize}
    \item \textbf{Protein Complexes}: AlphaFold2 excels at predicting the structures of individual proteins, but it is less accurate when predicting the structures of protein complexes (i.e., how multiple proteins interact to form a larger structure).
    \item \textbf{Dynamic Proteins}: AlphaFold2 predicts static structures, but many proteins are dynamic and undergo conformational changes. Predicting these dynamic behaviors remains a challenge.
    \item \textbf{Limited Accuracy for Disordered Proteins}: AlphaFold2 struggles with predicting the structures of intrinsically disordered proteins, which do not adopt a well-defined 3D structure.
\end{itemize}

\subsection{Future Directions for AlphaFold and Protein Prediction}

The success of AlphaFold2 has set the stage for further advancements in protein prediction. Some of the future directions for this research include:

\begin{itemize}
    \item \textbf{Predicting Protein Complexes}: Developing models that can accurately predict the interactions between multiple proteins will be a key area of research.
    \item \textbf{Incorporating Dynamics}: Future models may incorporate information about the dynamic nature of proteins, enabling predictions of how proteins change shape in response to different conditions.
    \item \textbf{Improving Speed and Accessibility}: As computational power continues to increase, the goal will be to make tools like AlphaFold more accessible to researchers worldwide, allowing them to predict protein structures more quickly and at lower costs.
\end{itemize}

\subsection{Conclusion}

AlphaFold and AlphaFold2 represent a major leap forward in the field of protein structure prediction. By leveraging deep learning techniques, these models have solved many of the challenges associated with predicting 3D protein structures from amino acid sequences. With applications ranging from drug discovery to synthetic biology, AlphaFold has already made a significant impact on biological research and will continue to be a critical tool for scientists in the years to come. As we continue to improve our understanding of protein folding and develop more advanced models, the future of protein structure prediction looks brighter than ever.

\section{The Improvements and Breakthroughs of AlphaFold2}

\textbf{AlphaFold2} represents one of the most significant breakthroughs in the field of protein structure prediction. Building on the success of AlphaFold1, AlphaFold2 introduced a series of key improvements that greatly enhanced the accuracy, speed, and reliability of protein structure predictions. These advancements allowed AlphaFold2 to achieve near-experimental accuracy in predicting the three-dimensional (3D) structures of proteins, setting new standards in computational biology. In this section, we will explore the critical improvements and breakthroughs that AlphaFold2 brought to the field, explaining them step by step to ensure that beginners can follow the concepts easily.

\subsection{Key Improvements in AlphaFold2}

AlphaFold2's success stems from several critical improvements over its predecessor. Each of these improvements addresses specific challenges in protein structure prediction, leading to more accurate and reliable results. Below, we will explore the major enhancements of AlphaFold2, including its end-to-end training, use of the attention mechanism, the recycling process, and the introduction of equivariant neural networks.

\subsubsection{End-to-End Training}

One of the most important improvements in AlphaFold2 is its \textbf{end-to-end training} approach. In contrast to AlphaFold1, which relied on separate models for different stages of prediction (such as distance prediction and structure optimization), AlphaFold2 directly learns to predict 3D structures from protein sequences in an end-to-end manner. This means that AlphaFold2 does not rely on intermediate predictions like contact maps or distance matrices. Instead, the model is trained to output the final 3D coordinates of the protein atoms from the raw input sequence in a single pipeline.

\begin{itemize}
    \item \textbf{Benefits of End-to-End Training}:
    \begin{itemize}
        \item Directly optimizes the final structure prediction, improving accuracy.
        \item Reduces the complexity of the model by eliminating the need for separate stages.
        \item Leads to more efficient training, allowing the model to learn better representations of the protein sequence.
    \end{itemize}
\end{itemize}

This approach allows the model to focus on the end goal-accurately predicting the 3D structure-without getting constrained by intermediate steps that may introduce errors.

\subsubsection{Attention Mechanism: Capturing Long-Range Interactions}

Another major breakthrough in AlphaFold2 is the use of the \textbf{attention mechanism}, borrowed from the field of natural language processing (NLP). In proteins, amino acids often interact with each other across long distances in the sequence. These long-range interactions are crucial for the folding process, as they help determine how the protein adopts its 3D structure.

AlphaFold2 uses a \textbf{transformer model} with attention layers to capture these long-range dependencies between amino acids. The attention mechanism allows the model to focus on specific parts of the protein sequence that are important for folding, even if they are far apart in the linear sequence.

\textbf{Example: Attention Mechanism in Protein Folding}

We can visualize the attention mechanism using a simple diagram to show how it allows the model to focus on interactions between amino acids:

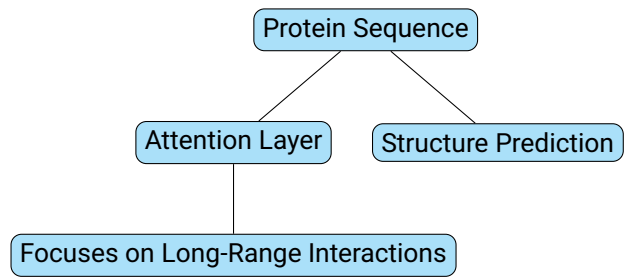
\begin{center}
\begin{tikzpicture}[sibling distance=10em, every node/.style = {shape=rectangle, rounded corners, draw, align=center,fill=cyan!30}]
  \node {Protein Sequence}
    child { node {Attention Layer} 
      child { node {Focuses on Long-Range Interactions} }
    }
    child { node {Structure Prediction} };
\end{tikzpicture}
\end{center}

In this diagram, the attention mechanism helps the model understand which parts of the sequence interact with each other, allowing it to make more accurate predictions about how the protein will fold.

\textbf{Python Example: Implementing an Attention Mechanism}

Here is a simplified Python example that demonstrates how the attention mechanism is used to capture dependencies between amino acids:

\begin{lstlisting}[style=python]
import tensorflow as tf
from tensorflow.keras import layers

# Define a simple transformer model with an attention layer
input_sequence = layers.Input(shape=(100, 20))  # Example: sequence length 100, one-hot encoded
attention_layer = layers.MultiHeadAttention(num_heads=4, key_dim=20)(input_sequence, input_sequence)
output = layers.Dense(3, activation='softmax')(attention_layer)  # Output: 3D structure prediction

# Compile the model
model = tf.keras.Model(inputs=input_sequence, outputs=output)
model.compile(optimizer='adam', loss='mse')

# Summary of the model
model.summary()
\end{lstlisting}

In this example, we define a transformer model with an attention layer that helps capture long-range dependencies in the input sequence. The attention layer processes the input sequence and outputs predictions about the 3D structure of the protein.

\subsubsection{Recycling Mechanism: Iterative Refinement of Predictions}

AlphaFold2 also introduces a novel \textbf{recycling mechanism}, which allows the model to iteratively refine its predictions \cite{adiyaman2023improvement}. After predicting the structure of the protein in the first pass, the model evaluates the accuracy of its prediction and feeds the predicted structure back into the model for further refinement.

This recycling mechanism allows the model to correct mistakes and improve its predictions in multiple iterations. As a result, the final predicted structure is more accurate and reliable.

\begin{itemize}
    \item \textbf{Benefits of Recycling}:
    \begin{itemize}
        \item Allows the model to refine its understanding of the protein's structure.
        \item Reduces errors by reprocessing the initial predictions.
        \item Leads to more accurate final structures, especially for complex proteins.
    \end{itemize}
\end{itemize}

\textbf{Example: Recycling Process in AlphaFold2}

The recycling process in AlphaFold2 can be visualized as a feedback loop, where the model revisits its predictions and makes iterative improvements:

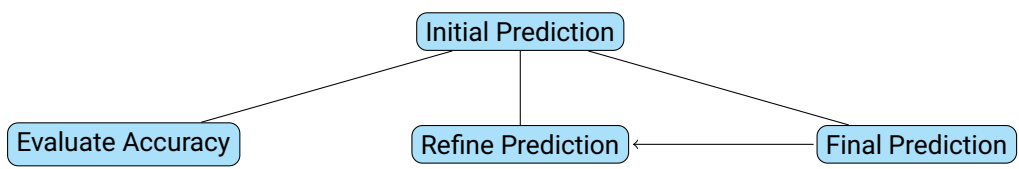
\begin{center}
\begin{tikzpicture}[sibling distance=15em, every node/.style = {shape=rectangle, rounded corners, draw, align=center,fill=cyan!30}]
  \node {Initial Prediction}
    child { node {Evaluate Accuracy} }
    child { node {Refine Prediction} }
    child { node {Final Prediction} };
  \draw[->] (3.9,-1.5) -- (1.5,-1.5);  
\end{tikzpicture}
\end{center}

\subsubsection{Equivariant Neural Networks}

AlphaFold2 also uses \textbf{equivariant neural networks} \cite{celledoni2021equivariant}, a type of network that ensures that the predicted structure is invariant to rotation and translation. This is important because a protein's function is determined by its shape, not by its orientation in space. Equivariant networks ensure that AlphaFold2 predicts the same structure regardless of how the input data is rotated or shifted.

Equivariant neural networks are particularly well-suited for 3D structure prediction because they enforce geometric consistency in the model's predictions, making them more physically realistic.

\begin{itemize}
    \item \textbf{Advantages of Equivariant Neural Networks}:
    \begin{itemize}
        \item Ensure that the model's predictions are consistent regardless of rotation or translation.
        \item Improve the physical realism of the predicted protein structures.
        \item Capture the underlying geometry of the protein more effectively.
    \end{itemize}
\end{itemize}

\subsection{Breakthroughs in Accuracy and Speed}

The improvements in AlphaFold2 have led to remarkable breakthroughs in both the accuracy and speed of protein structure predictions. Below, we will discuss these breakthroughs in more detail.

\subsubsection{Accuracy: Near-Experimental Precision}

One of the most significant achievements of AlphaFold2 is its ability to achieve \textbf{near-experimental accuracy} \cite{heo2019driven}. In CASP14, AlphaFold2 demonstrated a median global distance test (GDT) score of 92.4, which is comparable to the accuracy of experimental methods like X-ray crystallography. This means that AlphaFold2 can predict protein structures with a level of precision that was previously unimaginable using computational methods alone.

\textbf{Impact on Protein Science}:
\begin{itemize}
    \item Researchers can now obtain accurate protein structures for proteins that are difficult to study using experimental methods.
    \item AlphaFold2 accelerates the process of structure determination, enabling faster discoveries in fields like drug design and enzyme engineering.
\end{itemize}

\subsubsection{Speed: Faster Predictions with Fewer Resources}

Another major breakthrough in AlphaFold2 is its improved computational efficiency. Thanks to the end-to-end training approach and the use of advanced neural network architectures, AlphaFold2 can make predictions faster and with fewer computational resources compared to previous methods.

\textbf{Example: Computational Efficiency of AlphaFold2}

Here is an example that demonstrates how AlphaFold2 reduces computational requirements compared to older methods:

\begin{lstlisting}[style=cmd]
# Run AlphaFold2 on a protein sequence
python run_alphafold.py --fasta_path=protein_sequence.fasta --output_dir=./output --max_recycles=3
\end{lstlisting}

In this example, we use the \texttt{--max\_recycles=3} parameter to limit the number of recycling iterations, which reduces the computational cost while still producing accurate predictions. AlphaFold2's flexible design allows users to balance speed and accuracy depending on their needs.

\subsection{Applications of AlphaFold2's Breakthroughs}

The improvements and breakthroughs in AlphaFold2 have far-reaching applications in various fields of biological research and industry. Some of the most notable applications include:

\begin{itemize}
    \item \textbf{Drug Discovery}: AlphaFold2 enables rapid and accurate prediction of protein targets, speeding up the drug discovery process.
    \item \textbf{Enzyme Design}: Researchers can use AlphaFold2 to design enzymes with improved catalytic activity by predicting how changes in the sequence will affect the protein's 3D structure.
    \item \textbf{Protein-Protein Interactions}: AlphaFold2's high accuracy makes it useful for studying how proteins interact with each other, which is essential for understanding cellular processes and developing therapeutics.
    \item \textbf{Synthetic Biology}: By predicting how designed proteins will fold, AlphaFold2 helps researchers in synthetic biology create novel proteins with specific functions.
\end{itemize}

\subsection{Future Directions for AlphaFold2}

While AlphaFold2 has achieved remarkable success, there are still areas where future improvements can be made. Some of the potential future directions for AlphaFold2 and similar models include:

\begin{itemize}
    \item \textbf{Predicting Protein Complexes}: Developing models that can accurately predict the structures of multi-protein complexes remains a key challenge.
    \item \textbf{Incorporating Dynamics}: Future models could include information about protein dynamics, allowing for predictions of how proteins change shape in different environments or under different conditions.
    \item \textbf{Improved Accuracy for Disordered Proteins}: AlphaFold2 struggles with intrinsically disordered regions of proteins, and improving predictions in these regions could enhance its utility in studying protein folding diseases.
\end{itemize}

\subsection{Conclusion}

AlphaFold2's improvements and breakthroughs have fundamentally changed how we approach protein structure prediction. Its innovations, such as end-to-end training, the attention mechanism, recycling, and equivariant neural networks, have allowed it to achieve near-experimental accuracy at a fraction of the time and cost of traditional methods. As the model continues to evolve, it will play an increasingly important role in fields ranging from drug discovery to synthetic biology, opening new avenues for scientific discovery and innovation.

\section{Comparison with Traditional Methods}

AlphaFold and AlphaFold2 represent a significant leap forward in protein structure prediction, but to fully appreciate their impact, it's important to understand how these deep learning models compare to traditional methods. Traditional protein structure prediction methods, such as homology modeling, threading, and molecular dynamics (MD) simulations, have been the cornerstone of protein science for decades. Each of these methods has its strengths and limitations, and AlphaFold2 has introduced new capabilities that address many of the challenges faced by these older techniques. In this section, we will provide a detailed comparison between AlphaFold2 and traditional methods, exploring how deep learning has transformed protein structure prediction.

\subsection{Traditional Methods for Protein Structure Prediction}

Before the rise of deep learning, several traditional approaches were used to predict protein structures. These methods include:

\begin{itemize}
    \item \textbf{Homology Modeling}
    \item \textbf{Threading}
    \item \textbf{Molecular Dynamics (MD) Simulations}
    \item \textbf{Energy-Based Folding Methods}
\end{itemize}

Each of these methods operates on different principles, and their effectiveness varies depending on the availability of data and the complexity of the protein being studied. Let's explore these methods in more detail before comparing them to AlphaFold2.

\subsubsection{Homology Modeling}

\textbf{Homology modeling}, also known as comparative modeling, is based on the principle that proteins with similar amino acid sequences tend to have similar structures. If a protein's structure is unknown, and a homologous protein with a known structure exists, homology modeling can be used to predict the unknown structure by aligning the sequences and transferring the structural information.

\textbf{How Homology Modeling Works:}
\begin{itemize}
    \item A sequence alignment is performed between the target protein and a known template protein.
    \item The template's 3D structure is used as a scaffold, and the target sequence is modeled onto it.
    \item Refinement steps are taken to account for differences between the sequences, ensuring that the predicted structure folds appropriately.
\end{itemize}

While homology modeling can be highly accurate when a close homolog is available, it struggles when the sequence similarity between the target and template is low. This makes it less effective for proteins with no known homologs in structural databases.

\subsubsection{Threading (Fold Recognition)}

\textbf{Threading}, or fold recognition, is used when homology modeling fails due to low sequence similarity between the target and known structures. Threading attempts to align the target sequence to a database of known protein folds, even if there is little sequence similarity. It focuses on matching the overall 3D fold rather than individual residues.

\textbf{How Threading Works:}
\begin{itemize}
    \item The target sequence is threaded through known structural folds to find the best fit.
    \item Scoring functions evaluate the compatibility of the sequence with the fold based on factors like residue interactions and structural constraints.
    \item The best-fitting fold is used to predict the protein's structure.
\end{itemize}

Threading can be useful for predicting novel folds, but it is limited by the availability of known structures and the quality of the scoring functions, which may not always accurately capture subtle structural features.

\subsubsection{Molecular Dynamics Simulations}

\textbf{Molecular dynamics (MD) simulations} take a different approach by simulating the physical behavior of proteins over time. MD simulations model the movements of individual atoms based on physical laws, allowing researchers to study protein folding and conformational changes in atomic detail.

\textbf{How MD Simulations Work:}
\begin{itemize}
    \item A protein structure, either experimental or predicted, is placed in a simulated environment (usually in water or another solvent).
    \item The interactions between atoms are calculated using a force field, and the atoms are moved according to Newton's equations of motion.
    \item The simulation runs over time, showing how the protein's structure changes and folds.
\end{itemize}

MD simulations are extremely powerful for studying protein dynamics, but they are computationally expensive and typically require a known starting structure. Predicting the structure of a protein from scratch using MD simulations alone is infeasible for most proteins due to the sheer amount of computational power required.

\subsubsection{Energy-Based Folding Methods}

Traditional \textbf{energy-based folding methods} attempt to predict protein structure by minimizing the protein's free energy \cite{wu2021ebm}. Proteins tend to fold into the lowest-energy state, so by calculating the free energy for different conformations, these methods aim to find the structure with the lowest energy.

\textbf{How Energy-Based Methods Work:}
\begin{itemize}
    \item A potential energy function is used to model the forces acting between atoms in the protein.
    \item The protein is represented as a set of atomic coordinates, and the algorithm searches for the conformation that minimizes the total energy of the system.
    \item Techniques like Monte Carlo simulations or gradient descent are often used to search the energy landscape.
\end{itemize}

These methods can be accurate but are often computationally expensive. Additionally, finding the global minimum energy is difficult, as proteins often fold via complex, multi-step pathways that may involve local energy minima.

\subsection{How AlphaFold2 Differs from Traditional Methods}

AlphaFold2 introduced several breakthroughs that address many of the limitations of traditional methods. Below, we'll compare AlphaFold2 to each traditional method and explain the advantages that deep learning brings to protein structure prediction.

\subsubsection{Comparison with Homology Modeling}

\textbf{Key Differences:}
\begin{itemize}
    \item \textbf{Template Independence}: While homology modeling relies on a known template with high sequence similarity, AlphaFold2 can predict structures without needing a homologous template. It uses deep learning to model relationships between residues based on evolutionary data and physical constraints, allowing it to make accurate predictions even for proteins with no close homologs.
    \item \textbf{Accuracy}: AlphaFold2's predictions are often more accurate than homology models, even for proteins with known homologs. Its end-to-end training approach directly optimizes for structure accuracy, allowing it to outperform traditional homology models.
\end{itemize}

\textbf{Example of Homology Modeling vs. AlphaFold2:}

Consider a case where a protein has 30\% sequence similarity to a known structure. In homology modeling, the low similarity would likely result in an unreliable model, especially for regions that differ between the two sequences. In contrast, AlphaFold2 can use evolutionary and structural information to make a much more accurate prediction, even in regions of low similarity.

\subsubsection{Comparison with Threading}

\textbf{Key Differences:}
\begin{itemize}
    \item \textbf{Novel Folds}: AlphaFold2 can predict entirely new folds that have never been seen before, while threading is limited to selecting from a database of known folds. This gives AlphaFold2 a distinct advantage in predicting structures for proteins with novel architectures.
    \item \textbf{Scoring Functions}: Threading relies on scoring functions to evaluate how well a sequence fits a known fold. These scoring functions may not capture all aspects of protein folding, leading to suboptimal predictions. AlphaFold2, on the other hand, learns directly from data and is trained to optimize the final structure prediction.
\end{itemize}

Threading is useful when no close homologs are available, but it struggles with proteins that don't match any known fold. AlphaFold2 is not constrained by the need for known folds, allowing it to make predictions for a wider range of proteins.

\subsubsection{Comparison with Molecular Dynamics Simulations}

\textbf{Key Differences:}
\begin{itemize}
    \item \textbf{Speed}: AlphaFold2 can predict protein structures in a matter of hours, while molecular dynamics simulations often take days or weeks, even on high-performance computing clusters. MD simulations are focused on dynamics and atom-level movements, whereas AlphaFold2 predicts static structures, which is much faster.
    \item \textbf{Prediction from Sequence}: While MD simulations require a starting structure to simulate folding, AlphaFold2 predicts the structure directly from the amino acid sequence, making it a more versatile tool for structure prediction.
    \item \textbf{Application Focus}: MD simulations are best suited for studying the dynamics and flexibility of proteins, while AlphaFold2 excels at predicting static structures. These approaches are complementary, with AlphaFold2 providing the initial structure that MD simulations can refine or analyze over time.
\end{itemize}

\textbf{Example: Speed and Feasibility Comparison}

Suppose a researcher wants to predict the structure of a protein that has no known homolog and is suspected to have a novel fold. Using MD simulations alone to predict the fold would be prohibitively time-consuming, requiring months of computation. AlphaFold2, by contrast, can produce a highly accurate prediction in a matter of hours, providing a much more practical solution for initial structure determination.

\subsubsection{Comparison with Energy-Based Methods}

\textbf{Key Differences:}
\begin{itemize}
    \item \textbf{Global Energy Optimization vs. Learned Representations}: Traditional energy-based methods attempt to find the global minimum energy conformation, which can be computationally challenging due to the complex energy landscape of proteins. AlphaFold2 avoids the need for explicit energy minimization by learning how proteins fold from data, allowing it to predict structures directly.
    \item \textbf{Speed and Scalability}: Energy-based folding methods can be very slow, especially for large proteins, because they involve complex optimization algorithms. AlphaFold2's deep learning approach is much faster and can scale to predict large proteins with many residues.
\end{itemize}

Energy-based methods are powerful for studying the thermodynamics of protein folding, but they are not well-suited for large-scale structure prediction tasks where speed and efficiency are essential. AlphaFold2 provides a more scalable solution for predicting structures on a large scale.

\subsection{Visualization of Method Comparisons}

The following diagram summarizes the key differences between AlphaFold2 and traditional methods, highlighting the strengths and weaknesses of each approach.

\begin{itemize}
    \item Protein Structure Prediction
    \begin{itemize}
        \item Traditional Methods
        \begin{itemize}
            \item Homology Modeling
            \item Threading
            \item Molecular Dynamics
            \item Energy-Based Methods
        \end{itemize}
        \item AlphaFold2
        \begin{itemize}
            \item End-to-End Training
            \item Attention Mechanism
            \item Fast and Accurate
        \end{itemize}
    \end{itemize}
\end{itemize}

\subsection{Conclusion}

AlphaFold2 represents a significant advancement over traditional protein structure prediction methods. It offers greater accuracy, speed, and versatility, addressing many of the limitations of homology modeling, threading, molecular dynamics simulations, and energy-based methods. While traditional methods are still useful in certain contexts-such as studying protein dynamics or simulating protein-ligand interactions-AlphaFold2 provides a powerful tool for generating accurate protein structures from sequence data in a fraction of the time. Its ability to predict novel folds, combined with its speed and scalability, has transformed the field of protein structure prediction, opening up new possibilities for research in biology, drug discovery, and protein engineering.

\section{The Improvements and Breakthroughs of AlphaFold3}

\textbf{AlphaFold3} \cite{abramson2024accurate} represents the next significant advancement in the field of protein structure prediction and biomolecular interactions. Building on the successes of AlphaFold2, AlphaFold3 introduces several new features and breakthroughs that expand its capabilities, particularly in handling complex biological systems, including proteins, nucleic acids, small molecules, ions, and modified residues. In this section, we will delve into the key improvements of AlphaFold3, explaining how these developments enhance the accuracy, versatility, and applicability of the model.

\subsection{Key Improvements in AlphaFold3}

The development of AlphaFold3 addresses several limitations observed in previous versions and expands the range of biomolecular interactions it can accurately predict. Some of the most notable improvements include:

\subsubsection{Introduction of Diffusion-Based Architecture}

One of the core innovations in AlphaFold3 is the introduction of a \textbf{diffusion-based architecture} \cite{karras2022elucidating} for predicting atomic coordinates directly from raw input data. This represents a significant departure from the structure module used in AlphaFold2, which relied on torsion angles and backbone frames. By employing a generative diffusion model, AlphaFold3 can now handle a much wider variety of molecular entities and predict their interactions more efficiently.

\begin{itemize}
    \item \textbf{How it Works:} The diffusion model is trained to predict atomic positions by progressively denoising a random input, generating the structure in stages. At each step, the model refines its predictions, focusing on both local stereochemistry and larger-scale molecular features.
    \item \textbf{Benefits:} This architecture allows AlphaFold3 to predict highly accurate structures across a wide range of molecular systems, from small molecules to large protein complexes. The diffusion process also eliminates the need for complex stereochemical constraints, making the model simpler and more versatile.
\end{itemize}

\textbf{Example: Diffusion Process in AlphaFold3}

The diffusion process in AlphaFold3 can be visualized as follows:

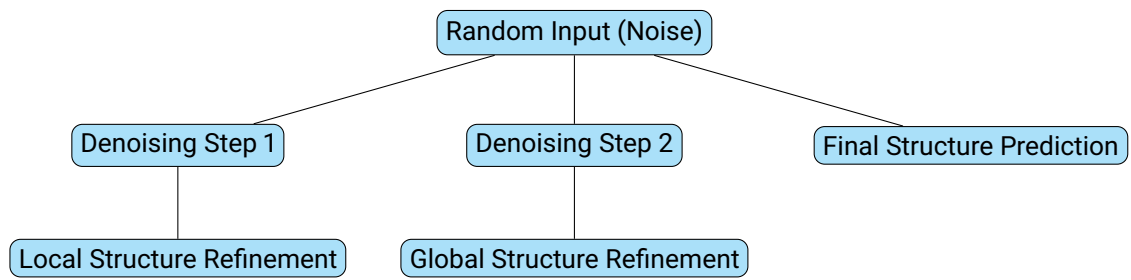
\begin{center}
\begin{tikzpicture}[sibling distance=15em, every node/.style = {shape=rectangle, rounded corners, draw, align=center,fill=cyan!30}]
  \node {Random Input (Noise)}
    child { node {Denoising Step 1} 
      child { node {Local Structure Refinement} }
    }
    child { node {Denoising Step 2} 
      child { node {Global Structure Refinement} }
    }
    child { node {Final Structure Prediction} };
\end{tikzpicture}
\end{center}

The diffusion model refines its predictions iteratively, gradually improving the accuracy of the predicted structure at each step.

\subsubsection{Pairformer Module Replacing Evoformer}

In AlphaFold3, the \textbf{pairformer module} \cite{abramson2024accurate} replaces the evoformer module from AlphaFold2. The pairformer is responsible for processing the relationships between atoms in a biomolecular complex, but it does so more efficiently and with greater flexibility.

\begin{itemize}
    \item \textbf{Key Change:} The pairformer operates directly on pairwise representations between entities in the complex, such as between protein residues or between a protein and a ligand. It reduces the reliance on multiple sequence alignments (MSAs), making the model less dependent on evolutionary data.
    \item \textbf{Benefits:} This modification allows AlphaFold3 to handle a broader range of biomolecular interactions, including protein-ligand and protein-nucleic acid complexes, without needing extensive MSA data.
\end{itemize}

\subsubsection{Handling of Small Molecules, Nucleic Acids, and Ions}

A major breakthrough in AlphaFold3 is its ability to predict the structure of complexes that include \textbf{small molecules \cite{thomas2008targeting}, nucleic acids \cite{fleck1966determination}, and modified residues \cite{motorin2007identification}}. Previous versions of AlphaFold focused primarily on protein structures, but AlphaFold3 broadens its scope to include virtually any type of biomolecular complex found in the \textbf{Protein Data Bank (PDB)} \cite{burley2017protein}.

\textbf{Examples of New Capabilities:}
\begin{itemize}
    \item \textbf{Protein-Ligand Interactions}: AlphaFold3 shows far greater accuracy in predicting how small molecules bind to proteins compared to traditional docking methods.
    \item \textbf{Protein-Nucleic Acid Interactions}: It outperforms nucleic-acid-specific predictors, providing highly accurate models of protein-DNA and protein-RNA complexes.
    \item \textbf{Ions and Modified Residues}: AlphaFold3 can now predict interactions involving ions and covalently modified residues, which are critical for understanding enzyme functions and post-translational modifications.
\end{itemize}

\subsubsection{Improved Accuracy for Antibody-Antigen Interactions}

AlphaFold3 significantly improves the accuracy of \textbf{antibody-antigen} \cite{wilson1993antibody}interaction predictions. This is crucial for applications in immunotherapy, where understanding how antibodies bind to specific antigens is essential for designing effective treatments.

\begin{itemize}
    \item \textbf{Improved Prediction Metrics}: Compared to AlphaFold-Multimer v2.3, AlphaFold3 demonstrates a marked improvement in DockQ scores (a measure of protein-protein interface prediction accuracy) for antibody-antigen interactions.
    \item \textbf{Multiple Seeds for Improved Results}: AlphaFold3 uses multiple model seeds and diffusion samples to generate highly accurate predictions for these complex interactions.
\end{itemize}

\subsubsection{Generative Model for Multiple Conformations}

AlphaFold3 also introduces a generative model that can predict \textbf{multiple conformations} \cite{wayment2024predicting} of biomolecular complexes. This is especially important for proteins and enzymes that adopt different conformations depending on their binding partners or environmental conditions.

\begin{itemize}
    \item \textbf{Handling Dynamic Proteins}: The generative model allows AlphaFold3 to explore the conformational space of dynamic proteins, predicting not just one static structure but multiple plausible conformations.
    \item \textbf{Avoiding Hallucinations}: A cross-distillation method was introduced to prevent the model from generating "hallucinations" or erroneous structures in disordered regions of the protein.
\end{itemize}

\subsection{Breakthroughs in Performance and Speed}

AlphaFold3 represents a significant leap in both \textbf{performance and speed}. Its diffusion-based architecture allows for faster predictions while maintaining or even improving upon the accuracy of previous models. Below are some key performance breakthroughs:

\subsubsection{Increased Prediction Speed}

Due to its streamlined architecture and reduced reliance on MSA processing, AlphaFold3 can predict complex structures in a shorter time frame compared to AlphaFold2.

\subsubsection{Higher Accuracy Across Biomolecular Space}

AlphaFold3's generalist approach allows it to outperform specialized tools in many domains. For example, it achieves higher accuracy than state-of-the-art docking tools for protein-ligand interactions and outperforms nucleic acid-specific predictors for protein-DNA/RNA complexes.

\begin{itemize}
    \item AlphaFold3
    \begin{itemize}
        \item Improved Accuracy
        \begin{itemize}
            \item Protein-Ligand Interactions
            \item Protein-Nucleic Acid Complexes
            \item Antibody-Antigen Interactions
        \end{itemize}
        \item Faster Predictions
        \begin{itemize}
            \item Reduced MSA Dependence
            \item Diffusion-Based Architecture
        \end{itemize}
    \end{itemize}
\end{itemize}

\subsection{Applications and Future Directions}

The breakthroughs in AlphaFold3 have broad applications across many fields, from drug discovery to synthetic biology and immunotherapy. Its ability to accurately model biomolecular complexes opens new avenues for rational drug design, enzyme engineering, and the study of protein-protein and protein-nucleic acid interactions.

\begin{itemize}
    \item \textbf{Drug Discovery}: AlphaFold3's ability to predict protein-ligand interactions with high accuracy makes it a powerful tool for designing new drugs that target specific proteins.
    \item \textbf{Synthetic Biology}: The generative capabilities of AlphaFold3 enable the design of novel proteins and enzymes with specific functions.
    \item \textbf{Immunotherapy}: By accurately modeling antibody-antigen interactions, AlphaFold3 can aid in the design of new therapeutic antibodies for diseases such as cancer and autoimmune disorders.
\end{itemize}

\subsubsection{Future Improvements}

Despite its remarkable advancements, AlphaFold3 still has areas for potential improvement. Future versions could focus on:

\begin{itemize}
    \item \textbf{Improved Dynamics}: Currently, AlphaFold3 predicts static structures. Incorporating dynamic behavior into future models would allow for better predictions of proteins that undergo conformational changes.
    \item \textbf{Handling Larger Complexes}: While AlphaFold3 can predict large biomolecular complexes, further optimizations are needed to efficiently model systems with thousands of residues or multiple interacting partners.
\end{itemize}

\subsection{Conclusion}

AlphaFold3 builds on the success of AlphaFold2, introducing new capabilities such as a diffusion-based architecture, pairformer module, and support for a wider range of biomolecular interactions. These improvements result in greater accuracy, faster predictions, and expanded applicability across biological research and drug development. As the field of protein structure prediction continues to evolve, AlphaFold3 represents a significant step forward, enabling new discoveries in biology and medicine.
\chapter{Modules of AlphaFold2}
\section{Structure Generation Module}

The \textbf{structure generation module} is one of the most crucial components of AlphaFold2, responsible for predicting the three-dimensional (3D) structure of a protein based on its amino acid sequence. This module operates at the heart of AlphaFold2's prediction pipeline, transforming the processed sequence and relational information into the final 3D coordinates of the protein. Understanding how the structure generation module works, step by step, is critical for grasping how AlphaFold2 achieves such high accuracy in protein structure prediction. In this section, we will explore the different elements of the structure generation module in detail, explaining its mechanisms in a way that is accessible to beginners.

\subsection{Overview of the Structure Generation Module}

The structure generation module takes the processed information from the earlier stages of AlphaFold2, such as the multiple sequence alignments (MSAs) and pairwise representations, and converts them into accurate predictions of a protein's 3D coordinates \cite{yang2023alphafold2}. The goal of this module is to predict the precise atomic positions of the protein's backbone and side chains while maintaining physical and chemical realism. This process is built on several key innovations, including the use of neural networks that are tailored for 3D geometry and specific modules designed to refine the predictions iteratively.

\subsubsection{The Role of Neural Networks in Structure Prediction}

AlphaFold2's structure generation module relies on deep neural networks to predict the spatial arrangement of atoms in a protein. Unlike traditional methods that use physical simulations or energy minimization, AlphaFold2's approach uses learned representations of protein structure and folding patterns \cite{garrido2024analysis}. This makes it possible for the model to make predictions that are not only faster but also more accurate than older computational methods.

The neural networks in this module learn from vast datasets of known protein structures and are trained to recognize patterns of folding and spatial interactions that occur in proteins with similar sequences. This learning enables the module to predict the following:

\begin{itemize}
    \item \textbf{Backbone structure}: The positions of the main chain atoms (N, C$_\alpha$, C) that form the protein's backbone.
    \item \textbf{Side chain positioning}: The arrangement of atoms in the side chains of each amino acid, which often determine the protein's function and interactions.
    \item \textbf{Torsion angles}: The angles between adjacent planes in the protein backbone (phi, psi, and omega angles), which define the protein's secondary and tertiary structures.
\end{itemize}

\subsection{Backbone Prediction: Building the Scaffold of the Protein}

The first step in the structure generation module is predicting the backbone structure of the protein \cite{goverde2023novo}. The backbone forms the core scaffold of the protein and dictates the overall 3D shape. Predicting the backbone is a critical first step because it provides a framework upon which the side chains and other structural elements can be built.

\subsubsection{Torsion Angles and Frame Prediction}

Proteins have a flexible backbone that can adopt different shapes based on the torsion angles between amino acids. These angles, known as phi ($\phi$), psi ($\psi$), and omega ($\omega$), define how the protein chain twists and folds in space.

\begin{itemize}
    \item \textbf{Phi ($\phi$) and Psi ($\psi$)}: These angles describe the rotation of the backbone around the N-C$_\alpha$ bond and C$_\alpha$-C bond, respectively. They are key determinants of secondary structure elements like $\alpha$-helices and $\beta$-sheets.
    \item \textbf{Omega ($\omega$)}: This angle typically adopts a planar conformation, as it corresponds to the peptide bond between the C and N atoms of adjacent residues.
\end{itemize}

In AlphaFold2, the structure generation module uses deep learning to predict these angles for each residue in the protein sequence. These predicted torsion angles allow the model to build a backbone that closely resembles the native structure of the protein.

\textbf{Example: Torsion Angles in Protein Folding}

The diagram below shows the relationship between phi and psi angles and how they define the backbone conformation:

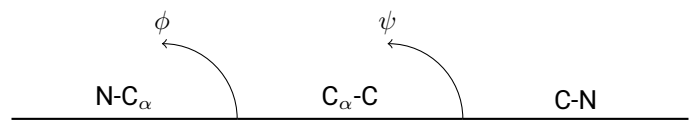
\begin{center}
\begin{tikzpicture}
    \draw[thick] (0,0) -- (3,0) node[midway,above] {N-C$_\alpha$};
    \draw[thick] (3,0) -- (6,0) node[midway,above] {C$_\alpha$-C};
    \draw[thick] (6,0) -- (9,0) node[midway,above] {C-N};
    \draw[->] (3,0) arc (0:90:1) node[above] {$\phi$};
    \draw[->] (6,0) arc (0:90:1) node[above] {$\psi$};
\end{tikzpicture}
\end{center}

In this simplified diagram, the angles $\phi$ and $\psi$ dictate the rotation of the backbone, enabling the prediction of secondary structures.

\subsubsection{Frame Representations}

AlphaFold2 employs a frame-based representation of the backbone to model its 3D coordinates. Each residue in the sequence is assigned a local coordinate frame, which describes the spatial orientation of the atoms around that residue. The frames for consecutive residues are connected to build the overall backbone structure.

This approach provides a flexible way to model protein backbones, as the frames can be adjusted iteratively to refine the predicted structure. By predicting the local frames for each residue, the model can ensure that the backbone forms a physically plausible structure that is compatible with known protein folds.

\subsection{Side Chain Prediction: Adding the Functional Elements}

Once the backbone of the protein has been predicted, the next step is to predict the positions of the side chains. Side chains are crucial for protein function, as they determine how the protein interacts with other molecules, including ligands, substrates, and other proteins \cite{holm2023dali}. AlphaFold2's structure generation module uses learned representations of side chain geometry to accurately place the atoms in each side chain.

\subsubsection{Challenges in Side Chain Prediction}

Predicting the side chain positions is more challenging than predicting the backbone because side chains can adopt many different conformations depending on the local environment and interactions within the protein. Factors that influence side chain positioning include:

\begin{itemize}
    \item \textbf{Steric hindrance}: Side chains must avoid clashing with other atoms in the protein.
    \item \textbf{Hydrophobicity}: Hydrophobic side chains tend to be buried in the interior of the protein, while hydrophilic side chains are often exposed to the solvent.
    \item \textbf{Electrostatic interactions}: Charged side chains can form salt bridges with other charged residues or interact with ions and ligands.
\end{itemize}

AlphaFold2's neural networks are trained to predict the side chain conformations by taking into account these factors, along with the predicted backbone structure. The model uses a combination of geometric reasoning and learned patterns from large datasets to place the side chains in their most likely positions.

\subsection{Iterative Refinement: Improving Prediction Accuracy}

The structure generation module in AlphaFold2 operates iteratively, meaning that the predicted structure is refined over several passes through the network. This \textbf{recycling} process allows the model to adjust the predicted coordinates, improving the accuracy of the final structure. Each pass through the network refines the backbone and side chain positions based on the predictions from the previous pass.

\textbf{Benefits of Iterative Refinement:}
\begin{itemize}
    \item \textbf{Error Correction}: By revisiting the structure multiple times, the model can correct errors in the initial predictions, leading to a more accurate final structure.
    \item \textbf{Consistency with Known Structures}: The iterative process ensures that the predicted structure is consistent with known protein folds and does not contain unrealistic features.
\end{itemize}

\subsubsection{Evaluating the Refined Structure}

After several rounds of refinement, the final predicted structure is evaluated based on its consistency with physical and biological constraints. AlphaFold2 ensures that the structure is stereochemically plausible, meaning that bond lengths, angles, and torsion angles fall within acceptable ranges for proteins. The model also checks for clashes between atoms and ensures that side chains adopt physically realistic conformations.

\subsection{Summary of the Structure Generation Module}

The structure generation module in AlphaFold2 is responsible for transforming sequence information into accurate 3D protein structures. This is achieved through a combination of deep learning techniques that predict torsion angles, backbone frames, and side chain conformations, followed by iterative refinement to improve accuracy. Key aspects of this process include:

\begin{itemize}
    \item \textbf{Backbone prediction}: Using predicted torsion angles and local coordinate frames to model the protein's backbone.
    \item \textbf{Side chain positioning}: Predicting the conformations of side chains based on geometric and chemical factors.
    \item \textbf{Iterative refinement}: Recycling the predicted structure to correct errors and improve accuracy.
\end{itemize}

By leveraging these techniques, AlphaFold2 can predict protein structures with a level of accuracy that was previously only achievable through experimental methods. This module is a key reason for AlphaFold2's success in transforming the field of protein structure prediction.

\section{Evaluation and Scoring Module}

The \textbf{Evaluation and Scoring Module} in AlphaFold2 plays a crucial role in assessing the accuracy and reliability of the predicted protein structures \cite{akdel2022structural}. After the structure generation module has created a 3D model of a protein, it is essential to evaluate this model to ensure that it meets biological and physical standards. The evaluation process involves scoring the predicted structure based on various metrics that reflect how well the predicted model aligns with known principles of protein folding, stereochemistry, and empirical data from experimentally determined structures. In this section, we will break down the evaluation process in detail, explaining the scoring mechanisms used by AlphaFold2, how they work, and why they are important for ensuring the model's reliability.

\subsection{Overview of the Evaluation Process}

The purpose of the evaluation and scoring module is to determine how closely the predicted structure matches the expected or true structure of the protein. This is done by comparing the predicted structure against known structural features, such as bond angles, torsion angles, and overall folding patterns. Additionally, when experimental data is available (e.g., from X-ray crystallography or cryo-electron microscopy), AlphaFold2 can assess how well its prediction agrees with this data.

The evaluation process is divided into two main stages:

\begin{itemize}
    \item \textbf{Internal evaluation}: This focuses on ensuring that the predicted structure adheres to fundamental physical and biological principles, such as stereochemistry and residue interactions.
    \item \textbf{External evaluation}: This involves comparing the predicted structure with known experimental data or using scoring metrics designed to gauge how well the predicted structure would perform if it were experimentally validated.
\end{itemize}

\subsubsection{Internal Evaluation: Ensuring Physical Realism}

One of the key challenges in protein structure prediction is ensuring that the predicted model is physically realistic. Proteins must follow certain rules dictated by chemistry and physics, such as correct bond lengths and angles, avoidance of steric clashes, and the proper arrangement of side chains. AlphaFold2's internal evaluation step assesses these factors.

\textbf{Metrics Used in Internal Evaluation:}

\begin{itemize}
    \item \textbf{Stereochemistry} \cite{coskun2023using}: The model is checked for the correct stereochemistry of atoms, meaning that bond angles and torsion angles fall within expected ranges. AlphaFold2 uses known stereochemical rules for amino acids to validate that the predicted structure makes sense chemically.
    \item \textbf{Bond Lengths and Angles} \cite{agarwal2024power}: The predicted structure is evaluated for correct bond lengths and angles between atoms. Any deviations from expected values can indicate that the structure is unrealistic or physically unstable.
    \item \textbf{Clash Detection} \cite{al2023investigating}: AlphaFold2 looks for steric clashes, where atoms in the predicted structure are too close to each other, violating the principles of atomic repulsion. A structure with significant clashes would not be physically plausible in a biological system.
    \item \textbf{Ramachandran Plot Evaluation} \cite{alhumaid2024reliability}: The Ramachandran plot is a two-dimensional plot that visualizes the phi ($\phi$) and psi ($\psi$) angles of the protein's backbone. In proteins, most of the backbone torsion angles fall within certain allowed regions on this plot. AlphaFold2 evaluates the predicted structure to ensure that its torsion angles fall within these allowed regions.
\end{itemize}

\textbf{Example: Ramachandran Plot Analysis}

The Ramachandran plot is a commonly used tool to evaluate the backbone torsion angles of proteins. The allowed regions on the plot correspond to stable secondary structures, such as $\alpha$-helices and $\beta$-sheets. A well-predicted protein structure should have the majority of its torsion angles falling within these allowed regions.

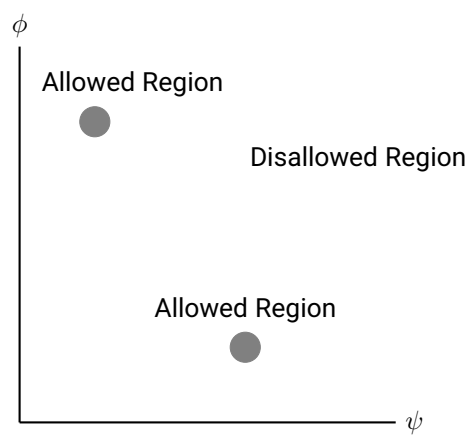
\begin{center}
\begin{tikzpicture}
    \draw[thick] (0,0) -- (5,0) node[right] {$\psi$};  
    \draw[thick] (0,0) -- (0,5) node[above] {$\phi$};  
    \filldraw[gray] (1,4) circle [radius=0.2];
    \node at (1.5,4.5) {Allowed Region};  
    \filldraw[gray] (3,1) circle [radius=0.2];
    \node at (3,1.5) {Allowed Region};  
    \node at (4.5,3.5) {Disallowed Region};  
\end{tikzpicture}
\end{center}

\textbf{Importance of Stereochemical Evaluation:}

Ensuring that the structure adheres to stereochemical rules is important because biologically functional proteins must follow these rules. Errors in stereochemistry can lead to incorrect folding or functional impairment in the real-world protein.

\subsubsection{External Evaluation: Comparison with Experimental Data}

Once internal evaluation confirms that the structure is physically plausible, AlphaFold2's evaluation module moves on to \textbf{external evaluation}, where the predicted structure is compared with experimentally determined structures if available. This process is particularly important for validating the accuracy of AlphaFold2's predictions in cases where the true structure has been experimentally resolved using techniques like X-ray crystallography, NMR spectroscopy, or cryo-electron microscopy (cryo-EM).

\textbf{Metrics Used in External Evaluation:}

\begin{itemize}
    \item \textbf{Root-Mean-Square Deviation (RMSD)} \cite{damm2013csar}: RMSD is one of the most common metrics used to assess how closely two structures align. It measures the average distance between corresponding atoms in the predicted structure and the experimentally determined structure. Lower RMSD values indicate a closer match between the two structures.
    \item \textbf{Global Distance Test (GDT)} \cite{olechnovivc2013cad}: The GDT score measures the percentage of residues in the predicted structure that are within a certain distance of the corresponding residues in the experimental structure. The GDT score ranges from 0 to 100, with higher scores indicating better agreement with the experimental structure.
    \item \textbf{Template Modeling Score (TM-Score)} \cite{zhang2004scoring}: The TM-Score is another metric used to assess the similarity between two protein structures. It ranges from 0 to 1, with scores above 0.5 indicating a reasonably accurate prediction.
\end{itemize}

\textbf{Example: Evaluating a Predicted Structure with RMSD}

Suppose we have an experimentally determined protein structure and a predicted structure from AlphaFold2. The RMSD value between these two structures can be calculated by comparing the positions of corresponding atoms:

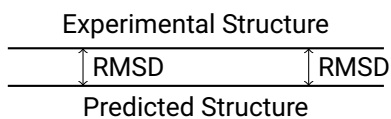
\begin{center}
\begin{tikzpicture}
    \draw[thick] (0,0) -- (5,0) node[midway,below] {Predicted Structure};
    \draw[thick] (0,0.5) -- (5,0.5) node[midway,above] {Experimental Structure};
    \draw[<->] (1,0) -- (1,0.5) node[midway,right] {RMSD};
    \draw[<->] (4,0) -- (4,0.5) node[midway,right] {RMSD};
\end{tikzpicture}
\end{center}

The RMSD value is calculated as the square root of the average squared deviations between the predicted and experimental atom positions.

\subsection{Scoring Mechanisms: Assessing Confidence in Predictions}

AlphaFold2 also includes a built-in \textbf{scoring mechanism} that assesses the model's confidence in its own predictions. This is done using a metric called the \textbf{pLDDT score} \cite{david2022alphafold} (predicted Local Distance Difference Test). The pLDDT score provides a per-residue confidence score, allowing researchers to evaluate which parts of the structure are predicted with high certainty and which parts may be less reliable.

\textbf{How pLDDT Works:}

The pLDDT score ranges from 0 to 100, with higher values indicating greater confidence in the prediction. This score is computed based on how well the model predicts the distances between pairs of atoms within each residue. Residues with well-defined, stable structures tend to have higher pLDDT scores, while flexible or disordered regions may have lower scores.

\begin{itemize}
    \item \textbf{High pLDDT Score ($>$90)}: Indicates high confidence in the predicted structure for that region.
    \item \textbf{Moderate pLDDT Score (70-90)}: The structure is predicted with reasonable confidence, but there may be some uncertainties.
    \item \textbf{Low pLDDT Score ($<$70)}: Indicates that the predicted structure in this region is uncertain, and experimental validation may be necessary.
\end{itemize}

\textbf{Visualization of pLDDT Confidence Scores}

The pLDDT scores can be visualized on the protein structure by coloring different regions based on their confidence levels. This allows researchers to easily identify areas of high and low confidence:

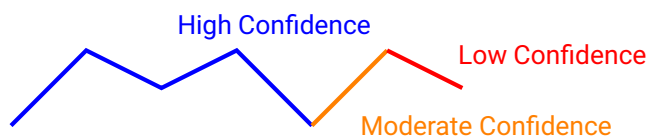
\begin{center}
\begin{tikzpicture}
    \draw[ultra thick,blue] (0,0) -- (1,1) -- (2,0.5) -- (3,1) -- (4,0) node[midway,above,xshift=0.0cm,yshift=0.5cm] {High Confidence};
    \draw[ultra thick,orange] (4,0) -- (5,1) node[midway,right,xshift=0.0cm,yshift=-0.5cm] {Moderate Confidence};
    \draw[ultra thick,red] (5,1) -- (6,0.5) node[midway,right,xshift=0.3cm,yshift=0.2cm] {Low Confidence};
\end{tikzpicture}
\end{center}

In this visualization, the blue regions represent areas where the model has high confidence in its prediction, while the red regions indicate areas of uncertainty.

\subsubsection{Importance of pLDDT Scores}

The pLDDT score is important for assessing the reliability of different regions of the predicted structure. In many cases, researchers may focus on the high-confidence regions for downstream applications, such as drug design or protein engineering, while treating the low-confidence regions with caution or subjecting them to further experimental validation.

\subsection{Importance of the Evaluation and Scoring Module}

The evaluation and scoring module is a critical part of AlphaFold2's architecture, as it ensures that the predicted structures are not only physically plausible but also aligned with experimental observations. By providing confidence scores and comparing predicted structures to known data, this module helps researchers assess the reliability of the predictions, making AlphaFold2 a valuable tool for scientific research and practical applications.

\textbf{Summary of the Evaluation and Scoring Module:}
\begin{itemize}
    \item \textbf{Internal Evaluation}: Ensures that the predicted structure adheres to physical and chemical principles, such as correct bond angles, stereochemistry, and avoidance of atomic clashes.
    \item \textbf{External Evaluation}: Compares the predicted structure to experimental data, using metrics like RMSD, GDT, and TM-Score to gauge accuracy.
    \item \textbf{pLDDT Scoring}: Provides a confidence score for each residue, helping researchers identify areas of high and low certainty in the predicted structure.
\end{itemize}

This thorough evaluation process ensures that AlphaFold2's predictions are not only fast but also reliable and accurate, making it one of the most advanced tools for protein structure prediction available today.

\section{Role of Multiple Sequence Alignment (MSA)}

One of the key innovations that enables AlphaFold2 to achieve its remarkable accuracy in protein structure prediction is the use of \textbf{Multiple Sequence Alignment} (MSA). MSAs play a crucial role in AlphaFold2's architecture by providing evolutionary information that helps the model understand the relationships between amino acids in a protein sequence. This evolutionary context is critical for predicting how a protein folds, as amino acids that are conserved across different species are often important for maintaining the structure and function of a protein. In this section, we will explore the concept of MSA in detail, explain how it contributes to AlphaFold2's predictions, and discuss its importance for both beginners and experts in the field of protein design and prediction.

\subsection{What is Multiple Sequence Alignment (MSA)?}

\textbf{Multiple Sequence Alignment (MSA)} is a method used to align three or more protein or nucleotide sequences in such a way that regions of similarity are highlighted. These regions of similarity often indicate evolutionary relationships and shared functional or structural roles between the sequences. By comparing a target protein sequence to a set of related sequences from other organisms, MSAs can provide insights into which amino acids are important for the structure or function of the protein.

\textbf{Example: Aligning Sequences Using MSA}

Suppose we have several protein sequences that are homologous to a target protein. The MSA process will align these sequences to reveal conserved residues:

\begin{center}
\begin{lstlisting}[style=cmd]
Sequence 1: ATGC--TACGAT
Sequence 2: ATGCTGTA-GAT
Sequence 3: ATGCCT--GGAT
Sequence 4: ATGCGT-ACGAT
\end{lstlisting}
\end{center}

In this example, the MSA reveals that some residues, such as the positions containing "ATGC" at the start, are conserved across all four sequences. These conserved residues may play a crucial role in the structure or function of the protein.

\subsection{How AlphaFold2 Uses MSA Information}

AlphaFold2 utilizes MSAs to gather \textbf{evolutionary couplings} \cite{weinreb20163d} between amino acids in a target protein. Evolutionary couplings provide important clues about which amino acids are likely to be close to each other in the protein's final 3D structure, even if they are far apart in the linear sequence. By analyzing MSAs, AlphaFold2 can learn which parts of the protein sequence have co-evolved, suggesting that these regions form important structural or functional units.

\subsubsection{Evolutionary Couplings}

Evolutionary couplings refer to pairs of amino acids that tend to mutate together over evolutionary time. This co-evolution suggests that the amino acids are functionally or structurally related, often because they form part of the same structural motif or are involved in the same biochemical interaction. For example, if two residues are frequently observed to mutate in tandem across different species, it is likely that they are spatially close in the 3D structure and need to maintain their interaction to preserve the protein's stability or function.

\textbf{How AlphaFold2 Learns from Evolutionary Couplings:}
\begin{itemize}
    \item AlphaFold2 extracts evolutionary couplings from the MSA by examining which pairs of residues are conserved or co-vary across multiple sequences.
    \item These couplings are used to predict the distances between pairs of amino acids, helping AlphaFold2 form accurate models of the protein's tertiary structure.
    \item By incorporating this evolutionary information, AlphaFold2 can better predict long-range interactions between residues, which are often difficult to infer from the primary sequence alone.
\end{itemize}

\textbf{Example: Evolutionary Couplings in MSA}

Consider the following MSA, where the highlighted residues tend to mutate together across different sequences:

\begin{center}
\begin{lstlisting}[style=cmd]
Sequence 1: ATGC**TA**CGAT
Sequence 2: ATG**C**TG**TA**GAT
Sequence 3: ATGC**C**T**--**GGAT
Sequence 4: ATG**CGT**ACGAT
\end{lstlisting}
\end{center}

The conserved regions, shown in bold, indicate evolutionary couplings between the highlighted residues. These residues may be spatially close in the final 3D structure, and AlphaFold2 uses this information to predict their proximity.

\subsubsection{Incorporating MSA Data into AlphaFold2's Prediction Pipeline}

In AlphaFold2, MSAs are incorporated into the \textbf{Evoformer module} \cite{hu2022exploring}, a specialized deep learning model that processes both sequence information and pairwise relationships between residues. The Evoformer is designed to extract meaningful patterns from the MSA data, which it then uses to build accurate representations of the protein's structure.

\textbf{Steps in Incorporating MSA Data:}
\begin{enumerate}
    \item \textbf{Input Sequences}: AlphaFold2 first collects a set of homologous sequences from databases such as UniRef, using tools like JackHMMER to identify sequences that are evolutionarily related to the target protein.
    \item \textbf{MSA Creation}: These sequences are aligned to form an MSA. The resulting MSA captures the evolutionary relationships between the residues in the target protein and those in related sequences.
    \item \textbf{Feature Extraction}: The Evoformer module processes the MSA, extracting features that describe the evolutionary couplings between residues. These features provide critical information about which residues are likely to be close in 3D space.
    \item \textbf{Structure Prediction}: The features derived from the MSA are used to predict inter-residue distances and build the 3D model of the protein. The model combines this evolutionary information with other structural constraints to generate an accurate prediction.
\end{enumerate}

\subsection{The Importance of MSAs in Improving Prediction Accuracy}

MSAs are essential for improving the accuracy of AlphaFold2's predictions because they provide rich evolutionary information that can help infer structural relationships. Proteins are highly conserved across different organisms, and the conservation patterns revealed by MSAs allow AlphaFold2 to make more informed predictions about how a protein will fold.

\subsubsection{Handling Long-Range Interactions}

One of the main challenges in protein structure prediction is accurately predicting long-range interactions-those between amino acids that are far apart in the sequence but close in the final 3D structure. MSAs help address this challenge by revealing which residues tend to co-evolve, providing clues about these long-range contacts. By using MSAs, AlphaFold2 can predict long-range interactions more accurately than models that rely solely on the primary sequence.

\textbf{Example: Long-Range Interactions from Evolutionary Couplings}

In the MSA example below, the residues at positions 2 and 12 are highly conserved across multiple sequences, indicating a potential long-range interaction:

\begin{center}
\begin{lstlisting}[style=cmd]
Sequence 1: A---G----R---
Sequence 2: A---G----R---
Sequence 3: A---G----R---
Sequence 4: A---G----R---
\end{lstlisting}
\end{center}

Even though these residues are far apart in the sequence, their evolutionary conservation suggests that they may be close in the folded structure.

\subsubsection{Improving Predictions for Difficult Cases}

AlphaFold2 is particularly effective for proteins that have few or no homologous structures available in databases. In such cases, traditional homology modeling methods would struggle to make accurate predictions. However, by leveraging the power of MSAs, AlphaFold2 can still infer useful information from evolutionary data, even when no closely related structures are available. This makes AlphaFold2 more versatile and effective than traditional methods.

\subsection{Limitations of MSAs in AlphaFold2}

While MSAs are extremely powerful, they do have limitations. For example, if a target protein has few homologous sequences available, the MSA may not provide enough evolutionary information to accurately predict the structure \cite{meng2023improved}. Additionally, MSAs may struggle with intrinsically disordered proteins, which do not have well-defined structures and may not show clear evolutionary conservation.

\subsubsection{Low MSA Depth}

The effectiveness of an MSA depends on the number and diversity of sequences included in the alignment, often referred to as \textbf{MSA depth}. A shallow MSA, with only a few homologous sequences, may not capture enough evolutionary information to accurately predict the structure. Conversely, a deep MSA, with many diverse sequences, provides much more information about evolutionary couplings and is more likely to lead to accurate predictions.

\textbf{Example: Shallow vs. Deep MSA}

\begin{itemize}
    \item \textbf{Shallow MSA:} 
    \begin{lstlisting}[style=cmd]
    Sequence 1: ATGC--TACGAT
    Sequence 2: ATGCTGTA-GAT
    \end{lstlisting}
    With only two sequences, this MSA provides limited information about evolutionary relationships.
    
    \item \textbf{Deep MSA:} 
    \begin{lstlisting}[style=cmd]
    Sequence 1: ATGC--TACGAT
    Sequence 2: ATGCTGTA-GAT
    Sequence 3: ATGCCT--GGAT
    Sequence 4: ATGCGT-ACGAT
    Sequence 5: ATGCTGTCAGAT
    \end{lstlisting}
    A deeper MSA with more sequences provides a richer source of evolutionary information, improving the accuracy of the prediction.
\end{itemize}

\subsection{The Future of MSA-Free Predictions}

While MSAs are currently central to AlphaFold2's accuracy, there is ongoing research into \textbf{MSA-free prediction models} \cite{fang2022helixfold}, which aim to predict protein structures without relying on evolutionary data. These models would be particularly useful for predicting the structures of novel proteins that do not have homologous sequences in existing databases.

AlphaFold2 has already made strides toward reducing its reliance on MSAs, especially for more difficult cases where few homologous sequences are available. Future versions of AlphaFold and other deep learning models may further reduce the need for MSAs by using more advanced neural networks that can directly predict protein structures from the primary sequence alone.

\subsection{Conclusion}

The role of Multiple Sequence Alignment (MSA) in AlphaFold2 is fundamental to its ability to accurately predict protein structures. MSAs provide valuable evolutionary information that helps AlphaFold2 infer long-range interactions and structural constraints. By analyzing patterns of sequence conservation, AlphaFold2 can make informed predictions about how a protein folds, even in cases where little experimental data is available. While MSAs have limitations, especially for proteins with few homologous sequences, they remain a powerful tool for improving the accuracy of protein structure prediction.

As the field of protein design continues to evolve, we may see further developments in MSA-free models, but for now, MSAs remain an essential component of AlphaFold2's success in revolutionizing protein structure prediction.

\section{Network Architecture and Attention Mechanisms}

The \textbf{network architecture} of AlphaFold2, along with its innovative use of \textbf{attention mechanisms}, is central to its success in protein structure prediction. AlphaFold2 leverages a deep learning architecture that is specifically designed to capture the complex relationships between amino acids in a protein sequence, using attention mechanisms to identify critical interactions between residues that ultimately guide the protein's folding. This section will walk through the architecture of AlphaFold2, the role of attention mechanisms, and how they work together to provide accurate predictions of protein structures. We will explain these concepts in a clear and step-by-step manner, making them accessible to beginners.

\subsection{Overview of AlphaFold2's Network Architecture}

At its core, AlphaFold2's architecture is a combination of neural networks that work together to predict the three-dimensional structure of proteins. Unlike traditional methods that rely on physical simulations or fixed templates, AlphaFold2 uses a data-driven approach where the model learns from vast datasets of known protein sequences and structures. This enables it to predict the folding patterns of proteins with remarkable accuracy, even for proteins that lack known homologs in structural databases.

The architecture of AlphaFold2 is built around two main components:

\begin{itemize}
    \item \textbf{Evoformer module}: This module processes the multiple sequence alignments (MSAs) and pairwise relationships between residues, capturing the evolutionary information and interaction patterns needed for structure prediction.
    \item \textbf{Structure module}: The structure module uses the output of the Evoformer to predict the 3D coordinates of the protein atoms. It employs a combination of geometric reasoning and iterative refinement to ensure that the predicted structure is physically realistic.
\end{itemize}

These two components work together to generate accurate predictions, but the key to AlphaFold2's success lies in its use of \textbf{attention mechanisms}, which allow the model to focus on critical interactions between residues during the folding process.

\subsection{Evoformer: Capturing Evolutionary and Structural Relationships}

The \textbf{Evoformer module} \cite{hu2022exploring} is responsible for processing the MSA and pairwise relationships between residues. It is designed to extract features that describe the evolutionary and structural relationships between amino acids in the protein sequence. These features include information about which residues are conserved across species, which are likely to interact in the folded structure, and which are involved in important structural motifs.

\subsubsection{Attention Mechanisms in the Evoformer}

One of the most innovative aspects of the Evoformer module is its use of \textbf{attention mechanisms}. These mechanisms allow the model to selectively focus on the parts of the sequence and MSA that are most relevant for predicting the protein's structure. In AlphaFold2, attention mechanisms are used to identify important long-range interactions between residues that may be far apart in the sequence but close in the final folded structure.

\textbf{How Attention Works in the Evoformer:}

\begin{itemize}
    \item \textbf{Self-Attention}: Self-attention is used to focus on relationships within the sequence itself. The model learns which residues in the sequence are most important for folding by computing attention weights that highlight these relationships. For example, self-attention might identify that residues near the start of the sequence need to form interactions with residues near the end, despite their distance in the linear sequence.
    \item \textbf{Pairwise Attention}: Pairwise attention mechanisms are used to analyze the relationships between pairs of residues. These mechanisms focus on the interactions between specific pairs of amino acids, helping the model predict which residues are likely to form contacts in the 3D structure.
\end{itemize}

\textbf{Example: Self-Attention in Protein Sequences}

In the following simplified diagram, self-attention allows the model to focus on key interactions between residues in a protein sequence, even if they are far apart:

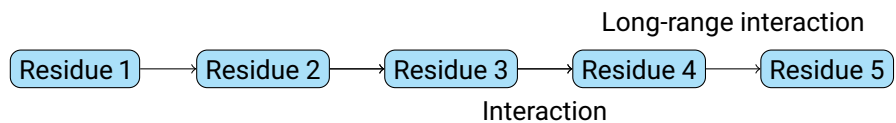
\begin{center}
\begin{tikzpicture}[node distance=2.5cm]
    \node (seq1) [rectangle,fill=cyan!30, rounded corners, draw] {Residue 1};
    \node (seq2) [rectangle,fill=cyan!30, rounded corners,draw, right of=seq1] {Residue 2};
    \node (seq3) [rectangle, fill=cyan!30,rounded corners, draw, right of=seq2] {Residue 3};
    \node (seq4) [rectangle, fill=cyan!30,rounded corners,draw, right of=seq3] {Residue 4};
    \node (seq5) [rectangle, fill=cyan!30,rounded corners, draw, right of=seq4] {Residue 5};
    
    \draw[->] (seq1) -- (seq2);
    \draw[->] (seq2) -- (seq3);
    \draw[->] (seq3) -- (seq4);
    \draw[->] (seq4) -- (seq5) node[midway, above, yshift=0.3cm] {Long-range interaction};
    
    \draw[->] (seq2) -- (seq3);
    \draw[->] (seq3) -- (seq4) node[midway, below, yshift=-0.3cm] {Interaction};
    
\end{tikzpicture}
\end{center}

This diagram shows how attention mechanisms can identify important interactions that are not immediately apparent from the linear sequence.

\subsubsection{MSA Attention}

In addition to self-attention and pairwise attention, the Evoformer also employs \textbf{MSA attention}, which allows the model to process the multiple sequence alignments. The goal of MSA attention is to capture evolutionary relationships by focusing on conserved residues and identifying co-evolved pairs of amino acids. These co-evolved pairs are often spatially close in the final 3D structure, and the MSA attention mechanism helps the model learn these relationships.

\textbf{Example: MSA Attention}

Consider the following MSA where the attention mechanism highlights important evolutionary couplings:

\begin{center}
\begin{lstlisting}[style=cmd]
Sequence 1: ATGC--TACGAT
Sequence 2: ATGCTGTA-GAT
Sequence 3: ATGCCT--GGAT
Sequence 4: ATGCGT-ACGAT
\end{lstlisting}
\end{center}

In this example, MSA attention helps the model focus on conserved regions (e.g., "ATGC" and "GAT"), which are likely to be structurally important.

\subsection{Structure Module: Predicting 3D Coordinates}

Once the Evoformer module has processed the evolutionary and structural information, the \textbf{Structure module} takes over to predict the 3D coordinates of the protein atoms. This module uses the information from the Evoformer to construct the final structure, employing a combination of geometric reasoning and iterative refinement to ensure that the predicted structure is accurate.

\subsubsection{Attention Mechanisms in the Structure Module}

The structure module also uses attention mechanisms to refine its predictions. In this case, attention is used to focus on the spatial relationships between residues, ensuring that the predicted 3D structure is physically plausible. The model predicts the distances and angles between residues, iteratively adjusting these predictions until the final structure is achieved.

\textbf{Example: Refining 3D Structure with Attention}

The structure module might initially predict that two residues are too far apart, leading to a physically unrealistic structure. Attention mechanisms help the model refine this prediction by focusing on the spatial relationships between these residues and adjusting the distance between them.

\subsubsection{Iterative Refinement}

AlphaFold2 uses a process called \textbf{recycling}, where the predicted structure is fed back into the model for further refinement. The structure module continuously updates its predictions, adjusting the 3D coordinates based on the attention mechanisms and geometric constraints. This iterative process ensures that the final structure is as accurate as possible.

\textbf{Benefits of Iterative Refinement:}
\begin{itemize}
    \item \textbf{Error Correction}: The recycling process allows the model to correct errors in the initial predictions, leading to a more accurate final structure.
    \item \textbf{Consistency}: Iterative refinement ensures that the predicted structure is consistent with known physical and chemical properties of proteins, such as correct bond angles and avoidance of steric clashes.
\end{itemize}

\subsection{Attention Mechanisms: A Key Innovation}

The use of attention mechanisms in AlphaFold2 represents a key innovation that sets it apart from previous protein structure prediction methods. Attention mechanisms allow the model to selectively focus on important interactions, both within the sequence and between pairs of residues, enabling AlphaFold2 to capture long-range interactions that are critical for accurate structure prediction.

\subsubsection{Global vs. Local Attention}

In AlphaFold2, attention mechanisms operate at both a \textbf{global} and \textbf{local} level:

\begin{itemize}
    \item \textbf{Global attention}: This focuses on identifying long-range interactions between residues that are far apart in the sequence but close in the 3D structure. These interactions are crucial for predicting the overall fold of the protein.
    \item \textbf{Local attention}: Local attention focuses on short-range interactions, such as those between adjacent residues in the sequence. These interactions help the model predict local secondary structures like $\alpha$-helices and $\beta$-sheets.
\end{itemize}

By combining both global and local attention, AlphaFold2 is able to predict the entire structure of the protein, from local motifs to the overall fold.

\subsubsection{Benefits of Attention Mechanisms}

\begin{itemize}
    \item \textbf{Long-Range Interactions}: Attention mechanisms help the model capture long-range interactions, which are often difficult to infer from the sequence alone. This improves the accuracy of the predicted structure.
    \item \textbf{Flexibility}: Attention mechanisms are highly flexible, allowing AlphaFold2 to adapt to a wide range of protein sequences and structures. This flexibility is one of the reasons AlphaFold2 can predict the structure of proteins with diverse folding patterns.
    \item \textbf{Scalability}: The attention mechanism scales well to large proteins with many residues, making it possible to predict the structure of complex proteins with thousands of amino acids.
\end{itemize}

\subsection{Conclusion}

The network architecture and attention mechanisms of AlphaFold2 are central to its ability to predict protein structures with such high accuracy. By employing attention mechanisms at multiple stages of the prediction pipeline, AlphaFold2 is able to capture both local and long-range interactions, making it one of the most effective tools for protein structure prediction. The Evoformer module processes evolutionary relationships, while the structure module refines the 3D coordinates using attention-based mechanisms and iterative refinement. Together, these components allow AlphaFold2 to generate accurate, physically realistic models of protein structures, revolutionizing the field of structural biology.

\section{Limitations and Future Prospects of AlphaFold2}

While \textbf{AlphaFold2} has revolutionized the field of protein structure prediction by achieving unprecedented accuracy, it is important for beginners to understand its limitations and the areas where future improvements are possible. This section will delve into the current limitations of AlphaFold2, providing detailed explanations and examples, and discuss the prospects for future developments in protein structure prediction and design.

\subsection{Limitations of AlphaFold2}

Despite its groundbreaking success, AlphaFold2 has certain limitations that are important to recognize:

\subsubsection{Static Structure Predictions}

AlphaFold2 predicts a \textbf{single, static structure} for each protein sequence. Proteins are dynamic molecules that can adopt multiple conformations under different physiological conditions. This dynamic behavior is crucial for many proteins' functions, such as enzymes that change shape during catalysis or channels that open and close in response to signals.

\begin{itemize}
    \item \textbf{Example: Conformational Flexibility in Enzymes}
    
    Consider the enzyme hexokinase, which undergoes a significant conformational change upon binding to glucose. AlphaFold2 predicts only one static structure and may not capture both the open (unbound) and closed (substrate-bound) states.
\end{itemize}

\subsubsection{Limited Prediction of Protein Complexes}

AlphaFold2 primarily focuses on predicting the structures of individual proteins. While it has some capability to predict \textbf{homomeric complexes} \cite{pereira2007evolution}(assemblies of identical protein units), its ability to accurately predict \textbf{heteromeric protein-protein interactions} \cite{hou2017seeing} (different proteins interacting together) is limited.

\begin{itemize}
    \item \textbf{Example: Protein-Protein Interactions}
    
    The interaction between antibodies and antigens is critical for immune response. AlphaFold2 may not reliably predict the complex structure formed by an antibody and its antigen, which is essential for understanding immune recognition.
\end{itemize}

\subsubsection{Challenges with Disordered Regions}

Proteins often contain \textbf{intrinsically disordered regions} (IDRs) \cite{oldfield2014intrinsically} that do not adopt a fixed 3D structure. AlphaFold2 tends to predict these regions as structured, which may not reflect their true, flexible nature.

\begin{itemize}
    \item \textbf{Example: Intrinsically Disordered Proteins}
    
    The tumor suppressor protein p53 has disordered regions that are crucial for its function. AlphaFold2 might incorrectly model these regions as structured, potentially misleading interpretations of the protein's behavior.
\end{itemize}

\subsubsection{Post-Translational Modifications (PTMs)}

AlphaFold2 does not account for \textbf{post-translational modifications} \cite{ramazi2021post}, such as phosphorylation, glycosylation, or methylation, which can significantly alter a protein's structure and function.

\begin{itemize}
    \item \textbf{Example: Impact of Phosphorylation}
    
    Phosphorylation of certain residues can induce conformational changes that activate or deactivate a protein's function. Without considering PTMs, AlphaFold2's predictions may not represent the biologically active form of the protein.
\end{itemize}

\subsubsection{Membrane Proteins}

Predicting the structures of \textbf{membrane proteins} remains challenging for AlphaFold2 due to the complexity of the membrane environment and limited availability of high-resolution membrane protein structures in databases.

\begin{itemize}
    \item \textbf{Example: G-Protein Coupled Receptors (GPCRs) \cite{kroeze2003g}}
    
    GPCRs are a large class of membrane proteins involved in signal transduction. Their structures are difficult to predict accurately because of their complex topology and interactions with the lipid bilayer.
\end{itemize}

\subsubsection{Dependency on Multiple Sequence Alignments (MSAs)}

AlphaFold2 relies heavily on \textbf{deep MSAs} to extract evolutionary information. For proteins with few homologous sequences, the quality of the MSA is poor, which can reduce prediction accuracy.

\begin{itemize}
    \item \textbf{Example: Orphan Proteins}
    
    Proteins unique to a particular organism or with no known homologs (orphan proteins) may have shallow MSAs, leading to less accurate structure predictions by AlphaFold2.
\end{itemize}

\subsubsection{Computational Resources and Accessibility}

Running AlphaFold2 requires significant computational resources, including powerful GPUs and substantial memory. This can be a barrier for researchers with limited access to high-performance computing infrastructure.

\begin{itemize}
    \item \textbf{Example: Resource Requirements}
    
    Predicting a large protein or complex can take hours to days on a high-end GPU, making it impractical for some users or large-scale studies.
\end{itemize}

\subsubsection{Lack of Functional Insights}

While AlphaFold2 excels at predicting structures, it does not directly provide information about \textbf{protein function}, dynamics, or interactions with small molecules, nucleic acids, or other ligands.

\begin{itemize}
    \item \textbf{Example: Enzyme Active Sites \cite{adam2004mapping}}
    
    AlphaFold2 may predict the overall fold of an enzyme accurately but does not identify the catalytic residues or provide insights into the mechanism of action.
\end{itemize}

\subsection{Future Prospects of AlphaFold2 and Protein Structure Prediction}

The limitations of AlphaFold2 highlight areas where future research and development can focus to enhance protein structure prediction and its applications.

\subsubsection{Incorporating Protein Dynamics}

Future models may integrate \textbf{molecular dynamics simulations} or other computational methods to predict multiple conformations of proteins, capturing their dynamic behavior.

\begin{itemize}
    \item \textbf{Potential Developments}
    
    \begin{itemize}
        \item Predicting \textbf{conformational ensembles} rather than single structures.
        \item Modeling \textbf{allosteric changes} and \textbf{conformational transitions}.
    \end{itemize}
\end{itemize}

\subsubsection{Improved Prediction of Protein Complexes}

Enhancing the ability to predict \textbf{protein-protein interactions} and complexes is a key area for future improvement.

\begin{itemize}
    \item \textbf{Potential Developments}
    
    \begin{itemize}
        \item Developing specialized algorithms for \textbf{docking} proteins together.
        \item Incorporating \textbf{co-evolutionary data} from interacting proteins.
    \end{itemize}
\end{itemize}

\subsubsection{Accounting for Post-Translational Modifications}

Future models might include the capability to predict structures with \textbf{post-translational modifications}.

\begin{itemize}
    \item \textbf{Potential Developments}
    
    \begin{itemize}
        \item Integrating data on common PTMs and their structural impact.
        \item Allowing users to specify modifications as input parameters.
    \end{itemize}
\end{itemize}

\subsubsection{Improved Handling of Membrane Proteins}

Advancements in modeling the membrane environment could enhance predictions for \textbf{membrane proteins}.

\begin{itemize}
    \item \textbf{Potential Developments}
    
    \begin{itemize}
        \item Incorporating \textbf{membrane mimetic models} into the prediction pipeline.
        \item Utilizing \textbf{lipid-specific interactions} and constraints.
    \end{itemize}
\end{itemize}

\subsubsection{Reducing Dependency on MSAs}

Developing methods that reduce reliance on MSAs could improve predictions for proteins with few homologs.

\begin{itemize}
    \item \textbf{Potential Developments}
    
    \begin{itemize}
        \item Utilizing \textbf{protein language models} trained on large sequence databases to capture evolutionary information implicitly.
        \item Implementing \textbf{one-shot} or \textbf{few-shot learning} approaches.
    \end{itemize}
\end{itemize}

\subsubsection{Integration with Functional Annotation Tools}

Combining structure prediction with functional annotation could provide more comprehensive insights.

\begin{itemize}
    \item \textbf{Potential Developments}
    
    \begin{itemize}
        \item Predicting \textbf{active sites}, \textbf{binding pockets}, and \textbf{interaction interfaces}.
        \item Linking structural predictions with \textbf{functional databases} and \textbf{pathway analysis}.
    \end{itemize}
\end{itemize}

\subsubsection{Enhancing Computational Efficiency and Accessibility}

Improving the computational efficiency of structure prediction algorithms will make them more accessible.

\begin{itemize}
    \item \textbf{Potential Developments}
    
    \begin{itemize}
        \item Optimizing algorithms for \textbf{parallel processing} and \textbf{distributed computing}.
        \item Developing \textbf{cloud-based platforms} to make tools available without the need for local high-performance computing resources.
    \end{itemize}
\end{itemize}

\subsubsection{Predicting Protein-Ligand Interactions}

Extending predictions to include \textbf{small molecule binding} could have significant implications for drug discovery.

\begin{itemize}
    \item \textbf{Potential Developments}
    
    \begin{itemize}
        \item Integrating \textbf{ligand docking} algorithms with structure prediction.
        \item Predicting \textbf{binding affinities} and \textbf{specificity}.
    \end{itemize}
\end{itemize}

\subsection{Examples of Future Applications}

\subsubsection{Drug Discovery and Design}

Improved protein structure prediction can accelerate \textbf{drug discovery} by identifying novel targets and optimizing lead compounds.

\begin{itemize}
    \item \textbf{Example: Targeting Undruggable Proteins \cite{zhang2022strategies}}
    
    Some proteins are considered "undruggable" due to a lack of structural information. Enhanced prediction models could reveal binding pockets and enable the development of inhibitors.
\end{itemize}

\subsubsection{Understanding Disease Mechanisms}

Structural insights can elucidate how \textbf{mutations} affect protein function, aiding in understanding genetic diseases.

\begin{itemize}
    \item \textbf{Example: Cancer-associated Mutations \cite{risques2018aging}}
    
    Predicting the structural impact of mutations in oncogenes or tumor suppressors can reveal mechanisms of oncogenesis and potential therapeutic interventions.
\end{itemize}

\subsubsection{Synthetic Biology and Protein Engineering}

Accurate structure prediction facilitates the design of \textbf{novel proteins} with desired functions.

\begin{itemize}
    \item \textbf{Example: Enzyme Engineering \cite{sharma2021enzyme}}
    
    Designing enzymes with enhanced catalytic activity or altered substrate specificity can be guided by accurate structural models.
\end{itemize}

\subsection{Conclusion}

AlphaFold2 has significantly advanced the field of protein structure prediction, providing accurate models that were previously unattainable computationally. However, recognizing its limitations is crucial for appropriate application and for guiding future research. By addressing these limitations, future developments can expand the capabilities of computational protein modeling, opening new horizons in biology, medicine, and biotechnology.

\subsection{Summary}

\begin{itemize}
    \item AlphaFold2 predicts static structures and may not capture protein dynamics.
    \item It has limited ability to predict protein complexes and interactions.
    \item Disordered regions and post-translational modifications present challenges.
    \item Dependency on MSAs can limit accuracy for proteins with few homologs.
    \item Future prospects include incorporating dynamics, improving complex predictions, and integrating functional insights.
    \item Advancements can impact drug discovery, disease understanding, and synthetic biology.
\end{itemize}
\chapter{Practical Applications in Protein Structure Prediction and Design}

\section{How to Use AlphaFold2 for Prediction}

In this section, we will explore the practical steps involved in using \textbf{AlphaFold2} to predict protein structures. This guide is designed for beginners and those new to computational biology, walking through the process in a clear and detailed manner. AlphaFold2 has become a widely used tool due to its ability to predict protein structures with high accuracy, and understanding how to run the model will enable you to harness its power for a wide range of applications in research, drug discovery, and protein design.

\subsection{Overview of the Prediction Process}

Before diving into the detailed steps, it's important to understand the overall workflow for using AlphaFold2. The process involves preparing input data, running the prediction, and analyzing the output structures. Below is a high-level overview of the workflow:

\begin{enumerate}
    \item \textbf{Input Preparation:} Prepare the amino acid sequence in FASTA format and gather additional required data (e.g., homologous sequences for Multiple Sequence Alignment).
    \item \textbf{Running AlphaFold2:} Set up the AlphaFold2 environment, install dependencies, and execute the prediction with the appropriate command.
    \item \textbf{Analyzing the Output:} Analyze the predicted structure using visualization tools and assess the confidence scores provided by AlphaFold2.
\end{enumerate}

Now, let's walk through each of these steps in more detail.

\subsection{Step 1: Preparing the Input Data}

To run AlphaFold2, you need to start by preparing the \textbf{amino acid sequence} \cite{guo2022alphafold2} of the protein you want to predict. This sequence must be saved in a text file using the FASTA format, which is commonly used to represent nucleotide or protein sequences. The FASTA format consists of a header line that starts with a ">" symbol followed by the sequence on subsequent lines.

\textbf{Example of a Protein Sequence in FASTA Format:}

\begin{lstlisting}[style=cmd]
>Example_Protein
MVLSPADKTNVKAAWGVLGAFSDGLAHLDNLKGTFATLSELHCDKLHVDPENFRLLGNVLVCVLAHHFGKEFTPPVQAAYQKVVAGVANALAHKYH
\end{lstlisting}

In this example, the first line is the header, which contains the name of the protein ("Example\_Protein"). The second line contains the amino acid sequence itself. This sequence is what AlphaFold2 will use to predict the structure.

\subsubsection{Importance of Sequence Quality}

The accuracy of AlphaFold2's predictions is highly dependent on the quality of the input sequence. Ensure that your sequence is complete and accurate, as any missing or incorrectly entered residues can lead to inaccurate structural predictions.

\subsubsection{Additional Data: Multiple Sequence Alignment (MSA)}

AlphaFold2 requires an MSA as part of its prediction process. MSAs are used to gather evolutionary information by comparing the target protein sequence to sequences from related proteins across different species. To generate the MSA, AlphaFold2 uses tools such as JackHMMER, which is included in the AlphaFold2 pipeline.

While AlphaFold2 automatically generates the MSA when you run the prediction, it's helpful to understand that this step is crucial for extracting information about how certain residues have evolved together, which informs the prediction model about which regions are likely to be structurally conserved.

\subsection{Step 2: Setting Up AlphaFold2 and Running the Prediction}

Once the input data is prepared, the next step is to run AlphaFold2. This involves setting up the environment, installing the necessary dependencies, and executing the prediction.

\subsubsection{System Requirements}

AlphaFold2 is computationally intensive and requires a system with a powerful \textbf{GPU} (Graphics Processing Unit). The following are the typical hardware requirements for running AlphaFold2:

\begin{itemize}
    \item \textbf{GPU:} A modern NVIDIA GPU with at least 16GB of memory (e.g., NVIDIA Tesla V100, RTX 2080, or newer).
    \item \textbf{CPU:} A multi-core CPU for running the MSA generation and other preprocessing steps.
    \item \textbf{Memory:} At least 64GB of RAM, especially for predicting large proteins or protein complexes.
    \item \textbf{Storage:} Several terabytes of disk space for storing databases and temporary files.
\end{itemize}

\subsubsection{Software Installation}

Before running AlphaFold2, you need to install the software and set up the required dependencies. Follow these steps to install AlphaFold2 on a Linux system:

\begin{lstlisting}[style=cmd]
# Step 1: Clone the AlphaFold2 repository from GitHub
git clone https://github.com/deepmind/alphafold.git
cd alphafold

# Step 2: Set up the Python environment
# Install Docker if necessary (optional but recommended)
sudo apt-get install docker-ce docker-ce-cli containerd.io

# Step 3: Download the AlphaFold2 databases (this step requires significant storage space)
bash scripts/download_all_data.sh /path/to/download/directory

# Step 4: Install Python dependencies (if using Python directly)
pip install -r requirements.txt
\end{lstlisting}

\textbf{Note:} You can also run AlphaFold2 using Docker, which simplifies the setup process by providing a pre-configured environment. Docker is especially useful if you want to avoid manually managing dependencies.

\subsubsection{Running the AlphaFold2 Prediction}

Once the environment is set up, you are ready to run AlphaFold2. The command below shows how to run AlphaFold2 on a protein sequence, specifying the path to the input sequence and the output directory.

\begin{lstlisting}[style=cmd]
python run_alphafold.py \
  --fasta_paths=./input/Example_Protein.fasta \
  --output_dir=./output/Example_Protein/ \
  --max_template_date=2022-12-31 \
  --model_preset=monomer
\end{lstlisting}

\textbf{Explanation of the Parameters:}
\begin{itemize}
    \item \textbf{--fasta\_paths}: Specifies the path to the input FASTA file containing the protein sequence.
    \item \textbf{--output\_dir}: Specifies the directory where the predicted structure and related files will be saved.
    \item \textbf{--max\_template\_date}: Specifies the latest date for using template structures from the Protein Data Bank (PDB). This parameter ensures that no future data is used during prediction.
    \item \textbf{--model\_preset}: Determines the type of model to run (e.g., "monomer" for single proteins, "multimer" for protein complexes).
\end{itemize}

Once you run the command, AlphaFold2 will begin processing the input sequence. This involves several stages, including MSA generation, feature extraction, and structure prediction. Depending on the size of the protein and the hardware available, this process can take anywhere from a few minutes to several hours.

\subsection{Step 3: Analyzing the Output}

After AlphaFold2 has completed the prediction, the results will be saved in the specified output directory. The primary output is a \textbf{PDB file} \cite{berman2002protein} (Protein Data Bank format) that contains the predicted 3D structure of the protein. Additional outputs include:

\begin{itemize}
    \item \textbf{pLDDT Scores}: AlphaFold2 generates per-residue confidence scores known as \textbf{pLDDT} \cite{wilson2022alphafold2} (predicted Local Distance Difference Test) scores, which indicate the model's confidence in the prediction for each residue.
    \item \textbf{Predicted Aligned Error (PAE) Plot \cite{schwartz2000pipmaker}}: This plot provides information about the predicted alignment error between different regions of the protein. It is useful for understanding the relative accuracy of different parts of the predicted structure.
\end{itemize}

\subsubsection{Visualizing the Predicted Structure}

To visualize the predicted structure, you can use software tools such as \textbf{PyMOL} \cite{delano2002pymol}, \textbf{ChimeraX} \cite{pettersen2021ucsf}, or \textbf{UCSF Chimera} \cite{pettersen2004ucsf}. These tools allow you to load the PDB file and explore the 3D structure of the protein.

\textbf{Example of Visualizing a PDB File in PyMOL:}

\begin{lstlisting}[style=cmd]
# Open PyMOL and load the predicted structure
pymol Example_Protein.pdb

# Display the structure as a cartoon representation
show cartoon
color blue, Example_Protein

# Highlight specific regions of the protein (e.g., alpha helices)
select alpha_helices, ss h
color red, alpha_helices
\end{lstlisting}

In this example, PyMOL is used to load the PDB file and visualize the predicted structure. You can apply different visual representations, such as cartoons, surface views, or ribbon diagrams, to highlight various aspects of the protein's structure.

\subsubsection{Interpreting the Confidence Scores (pLDDT)}

AlphaFold2 provides confidence scores for each residue, represented as pLDDT values, which range from 0 to 100. These scores help you assess which parts of the structure are predicted with high confidence and which regions are less certain.

\begin{itemize}
    \item \textbf{High pLDDT (90–100)}: Indicates high confidence in the prediction. These regions are likely to be structurally accurate.
    \item \textbf{Moderate pLDDT (70–90)}: Indicates reasonable confidence, but further validation may be required for experimental studies.
    \item \textbf{Low pLDDT ($<$70)}: Indicates low confidence. These regions might be disordered or inaccurately predicted, requiring caution in interpretation.
\end{itemize}

\subsubsection{Using the Predicted Aligned Error (PAE) Plot}

The PAE plot provides insights into the relative positioning of different regions within the protein. This plot can be particularly useful when interpreting large or multi-domain proteins. For example, a high PAE score between two regions indicates uncertainty in their relative positioning, even if each individual region is predicted with high confidence.

\subsection{Practical Tips for Using AlphaFold2}

Here are some practical tips to keep in mind when using AlphaFold2 for protein structure prediction:

\begin{itemize}
    \item \textbf{Run in Multimer Mode for Complexes}: If you are predicting the structure of a protein complex, use the \texttt{multimer} preset, which is optimized for multi-chain predictions.
    \item \textbf{Check for Template Structures}: AlphaFold2 can use structural templates from the PDB if available. These templates can improve prediction accuracy for proteins with known homologs.
    \item \textbf{Use Cloud-Based Solutions}: If you do not have access to high-performance computing resources, consider using cloud-based platforms like Google Cloud or AWS to run AlphaFold2.
    \item \textbf{Validate Predictions Experimentally}: While AlphaFold2 provides highly accurate predictions, it is important to validate these structures through experimental techniques such as X-ray crystallography or cryo-EM when possible.
\end{itemize}

\subsection{Conclusion}

Using AlphaFold2 for protein structure prediction involves several steps, from preparing the input sequence to running the prediction and analyzing the output. By following the detailed instructions provided in this section, even beginners can successfully use AlphaFold2 to predict protein structures and leverage its powerful capabilities for research and design. As you become more familiar with the tool, you can apply it to increasingly complex projects, from studying protein interactions to designing novel proteins for biotechnology applications.

\section{Other Open-Source Tools (e.g., RoseTTAFold, OmegaFold)}

While AlphaFold2 has revolutionized protein structure prediction, it is not the only tool available to researchers. Several other open-source tools, such as \textbf{RoseTTAFold} \cite{lee2022comparative} and \textbf{OmegaFold} \cite{wu2022high}, also offer powerful capabilities for predicting protein structures, with some unique features and advantages. In this section, we will provide a detailed overview of these tools, explaining how they work, what makes them different, and how they can be used in protein structure prediction. This section is written for beginners, so we will go step by step, providing practical examples where appropriate.

\subsection{RoseTTAFold: A Hybrid Approach to Protein Structure Prediction}

\textbf{RoseTTAFold} is a deep learning-based protein structure prediction tool developed by the Baker lab at the University of Washington. It uses a \textbf{three-track architecture} that integrates sequence, distance, and structure information simultaneously, making it a hybrid method that leverages both deep learning and traditional structural biology principles. RoseTTAFold was inspired by the success of AlphaFold2, but it incorporates some unique features that make it an attractive alternative for certain applications.

\subsubsection{Key Features of RoseTTAFold}

\begin{itemize}
    \item \textbf{Three-Track Architecture}: Unlike AlphaFold2, which uses a two-track model to process sequence and structure information separately, RoseTTAFold uses a three-track system where sequence, pairwise distances between residues, and the 3D structure are processed in parallel. This allows the model to integrate these different data types in a more unified way.
    \item \textbf{Flexibility in Prediction}: RoseTTAFold is flexible enough to predict the structures of individual proteins as well as protein-protein complexes. It can also be used to model protein-ligand interactions, making it useful for a wide range of applications.
    \item \textbf{Open-Source and Extensible}: Like AlphaFold2, RoseTTAFold is fully open-source, and it is designed to be extensible. Researchers can modify the source code and integrate new features as needed.
\end{itemize}

\subsubsection{How RoseTTAFold Works}

RoseTTAFold operates through an iterative process in which the three tracks (sequence, distance, and structure) are refined together over multiple cycles. Each cycle updates the distance matrix and the 3D structure representation based on the sequence data, allowing the model to gradually converge on an accurate structure. This integration of different data types enables RoseTTAFold to predict complex protein folds, even in cases where sequence information alone may not be sufficient.

\textbf{Example Workflow for Using RoseTTAFold}

The basic steps for using RoseTTAFold are similar to those for AlphaFold2, but with some differences in terms of how the model handles input and output data:

\begin{itemize}
    \item \textbf{Input}: The input to RoseTTAFold is typically a protein sequence in FASTA format. It also benefits from using Multiple Sequence Alignments (MSAs) and template structures if available.
    \item \textbf{Processing}: RoseTTAFold runs iterative cycles where the sequence, distance predictions, and structure predictions are updated in parallel.
    \item \textbf{Output}: The output is a 3D structure of the protein in PDB format, along with confidence scores for each residue in the structure.
\end{itemize}

\begin{lstlisting}[style=cmd]
# Example command to run RoseTTAFold
python run_rosettafold.py --fasta=input.fasta --output_dir=output/
\end{lstlisting}

\textbf{Advantages of RoseTTAFold:}
\begin{itemize}
    \item More flexible with protein-protein complexes and protein-ligand interactions.
    \item Efficient with fewer computational resources compared to AlphaFold2.
    \item Integrated with the broader Rosetta suite for additional functionality in protein design and docking.
\end{itemize}

\subsubsection{When to Use RoseTTAFold}

RoseTTAFold is particularly useful when you are working on protein-protein interactions, protein-ligand docking, or other cases where AlphaFold2 may not perform as well. It is also a great option for researchers already familiar with the Rosetta software suite, as it integrates seamlessly with other Rosetta tools.

\subsection{OmegaFold: Structure Prediction Without MSAs}

\textbf{OmegaFold} is another open-source protein structure prediction tool, designed to work \textbf{without relying on Multiple Sequence Alignments (MSAs)}. This makes OmegaFold particularly valuable for predicting the structures of proteins that have few or no homologous sequences, where traditional MSA-based methods, like AlphaFold2, might struggle. OmegaFold is developed with the goal of expanding the range of proteins that can be accurately predicted, even when evolutionary data is scarce.

\subsubsection{Key Features of OmegaFold}

\begin{itemize}
    \item \textbf{MSA-Free Prediction}: OmegaFold does not require MSAs, making it ideal for orphan proteins or sequences with very few homologs. Instead of relying on evolutionary data, it uses advanced deep learning techniques to predict the structure directly from the primary sequence.
    \item \textbf{Fast Prediction Times}: Without the need to generate MSAs, OmegaFold can predict structures more quickly than AlphaFold2, making it a good option when time is a critical factor.
    \item \textbf{Open-Source and Lightweight}: OmegaFold is designed to be lightweight and easy to use, requiring less computational power than AlphaFold2, making it more accessible to researchers without access to high-end computational resources.
\end{itemize}

\subsubsection{How OmegaFold Works}

OmegaFold uses a deep learning model that is trained on a large dataset of protein sequences and their corresponding structures, but it does not rely on evolutionary relationships between proteins. Instead, the model learns directly from the sequence to predict distances between residues and ultimately builds the 3D structure based on these distance predictions.

\textbf{Example Workflow for Using OmegaFold}

The process for running OmegaFold is simpler than AlphaFold2 because it skips the MSA generation step. Here's an example of how to run OmegaFold:

\begin{lstlisting}[style=cmd]
# Example command to run OmegaFold
python run_omegafold.py --fasta=input.fasta --output_dir=output/
\end{lstlisting}

\textbf{Advantages of OmegaFold:}
\begin{itemize}
    \item Fast structure predictions without the need for MSAs.
    \item Ideal for proteins with few homologous sequences or unique evolutionary histories.
    \item Lower computational requirements, making it accessible for labs with limited resources.
\end{itemize}

\subsubsection{When to Use OmegaFold}

OmegaFold is particularly suited for situations where you are dealing with \textbf{orphan proteins} \cite{tautz2011evolutionary}, sequences that lack significant evolutionary data, or when you need fast predictions. It is also useful for smaller proteins or projects where computational resources are limited. While it may not achieve the same level of accuracy as AlphaFold2 in cases where deep MSAs are available, OmegaFold excels in scenarios where MSA generation is challenging or impossible.

\subsection{Comparison of AlphaFold2, RoseTTAFold, and OmegaFold}

\begin{itemize}
    \item Protein Structure Prediction Tools
    \begin{itemize}
        \item AlphaFold2
        \begin{itemize}
            \item High Accuracy for Individual Proteins
            \item Requires MSAs
        \end{itemize}
        \item RoseTTAFold
        \begin{itemize}
            \item Three-Track Model
            \item Supports Protein Complexes
        \end{itemize}
        \item OmegaFold
        \begin{itemize}
            \item MSA-Free Prediction
            \item Fast Predictions for Orphan Proteins
        \end{itemize}
    \end{itemize}
\end{itemize}

\subsubsection{AlphaFold2 vs. RoseTTAFold}

While both AlphaFold2 and RoseTTAFold offer high-accuracy predictions, AlphaFold2 is generally more accurate for single proteins where deep MSAs are available. RoseTTAFold, on the other hand, offers more flexibility for predicting \textbf{protein-protein complexes} \cite{jones1996principles} and is integrated with the Rosetta software suite, which makes it a powerful tool for protein design and protein-ligand interactions.

\subsubsection{AlphaFold2 vs. OmegaFold}

OmegaFold's key advantage is its ability to predict protein structures without relying on MSAs. This makes it especially useful for proteins that lack evolutionary homologs. However, for well-studied proteins with large MSAs, AlphaFold2 typically offers higher accuracy due to its use of evolutionary information.

\subsubsection{RoseTTAFold vs. OmegaFold}

RoseTTAFold and OmegaFold serve different purposes. RoseTTAFold is better suited for complex protein-protein interactions and cases where evolutionary data is available. OmegaFold excels when you need fast predictions without the need for MSAs, particularly for novel or less-studied proteins.

\subsection{Conclusion}

In the field of protein structure prediction, AlphaFold2, RoseTTAFold, and OmegaFold each offer unique strengths and can be applied to different research scenarios. AlphaFold2 remains the gold standard for individual protein predictions when MSAs are available, but RoseTTAFold's flexibility and OmegaFold's MSA-free approach make them valuable tools for specific cases, such as protein complexes or orphan proteins. By understanding the strengths and limitations of each tool, researchers can choose the best method for their specific applications, whether that is designing new proteins, studying complex interactions, or predicting the structure of less-characterized proteins.

\section{Case Studies and Interpretation of Results}

In this section, we will explore practical case studies that demonstrate the application of protein structure prediction tools, such as AlphaFold2, and provide a detailed guide on how to interpret the results. These case studies are designed for beginners and will walk through each step, providing real-world examples and clear explanations. By the end of this section, you will have a solid understanding of how to use predicted protein structures for practical applications in research and design.

\subsection{Case Study 1: Predicting the Structure of a Single Protein}

Let's begin with a straightforward case where AlphaFold2 is used to predict the structure of a single, well-studied protein. For this example, we will use the enzyme \textbf{lysozyme} \cite{lesnierowski2007lysozyme}, which is involved in breaking down bacterial cell walls and has been widely studied due to its antimicrobial properties.

\subsubsection{Step 1: Input Preparation}

The amino acid sequence of lysozyme is readily available in FASTA format. Here is an example of the sequence used for the prediction:

\begin{lstlisting}[style=cmd]
>Lysozyme
KVFERCELARTLKRLGMDGYRGISLANWMCLAKWESGYNTRATNYNAGDRSTDYGIFQINSRYWCNDG...
\end{lstlisting}

The sequence is saved as a FASTA file (e.g., \texttt{lysozyme.fasta}) and will be used as the input for AlphaFold2.

\subsubsection{Step 2: Running the Prediction}

After preparing the input, we run AlphaFold2 using the following command:

\begin{lstlisting}[style=cmd]
python run_alphafold.py \
  --fasta_paths=./input/lysozyme.fasta \
  --output_dir=./output/lysozyme/ \
  --model_preset=monomer
\end{lstlisting}

The \texttt{model\_preset=monomer} argument tells AlphaFold2 that we are predicting the structure of a single protein.

\subsubsection{Step 3: Output Files}

Once AlphaFold2 completes the prediction, it generates several output files:

\begin{itemize}
    \item \textbf{PDB file}: The main output is the predicted 3D structure of lysozyme in PDB format (\texttt{lysozyme.pdb}). This file contains the atomic coordinates for the protein's structure.
    \item \textbf{pLDDT Scores}: These scores are included to indicate the model's confidence in the predicted structure, with values ranging from 0 to 100.
    \item \textbf{Predicted Aligned Error (PAE)}: The PAE plot helps visualize the model's uncertainty in the relative positioning of different regions within the protein.
\end{itemize}

\subsubsection{Step 4: Visualizing and Interpreting the Results}

Using a molecular visualization tool like PyMOL or Chimera, you can open the \texttt{lysozyme.pdb} file to inspect the predicted structure. Below is a simple command to visualize the structure in PyMOL:

\begin{lstlisting}[style=cmd]
pymol lysozyme.pdb
show cartoon
color blue, all
\end{lstlisting}

\textbf{Interpreting pLDDT Scores:}

\begin{itemize}
    \item \textbf{High confidence (90–100)}: Regions with these scores are predicted with high confidence. For lysozyme, the active site and core structural motifs (like alpha helices and beta sheets) should have high pLDDT scores, indicating a reliable prediction.
    \item \textbf{Moderate confidence (70–90)}: Some loops and surface-exposed regions may have lower confidence. These regions are typically more flexible and less structured in real proteins.
    \item \textbf{Low confidence ($<$70)}: Any region with a low pLDDT score should be interpreted with caution, as the prediction may be inaccurate or correspond to a disordered region.
\end{itemize}

\textbf{Example of Result Interpretation:}

For lysozyme, the active site is predicted with high confidence, which is expected given its conserved catalytic role. However, some loop regions may have lower confidence due to their inherent flexibility. These results align with experimental observations, as lysozyme's structure has been extensively studied, and flexible regions often exhibit less structural stability.

\subsection{Case Study 2: Predicting a Protein-Protein Complex}

In this case study, we will use RoseTTAFold to predict the structure of a protein-protein complex. Specifically, we will look at the interaction between an antibody and an antigen. Understanding how proteins interact is crucial for applications such as drug development and immunotherapy.

\subsubsection{Step 1: Input Preparation}

To predict a protein-protein complex, we need the sequences of both the antibody and antigen in FASTA format. Below is an example of two sequences saved in separate FASTA files.

\textbf{Antibody Sequence (FASTA format):}
\begin{lstlisting}[style=cmd]
>Antibody
EVQLVESGGGLVQPGGSLRLSCAASGFTFSSYGMHWVRQAPGKGLEWVSG...
\end{lstlisting}

\textbf{Antigen Sequence (FASTA format):}
\begin{lstlisting}[style=cmd]
>Antigen
MKLGAIAIVIILGGFVLFIGLVGITGGKVGFAIGSSGLKLAMMGNAG...
\end{lstlisting}

\subsubsection{Step 2: Running RoseTTAFold for Complex Prediction}

To predict the interaction between these two proteins, we use RoseTTAFold with the following command:

\begin{lstlisting}[style=cmd]
python run_rosettafold.py \
  --fasta_paths=./input/antibody.fasta,./input/antigen.fasta \
  --output_dir=./output/complex/
\end{lstlisting}

\subsubsection{Step 3: Output Files}

RoseTTAFold will generate several files, including:

\begin{itemize}
    \item \textbf{Complex PDB file}: This file contains the predicted structure of the antibody-antigen complex, showing how the two proteins interact.
    \item \textbf{Confidence Scores}: These scores indicate the reliability of the predicted interaction interface.
\end{itemize}

\subsubsection{Step 4: Visualizing and Interpreting the Complex}

Load the complex structure into PyMOL to explore how the antibody binds to the antigen:

\begin{lstlisting}[style=cmd]
pymol complex.pdb
color cyan, antibody
color green, antigen
show cartoon
\end{lstlisting}

\textbf{Interpreting the Interface:}

The antibody-antigen interface is a critical region to examine. In a typical immune response, antibodies bind to specific antigens via their variable regions. By examining the predicted structure, we can identify key residues that form contacts between the two proteins.

\begin{itemize}
    \item \textbf{High-confidence interface}: If the interface residues have high confidence scores, we can be reasonably certain that these regions accurately represent the interaction.
    \item \textbf{Potential binding hotspots}: The structure might reveal "hotspots" where critical residues drive the binding interaction. These can be targets for drug development or therapeutic design.
\end{itemize}

\textbf{Example of Result Interpretation:}

If the antibody's complementarity-determining regions (CDRs) show strong interactions with the antigen, the predicted structure provides valuable insights into the immune recognition process. This can inform further experimental validation or even guide therapeutic antibody design.

\subsection{Case Study 3: Predicting a Disordered Protein Region with OmegaFold}

Next, we will use OmegaFold to predict the structure of a protein with a disordered region. Disordered proteins, or proteins with intrinsically disordered regions (IDRs), do not adopt a fixed structure under physiological conditions, but they play crucial roles in signaling, regulation, and interactions.

\subsubsection{Step 1: Input Preparation}

We will use a protein sequence known to contain an IDR:

\begin{lstlisting}[style=cmd]
>Disordered_Protein
MASMTGGQQMGSTTLMGEQLSTGGGGGSQGTGQMSGAQDRSM...
\end{lstlisting}

\subsubsection{Step 2: Running OmegaFold}

Since OmegaFold can predict structures without relying on MSAs, it is a good choice for disordered proteins:

\begin{lstlisting}[style=cmd]
python run_omegafold.py \
  --fasta=./input/disordered_protein.fasta \
  --output_dir=./output/disordered_protein/
\end{lstlisting}

\subsubsection{Step 3: Output Files}

OmegaFold generates a PDB file that contains the predicted structure of the protein, including any structured and disordered regions.

\subsubsection{Step 4: Visualizing and Interpreting Disordered Regions}

When visualizing the structure, you may notice that certain regions are predicted with low confidence, which could correspond to the disordered segments. These regions may not form stable secondary or tertiary structures but could still be functionally significant.

\begin{itemize}
    \item \textbf{Disordered regions}: Look for regions with low pLDDT scores or higher flexibility. These regions may lack regular secondary structures like alpha helices or beta sheets.
    \item \textbf{Functional significance}: Despite the lack of structure, IDRs are often involved in regulatory functions or transient interactions. The prediction might highlight flexible loops or regions that facilitate protein-protein interactions.
\end{itemize}

\textbf{Example of Result Interpretation:}

In this case, OmegaFold might predict a largely disordered protein, with only a few structured motifs. The disordered regions might still play key roles in signal transduction or molecular recognition. These predictions can guide further investigation into the protein's biological functions.

\subsection{Conclusion}

These case studies illustrate how different open-source protein structure prediction tools like AlphaFold2, RoseTTAFold, and OmegaFold can be applied to various research scenarios. From predicting single protein structures to exploring complex interactions and disordered regions, these tools provide invaluable insights that can drive experimental validation, drug discovery, and protein engineering. By carefully interpreting the predicted structures and associated confidence scores, researchers can make informed decisions about the biological relevance and potential applications of their results.

\section{Practical Protein Design}

Protein design is a powerful tool that allows researchers to create proteins with desired properties and functions. Using computational tools such as AlphaFold2, RoseTTAFold, and OmegaFold, it is possible to predict how a protein will fold and behave in a biological system. However, protein design goes beyond just predicting structures-it involves designing sequences that perform specific functions, optimizing these sequences for stability and efficiency, and ultimately validating them experimentally. In this section, we will guide beginners through the entire process of practical protein design, from defining the target function to sequence generation, optimization strategies, and experimental validation.

\subsection{From Target Function to Sequence Generation}

The first step in practical protein design is defining the \textbf{target function} \cite{hubbard1993target} of the protein. This function could range from catalyzing a chemical reaction (in the case of enzymes), binding a specific target molecule (e.g., antibodies), or interacting with other proteins to regulate cellular processes.

\subsubsection{Defining the Target Function}

Before designing a protein, it is essential to clearly define what the protein is supposed to do. This could involve enhancing an existing protein's function or creating an entirely new functionality. For example:

\begin{itemize}
    \item \textbf{Enzyme Design}: If you are designing an enzyme, the goal might be to create a protein that catalyzes a specific chemical reaction more efficiently than natural enzymes.
    \item \textbf{Antibody Design}: In antibody design, the target function may involve binding to a specific antigen with high affinity, such as designing an antibody that binds to a viral protein.
    \item \textbf{Protein-Protein Interaction Design}: You may be designing a protein to interact with another protein in a specific way, either to activate or inhibit a biological pathway.
\end{itemize}

\textbf{Example: Designing a New Enzyme}

Let's say you want to design an enzyme that catalyzes the conversion of a substrate that currently lacks an efficient natural enzyme. The first step is to define the desired reaction mechanism and the types of chemical bonds that the enzyme needs to break or form. Understanding this will guide you in designing an active site that can accommodate and catalyze the reaction.

\subsubsection{Generating a Sequence to Achieve the Target Function}

Once the function is defined, the next step is to generate a protein sequence that is likely to fold into a structure that can perform the desired function. This process is known as \textbf{sequence design} \cite{anand2022protein}. There are several approaches to sequence design, including:

\begin{itemize}
    \item \textbf{Homology-Based Design}: One approach is to start with an existing protein that has a similar function and modify its sequence to optimize or change its activity. Tools like AlphaFold2 and RoseTTAFold can help predict how the modified sequence will fold.
    \item \textbf{De Novo Design}: In cases where no homologous protein exists, it may be necessary to design a protein \textit{de novo} (from scratch). This involves generating new sequences and predicting how they fold into stable structures.
\end{itemize}

\textbf{Example: Homology-Based Design of an Antibody}

Suppose you are designing an antibody to bind a novel antigen. You might begin by identifying an existing antibody with a similar antigen-binding region and modifying the complementarity-determining regions (CDRs) to target the new antigen \cite{zhang2025u}. Sequence design tools can help you make these modifications while maintaining the overall stability of the antibody structure.

\textbf{Python Example: Generating a Sequence}

Using Python and a sequence design tool like \texttt{PyRosetta}, you can generate new protein sequences:

\begin{lstlisting}[style=python]
from pyrosetta import init, pose_from_sequence
init()

# Define an example protein sequence
sequence = "MKTAYIAKQRQISFVKSHFSRQLEERLGLIEVQAAEQLQIR"

# Generate a protein pose based on the sequence
pose = pose_from_sequence(sequence)
print(pose)
\end{lstlisting}

In this example, a sequence is provided, and the PyRosetta library is used to generate a 3D pose \cite{zhang2024vocapter}, which represents the structure that the sequence is likely to adopt.

\subsection{Considerations and Optimization Strategies}

Once a protein sequence has been generated, the next step is to optimize the sequence for the target function. This involves several key considerations, including the stability of the protein, the efficiency of its function, and the feasibility of its expression and purification in a laboratory setting.

\subsubsection{Stability Optimization}

One of the most important aspects of protein design is ensuring that the protein will fold into a stable structure under physiological conditions. A protein that is unstable may misfold, aggregate, or degrade, which can severely limit its functionality. Several strategies are used to optimize protein stability:

\begin{itemize}
    \item \textbf{Thermodynamic Stability}: The protein should be thermodynamically stable, meaning that its folded state should have a lower free energy than its unfolded state. This can be achieved by designing favorable interactions, such as hydrogen bonds, hydrophobic core packing, and salt bridges.
    \item \textbf{Avoiding Aggregation}: Proteins must avoid aggregation, which occurs when misfolded proteins clump together. Avoiding hydrophobic surface patches and ensuring proper solubility can reduce aggregation.
    \item \textbf{Disulfide Bonds}: Introducing disulfide bonds between cysteine residues can help stabilize the protein by linking distant parts of the structure.
\end{itemize}

\textbf{Example: Adding Disulfide Bonds}

If the designed protein is predicted to have regions that might be unstable, adding disulfide bonds can increase stability. You can identify pairs of cysteine residues and model the impact of introducing a covalent bond between them using computational tools.

\subsubsection{Functional Optimization}

In addition to stability, the protein must be optimized for function. Functional optimization ensures that the protein's active site or binding interface is correctly positioned to carry out its intended role.

\begin{itemize}
    \item \textbf{Active Site Design}: For enzymes, the active site must be designed to properly accommodate the substrate and catalyze the reaction efficiently. This might involve modifying the shape, charge, or hydrophobicity of the active site.
    \item \textbf{Binding Affinity}: For proteins that interact with other molecules (such as antibodies or receptors), the binding affinity must be optimized. This can be done by fine-tuning the interface to maximize favorable interactions (e.g., hydrogen bonds, van der Waals forces) and minimize unfavorable ones.
\end{itemize}

\textbf{Example: Optimizing an Enzyme Active Site}

Suppose you are optimizing an enzyme to improve its catalytic efficiency. You might introduce mutations that increase the proximity of key residues to the substrate or modify the electrostatic environment to facilitate the reaction.

\subsubsection{Experimental Considerations}

Even if a protein is designed in silico (using computational models), it must be experimentally feasible to express and purify the protein in a laboratory setting. Several practical considerations include:

\begin{itemize}
    \item \textbf{Codon Optimization}: The DNA sequence coding for the protein must be optimized for the expression system used (e.g., bacterial, yeast, or mammalian cells). Codon optimization ensures that the host organism can efficiently translate the protein \cite{fu2020codon}.
    \item \textbf{Solubility}: The protein must be soluble under physiological conditions. If the designed protein is insoluble, it may form inclusion bodies or aggregates, which can impede functional studies.
    \item \textbf{Expression and Purification}: The protein must be easy to express and purify. Adding tags (e.g., His-tags \cite{meredith2004targeted}) can simplify purification, but these tags may need to be removed later to ensure proper function.
\end{itemize}

\textbf{Example: Codon Optimization for Expression}

Suppose you have designed a protein intended for expression in \textit{Escherichia coli}. The codon usage in the designed sequence may not be optimal for \textit{E. coli}, leading to poor expression. Codon optimization tools can help modify the sequence without changing the amino acids to enhance expression levels.

\subsection{From Virtual Design to Experimental Validation}

Once a protein has been designed and optimized computationally, it is essential to validate the design through experimental methods. This step is crucial to ensure that the designed protein folds correctly and performs its intended function in a real biological system.

\subsubsection{Protein Expression and Purification}

The first step in experimental validation is to express the designed protein in a suitable host organism (such as bacteria, yeast, or mammalian cells) and purify it for further analysis. Common methods for protein expression include:

\begin{itemize}
    \item \textbf{Bacterial Expression}: \textit{E. coli} \cite{nataro1998diarrheagenic} is one of the most widely used organisms for expressing recombinant proteins. It offers fast growth, high yields, and cost-effectiveness.
    \item \textbf{Yeast Expression}: Yeast systems, such as \textit{Pichia pastoris}, are used when post-translational modifications (e.g., glycosylation \cite{haltiwanger2004role}) are required.
    \item \textbf{Mammalian Expression}: Mammalian cells are used for expressing proteins that require complex post-translational modifications and proper folding.
\end{itemize}

\subsubsection{Functional Assays and Structural Validation}

After purifying the protein, the next step is to test whether the protein functions as intended. This can involve:

\begin{itemize}
    \item \textbf{Enzymatic Assays}: For enzymes, you can measure the reaction rate or substrate turnover to determine whether the enzyme is catalyzing the reaction efficiently.
    \item \textbf{Binding Assays}: For proteins designed to bind specific targets (e.g., antibodies \cite{forthal2014functions}), you can use binding assays (e.g., ELISA \cite{hnasko2015elisa}, surface plasmon resonance \cite{hou2013review}) to measure binding affinity and specificity.
    \item \textbf{Structural Validation}: Techniques such as X-ray crystallography, NMR spectroscopy, or cryo-EM can be used to experimentally determine the 3D structure of the protein and confirm that it matches the predicted design.
\end{itemize}

\textbf{Example: Validating an Antibody Design}

Suppose you designed an antibody to bind a viral antigen. After expressing and purifying the antibody, you can perform an ELISA to confirm that the antibody binds to the antigen with high affinity. You can also use structural techniques like X-ray crystallography to confirm the antibody's binding mode.

\subsubsection{Iterative Design and Improvement}

Protein design is often an iterative process. After the initial design and experimental validation, you may need to return to the computational design phase to make further optimizations based on experimental data. For example, if the protein is not stable enough or does not bind as strongly as expected, you can modify the sequence and repeat the process.

\subsection{Conclusion}

Practical protein design is a multifaceted process that combines computational tools with experimental validation. By defining the target function, generating a sequence, optimizing the design, and validating the protein experimentally, researchers can create proteins with tailored properties for use in research, industry, and medicine. With the right tools and strategies, even beginners can start designing functional proteins and contributing to advancements in biotechnology and synthetic biology.
\chapter{Modules of AlphaFold3}

\section{Introduction}

AlphaFold3 represents the latest evolution in protein structure prediction, building on the successes and breakthroughs of AlphaFold2. One of the key improvements in AlphaFold3 is the incorporation of several new modules that enhance its ability to predict protein structures with higher accuracy, faster inference, and the ability to handle more complex biomolecular systems. These modules work together in a unified architecture to process inputs such as amino acid sequences, evolutionary data, and structural templates, ultimately predicting three-dimensional (3D) protein structures.

In this chapter, we will explore the different modules of AlphaFold3, focusing on their specific roles and contributions to the overall prediction process. We will start with an in-depth look at the \textbf{Pairformer Module} \cite{abramson2024accurate}, one of the most significant innovations in AlphaFold3, which enhances the model's ability to predict protein-protein interactions and handle large biomolecular complexes.

\section{Pairformer Module}

The \textbf{Pairformer Module} is a core component of AlphaFold3, replacing the Evoformer module that was central to AlphaFold2. The Pairformer Module is responsible for processing pairwise relationships between residues in a protein sequence, making it particularly well-suited for predicting how multiple proteins or protein domains interact in complex biological systems.

\subsection{Overview of the Pairformer Module}

The Pairformer Module focuses on learning the pairwise interactions between residues. It takes as input both the amino acid sequence and the structural templates (if available), then processes this information to predict the spatial relationships between residues, particularly those that are crucial for maintaining the overall 3D structure of the protein or complex. The module operates on both intra-protein and inter-protein interactions, enabling AlphaFold3 to predict the structures of large protein complexes and multimeric assemblies with much greater accuracy than AlphaFold2.

\textbf{Key Features of the Pairformer Module:}
\begin{itemize}
    \item \textbf{Pairwise Attention Mechanism}: The Pairformer Module uses a novel attention mechanism that is optimized for pairwise residue interactions. This mechanism allows the model to focus on critical contacts between residues, particularly those that are distant in the sequence but close in 3D space.
    \item \textbf{Reduced Reliance on MSAs}: Unlike AlphaFold2, which heavily relied on multiple sequence alignments (MSAs) to gather evolutionary information, the Pairformer Module can function more effectively even with limited MSA depth. This makes AlphaFold3 more robust for proteins that have few homologs.
    \item \textbf{Handling Protein Complexes}: The Pairformer is designed to excel at predicting protein-protein interactions, making it ideal for handling multimeric proteins, enzyme-substrate complexes, and large biomolecular assemblies.
\end{itemize}

\subsection{Pairwise Attention Mechanism}

The attention mechanism is the backbone of the Pairformer Module. In a protein structure, many residues that are distant from each other in the linear sequence are actually close in the 3D folded form. Identifying and learning these relationships is crucial for accurate structure prediction. The Pairformer's attention mechanism allows AlphaFold3 to efficiently learn these interactions by focusing on pairwise residue relationships rather than relying solely on MSAs.

\textbf{Example of Pairwise Attention:}
Let's consider a hypothetical protein where residues at positions 5 and 100 are far apart in the linear sequence but come close together in the folded 3D structure. The Pairformer Module identifies this long-range interaction through its pairwise attention mechanism and adjusts the model to reflect this proximity in the final predicted structure.

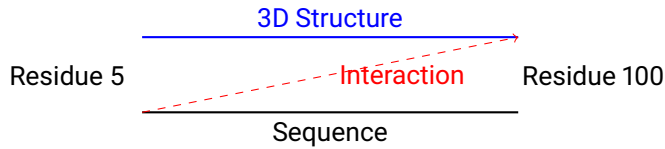
\begin{center}
\begin{tikzpicture}
    \draw[thick] (0,0) -- (5,0) node[midway,below] {Sequence};
    \draw[thick,blue] (0,1) -- (5,1) node[midway,above] {3D Structure};
    \draw[dashed,red,->] (0,0) -- (5,1) node[midway,right] {Interaction};
    \node at (-1.0,0.5) {Residue 5};
    \node at (6.0,0.5) {Residue 100};
\end{tikzpicture}
\end{center}

\textbf{Benefits of Pairwise Attention:}
\begin{itemize}
    \item \textbf{Improved Long-Range Interaction Prediction}: By focusing on long-range interactions, the Pairformer Module helps AlphaFold3 accurately predict tertiary and quaternary structures, even in cases where residues are far apart in the sequence.
    \item \textbf{Efficient Data Utilization}: Since it reduces reliance on deep MSAs, the Pairformer can generate accurate predictions even with limited evolutionary data, making AlphaFold3 more flexible for orphan proteins or proteins with few homologs.
\end{itemize}

\subsection{Handling Protein-Protein Interactions}

One of the most significant challenges in structural biology is predicting how proteins interact with each other to form functional complexes. Protein-protein interactions are critical for nearly every biological process, from signal transduction to enzymatic activity. AlphaFold3's Pairformer Module is designed to tackle this challenge head-on.

\textbf{Example: Antibody-Antigen Interaction}
Consider an antibody binding to an antigen. The antibody's variable region recognizes and binds to specific epitopes on the antigen surface. The Pairformer Module can predict the structure of the entire antibody-antigen complex by focusing on the pairwise interactions between residues on both proteins. This allows AlphaFold3 to predict the precise binding interface between the two molecules.

\textbf{Application of the Pairformer in Drug Design:}
Accurate prediction of protein-protein interfaces is critical for drug discovery. For example, understanding how a drug binds to its target protein can guide the optimization of drug candidates. The Pairformer's ability to handle protein-protein interactions enables AlphaFold3 to predict drug binding sites and suggest modifications to improve binding affinity.

\subsection{Reduced Dependence on MSAs}

While MSAs are still useful for gathering evolutionary information, AlphaFold3's Pairformer Module reduces the dependence on them. This is particularly advantageous in cases where MSAs are sparse or non-existent, such as orphan proteins or rapidly evolving proteins with few known homologs. By focusing on pairwise interactions directly, AlphaFold3 can produce high-quality predictions even with limited evolutionary data.

\textbf{Example: Predicting the Structure of an Orphan Protein}
Orphan proteins, which have few or no homologs in current databases, are difficult to predict using traditional MSA-based methods. AlphaFold3, with its Pairformer Module, can overcome this limitation by leveraging structural information directly from the sequence, without needing deep evolutionary data.

\subsection{Pairformer's Role in Multi-Domain and Multi-Chain Predictions}

Multi-domain proteins and multi-chain complexes present additional challenges due to the complexity of inter-domain or inter-chain interactions. The Pairformer Module excels at these predictions by focusing on interactions both within and between chains. This capability allows AlphaFold3 to predict the quaternary structure of complexes, such as homomeric and heteromeric assemblies.

\textbf{Example: Multi-Chain Protein Complex Prediction}
In the case of a multi-chain enzyme complex, where multiple protein subunits work together to perform a biological function, the Pairformer Module can predict how each subunit interacts with the others. This is essential for understanding the full functionality of the enzyme and designing inhibitors or activators that target specific subunits.

\subsection{Conclusion}

The Pairformer Module represents a significant advancement in AlphaFold3's architecture. By focusing on pairwise residue interactions, it enhances the model's ability to predict long-range contacts, protein-protein interfaces, and complex multimeric structures. The reduction in reliance on MSAs further improves the versatility of AlphaFold3, making it an invaluable tool for predicting orphan proteins and other challenging cases. In the following sections, we will explore additional modules that contribute to AlphaFold3's overall performance and accuracy.

\section{Diffusion Module}

The \textbf{Diffusion Module} in AlphaFold3 represents a significant departure from the traditional methods used in AlphaFold2 for generating protein structures. This module introduces a diffusion-based approach that operates directly on atomic coordinates, allowing the model to iteratively refine the protein structure. The diffusion process is key to how AlphaFold3 predicts highly complex protein structures, including biomolecular complexes and protein-ligand interactions, with increased accuracy.

\subsection{What is Diffusion in Protein Structure Prediction?}

In the context of deep learning and generative models, diffusion refers to a process where random noise is gradually added to the input data (in this case, the protein's atomic coordinates), and the model is trained to remove this noise to recover the original structure \cite{wu2024protein}. This process allows the model to learn how to progressively refine a coarse structure into a highly detailed and accurate one. Diffusion-based models are often used in image generation, but in AlphaFold3, this concept has been adapted for protein structure prediction.

\textbf{How it works:}

The diffusion model starts with a highly noised version of the atomic coordinates of a protein or biomolecular complex. The noise can be thought of as a random disturbance that disrupts the initial structure. The goal of the model is to predict the true, noiseless coordinates by iteratively refining the structure over multiple steps.

At each step of the diffusion process, the model improves different aspects of the structure:

\begin{itemize}
    \item \textbf{Early steps} focus on large-scale features, such as the overall shape and fold of the protein.
    \item \textbf{Later steps} refine smaller details, such as side-chain orientations and the precise positioning of atoms.
\end{itemize}

\subsubsection{Generative Training Process}

The diffusion module is trained using a generative process where it learns to denoise atomic coordinates at different levels of noise. This generative approach means that AlphaFold3 can produce a range of plausible structures rather than a single deterministic output. This is useful when dealing with flexible proteins or regions that lack a well-defined structure.

\textbf{Example:}
Imagine predicting the structure of a protein-ligand complex where the ligand is bound deep within the protein. The diffusion module can begin with an approximate structure of the complex and gradually refine it, paying attention to how the protein's binding pocket forms around the ligand.

\subsection{Key Features of the Diffusion Module}

The Diffusion Module brings several advantages over traditional structure prediction approaches:

\begin{itemize}
    \item \textbf{Direct prediction of atomic coordinates}: Unlike AlphaFold2, which relied on intermediate representations such as backbone frames and torsion angles, AlphaFold3's Diffusion Module directly predicts the positions of atoms in space.
    \item \textbf{Handling complex biomolecular interactions}: The diffusion approach is particularly effective for predicting the structures of complex assemblies, such as protein-protein interactions, protein-ligand complexes, and even protein-nucleic acid interactions.
    \item \textbf{Generative nature}: Because the diffusion process is generative, it can produce multiple plausible structures for proteins with flexible regions or multiple conformations, providing a more comprehensive understanding of the protein's potential states.
\end{itemize}

\subsection{Comparison with AlphaFold2's Structure Module}

The Diffusion Module replaces the structure module of AlphaFold2, which relied heavily on specific parameterizations of protein residues, such as backbone frames and torsion angles. While AlphaFold2 used these parameterizations to enforce stereochemical constraints, AlphaFold3's diffusion approach allows for a more flexible and general representation of the protein's atomic structure. This results in several improvements:

\begin{itemize}
    \item \textbf{Stereochemical constraints are naturally enforced}: AlphaFold3 does not require specific loss functions to penalize stereochemical violations, as the diffusion process inherently learns to predict chemically plausible structures.
    \item \textbf{Elimination of torsion-based parameterization}: The reliance on torsion angles in AlphaFold2 added complexity to the model. AlphaFold3 simplifies this by predicting atomic coordinates directly, which is more general and applicable to a wider range of biomolecular structures.
\end{itemize}

\subsection{Applications of the Diffusion Module}

The diffusion approach is especially useful in scenarios where traditional methods struggle, such as predicting the structure of biomolecular complexes that include multiple types of molecules (e.g., proteins, nucleic acids, and small molecules). Below are a few key applications:

\subsubsection{Protein-Ligand Interactions}

AlphaFold3 has demonstrated significant improvements in predicting protein-ligand interactions, surpassing traditional docking methods. The Diffusion Module is able to predict the binding pose of ligands with high accuracy, even in cases where the ligand undergoes conformational changes upon binding.

\textbf{Example:}
In a drug discovery context, researchers can use AlphaFold3 to predict how a small molecule drug interacts with its protein target. The Diffusion Module refines the protein's binding pocket and the ligand's position to predict a highly accurate structure, which can then be used to guide drug optimization.

\subsubsection{Protein-Protein and Protein-Nucleic Acid Complexes}

The ability to predict large biomolecular complexes is another major advantage of AlphaFold3. The Diffusion Module can handle interactions between proteins and nucleic acids, such as those seen in transcription factors binding to DNA, or between proteins in a multimeric assembly.

\textbf{Example:}
In the case of a transcription factor bound to a segment of DNA, AlphaFold3 predicts both the protein structure and the DNA structure simultaneously. The diffusion process ensures that the predicted structure maintains biologically relevant interactions between the protein and the DNA.

\subsection{Challenges and Limitations}

While the Diffusion Module offers significant improvements, there are some challenges associated with its use:

\begin{itemize}
    \item \textbf{Generative noise}: One challenge of generative diffusion models is the possibility of producing hallucinated structures, especially in regions where the model is uncertain. To counteract this, AlphaFold3 employs cross-distillation techniques to ensure that the generated structures remain realistic.
    \item \textbf{Computational cost}: The iterative nature of the diffusion process requires more computational resources compared to deterministic models. Generating multiple samples to account for structural variability can also increase the time required for predictions.
\end{itemize}

\subsection{Conclusion}

The Diffusion Module is a powerful innovation in AlphaFold3, allowing it to predict highly accurate protein structures across a wide range of biomolecular systems. By directly predicting atomic coordinates and using a generative approach to refine structures, AlphaFold3 offers improved accuracy, especially for protein-ligand interactions and large biomolecular complexes. While there are challenges related to computational cost and noise, the Diffusion Module's ability to handle complex biological systems makes it a critical component of AlphaFold3's architecture.

\section{Template Module}

The \textbf{Template Module} \cite{abramson2024accurate} in AlphaFold3 plays a crucial role in improving the accuracy of protein structure predictions by incorporating known structural information from homologous proteins or similar sequences. This approach allows AlphaFold3 to use both evolutionary information and structural knowledge, especially when predicting novel proteins or protein complexes. Templates are particularly useful in cases where multiple sequence alignments (MSAs) do not provide enough information due to limited evolutionary data. The Template Module, therefore, acts as a complementary source of structural data, ensuring that AlphaFold3 predictions are as accurate as possible, even in challenging cases.

\subsection{Overview of the Template Module}

The Template Module works by searching for known structures in databases such as the Protein Data Bank (PDB). When a similar structure is found, AlphaFold3 incorporates this structural information as a template, using it to guide the prediction process. This is particularly important for proteins that have homologs with well-characterized structures, as the template provides a reliable scaffold upon which AlphaFold3 can build its predictions.

The template-based approach in AlphaFold3 differs from traditional homology modeling in that it integrates template information into the deep learning framework rather than relying entirely on templates to predict the structure. This integration allows AlphaFold3 to combine template information with its powerful attention-based architecture to refine and improve the final structure prediction.

\subsubsection{How Template Search Works}

The first step of the Template Module involves searching a structural database for potential templates. AlphaFold3 uses a similarity-based approach to identify templates that share significant sequence or structural similarity with the target protein. Once a suitable template is found, the information is extracted and processed by the model.

\textbf{Example:}
Imagine you are predicting the structure of a protein with limited sequence homologs. The Template Module would search the PDB for proteins with similar domains or regions, such as a known enzyme with a well-characterized catalytic domain. This template would then be used to help guide the structure prediction of the unknown protein.

In Python, you can use tools like \texttt{Biopython} or \texttt{hmmer} to perform sequence searches against databases like PDB:

\begin{lstlisting}[style=python]
from Bio import SearchIO
# Load the HMMER results
result = SearchIO.read("example_hmmer_output.txt", "hmmer3-text")
# Parse and extract the best hit from the template search
best_template = result.hits[0]
print("Best template:", best_template.id)
\end{lstlisting}

This simple example demonstrates how you might search for similar sequences in a database to find a potential structural template.

\subsection{Key Features of the Template Module}

The Template Module provides several key advantages in structure prediction:

\begin{itemize}
    \item \textbf{Improved Prediction for Novel Proteins}: For proteins that lack sufficient homologs or have novel domains, the Template Module can find related structures to provide additional structural insights. This is particularly useful in cases where evolutionary information from MSAs is sparse.
    \item \textbf{Combining Structural and Evolutionary Data}: The Template Module allows AlphaFold3 to combine known structural information with the powerful pattern recognition capabilities of the deep learning model. This combination enhances the accuracy of predictions, especially for proteins with known structural motifs.
    \item \textbf{Increased Accuracy for Complexes}: When predicting the structure of protein complexes, the Template Module can provide critical information about how individual subunits of the complex interact with each other, based on previously solved structures of similar complexes.
\end{itemize}

\subsection{Incorporation of Templates into the Prediction Process}

The Template Module in AlphaFold3 is not just used at the beginning of the prediction process but is integrated throughout the model. After the initial template search, the structural information is processed and passed through the attention layers of AlphaFold3's architecture. This allows the model to refine the template data and ensure that it fits well with the other data sources, such as MSAs and pairwise attention mechanisms.

The template information helps the model make corrections where the evolutionary data alone might lead to errors. For instance, if the MSA suggests a certain fold, but the template reveals a more accurate configuration, AlphaFold3 will incorporate the template's structure to refine the prediction.

\subsubsection{Example: Template Use in a Protein-Protein Complex}

Consider the prediction of a protein-protein complex where one protein has a known structure, but the other does not. The Template Module would use the known structure as a guide for predicting the unknown protein's binding conformation. This is particularly useful in drug discovery, where understanding how a protein interacts with other proteins or ligands is crucial for developing effective treatments.

In this case, the known structure (template) provides a reliable starting point for the prediction of the unknown partner in the complex, allowing the model to predict the interaction interface more accurately.

\subsection{Challenges with Template-Based Predictions}

While the Template Module significantly enhances the accuracy of AlphaFold3, there are challenges associated with relying on templates:

\begin{itemize}
    \item \textbf{Template Availability}: The effectiveness of the Template Module depends on the availability of similar structures in the database. For proteins with novel folds or no known homologs, finding an appropriate template may be difficult.
    \item \textbf{Template Selection}: Selecting the most appropriate template is a non-trivial task, as some templates may share sequence similarity but have different structural characteristics. AlphaFold3 mitigates this by integrating the template information with its deep learning-based predictions, but template selection can still affect the quality of the final prediction.
    \item \textbf{Template Bias}: There is a risk that reliance on templates can introduce bias into the model, leading to over-reliance on structural motifs from the templates rather than novel predictions. AlphaFold3 addresses this by carefully balancing the template information with other data sources.
\end{itemize}

\subsection{Improving the Performance of Template-Based Models}

To maximize the utility of templates in AlphaFold3, several optimization strategies can be employed:

\begin{itemize}
    \item \textbf{Template Pruning}: Carefully selecting templates based on sequence identity, structural quality, and functional similarity can improve prediction accuracy. For example, filtering templates to exclude those that are too distantly related or have unresolved regions can help refine the predictions.
    \item \textbf{Combining Multiple Templates}: In cases where multiple templates are available, combining them into a consensus model can improve prediction accuracy. This approach allows AlphaFold3 to leverage information from multiple sources to build a more complete picture of the protein's structure.
    \item \textbf{Template Refinement}: After incorporating a template, further refinement using AlphaFold3's deep learning architecture can improve the accuracy of the predicted structure. This refinement step is crucial for ensuring that the template fits well with the rest of the model's predictions.
\end{itemize}

\textbf{Example:}
In the case of predicting the structure of a viral protein with limited homologs, the Template Module might find a distant relative in the PDB. By refining the initial template prediction using the deep learning model's attention mechanism, AlphaFold3 can generate a highly accurate structure despite the lack of close homologs.

\subsection{Conclusion}

The Template Module is a vital component of AlphaFold3, allowing it to leverage existing structural knowledge to improve protein structure predictions. By integrating template information into the deep learning framework, AlphaFold3 can make accurate predictions even for proteins with limited evolutionary data or complex protein-protein interactions. Although there are challenges associated with template selection and bias, careful optimization and refinement strategies help mitigate these issues, ensuring that the Template Module enhances, rather than limits, the predictive power of AlphaFold3.

\section{MSA Processing Module}

The \textbf{Multiple Sequence Alignment (MSA) Processing Module} in AlphaFold3 plays a crucial role in gathering evolutionary information from protein sequences. MSAs are a fundamental tool in bioinformatics that help identify conserved regions across protein sequences, which are often key to understanding protein structure and function. In AlphaFold3, MSAs contribute to the prediction of protein folding by revealing which residues are critical for maintaining the protein's structure across different species. However, in comparison to AlphaFold2, AlphaFold3 reduces the reliance on deep MSAs, which allows it to make accurate predictions even when limited evolutionary data are available.

\subsection{What is Multiple Sequence Alignment?}

MSA is a method used to align three or more biological sequences (proteins or nucleotides) so that similar residues across sequences are aligned in columns. The idea is that residues or motifs that are conserved across different species are likely to be structurally or functionally important.

Here's a simple example of a multiple sequence alignment for three hypothetical protein sequences:

\begin{lstlisting}[style=cmd]
Seq1:  M A L Q G L T K A 
Seq2:  M A L Q G - T K A 
Seq3:  M A - Q G L T K -
\end{lstlisting}

In this alignment, we can see that positions 1, 2, 4, 5, 7, and 8 are conserved across all three sequences, indicating that these residues are likely crucial for the protein's structure or function. The gaps introduced in sequences 2 and 3 represent insertions or deletions that have occurred during evolution.

\subsection{The Role of MSAs in AlphaFold3}

In AlphaFold2, MSAs were heavily relied upon to extract evolutionary constraints that guided the protein folding process. AlphaFold3, however, introduces significant improvements in how MSAs are handled, allowing it to function effectively even with shallow or incomplete MSAs. This is particularly useful for proteins that have few homologs, such as orphan proteins or those that evolve rapidly.

The MSA Processing Module in AlphaFold3 still plays an important role, but it is integrated more efficiently into the model's architecture. The model uses the MSA to build a profile of conserved residues, but it places less emphasis on the depth and breadth of the alignment, instead focusing on pairwise relationships between residues.

\textbf{Example of MSA Processing:}

Consider a scenario where you have a target protein with only a few known homologs. AlphaFold3 can still process this limited MSA and extract valuable information from the few available sequences. The model integrates this information with structural templates and its pairwise attention mechanism to accurately predict the protein's 3D structure.

\subsection{MSA Processing in AlphaFold3 vs. AlphaFold2}

One of the key advancements of AlphaFold3 is that it reduces the computational burden associated with deep MSA searches. In AlphaFold2, generating a deep MSA for a protein sequence could take significant computational time and resources, especially for proteins with many homologs. AlphaFold3 introduces a simplified approach where fewer MSA blocks are used, and pairwise representations take over much of the work that was previously done by MSAs.

\textbf{Improvements in AlphaFold3:}
\begin{itemize}
    \item \textbf{Reduced MSA Depth}: AlphaFold3 can generate accurate predictions even with shallow MSAs, making it more efficient and faster.
    \item \textbf{Pairwise Representations}: The focus shifts from deep evolutionary data to pairwise residue interactions, allowing AlphaFold3 to handle cases where MSA data is sparse or unavailable.
    \item \textbf{Efficiency Gains}: By reducing the complexity of the MSA processing, AlphaFold3 can perform structure predictions more quickly without sacrificing accuracy.
\end{itemize}

\subsection{Key Features of the MSA Processing Module}

While AlphaFold3 places less emphasis on MSAs compared to its predecessor, it still extracts useful evolutionary information through a streamlined process. Here are some of the key features:

\begin{itemize}
    \item \textbf{Shallow MSAs are sufficient}: AlphaFold3 can work effectively even with minimal sequence alignment data, making it suitable for proteins that have very few homologs.
    \item \textbf{Simplified MSA embedding}: Instead of using deep stacks of MSA blocks, AlphaFold3 uses a smaller, more efficient MSA embedding block. This reduces the computational overhead and speeds up the prediction process.
    \item \textbf{Focus on Pairwise Representations}: The pairwise relationships between residues are now more important, reducing the model's reliance on evolutionary data and making it more robust in challenging cases.
\end{itemize}

\subsection{Challenges and Limitations of MSA-Based Predictions}

While the MSA Processing Module in AlphaFold3 is more efficient than in previous versions, there are still some challenges associated with using MSAs in protein structure prediction:

\begin{itemize}
    \item \textbf{Limited Homologs}: For some proteins, especially those that evolve rapidly or have unique functions, there may be few or no homologs available. In such cases, the MSA may not provide enough information for accurate structure prediction.
    \item \textbf{Conservation Bias}: MSAs rely on conserved residues, which can sometimes introduce bias. For example, highly conserved regions might dominate the prediction, while less conserved but structurally important regions might be underrepresented.
    \item \textbf{Computational Cost}: While AlphaFold3 has reduced the reliance on MSAs, generating even a shallow MSA can still be computationally expensive for very large proteins or complexes.
\end{itemize}

\subsection{Applications of the MSA Processing Module}

The MSA Processing Module is crucial in several applications of protein structure prediction, especially for proteins where evolutionary data provides important insights into their function or stability.

\subsubsection{Predicting Protein Function}

MSAs are particularly useful for identifying functionally important residues. By analyzing conserved regions across multiple species, researchers can identify key active sites, binding residues, and other functional motifs. This information can be combined with AlphaFold3's structural predictions to generate hypotheses about how the protein functions at the molecular level.

\textbf{Example: Enzyme Active Site Prediction}
Consider an enzyme that catalyzes a biochemical reaction. By aligning the enzyme's sequence with homologs from other species, conserved residues that form the active site can be identified. AlphaFold3's MSA Processing Module would help predict the structure of the enzyme, highlighting these functionally important residues in the final 3D structure.

\subsubsection{Predicting Protein Stability}

Evolutionary conservation can also be used to predict regions of a protein that are important for maintaining structural stability. By identifying conserved regions that are critical for the protein's fold, the MSA Processing Module helps AlphaFold3 predict which parts of the protein are less likely to mutate and more likely to maintain a stable structure over evolutionary time.

\subsubsection{Drug Target Identification}

In drug discovery, MSAs are used to identify conserved regions in proteins that are potential drug targets. By focusing on residues that are conserved across species, researchers can identify parts of the protein that are likely to be critical for its function, making them attractive targets for small molecule drugs or antibodies.

\subsection{Conclusion}

The MSA Processing Module is an integral part of AlphaFold3, although its role has evolved compared to AlphaFold2. By reducing the emphasis on deep evolutionary data and incorporating more pairwise representations, AlphaFold3 is able to make accurate predictions even in cases where traditional MSA-based approaches would struggle. Despite these improvements, the module remains a powerful tool for understanding evolutionary conservation and predicting protein function and stability, making it indispensable in many applications of protein structure prediction.

\section{Confidence Module}

The \textbf{Confidence Module} \cite{abramson2024accurate} in AlphaFold3 is responsible for providing a measure of the reliability of its predicted protein structures. Confidence estimation is crucial because it helps users assess the quality of the predicted models, especially in regions where the predictions may be uncertain. This module builds upon the concepts introduced in AlphaFold2, where confidence scores such as pLDDT (predicted Local Distance Difference Test) and PAE (Predicted Aligned Error) were used to provide insight into the accuracy of different parts of the predicted protein structures. However, in AlphaFold3, the confidence evaluation has been enhanced through a combination of new techniques that are designed to work within its diffusion-based architecture.

\subsection{Role of the Confidence Module}

The primary role of the Confidence Module is to estimate how likely the predicted protein structure is to match the actual physical structure. This is especially important for regions where the model is uncertain, such as flexible loops, disordered regions, or regions with sparse evolutionary data. The Confidence Module provides scores that help users identify which parts of the structure can be trusted and which parts may require further experimental validation.

\textbf{Key Responsibilities of the Confidence Module:}
\begin{itemize}
    \item \textbf{Estimating per-residue confidence}: Each residue in the predicted protein structure receives a confidence score (pLDDT), which indicates the likelihood that its predicted position is accurate.
    \item \textbf{Predicting overall model accuracy}: In addition to per-residue confidence, the module also provides an overall assessment of the accuracy of the entire model, using metrics such as PAE.
    \item \textbf{Guiding model ranking}: The confidence scores are used to rank multiple predicted models, helping to select the most accurate one from a set of possible predictions.
\end{itemize}

\subsection{How Confidence is Estimated}

In AlphaFold2, confidence estimation was based on the output of the structure module, where the model was trained to predict local distance differences and alignments. In AlphaFold3, however, the diffusion-based architecture introduces new challenges because the structure is predicted iteratively across multiple steps, rather than in a single output. As a result, AlphaFold3 uses a new method called \textbf{diffusion rollout} \cite{abramson2024accurate} to estimate confidence throughout the prediction process.

\textbf{Diffusion Rollout Process:}
\begin{itemize}
    \item During the diffusion process, AlphaFold3 generates multiple intermediate structures at various noise levels. These intermediate predictions are used to estimate how well the model is refining the structure over time.
    \item At each stage of the diffusion process, the Confidence Module tracks how well the predicted structure aligns with the expected geometries, including bond angles, distances between atoms, and residue placements.
    \item The final confidence score is a composite measure, combining information from all stages of the diffusion process to provide a more accurate estimate of prediction reliability.
\end{itemize}

\subsubsection{pLDDT Score}

The \textbf{pLDDT score} \cite{ng2002predicting} (Predicted Local Distance Difference Test) is one of the most critical metrics provided by the Confidence Module. It assigns a confidence value between 0 and 100 to each residue in the protein structure. A higher pLDDT score indicates that the model is highly confident in the predicted position of the residue, while lower scores suggest that the prediction may be less reliable.

\textbf{Example:}
Imagine a protein with well-structured domains and flexible loops. The pLDDT score for residues in the structured domain would likely be high (e.g., 90 or above), indicating strong confidence in those regions. On the other hand, residues in flexible loops or disordered regions may have lower pLDDT scores (e.g., 50 or below), signaling that these areas should be interpreted with caution.

\begin{lstlisting}[style=python]
# Example of pLDDT scores for a predicted protein structure
predicted_structure = {"residue_1": 95, "residue_2": 90, "residue_3": 45, "residue_4": 30}
print("pLDDT scores:", predicted_structure)
\end{lstlisting}

\subsubsection{PAE Matrix}

The \textbf{Predicted Aligned Error} (PAE) matrix is another important output of the Confidence Module. It provides a pairwise confidence score that indicates the model's expected error in aligning pairs of residues in the predicted structure. The PAE matrix helps assess the accuracy of long-range interactions and overall protein fold.

\textbf{Example:}
In a multi-domain protein, the PAE matrix might reveal that the model is confident about the relative positioning of residues within a domain (low PAE values) but less certain about the relative alignment between two domains (high PAE values). This can guide experimentalists in focusing on regions where the predicted structure is less reliable.

\subsection{Applications of Confidence Scores}

Confidence scores are essential for several practical applications of AlphaFold3, especially when users need to make decisions based on the predicted structures. Here are some key scenarios where the Confidence Module plays a vital role:

\subsubsection{Model Ranking and Selection}

When AlphaFold3 generates multiple predictions for the same protein, confidence scores are used to rank these models. The model with the highest overall confidence score (based on pLDDT and PAE) is typically selected as the most reliable prediction. This process ensures that users can focus on the most accurate structure, reducing the need for manual inspection of multiple models.

\subsubsection{Experimental Validation}

In experimental structural biology, confidence scores guide researchers in deciding which regions of a predicted structure require further experimental validation. For example, if a predicted binding site has a low pLDDT score, researchers might prioritize experimental techniques such as X-ray crystallography or cryo-EM to resolve that region.

\subsubsection{Protein Design and Engineering}

In protein design, confidence scores are critical for assessing the feasibility of engineered modifications. For example, when designing a protein for a specific function, high-confidence regions can be targeted for modifications, while low-confidence regions may be left unchanged or subjected to additional refinement.

\subsection{Limitations and Challenges}

While the Confidence Module in AlphaFold3 represents a significant advancement, there are still limitations and challenges associated with its use:

\begin{itemize}
    \item \textbf{Over-reliance on confidence scores}: High pLDDT scores do not guarantee that the structure is functionally correct. For example, even though a predicted structure may be highly confident, it could still represent an inactive or non-functional state of the protein.
    \item \textbf{Uncertainty in flexible regions}: In cases where proteins contain intrinsically disordered regions, confidence scores may be lower, but this does not necessarily mean that the prediction is wrong. Instead, it reflects the model's uncertainty in predicting flexible or unstructured areas.
    \item \textbf{Generative nature of AlphaFold3}: Due to the generative nature of the diffusion model, the Confidence Module must deal with multiple plausible structures. This can sometimes make it challenging to determine whether low-confidence regions represent genuine structural variability or errors in the prediction.
\end{itemize}

\subsection{Conclusion}

The Confidence Module in AlphaFold3 plays a critical role in providing users with reliable measures of prediction accuracy. By combining metrics like pLDDT and PAE with advanced techniques such as diffusion rollout, AlphaFold3 offers a powerful toolset for evaluating the quality of predicted protein structures. This module not only helps guide experimental validation but also enhances the practical applications of AlphaFold3 in fields like protein design and drug discovery. However, it is important to remember the limitations of confidence estimation and interpret the results carefully, especially when working with flexible or disordered proteins.

\section{Training Regime}

AlphaFold3's training regime is designed to address the increased complexity of its architecture and the wide range of biomolecular interactions it predicts. The training process involves multiple stages, including initial training and fine-tuning phases, to ensure that the model learns both local and global structural features. AlphaFold3 introduces new components like the diffusion module and the pairformer, which require adjustments in the training strategy to accommodate the generative nature of the model and its ability to predict various types of biomolecular complexes, including proteins, nucleic acids, and ligands.

\subsection{Initial Training}

The \textbf{Initial Training} phase is crucial for establishing the foundation upon which AlphaFold3 builds its predictive capabilities. In this stage, the model is trained on a broad set of structural data using a fixed crop size to ensure it learns a robust representation of both single proteins and complex biomolecular interactions \cite{abramson2024accurate}. The training process involves using a mini-batch of input data that is passed through the network to update the weights and biases of the model.

\textit{Dataset and Preprocessing:} The initial training utilizes data from the Protein Data Bank (PDB), along with other databases containing protein-nucleic acid, protein-ligand, and protein-protein interaction data. Each structure is carefully preprocessed to include information about atomic coordinates, bonding patterns, and, where applicable, the presence of non-canonical amino acids or modifications. The training data also includes multiple sequence alignments (MSAs) to help the model understand evolutionary relationships between proteins, although MSAs play a smaller role in AlphaFold3 compared to AlphaFold2.

\textit{Example of Mini-Batch Training:} During the initial training, AlphaFold3 processes a mini-batch of 256 samples, each containing protein structures and their respective interaction data. Each sample goes through the following steps:

\begin{enumerate}
    \item The structure and interaction data are passed through the pairformer module to generate pairwise residue representations.
    \item The diffusion module processes atomic coordinates to refine the protein structure iteratively, starting from noised positions.
    \item The loss is computed by comparing the predicted structure to the true structure, using metrics like the Local Distance Difference Test (LDDT) and Predicted Aligned Error (PAE).
\end{enumerate}

\begin{lstlisting}[style=python]
# Example Python pseudo-code for training a batch of data
for epoch in range(num_epochs):
    for batch in dataloader:
        # Forward pass through the pairformer and diffusion modules
        pair_representation = model.pairformer(batch)
        predicted_structure = model.diffusion_module(pair_representation)
        
        # Compute the loss based on LDDT and PAE
        loss = compute_loss(predicted_structure, batch.true_structure)
        
        # Backpropagate and update model parameters
        optimizer.zero_grad()
        loss.backward()
        optimizer.step()
\end{lstlisting}

The goal of the initial training phase is to optimize the model's parameters to ensure that it can generalize well across a variety of biomolecular complexes. During this phase, the model rapidly learns local structural features like bond angles and torsion, as well as more complex global features, such as the relative orientation of different protein domains.

\subsubsection{Training Metrics}

AlphaFold3 uses several metrics to evaluate its performance during the initial training:

\begin{itemize}
    \item \textit{LDDT (Local Distance Difference Test)} \cite{mariani2013lddt}: This metric measures the accuracy of the predicted local geometry compared to the true structure. Higher LDDT scores indicate that the model is correctly predicting bond lengths and angles.
    \item \textit{PAE (Predicted Aligned Error)} \cite{elfmann2023pae}: PAE is used to assess the accuracy of long-range interactions between residues, helping the model learn how different parts of the protein fold relative to each other.
    \item \textit{Cross-entropy loss} \cite{mao2023cross}: This loss function is used to compare the predicted residue contacts and distances with the true values in the training set.
\end{itemize}

\subsection{Fine-tuning Stages}

Once the initial training is complete, AlphaFold3 undergoes two stages of \textit{fine-tuning} to further refine its predictions, especially for complex biomolecular interactions and multi-chain systems. Fine-tuning allows the model to adjust to more specific tasks, such as improving the accuracy of protein-ligand binding predictions or handling complex protein-nucleic acid interactions.

\textit{Fine-tuning Stage 1:} The first fine-tuning phase increases the crop size from 384 tokens to 640 tokens, allowing the model to process larger complexes and incorporate more detailed structural information. This stage focuses on improving the model's understanding of inter-domain and inter-protein interactions, which are critical for accurately predicting the structure of multi-protein complexes and assemblies.

\textit{Example: Fine-tuning Protein-Ligand Complexes} In this phase, AlphaFold3 may focus on specific tasks such as predicting how a ligand binds to a protein. The model uses both the pairformer and diffusion modules to refine the atomic coordinates of the ligand and the protein's binding pocket. The confidence module evaluates the prediction, providing feedback that guides further fine-tuning.

\textit{Fine-tuning Stage 2:} The second fine-tuning phase further increases the crop size to 768 tokens. This stage is particularly useful for handling very large complexes, such as ribosomes, transcription complexes, or viral assemblies. By increasing the amount of input data processed simultaneously, AlphaFold3 can better capture long-range interactions and subtle structural changes that occur in large systems.

\textit{Impact of Fine-tuning on Accuracy:} Fine-tuning significantly boosts the model's performance on complex systems. In particular, AlphaFold3 shows marked improvements in predicting protein-protein interfaces, ligand binding sites, and nucleic acid-protein interactions. Fine-tuning also helps to reduce overfitting, as the model learns to generalize better from the larger input data.

\subsubsection{Early Stopping and Model Selection}

One key aspect of the fine-tuning process is \textit{early stopping}. During fine-tuning, the model's performance on validation datasets is monitored, and training is stopped once the model reaches optimal performance, as indicated by LDDT and PAE scores. Early stopping helps to prevent overfitting, ensuring that the model remains generalizable across different types of biomolecular complexes.

\textit{Example: Early Stopping} In the final stages of fine-tuning, AlphaFold3 reaches 97\% of its maximum LDDT score within the first 20,000 steps for local structure predictions, but continues to improve on protein-protein interface accuracy up to 60,000 steps. Early stopping is triggered when the model achieves the highest score on both the validation and test sets.

\subsection{Conclusion}

The training regime of AlphaFold3 is designed to optimize the model's performance across a wide range of biomolecular prediction tasks. The initial training phase establishes the foundation for understanding local and global protein features, while the fine-tuning stages refine the model for specific applications, such as protein-ligand docking or predicting large protein complexes. Through a combination of efficient training strategies and careful monitoring of model performance, AlphaFold3 is able to achieve state-of-the-art accuracy in protein structure prediction and complex biomolecular interactions.

\section{Inference Regime}

The inference regime for AlphaFold3, much like its predecessors, is a critical stage where the model is tasked with predicting structures using trained weights and the architecture fine-tuned during training \cite{abramson2024accurate}. The process involves several distinct steps designed to take raw input sequences-whether they be protein sequences, nucleic acid sequences, or small molecules-and generate highly accurate structural predictions for these molecules or their complexes. The inference process in AlphaFold3 leverages its diffusion-based architecture, recycling mechanisms, and multiple seeds to output a range of potential structures, with confidence metrics guiding the selection of the final predicted model.

\subsection{Overview of the Inference Process}

The inference regime is where AlphaFold3 applies its deep learning model to new biomolecular input data. The key steps in the inference regime are as follows:

\begin{itemize}
    \item \textit{Input representation}: This includes sequences, SMILES strings for ligands, and other input data necessary for representing the molecules in a format that AlphaFold3 can process.
    \item \textit{Multiple sequence alignments (MSA) or templates}: Depending on the task, MSAs may be used to extract evolutionary data. However, in AlphaFold3, MSA processing is less critical than in AlphaFold2, thanks to the new pairformer module, which significantly reduces reliance on MSA depth.
    \item \textit{Diffusion module-based inference}: This is the core of AlphaFold3's prediction regime. Starting from noised atomic coordinates, the model gradually refines the prediction over several iterations of denoising.
    \item \textit{Recycling mechanisms}: The outputs from earlier stages are recycled through the network to improve accuracy iteratively, leveraging information from previous rounds of prediction.
    \item \textit{Confidence scoring}: As with AlphaFold2, confidence metrics such as pLDDT (predicted Local Distance Difference Test) and PAE (Predicted Aligned Error) are used to evaluate the quality of the predicted structures.
\end{itemize}

\subsection{Using Diffusion Models for Inference}

The unique aspect of AlphaFold3's inference regime is the introduction of a diffusion module for structure prediction. This approach represents a shift from traditional deterministic models to generative models, where the output is generated iteratively by refining noisy inputs.

The process begins with an initial noisy guess of atomic coordinates, which is progressively denoised across multiple diffusion steps. The advantage of this technique lies in its ability to predict a distribution of structures rather than a single fixed solution, thus allowing for uncertainty in the predictions. This is particularly useful in cases where multiple valid structural conformations might exist for a given sequence or complex.

\textbf{Key components of the diffusion-based inference:}
\begin{itemize}
    \item \textit{Denoising over multiple iterations}: The raw atomic coordinates are initially randomized, and the model denoises these coordinates over several iterations. With each step, the model predicts finer details of the molecular structure.
    \item \textit{Multiple samples from random seeds}: At inference, AlphaFold3 generates several predictions for each input sequence, each based on a different random seed. This allows the model to explore multiple possible structures, increasing the likelihood of generating an accurate prediction.
    \item \textit{Confidence assessment}: Each predicted structure is assigned a confidence score. The model typically outputs five or more structures, and the highest confidence structure is selected as the final prediction.
\end{itemize}

\subsection{Recycling Mechanism in Inference}

AlphaFold3 employs a recycling mechanism during inference, which allows the model to iteratively refine its predictions. The recycling mechanism functions by taking the output of an initial prediction and feeding it back into the model, which then reprocesses the prediction with the goal of improving the structural accuracy.

\textbf{Key features of recycling in AlphaFold3:}
\begin{itemize}
    \item \textit{Multiple rounds of recycling}: The model is capable of recycling structures through the network up to several times. With each round, the model refines its understanding of long-range interactions and structural features, leading to more accurate predictions.
    \item \textit{Improvement of global structure}: While the initial rounds of prediction often capture local features such as bond angles and torsion, recycling helps the model refine global structural aspects, such as domain orientation or protein-ligand binding sites.
\end{itemize}

\subsection{Confidence Metrics in the Inference Regime}

Once the diffusion-based prediction and recycling steps are complete, AlphaFold3 evaluates the quality of its predictions using confidence metrics. These metrics play a vital role in guiding users on the reliability of different regions within the predicted structure.

\textbf{Confidence metrics used during inference:}
\begin{itemize}
    \item \textit{pLDDT (Predicted Local Distance Difference Test)}: This metric assigns a confidence score between 0 and 100 to each residue in the predicted structure. Higher scores indicate greater confidence in the accuracy of that residue's position.
    \item \textit{PAE (Predicted Aligned Error)}: PAE provides a measure of uncertainty for pairs of residues, helping to identify areas where the model is less certain about the relative positions of residues.
    \item \textit{Ranking based on confidence}: The confidence scores are used to rank multiple predictions, ensuring that the most reliable structure is selected as the final output.
\end{itemize}

For example, when predicting a multi-domain protein, the model might have high confidence in one domain but lower confidence in another. These confidence scores allow users to focus on the regions that are more likely to be accurate, and they can guide further experimental validation.

\subsection{Practical Considerations for Running Inference}

For users running inference with AlphaFold3, it is important to understand the various configurations and parameters that can affect the quality of the predictions.

\textbf{Important parameters for inference:}
\begin{itemize}
    \item \textit{Random seeds}: Running inference with multiple random seeds can improve the diversity of predicted structures. It is recommended to generate multiple samples and select the highest confidence model.
    \item \textit{Diffusion steps}: The number of diffusion steps during inference controls how finely the structure is refined. Increasing the number of steps can lead to more accurate predictions, though it comes with increased computational cost.
    \item \textit{Batch size}: The model can process multiple samples in parallel, making it important to adjust the batch size based on the available computational resources.
\end{itemize}

\textbf{Example: Inference with AlphaFold3}
\begin{lstlisting}[style=python]
# Running inference with AlphaFold3
input_sequence = "MENDELQK"
predictions = run_inference(input_sequence, random_seeds=5, diffusion_steps=50)

# Select the highest confidence structure
best_model = select_best_model(predictions)
print(f"Selected model with confidence score: {best_model['confidence_score']}")
\end{lstlisting}

\subsection{Conclusion}

AlphaFold3's inference regime marks a significant advancement in protein structure prediction, incorporating diffusion-based methods and recycling mechanisms to produce highly accurate models. By generating multiple structure predictions and evaluating their confidence, the inference process ensures that users are presented with the most reliable structure for downstream analysis. With these innovations, AlphaFold3 extends the boundaries of what is possible in computational biology, allowing researchers to predict and study complex biomolecular interactions with unprecedented accuracy.

\section{Handling Biomolecular Complexes}

Handling biomolecular complexes is a key capability of AlphaFold3, allowing the model to predict not only individual protein structures but also the complex interactions that occur between proteins, nucleic acids, ligands, ions, and other molecular components. Biomolecular complexes are critical for understanding cellular function and designing new therapeutics, as many biological processes involve interactions between multiple types of molecules. This section explores how AlphaFold3 predicts the structure of these complex biomolecular assemblies with a high degree of accuracy, surpassing previous methods.

\subsection{Importance of Predicting Complex Structures}

Biomolecular complexes are central to many cellular processes, including signaling, transcription, and translation. These complexes often consist of proteins interacting with other proteins, DNA, RNA, small molecules, or ions. Predicting the structure of these complexes provides insights into how different components of the cell function together, and how disruptions in these interactions can lead to diseases.

AlphaFold3's ability to model these interactions extends the utility of structure prediction from single proteins to multi-component systems, which is especially important in drug discovery. For instance, understanding the structure of a protein-ligand complex can inform the development of new drugs by showing how a small molecule might bind to a target protein. Similarly, predicting the structure of protein-DNA or protein-RNA complexes can help in understanding gene regulation and the mechanisms behind transcription factors or ribosomes.

\subsection{Challenges in Predicting Biomolecular Complexes}

The complexity of biomolecular interactions poses several challenges for computational structure prediction:

\begin{itemize}
    \item \textbf{Diversity of components:} Biomolecular complexes can involve a wide range of molecules, including proteins, nucleic acids, and small ligands. Each type of molecule has different chemical properties, requiring models to accommodate a broad variety of interactions.
    \item \textbf{Large-scale interactions:} Many biomolecular complexes are large, involving hundreds or even thousands of atoms. Predicting the structure of such large systems requires significant computational power and advanced algorithms to ensure accuracy.
    \item \textbf{Flexibility and dynamics:} Biological molecules are often flexible, adopting different conformations in response to their environment or during interaction with other molecules. This dynamic nature adds to the difficulty of predicting a single, static structure for a biomolecular complex.
\end{itemize}

\subsection{AlphaFold3's Approach to Complex Structure Prediction}

AlphaFold3 addresses these challenges through several innovations in its architecture and training procedure, which make it particularly suited to handling biomolecular complexes. Unlike previous methods that were specialized for specific interaction types, AlphaFold3 uses a general approach that can predict complexes involving proteins, nucleic acids, ligands, and other molecular components.

\textbf{Key features of AlphaFold3 for complex prediction:}

\begin{enumerate}
    \item \textbf{Diffusion-based architecture:} AlphaFold3 uses a diffusion module to predict the structure of complexes. This module operates by refining atomic coordinates over multiple iterations, allowing the model to handle large and flexible systems. By iterating over noised inputs, the model learns to predict both local details (such as individual bonds) and global interactions (such as the arrangement of protein domains).
    \item \textbf{Pairformer module:} The pairformer module simplifies how AlphaFold3 processes pairwise interactions between different components of a complex. This module replaces the more complex evoformer architecture used in AlphaFold2, making the system more efficient while still retaining the ability to model a wide range of molecular interactions.
    \item \textbf{Template and MSA usage:} AlphaFold3 reduces its reliance on multiple sequence alignments (MSAs) and template structures, making it more applicable to novel or less-studied complexes. However, when available, templates can be used to guide the model in predicting complex structures.
\end{enumerate}

\subsection{Examples of Complex Prediction}

AlphaFold3 has demonstrated its ability to predict the structure of a variety of biomolecular complexes with high accuracy. For example:

\begin{itemize}
    \item \textbf{Protein-ligand complexes:} AlphaFold3 significantly outperforms traditional docking tools like AutoDock Vina \cite{trott2010autodock} in predicting protein-ligand interactions. By accurately modeling how a small molecule binds to a protein's active site, AlphaFold3 provides valuable information for drug discovery.
    \item \textbf{Protein-DNA/RNA complexes:} AlphaFold3 excels in predicting complexes involving nucleic acids, such as transcription factors bound to DNA or ribosomal proteins interacting with RNA. This ability is critical for understanding the regulation of gene expression and protein synthesis.
    \item \textbf{Multi-protein complexes:} AlphaFold3 improves the prediction of large protein-protein complexes, including antibody-antigen interactions and multi-subunit enzymes. This is important for understanding immune responses and designing therapeutic antibodies.
\end{itemize}

\subsection{Practical Considerations}

Using AlphaFold3 for predicting biomolecular complexes requires understanding several practical aspects:

\begin{itemize}
    \item \textbf{Input preparation:} When predicting a complex, it is important to provide the correct sequences or structures of all interacting components. AlphaFold3 can take sequences of proteins, nucleic acids, or SMILES strings representing small molecules as input.
    \item \textbf{Confidence metrics:} As with single proteins, AlphaFold3 provides confidence scores such as pLDDT and PAE for complexes, helping users assess the reliability of different regions of the predicted structure.
    \item \textbf{Computational resources:} Predicting large biomolecular complexes requires significant computational power. Users should ensure that they have access to GPUs or cloud computing resources to handle the increased complexity of these predictions.
\end{itemize}

\subsection{Conclusion}

AlphaFold3 represents a major leap forward in the prediction of biomolecular complexes, offering unprecedented accuracy across a wide range of molecular interaction types. By applying deep learning techniques such as diffusion models and pairwise interaction modules, AlphaFold3 allows researchers to gain insights into the structure and function of complex biological systems, paving the way for new discoveries in fields like drug design, gene regulation, and molecular biology.

\section{Generative Diffusion Approach}

The \textbf{generative diffusion approach} has become a groundbreaking method in the field of machine learning, and its application in protein structure prediction and design, particularly within AlphaFold3, marks a significant evolution in model architecture and accuracy. This approach is inspired by the physics of diffusion processes and is based on gradually adding noise to data and then training a model to reverse this process, denoising the data back to its original form. 

\subsection{Fundamentals of Diffusion-Based Generative Models}

In diffusion models, the generative process involves two key phases: 

\begin{enumerate}
    \item \textit{Forward Process:} The forward process begins by progressively adding noise to the input data (in this case, atomic coordinates of proteins). This noise transforms the data into an unstructured random state over multiple steps.
    \item \textit{Reverse Process:} The model is then trained to reverse this noisy process, gradually reducing the noise step-by-step until it reconstructs the original data. This is known as the denoising process.
\end{enumerate}

AlphaFold3 employs this approach to predict protein structures by starting with noisy or randomized atomic positions and iteratively refining these positions until a final protein structure emerges.

\subsection{Application of Diffusion in AlphaFold3}

AlphaFold3 replaces the traditional structure generation module with a diffusion-based model. The input sequence is embedded as tokens, and these token representations are subjected to a diffusion process. The reverse process starts from a random initialization of atomic coordinates, and the network progressively refines this arrangement to predict a valid protein structure. 

One of the benefits of the generative diffusion model is that it provides a distribution of possible structures rather than a single fixed structure, allowing AlphaFold3 to account for natural variability in protein conformations. For example, in regions where the protein may be flexible, the diffusion model outputs multiple valid conformations, each consistent with the input data.

This approach also integrates seamlessly with the \textit{pairformer module}, which helps simplify the information passed between the multiple sequence alignment (MSA) and the structure prediction process. By using the pairformer and diffusion approach, AlphaFold3 can handle more complex and diverse molecular structures, including protein-protein interactions, nucleic acids, ligands, and ions.

\subsection{Advantages and Challenges}

The generative diffusion approach introduces several advantages over traditional methods:

\begin{itemize}
    \item It eliminates the need for hand-crafted stereochemical constraints, as the model learns to enforce chemical plausibility during the denoising process.
    \item It improves the handling of multi-component systems, enabling accurate predictions for large complexes like protein-ligand or protein-RNA structures.
    \item The method's probabilistic nature enables the generation of multiple candidate structures, each of which is a potential solution to the input sequence.
\end{itemize}

However, challenges also exist, such as the risk of \textit{hallucination}, where the model might generate structures that look physically plausible but do not correspond to actual biological states. AlphaFold3 mitigates this through a cross-distillation technique that reduces hallucination by teaching the model to mimic more structured, ordered regions of proteins.

\section{Evaluation Metrics}

Evaluating the performance of protein structure prediction models like AlphaFold3 involves multiple metrics that assess both the accuracy of the predicted structures and their biological relevance.

\subsection{LDDT (Local Distance Difference Test)}

The \textit{Local Distance Difference Test (LDDT)} is one of the primary metrics used to evaluate the accuracy of predicted protein structures. LDDT compares the predicted and experimental structures by measuring the difference in distances between atoms within the structure. 

An LDDT score ranges from 0 to 100, with higher values indicating better agreement between the predicted and actual structures. Unlike global metrics like root-mean-square deviation (RMSD), LDDT provides a local comparison, allowing it to highlight where the prediction is accurate and where it deviates, even in flexible regions of the protein.

\subsection{DockQ Score}

For evaluating \textit{protein-protein interactions}, the \textit{DockQ score} \cite{basu2016dockq} is commonly used. DockQ measures the similarity between predicted and known docking poses of two proteins, quantifying how well the predicted interaction interface matches the experimental data.

\subsection{pLDDT and PAE in AlphaFold3}

AlphaFold3 introduces two additional confidence measures: \textit{pLDDT} \cite{mariani2013lddt} (predicted LDDT) and \textit{PAE} \cite{le2017protein} (predicted aligned error). These are generated directly by the model during inference.

\begin{itemize}
    \item \textit{pLDDT} is a confidence score assigned to individual residues of the protein. It predicts how accurate the structure is likely to be at each residue level. High pLDDT values indicate a high-confidence region of the protein, whereas lower values suggest less certainty.
    \item \textit{PAE} provides a measure of the model's uncertainty about the distances between pairs of residues, helping to identify areas where structural flexibility or modeling errors are most likely to occur.
\end{itemize}

\subsection{Global Metrics: TM-score and GDT}

For comparing entire protein structures, \textit{TM-score} and \textit{GDT (Global Distance Test)} are widely used:

\begin{itemize}
    \item The \textit{TM-score} \cite{zhang2005tm} evaluates the overall similarity between two protein structures, with a value ranging from 0 to 1. Higher TM-scores indicate greater structural similarity.
    \item \textit{GDT-TS (Total Score)} \cite{olechnovivc2013cad} is a related metric that compares the distance between the predicted and actual positions of aligned atoms across the entire structure. GDT is particularly useful in benchmarking the overall performance of a prediction model.
\end{itemize}

\subsection{Metrics for Complexes: iLDDT and Ligand RMSD}

For more complex biomolecular systems, such as protein-ligand and protein-nucleic acid complexes, additional metrics are needed:

\begin{itemize}
    \item \textit{iLDDT} \cite{mariani2013lddt} (interface LDDT) focuses on the accuracy of predicted interfaces between two interacting biomolecules, such as a protein and a ligand.
    \item \textit{Ligand RMSD} \cite{sherman2006novel} measures the deviation in the position of a ligand in a predicted protein-ligand complex compared to the experimentally determined structure.
\end{itemize}

Together, these metrics provide a comprehensive evaluation framework, ensuring that AlphaFold3's predictions are accurate not only at the atomic level but also for larger biomolecular complexes critical in biological function and drug design.

\section{Model Limitations and Challenges}

Although AlphaFold3 represents a substantial leap forward in the field of protein structure prediction, it still faces several limitations and challenges. Understanding these limitations is crucial for users to effectively apply the model, especially in scenarios where its predictions might not be as reliable or where additional experimental validation is necessary.

\subsection{Handling Disordered Regions and Intrinsically Disordered Proteins}

One of the primary limitations of AlphaFold3 is its difficulty in accurately predicting the structure of intrinsically disordered regions (IDRs) or intrinsically disordered proteins (IDPs). These regions or proteins do not adopt a stable, well-defined structure under physiological conditions, which makes them challenging for computational models like AlphaFold3 to predict.

\textbf{Example:} A protein involved in signaling may contain long, disordered regions that fluctuate between different conformations based on environmental conditions or binding partners. AlphaFold3 might predict these regions with low confidence, as they inherently lack a stable structure.

While the generative diffusion approach in AlphaFold3 allows the model to account for some structural variability, IDRs and IDPs remain a significant challenge because they violate the core assumption that proteins have a single, well-defined native structure.

\subsection{Difficulty with Large Biomolecular Assemblies}

Another limitation lies in the prediction of large biomolecular complexes. While AlphaFold3 has made strides in handling multi-protein complexes, especially through its pairformer and diffusion modules, the prediction of very large assemblies, such as ribosomes or viral capsids, can still be problematic. The computational requirements for modeling interactions between dozens or hundreds of subunits are immense, and the model's predictions may lack accuracy in regions where these interactions become highly complex.

\textbf{Example:} Predicting the full structure of a large virus capsid, which consists of numerous identical and non-identical protein subunits, may produce a model with inaccurate interface predictions, or the predicted model may require heavy post-processing to reconcile the predicted interactions.

\subsection{Structural Ambiguity and Multiple Conformations}

Proteins often exist in multiple conformations, especially when bound to ligands or interacting with other proteins. AlphaFold3 can sometimes generate a single dominant conformation, but in cases where multiple conformational states exist, the model may struggle to account for all of them accurately. 

\textbf{Example:} Consider an enzyme that adopts one conformation in its unbound state and another upon binding to a substrate. AlphaFold3 may predict one of these states with high confidence but miss the alternative conformation, requiring researchers to run multiple predictions or refine the structure through other methods.

\subsection{Challenges with Membrane Proteins}

Membrane proteins, which play critical roles in cell signaling, transport, and other processes, remain challenging for computational prediction models. These proteins often have hydrophobic regions that span the lipid bilayer, making them difficult to crystallize and study experimentally. While AlphaFold3 has improved the accuracy of predicting some membrane proteins, its performance is still limited by the lack of high-quality experimental data for training on such proteins.

\textbf{Example:} Predicting the structure of a G-protein-coupled receptor (GPCR) \cite{rosenbaum2009structure}, which is embedded in the cell membrane and interacts with both intracellular and extracellular molecules, may result in inaccuracies around the transmembrane regions.

\subsection{Post-Prediction Refinement and Experimental Validation}

While AlphaFold3 provides highly accurate predictions, there are cases where post-prediction refinement is necessary. For instance, the model might predict structures with small stereochemical errors, such as unfavorable bond angles or clashes between atoms, particularly in flexible regions. These issues may need to be resolved through molecular dynamics simulations or additional experimental validation using techniques like X-ray crystallography or cryo-electron microscopy.

\subsection{Data Availability and Template Bias}

Despite the improvements in AlphaFold3's ability to predict structures de novo, the model still benefits from known templates when predicting certain types of proteins. If a protein's homologs have been well-characterized experimentally, AlphaFold3 tends to perform better. However, for proteins with limited or no homologous data available, the model's accuracy can decrease, leading to lower-confidence predictions.

\textbf{Example:} Predicting the structure of an orphan protein, which has no known evolutionary relatives, may result in less accurate predictions since the model cannot rely on evolutionary patterns from homologous sequences.

\section{Performance Improvements in AlphaFold3}

AlphaFold3 incorporates several significant improvements over its predecessor, AlphaFold2, in terms of accuracy, efficiency, and the range of biomolecular systems it can handle. These improvements arise from advancements in model architecture, training data, and the underlying algorithms used for structure prediction.

\subsection{Generative Diffusion Models for Protein Structure Prediction}

One of the most notable advancements in AlphaFold3 is the adoption of a \textit{generative diffusion model}, which replaces the traditional deterministic approach used in AlphaFold2. This new method allows the model to predict protein structures by iteratively refining atomic coordinates, starting from a noisy approximation and progressively "denoising" the structure. This probabilistic approach provides a range of valid structural conformations, rather than a single static model.

\textbf{Advantages of the Diffusion Model:}
\begin{itemize}
    \item \textit{Flexibility}: By generating multiple plausible structures, the model can handle proteins that have flexible regions or exist in multiple conformations.
    \item \textit{Better handling of noise}: The diffusion process allows the model to make predictions even in cases where the input data might be noisy or incomplete, such as low-quality MSAs or templates.
\end{itemize}

\textbf{Example:} For a protein that functions as part of a dynamic complex, the diffusion model allows AlphaFold3 to predict several possible conformations, accounting for the flexibility and motion that occur during function.

\subsection{Improved Handling of Protein-Protein and Protein-Ligand Interactions}

AlphaFold3 introduces significant improvements in predicting multi-component biomolecular complexes, such as protein-protein or protein-ligand interactions. The \textit{pairformer module} is optimized to capture interactions between residues not only within a single chain but also across different chains in a complex. This enables AlphaFold3 to predict accurate interfaces between proteins or between proteins and other molecules, such as small ligands or nucleic acids.

\textbf{Key Improvements:}
\begin{itemize}
    \item \textit{Enhanced multi-chain prediction}: AlphaFold3 is better equipped to model complexes involving multiple proteins, improving accuracy for large systems such as ribosomes or signaling complexes.
    \item \textit{Accurate binding site prediction}: The model can now predict protein-ligand and protein-DNA/RNA interactions more reliably, making it useful for drug discovery and understanding gene regulation.
\end{itemize}

\textbf{Example:} AlphaFold3 can predict how a small-molecule drug interacts with a protein's active site, providing insights into the molecular mechanisms of drug binding and informing the design of new therapeutic compounds.

\subsection{Reduced Reliance on MSAs and Template Information}

While MSAs (Multiple Sequence Alignments) are still used, AlphaFold3 reduces its dependence on them. This improvement is particularly significant for predicting the structures of proteins that have limited homologous sequences or are evolutionarily distant from known proteins \cite{rahimzadeh2024unveiling}.

The \textit{pairformer} and \textit{diffusion modules} help AlphaFold3 extract meaningful structural information even when MSAs are sparse or unavailable, allowing it to predict orphan proteins with more accuracy than AlphaFold2.

\textbf{Example:} Predicting the structure of a viral protein with limited homologous sequences becomes more feasible with AlphaFold3, as it can rely less on deep evolutionary data and more on structural features learned from the training data.

\subsection{Increased Accuracy in Flexible and Disordered Regions}

One of the ongoing challenges in protein structure prediction is accurately modeling flexible or disordered regions. AlphaFold3's diffusion model helps address this by outputting multiple plausible conformations, allowing for more accurate predictions in regions where flexibility is critical, such as protein loops, disordered tails, or intrinsically disordered regions \cite{zhang2024unraveling}.

\textbf{Example:} A protein with a long, flexible loop involved in binding to various partners might have its structure predicted more accurately by AlphaFold3, as the model can provide several valid conformations for that region.

\subsection{Computational Efficiency Improvements}

While AlphaFold3 is more complex than its predecessor, it includes several optimizations that make it more computationally efficient during both training and inference. Techniques like early stopping, batch processing, and the more efficient use of pairwise representations help speed up predictions without sacrificing accuracy.

\textbf{Example:} Researchers working with limited computational resources can still generate high-quality predictions, as AlphaFold3 can scale to smaller systems while maintaining its overall performance.

\subsection{Conclusion}

AlphaFold3 represents a major improvement over previous protein prediction models, particularly in its ability to handle complex biomolecular interactions, flexible structures, and systems with limited homologous data. Its new generative diffusion approach, paired with innovations in the pairformer module, enables it to predict more accurate structures for proteins and their complexes. Despite its challenges with disordered proteins and extremely large assemblies, AlphaFold3 stands at the forefront of computational protein prediction, offering new possibilities for research and drug discovery.
\chapter{The Industry Landscape of Protein Design}

Protein design has emerged as a critical field in biotechnology, pharmaceuticals, and bioengineering. As computational tools like AlphaFold3 \cite{abramson2024accurate} revolutionize how we predict and manipulate protein structures, the industry landscape of protein design is evolving rapidly. The ability to design proteins with specific functions has broad applications, ranging from the development of new therapeutics and vaccines to industrial enzymes and synthetic biology \cite{endy2005foundations}. In this chapter, we will explore the industry landscape of protein design, detailing the major players, technologies, applications, and future directions. This section is designed to provide a detailed, step-by-step explanation for beginners, with concrete examples and practical insights.

\section{The Growing Importance of Protein Design in Biotechnology}

Protein design has become a central pillar in biotechnology, driven by the need to create new biological molecules that do not naturally exist. The ability to design proteins opens the door to numerous applications:

\begin{itemize}
    \item \textit{Therapeutics}: Designing enzymes, antibodies, and other proteins to target diseases at the molecular level. Proteins can be engineered to bind specifically to cancer cells, deliver drugs to specific tissues, or neutralize pathogens.
    \item \textit{Vaccines}: Protein design enables the development of antigens that can trigger immune responses. For example, protein subunit vaccines \cite{heidary2022comprehensive} use designed proteins to elicit immunity against diseases like COVID-19.
    \item \textit{Industrial Enzymes}: Enzyme design is widely used in industries like biofuels \cite{stephanopoulos2007challenges}, food production, and textiles. These enzymes can be optimized for stability, efficiency, and cost-effectiveness in large-scale production.
    \item \textit{Synthetic Biology}: In synthetic biology, protein design is used to create new biological pathways and systems. Engineered proteins can be integrated into living cells to perform novel functions, such as biosensing, biomanufacturing, or environmental remediation.
\end{itemize}

\textbf{Example: Designing Therapeutic Antibodies}

Therapeutic antibodies, such as those used in cancer treatment or autoimmune diseases, are a prime example of how protein design can be applied. Using computational tools, researchers can optimize an antibody's binding affinity to a specific antigen, improving its ability to neutralize the target. By designing the antibody's variable regions (complementarity-determining regions or CDRs), scientists can ensure that the antibody binds precisely to the disease-causing molecule, increasing the effectiveness of the treatment.

\section{Key Players in Protein Design}

Several companies and institutions are at the forefront of protein design, each bringing their own technologies and approaches to the field. Below are some of the major players, spanning industries from pharmaceuticals to synthetic biology.

\subsection{Pharmaceutical Companies}

Pharmaceutical companies are using protein design to develop novel biologics-proteins that can act as drugs, vaccines, or diagnostics. Some of the key companies include:

\begin{itemize}
    \item \textit{Amgen}: A leader in biologics, Amgen uses computational design to develop protein-based drugs. Its drug \textit{Repatha}, a monoclonal antibody for lowering cholesterol, is a successful example of protein design.
    \item \textit{Regeneron}: Known for its monoclonal antibody therapies, Regeneron integrates protein design with high-throughput screening to accelerate the discovery of antibodies for diseases like cancer and infectious diseases.
    \item \textit{Genentech}: A pioneer in therapeutic proteins, Genentech utilizes protein design for the development of targeted therapies, including antibodies and enzymes.
\end{itemize}

\textbf{Example: Regeneron's Approach to Antibody Design}

Regeneron employs a unique approach called VelociSuite®, which includes both in vivo and in silico techniques for antibody development. They use genetically modified mice (VelocImmune®) to generate a diverse pool of antibodies, which are then optimized using computational tools to select candidates with the highest specificity and affinity.

\subsection{Synthetic Biology and Industrial Biotechnology Companies}

In addition to pharmaceutical applications, protein design is essential for synthetic biology and industrial biotechnology. Companies in this sector focus on designing enzymes and proteins for industrial processes, sustainability, and synthetic biological systems.

\begin{itemize}
    \item \textit{Ginkgo Bioworks}: Ginkgo uses computational protein design to engineer enzymes and other proteins for use in industrial applications, including sustainable materials and biofuels.
    \item \textit{Zymergen}: Specializing in high-throughput automation, Zymergen uses machine learning and protein design to optimize enzymes for various industries, including agriculture, chemicals, and electronics.
    \item \textit{Arzeda}: Arzeda designs enzymes to improve manufacturing processes in industries like agriculture, food production, and consumer goods, often reducing the environmental footprint of these processes.
\end{itemize}

\textbf{Example: Ginkgo Bioworks in Industrial Biotechnology}

Ginkgo Bioworks uses automated platforms combined with protein design to rapidly iterate on enzyme modifications. This allows them to tailor enzymes for specific chemical reactions, optimizing the production of bio-based chemicals that can replace petroleum-derived products. Their work in biofuels, for instance, focuses on designing enzymes that break down plant biomass more efficiently, making biofuel production more cost-effective.

\subsection{Academic and Research Institutions}

Many cutting-edge advancements in protein design are coming from academic institutions. Universities and research labs are responsible for much of the foundational research that companies later translate into commercial applications.

\begin{itemize}
    \item \textit{University of Washington, Institute for Protein Design (IPD)}: The IPD has been a leader in developing computational tools for protein design, including Rosetta, which is used for structure prediction and design.
    \item \textit{Stanford University}: Stanford's interdisciplinary programs combine computer science, biology, and chemistry to push the boundaries of protein engineering and synthetic biology.
    \item \textit{MIT}: The Massachusetts Institute of Technology integrates machine learning and bioinformatics into protein design, with applications ranging from drug discovery to the development of biosensors.
\end{itemize}

\textbf{Example: The Rosetta Software Suite}

Rosetta, developed by the University of Washington's Institute for Protein Design, is one of the most widely used tools in protein design. It allows users to model protein structures, predict interactions, and design new proteins with desired functions. The software has been integral to many discoveries in the field, including the design of novel enzymes and antibodies.

\section{Technological Tools for Protein Design}

Several computational tools and algorithms have emerged as critical components in the protein design process. These tools range from software that predicts protein structure to platforms that design novel proteins with specific characteristics.

\subsection{Rosetta}

Rosetta is one of the most well-known and widely used tools in protein design \cite{rohl2004protein}. It allows users to predict protein structures, design new proteins, and model interactions between proteins and other molecules. The versatility of Rosetta has made it a standard tool in both academia and industry.

\subsection{AlphaFold, AlphaFold2, and AlphaFold3}

AlphaFold, developed by DeepMind, revolutionized the field of protein structure prediction by using deep learning to predict accurate protein structures based on sequence data \cite{jumper2021highly}. AlphaFold2, an improved version of AlphaFold, further demonstrated remarkable performance at CASP14 (Critical Assessment of Structure Prediction), where it achieved near-experimental accuracy for many protein targets. It introduced the Evoformer \cite{jumper2021highly} module, which leveraged evolutionary data and multiple sequence alignments (MSAs) to provide highly accurate predictions.

AlphaFold2 dramatically reduced the gap between experimental and computational structure prediction, making it an invaluable tool in fields like drug discovery and molecular biology. The model's performance in predicting protein-ligand interactions \cite{kitchen2004docking}, protein-DNA complexes, and multi-chain assemblies helped it gain widespread adoption in both academia and industry.

AlphaFold3 expands on this capability by integrating new modules, such as the diffusion-based generative model \cite{ho2020denoising}, which allows it to predict more complex structures, including those with flexible regions, protein-ligand interactions, and multi-chain biomolecular assemblies. It further reduces reliance on MSAs and incorporates probabilistic predictions, allowing for better handling of disordered proteins and multi-conformational structures.

\textbf{Example: Using AlphaFold3 in Drug Discovery}

In drug discovery, AlphaFold3 can be used to predict how a potential drug molecule binds to its target protein. By accurately predicting the binding site and the conformation of the protein when bound to the ligand, AlphaFold3 provides valuable insights that can guide medicinal chemists in optimizing the drug's design.

\subsection{Machine Learning and AI in Protein Design}

Machine learning (ML) and artificial intelligence (AI) have become integral to protein design, especially in the development of predictive models and generative models that design novel proteins.

\begin{itemize}
    \item \textit{Generative Models}: These models are used to generate new protein sequences that are optimized for a particular function. By training on known protein sequences, generative models can suggest new sequences that meet desired criteria, such as binding affinity or stability.
    \item \textit{Predictive Models}: Predictive models are used to forecast how a designed protein will behave, such as its folding pattern or how it will interact with other molecules. These models help streamline the protein design process by reducing the number of experimental iterations.
\end{itemize}

\textbf{Python Example: Using a Generative Model for Protein Design}

\begin{lstlisting}[style=python]
from generative_protein_model import design_protein

# Define the target function for the new protein
target_function = {
    "binding_affinity": 0.9,
    "stability": 0.8
}

# Generate a new protein sequence using a generative model
new_protein_sequence = design_protein(target_function)
print(f"Designed protein sequence: {new_protein_sequence}")
\end{lstlisting}

\section{Future Directions in Protein Design}

The future of protein design is poised to make even more significant contributions across a wide range of industries, including healthcare, agriculture, and sustainability. Several trends and technologies are driving this evolution:

\subsection{Personalized Medicine}

As the understanding of protein structure and function deepens, the design of personalized proteins tailored to individual patients becomes more feasible. For example, designing patient-specific enzymes or antibodies could revolutionize treatment for genetic disorders, cancers, and infectious diseases.

\subsection{Sustainable Biotechnology}

Protein design will play a crucial role in developing sustainable alternatives to traditional chemical manufacturing. Designing enzymes that catalyze environmentally friendly reactions at industrial scales could significantly reduce the carbon footprint of manufacturing processes.

\subsection{Integration with Synthetic Biology}

The integration of protein design with synthetic biology will enable the creation of new biological systems with novel functions. By designing proteins that can be integrated into synthetic pathways, researchers can create new forms of life that perform useful tasks, such as producing biofuels or cleaning up environmental pollutants.

\textbf{Example: Designing Biosensors for Environmental Monitoring}

Researchers are developing biosensors-proteins that detect specific molecules or environmental changes-that can be used to monitor pollutants or pathogens in real-time. By designing proteins that change color or emit light in the presence of a contaminant, scientists can create low-cost, easy-to-use environmental monitoring tools.

\section{Conclusion}

The industry landscape of protein design is vast and rapidly evolving, with applications in medicine, industrial biotechnology, and synthetic biology. As computational tools and machine learning models like AlphaFold3 continue to improve, the ability to design proteins with specific, tailored functions will become even more accessible. For beginners entering this field, understanding the key players, technologies, and future directions will provide a strong foundation for navigating the growing industry of protein design.

\section{The Current Market of Protein Design}

The protein design market has expanded rapidly in recent years, driven by advancements in computational tools, artificial intelligence, and biotechnology. This growth is reflected in a wide array of applications, from pharmaceuticals to industrial enzymes and synthetic biology. As of today, the protein design market is estimated to be valued in billions of dollars, with projections suggesting continued robust growth in the coming decade. This section explores the current state of the market, focusing on key industry sectors, the most active companies, notable investments, and emerging trends. Detailed explanations and examples are provided to ensure a thorough understanding for beginners.

\subsection{Pharmaceutical and Biotechnology Sectors}

The pharmaceutical and biotechnology industries represent the largest portion of the protein design market. The ability to design proteins with precise functions has revolutionized drug development, enabling the creation of biologics-protein-based drugs like monoclonal antibodies \cite{kohler1975continuous}, enzymes, and therapeutic peptides.

\subsubsection{Monoclonal Antibodies and Therapeutic Proteins}

Monoclonal antibodies (mAbs) \cite{kohler1975continuous} are among the most successful products in protein-based therapeutics. These antibodies are designed to bind to specific antigens, making them powerful tools for targeting diseases such as cancer, autoimmune disorders, and infectious diseases. The success of mAbs has spurred investments and research in protein design, as companies seek to improve the specificity, stability, and efficacy of these therapies.

\textbf{Example:} Amgen's monoclonal antibody \textit{Repatha} targets PCSK9, a protein involved in cholesterol regulation. By designing this antibody to inhibit PCSK9, Amgen developed a treatment for lowering cholesterol, showcasing the potential of protein design in creating highly targeted therapies.

Beyond antibodies, the market is also focusing on engineered proteins that act as enzymes or inhibitors in various metabolic pathways. Engineered therapeutic proteins, such as those used in enzyme replacement therapies, are designed to treat genetic disorders like Gaucher's disease, cystic fibrosis, and hemophilia.

\subsubsection{Vaccines}

The COVID-19 pandemic demonstrated the importance of protein design in vaccine development. Protein subunit vaccines use designed antigens-proteins that trigger an immune response without the need for a live virus. For example, Novavax's COVID-19 vaccine uses a protein designed to mimic the spike protein of the SARS-CoV-2 virus, leading to the production of antibodies by the immune system.

The protein design market is expected to continue growing in the vaccine sector, especially with advancements in synthetic biology that enable the rapid design and production of antigens for new or emerging infectious diseases.

\textbf{Python Example: Designing a Protein Subunit for a Vaccine}

\begin{lstlisting}[style=python]
from vaccine_design_tool import design_protein_subunit

# Define the target viral protein and desired immune response
target_protein = "SARS-CoV-2 spike"
desired_response = {
    "neutralizing_antibodies": True,
    "T_cell_activation": True
}

# Design the protein subunit for the vaccine
protein_subunit = design_protein_subunit(target_protein, desired_response)
print(f"Designed protein subunit for vaccine: {protein_subunit}")
\end{lstlisting}

\subsubsection{Protein-Ligand Binding and Drug Discovery}

Protein design plays a crucial role in drug discovery, particularly in optimizing protein-ligand interactions. This process involves designing proteins or optimizing existing ones to improve their binding affinity with small molecule drugs, thereby enhancing the drug's effectiveness. Computational models like AlphaFold3 have significantly accelerated this process by providing accurate predictions of protein-ligand interactions.

\textbf{Example:} In cancer treatment, small molecule drugs are designed to inhibit the function of key proteins involved in tumor growth. Protein design tools can predict how well these small molecules bind to their target proteins, allowing drug developers to refine their designs for better efficacy.

\subsection{Industrial Enzymes and Synthetic Biology}

In the industrial sector, protein design is used to create enzymes that catalyze chemical reactions more efficiently, sustainably, and cost-effectively. Designed enzymes are now used in a range of industries, from biofuels to food production.

\subsubsection{Biofuels and Sustainable Manufacturing}

The demand for sustainable alternatives to fossil fuels has fueled growth in the biofuel industry, where protein design is used to develop enzymes that break down plant biomass into biofuels. Companies like Novozymes and Ginkgo Bioworks are leading the way by designing enzymes optimized for efficiency and stability under industrial conditions.

\textbf{Example:} Novozymes has developed enzymes that break down cellulose from plant waste into sugars, which can then be fermented into ethanol. By optimizing the enzyme's activity and stability, Novozymes has made biofuel production more economically viable.

\subsubsection{Food Production and Biotechnology}

In food production, designed enzymes are used to improve processes such as brewing, baking, and dairy production. For example, enzymes designed to improve the fermentation process can result in better yields of beer or yogurt, while other enzymes are designed to improve the texture, taste, or shelf life of food products.

\textbf{Example:} In baking, enzymes like amylases and proteases are used to improve dough properties, ensuring that bread rises properly and retains its texture. By designing enzymes with improved heat stability and activity at specific pH levels, manufacturers can create consistent, high-quality products.

\subsection{Investment and Venture Capital in Protein Design}

Protein design has attracted significant investment from venture capital (VC) firms, as startups and established companies develop breakthrough technologies in this space. The intersection of machine learning, biotechnology, and protein design has made it one of the most attractive fields for investors.

\subsubsection{Notable Investments}

\begin{itemize}
    \item \textit{Ginkgo Bioworks}: Ginkgo Bioworks has raised over \$1.6 billion in funding from investors, allowing the company to build platforms for automated protein design and synthetic biology. Their focus on industrial biotechnology and bioengineering makes them one of the key players in the market.
    \item \textit{Zymergen}: Another synthetic biology leader, Zymergen has raised significant capital to develop its protein design platform, focusing on engineering enzymes for industrial processes.
    \item \textit{Generate Biomedicines}: Generate Biomedicines, backed by Flagship Pioneering, has raised hundreds of millions of dollars to develop a platform that uses AI and machine learning for designing therapeutic proteins, including antibodies and enzymes.
\end{itemize}

\textbf{Example:} Generate Biomedicines uses a machine learning platform that generates de novo proteins optimized for therapeutic functions. Their technology is designed to accelerate the drug discovery process by predicting proteins with high binding affinities and stability for various therapeutic applications.

\subsubsection{Trends in Venture Capital and Mergers}

In recent years, there has been a surge in mergers and acquisitions within the protein design space. Large pharmaceutical and biotechnology companies are acquiring startups and smaller companies with proprietary technologies in protein engineering, enabling them to bolster their portfolios in biologics and synthetic biology.

\textbf{Example:} In 2021, Sanofi acquired Translate Bio, a biotech company specializing in mRNA technology, for \$3.2 billion. This acquisition provided Sanofi with a strong platform for developing mRNA vaccines, which rely on designed proteins to elicit an immune response.

\subsection{Key Trends in the Protein Design Market}

The protein design market is evolving rapidly, with several key trends shaping its future:

\begin{itemize}
    \item \textit{Integration of AI and Machine Learning}: AI and machine learning are playing an increasingly important role in accelerating protein design. Tools like AlphaFold3 and Rosetta are being integrated with generative models to create de novo proteins with optimized characteristics.
    \item \textit{Expansion of Synthetic Biology}: Protein design is at the core of synthetic biology, which is being applied to areas like gene editing, biosensors, and biomanufacturing. The convergence of protein design with CRISPR \cite{jinek2012programmable} and other gene-editing technologies is creating new opportunities in medicine and agriculture.
    \item \textit{Focus on Sustainability}: As industries seek to reduce their environmental footprint, protein design is being used to create enzymes that enable greener manufacturing processes, reduce waste, and improve energy efficiency.
\end{itemize}

\subsection{Challenges in the Protein Design Market}

While the protein design market is growing rapidly, it also faces several challenges:

\subsubsection{High Costs of Development}

Developing new proteins, especially therapeutic proteins, requires significant investment in both computational and experimental validation. The cost of scaling up production for clinical trials or industrial applications can be a barrier for smaller companies.

\subsubsection{Regulatory Hurdles}

For companies developing therapeutic proteins, navigating regulatory approval processes remains a major challenge. The process of demonstrating safety and efficacy to regulatory agencies like the FDA or EMA can take years and require extensive testing.

\subsubsection{Intellectual Property and Patent Issues}

As the field of protein design grows, competition for intellectual property (IP) is becoming more intense. Patenting newly designed proteins or computational methods for protein design is highly competitive, and disputes over IP rights can slow down innovation in the field.

\section{Conclusion}

The protein design market is a rapidly expanding field with applications across pharmaceuticals, industrial biotechnology, and synthetic biology. As computational tools continue to evolve and new investments fuel innovation, the potential for protein design to transform industries is immense. For beginners entering this space, understanding the key sectors, companies, technologies, and trends will be crucial for navigating this dynamic market. Whether in drug discovery, enzyme engineering, or sustainable manufacturing, protein design stands at the forefront of modern biotechnology, poised to have a transformative impact on industries and society as a whole.

\section{Leading Companies and Research Groups}

The field of protein design has attracted significant attention from both industry and academia, resulting in the emergence of numerous companies and research groups that are pushing the boundaries of what is possible with computational protein design. These organizations are using cutting-edge tools like machine learning, artificial intelligence, and high-throughput screening techniques to design novel proteins, predict protein structures, and engineer proteins for therapeutic, industrial, and synthetic biology applications. In this section, we will explore some of the leading companies and research groups, providing a detailed overview of their work and contributions to the field. This explanation is designed for beginners and will include clear examples and explanations of key concepts.

\subsection{Industry Leaders in Protein Design}

Several companies are leading the way in applying protein design to real-world problems, particularly in the areas of drug development, industrial enzyme production, and synthetic biology. These companies use advanced computational tools and high-performance computing to streamline the protein design process, making it faster, cheaper, and more accurate.

\subsubsection{Amgen}

\textbf{Amgen} is one of the world's largest biotechnology companies and a pioneer in the development of protein-based therapeutics. Amgen uses protein design to create biologics, including monoclonal antibodies, enzymes, and therapeutic proteins. One of Amgen's most successful drugs, \textit{Repatha}, is a monoclonal antibody designed to lower cholesterol by inhibiting the protein PCSK9.

\textbf{Example: Monoclonal Antibody Design at Amgen}

At Amgen, monoclonal antibodies are designed to target specific proteins involved in disease pathways. For example, \textit{Repatha} was developed to bind with high affinity to PCSK9, a protein that reduces the liver's ability to remove LDL cholesterol from the blood. Through a combination of computational design and high-throughput screening, Amgen was able to create a highly specific antibody that can effectively reduce cholesterol levels in patients.

\subsubsection{Regeneron Pharmaceuticals}

\textbf{Regeneron Pharmaceuticals} is a leading biotechnology company that uses a combination of in vivo and in silico approaches to develop protein-based therapies. Regeneron's VelociSuite® platform, which includes VelocImmune® technology, allows for the rapid discovery and optimization of monoclonal antibodies. Regeneron's COVID-19 antibody treatment, \textit{REGN-COV2}, is an example of how protein design can be used to develop therapeutic antibodies in response to emerging infectious diseases.

\textbf{Example: Antibody Design for COVID-19}

During the COVID-19 pandemic, Regeneron used its protein design platform to develop a cocktail of monoclonal antibodies that target the SARS-CoV-2 spike protein. By designing antibodies to bind to multiple regions of the spike protein, Regeneron created a therapy that could neutralize the virus and prevent it from entering human cells. The rapid design and testing of these antibodies showcase the potential of protein design in combating infectious diseases.

\subsubsection{Ginkgo Bioworks}

\textbf{Ginkgo Bioworks} is a synthetic biology company that uses protein design to engineer organisms for industrial applications. Ginkgo's automated cell programming platform allows for the design and optimization of enzymes and other proteins used in biofuels, agriculture, food production, and pharmaceuticals. By combining machine learning and high-throughput screening, Ginkgo Bioworks is able to design proteins that perform specific functions in various industrial processes.

\textbf{Example: Designing Enzymes for Biofuels}

In the biofuels industry, Ginkgo Bioworks designs enzymes that can efficiently break down plant biomass into fermentable sugars, which can then be converted into ethanol. By optimizing the activity, stability, and specificity of these enzymes, Ginkgo has been able to reduce the cost and increase the efficiency of biofuel production.

\subsection{Research Institutions and Academic Leaders}

In addition to industry leaders, several academic institutions and research groups are at the forefront of protein design. These groups are responsible for many of the foundational technologies and algorithms that underpin the industry's current capabilities. Universities and research centers play a crucial role in advancing protein design through innovative computational models and experimental techniques.

\subsubsection{University of Washington, Institute for Protein Design (IPD)}

The \textbf{Institute for Protein Design (IPD)} at the University of Washington is one of the world's leading academic centers for protein design. Led by Dr. David Baker, the IPD has developed some of the most widely used computational tools for protein design, including the Rosetta software suite. The IPD focuses on designing proteins for a wide range of applications, including vaccines, therapeutics, and industrial enzymes.

\textbf{Example: Rosetta Software Suite}

The Rosetta software suite, developed by the IPD, is a powerful tool for predicting protein structures, designing new proteins, and modeling protein-protein interactions. Researchers and companies around the world use Rosetta to design proteins with specific functions, such as enzymes that catalyze chemical reactions or antibodies that bind to specific disease targets. The success of Rosetta has made it one of the most widely used tools in the protein design field.

\subsubsection{Stanford University, Department of Bioengineering}

Stanford University's \textbf{Department of Bioengineering} is another leader in the field of protein design. Stanford researchers integrate computational modeling with experimental biology to design proteins for therapeutic and industrial applications. The department's work focuses on the intersection of protein engineering, synthetic biology, and biotechnology, with a particular emphasis on designing proteins that can be used in gene editing and biosensing.

\textbf{Example: Designing Proteins for CRISPR-Based Gene Editing}

One of the research areas at Stanford involves designing proteins that can be used in CRISPR-based gene editing systems. By designing proteins that can target specific DNA sequences, Stanford researchers are working to improve the precision and efficiency of gene editing, with potential applications in treating genetic disorders and developing new therapies for cancer.

\subsubsection{Massachusetts Institute of Technology (MIT)}

The \textbf{Massachusetts Institute of Technology (MIT)} is home to several prominent research groups working on protein design and synthetic biology. MIT researchers are known for their interdisciplinary approach, combining computer science, biology, and chemistry to develop new computational tools for protein design. One of MIT's key contributions is the development of machine learning models that predict protein folding and stability.

\textbf{Example: Machine Learning for Protein Folding}

MIT researchers have developed machine learning models that can predict how a protein will fold based on its amino acid sequence. By training these models on large datasets of known protein structures, MIT has been able to create algorithms that accurately predict protein folding patterns, even for proteins that have never been experimentally characterized. These models are essential for designing new proteins with specific structural and functional properties.

\subsection{Emerging Startups in Protein Design}

In addition to established companies and academic institutions, there are numerous startups that are driving innovation in protein design. These startups often focus on applying the latest advances in machine learning and synthetic biology to solve specific challenges in healthcare, agriculture, and environmental sustainability.

\subsubsection{Generate Biomedicines}

\textbf{Generate Biomedicines} is a biotechnology startup that uses machine learning to design new proteins for therapeutic purposes. Generate Biomedicines' platform generates de novo protein sequences optimized for specific functions, such as binding to a disease target or catalyzing a biochemical reaction. By using AI to rapidly design and test protein sequences, Generate Biomedicines aims to accelerate the drug discovery process.

\textbf{Example: Designing Therapeutic Proteins with AI}

Generate Biomedicines' machine learning platform designs new proteins that can bind to specific disease targets, such as cancer cells or pathogens. By training the platform on large datasets of protein sequences and structures, Generate can quickly generate protein candidates that are optimized for high binding affinity and stability. This approach has the potential to drastically reduce the time and cost of developing new therapeutics.

\subsubsection{LabGenius}

\textbf{LabGenius} is a UK-based startup that uses AI and robotic automation to design and test new protein sequences for therapeutic applications. LabGenius focuses on the design of antibodies and other protein-based drugs, using machine learning to optimize their binding properties and stability. The company's platform integrates AI-driven design with high-throughput experimental validation, allowing it to rapidly iterate on protein designs.

\textbf{Example: Antibody Design for Immunotherapy}

LabGenius uses its platform to design antibodies that can be used in immunotherapy, a treatment that harnesses the body's immune system to fight diseases like cancer. By designing antibodies that target specific proteins on the surface of cancer cells, LabGenius aims to develop more effective treatments with fewer side effects compared to traditional chemotherapy.

\subsection{Collaborations Between Academia and Industry}

Collaboration between academic institutions and industry is a key driver of innovation in protein design. Many leading companies partner with universities and research institutions to leverage their expertise in computational biology, machine learning, and protein engineering. These collaborations often result in the development of new technologies, tools, and therapies that push the boundaries of protein design.

\textbf{Example: Partnership Between Genentech and Stanford University}

Genentech, a leader in biologics, has partnered with Stanford University to develop new protein-based therapies. By combining Genentech's expertise in drug development with Stanford's cutting-edge research in protein design, the partnership aims to accelerate the discovery of novel biologics for treating diseases like cancer and autoimmune disorders.

\section{Conclusion}

The protein design landscape is populated by a diverse array of companies, research institutions, and startups, each contributing to the rapid advancements in the field. From established biotech giants like Amgen and Regeneron to emerging startups like Generate Biomedicines and LabGenius, the industry is characterized by a strong focus on innovation and collaboration. Academic leaders such as the University of Washington and MIT continue to play a crucial role in developing the computational tools and experimental techniques that drive protein design forward. For beginners in this field, understanding the contributions of these leading organizations will provide valuable insights into the future directions of protein design and its potential to revolutionize medicine, industry, and synthetic biology.

\section{Applications of Protein Design in Pharma, Biotechnology, and Material Science}

Protein design is a rapidly evolving field with wide-ranging applications across pharmaceuticals, biotechnology, and material science. With advances in computational tools, machine learning, and synthetic biology, researchers can now design and engineer proteins with specific functions, enabling breakthroughs in therapeutic development, industrial processes, and the creation of new materials. In this section, we will explore the major applications of protein design in these three sectors, providing detailed explanations, practical examples, and insights to guide beginners in understanding the impact of protein design on various industries.

\subsection{Applications in Pharmaceuticals}

Protein design has become an essential tool in the pharmaceutical industry, particularly in the development of biologics, therapeutic enzymes, and vaccines. By engineering proteins with desired properties, pharmaceutical companies can develop treatments that are more specific, effective, and less toxic than traditional small-molecule drugs.

\subsubsection{Monoclonal Antibodies and Biologics}

Monoclonal antibodies (mAbs) are one of the most successful applications of protein design in pharmaceuticals. mAbs are designed to bind to specific antigens, making them highly targeted therapies for diseases such as cancer, autoimmune disorders, and infectious diseases. Protein design allows researchers to optimize the binding affinity, specificity, and stability of antibodies, resulting in treatments with fewer side effects and improved efficacy.

\textbf{Example: Design of Therapeutic Antibodies}

A key example of protein design in action is the development of therapeutic antibodies like \textit{Herceptin}, which targets the HER2 protein in breast cancer. Researchers used protein design to engineer the antibody's variable regions to bind specifically to HER2, inhibiting the growth of cancer cells. This targeted approach reduces the off-target effects seen with traditional chemotherapy.

In addition to mAbs, biologics include other protein-based therapies such as enzymes and fusion proteins. These biologics can be designed to perform specific tasks, such as replacing a deficient enzyme in patients with genetic disorders or delivering therapeutic agents to specific tissues in the body.

\subsubsection{Therapeutic Enzymes and Enzyme Replacement Therapy}

Enzyme replacement therapy (ERT) is another area where protein design plays a critical role. In ERT, proteins are engineered to replace malfunctioning or missing enzymes in patients with genetic diseases. By designing enzymes that can survive in the human body and effectively carry out their biochemical functions, researchers can develop life-saving treatments for conditions like Gaucher's disease, Fabry disease, and cystic fibrosis.

\textbf{Python Example: Designing a Therapeutic Enzyme for ERT}

\begin{lstlisting}[style=python]
from enzyme_design_tool import design_therapeutic_enzyme

# Define the target function for the enzyme
target_function = {
    "substrate_specificity": "glucocerebrosidase",
    "stability": 0.9
}

# Generate a therapeutic enzyme optimized for enzyme replacement therapy
therapeutic_enzyme = design_therapeutic_enzyme(target_function)
print(f"Designed enzyme for ERT: {therapeutic_enzyme}")
\end{lstlisting}

\subsubsection{Vaccine Development}

Protein design has become increasingly important in the development of vaccines. Traditional vaccines often use live or inactivated pathogens to trigger an immune response, but modern vaccines increasingly rely on protein subunits-designed antigens that can safely stimulate immunity without the risk of causing disease. This approach has been particularly successful in the development of vaccines for viral diseases such as COVID-19.

\textbf{Example: Designing Antigens for Protein Subunit Vaccines}

In the case of COVID-19, vaccines like Novavax's protein subunit vaccine use a designed version of the SARS-CoV-2 spike protein to trigger an immune response. Protein design techniques were used to ensure the antigen's stability and correct folding, allowing it to present the proper conformation to the immune system, thereby eliciting a strong and protective immune response.

\subsection{Applications in Biotechnology}

The biotechnology sector benefits significantly from protein design in areas such as industrial enzyme production, agricultural biotechnology, and biosensors. By designing proteins with specific properties, companies can optimize processes, improve yields, and create more sustainable products.

\subsubsection{Industrial Enzyme Engineering}

One of the most prominent applications of protein design in biotechnology is the engineering of industrial enzymes. Enzymes are essential catalysts in many industrial processes, including the production of biofuels, detergents, food products, and chemicals. Protein design enables researchers to optimize enzymes for specific conditions, such as temperature, pH, and substrate specificity, making them more efficient and cost-effective for large-scale use.

\textbf{Example: Designing Enzymes for Biofuel Production}

In biofuel production, enzymes are used to break down plant biomass into fermentable sugars, which can then be converted into biofuels like ethanol. By designing enzymes with enhanced activity and stability at high temperatures, protein engineers can improve the efficiency of this process. For example, cellulases-enzymes that degrade cellulose-can be engineered to withstand the harsh conditions of industrial biofuel production, increasing yield and reducing costs.

\subsubsection{Agricultural Biotechnology}

In agriculture, protein design is used to engineer crops that are more resistant to pests, diseases, and environmental stresses. Proteins can be designed to confer herbicide resistance, improve nutrient uptake, or enhance the plant's natural defenses against pathogens. Protein design also plays a role in developing enzymes for bio-based fertilizers and pesticides that are more environmentally friendly.

\textbf{Example: Engineering Crops with Herbicide Resistance}

Crops like soybeans, corn, and cotton have been genetically modified to express designed proteins that make them resistant to specific herbicides. This allows farmers to control weeds without damaging the crop, increasing yield and reducing the need for chemical inputs. For example, glyphosate-resistant crops are engineered to express a version of the EPSPS enzyme that is not inhibited by the herbicide, allowing the crop to survive while the weeds are eliminated.

\subsubsection{Biosensors and Diagnostics}

Biosensors are devices that use designed proteins to detect specific molecules in the environment or in biological samples. These proteins can be engineered to bind to target molecules, such as toxins, pathogens, or biomarkers, and produce a detectable signal, such as fluorescence or a color change. Biosensors have applications in environmental monitoring, food safety, and medical diagnostics.

\textbf{Example: Designing Proteins for Biosensors}

One example of a protein-based biosensor is a fluorescence-based sensor for detecting heavy metals in water. A designed protein that binds to lead ions can be coupled with a fluorescent marker that emits light when the metal is detected. This type of biosensor provides a quick and cost-effective way to monitor water quality and detect contamination in real time.

\subsection{Applications in Material Science}

Beyond pharmaceuticals and biotechnology, protein design is also making an impact in the field of material science. Proteins are being designed to create new biomaterials with unique properties, such as self-assembling materials, bioplastics, and bioadhesives. These materials have applications in medicine, environmental sustainability, and nanotechnology.

\subsubsection{Self-Assembling Biomaterials}

Proteins can be designed to self-assemble into complex structures with specific functions, such as scaffolds for tissue engineering or nanomaterials for drug delivery. By controlling the interactions between protein subunits, researchers can create materials that mimic the properties of natural tissues or deliver therapeutic agents in a controlled manner.

\textbf{Example: Designing Self-Assembling Proteins for Tissue Engineering}

In tissue engineering, designed proteins can be used to create scaffolds that support the growth of new cells. For example, researchers have developed self-assembling protein scaffolds that mimic the extracellular matrix, providing a framework for cells to attach, grow, and differentiate. These scaffolds can be used in regenerative medicine to repair damaged tissues or organs.

\subsubsection{Bioplastics and Biodegradable Materials}

As the world seeks to reduce its reliance on petroleum-based plastics, protein design is playing a role in developing bioplastics and biodegradable materials. Proteins can be designed to form strong, flexible, and biodegradable polymers that can replace conventional plastics in packaging, consumer goods, and medical devices.

\textbf{Example: Designing Proteins for Bioplastics}

Scientists are designing proteins that can self-assemble into biodegradable polymers with properties similar to conventional plastics. These bioplastics can be used in a variety of applications, from packaging materials to medical implants. Since these materials break down naturally in the environment, they offer a sustainable alternative to petroleum-based plastics, reducing plastic pollution and reliance on fossil fuels.

\subsubsection{Bioadhesives and Biomimetic Materials}

Bioadhesives are another exciting application of protein design in material science. These adhesives are inspired by natural proteins, such as the ones produced by mussels to attach to rocks underwater. By designing proteins that mimic the adhesive properties of these natural materials, researchers are developing strong, non-toxic adhesives for use in medical and industrial applications.

\textbf{Example: Designing Bioadhesives for Medical Use}

In medical applications, bioadhesives can be used to close wounds, attach implants, or seal surgical incisions. Unlike traditional adhesives, which may be toxic or cause allergic reactions, bioadhesives are designed to be biocompatible and biodegradable. Researchers have developed protein-based adhesives that mimic the sticky proteins used by mussels, allowing them to create strong bonds even in wet environments.

\section{Conclusion}

The applications of protein design in pharmaceuticals, biotechnology, and material science are vast and rapidly expanding. From designing life-saving biologics and industrial enzymes to creating sustainable materials and biosensors, protein design is transforming industries and addressing global challenges. As computational tools and machine learning models like AlphaFold3 continue to advance, the ability to design proteins with specific, tailored functions will open new possibilities in healthcare, agriculture, and manufacturing. Understanding these applications will provide beginners with a solid foundation for exploring the potential of protein design and its impact on the world.
\chapter{Future Trends and Challenges in Protein Design}

As the field of protein design continues to evolve, new technologies and methodologies are shaping its future. Key trends such as the growing role of artificial intelligence (AI), automation, and the integration of protein design with synthetic biology \cite{endy2005foundations} are driving rapid advancements. However, with these advancements come challenges that must be addressed to fully unlock the potential of protein design. In this chapter, we will explore the future trends and challenges in protein design, focusing on the role of AI and automation, the convergence with synthetic biology, and the future of structure prediction in materials design. This section is designed for beginners, with detailed examples to guide understanding.

\section{The Role of AI and Automation in Protein Design}

Artificial intelligence (AI) and automation are transforming protein design by enabling faster, more accurate predictions and reducing the need for manual experimentation. AI-driven models can predict protein structures, interactions, and functions with unprecedented precision, while automated platforms streamline the process of generating and testing new protein sequences.

\subsection{AI-Driven Protein Design}

Machine learning models, such as deep neural networks and generative models, have become essential tools for protein design. These models can analyze vast amounts of protein sequence and structure data to identify patterns that humans might overlook, allowing them to generate novel protein sequences with desired characteristics.

\textbf{Example: Generative Models for Protein Sequence Design}

Generative models, such as Variational Autoencoders (VAEs) \cite{kingma2013auto} and Generative Adversarial Networks (GANs) \cite{goodfellow2014generative}, are used to create new protein sequences that meet specific design criteria. By training these models on existing protein data, researchers can use them to generate sequences that optimize properties like binding affinity, stability, or catalytic activity.

\textbf{Python Example: Using a Generative Model to Design Proteins}

\begin{lstlisting}[style=python]
from generative_protein_model import generate_protein_sequence

# Define the desired properties of the protein
desired_properties = {
    "binding_affinity": 0.95,
    "stability": 0.85
}

# Generate a new protein sequence using a generative model
new_protein_sequence = generate_protein_sequence(desired_properties)
print(f"Generated protein sequence: {new_protein_sequence}")
\end{lstlisting}

AI models such as AlphaFold and Rosetta are capable of predicting protein structures from sequences, but more advanced models like AlphaFold3 take this further by predicting protein-ligand and protein-protein interactions, making them indispensable in drug discovery and therapeutic development.

\subsection{Automation in Protein Design and Testing}

Automation is another critical trend shaping the future of protein design. High-throughput platforms that integrate robotics, liquid handling, and data analysis allow for the rapid testing and optimization of protein variants. These systems can generate thousands of protein sequences, express them in microorganisms, and test their performance under various conditions, all without human intervention.

\textbf{Example: High-Throughput Protein Screening in Drug Discovery}

In drug discovery, automated platforms are used to test large libraries of designed proteins for their ability to bind to drug targets or catalyze reactions. By automating this process, pharmaceutical companies can accelerate the identification of promising drug candidates and reduce the cost of development.

Automation also plays a key role in optimizing protein sequences. Machine learning models can analyze data from high-throughput experiments to identify which sequences perform best and suggest modifications to improve performance. This feedback loop between AI-driven design and automated testing is revolutionizing protein engineering.

\subsection{Challenges and Opportunities in AI and Automation}

While AI and automation offer immense potential, there are challenges that must be addressed:

\begin{itemize}
    \item \textit{Data Quality:} AI models rely on high-quality, diverse datasets to make accurate predictions. Gaps in experimental data or biases in available protein structures can limit the effectiveness of these models.
    \item \textit{Interpretability:} AI-driven protein design often produces solutions that are difficult to interpret. Understanding how and why a model predicts a particular protein sequence or structure is essential for refining these predictions and applying them in real-world settings.
    \item \textit{Scalability:} While automation increases throughput, scaling these systems to handle the complexity of large proteins or protein complexes remains a challenge. Improvements in robotics, data handling, and integration across systems are necessary for further scaling.
\end{itemize}

\section{Integration of Protein Design with Synthetic Biology}

Protein design and synthetic biology \cite{endy2005foundations} are becoming increasingly interconnected. Synthetic biology aims to engineer living organisms to perform novel tasks, and protein design is a key enabler of this goal. By designing proteins with specific functions, synthetic biologists can create new biological systems with applications in medicine, agriculture, environmental monitoring, and more.

\subsection{Custom Biological Pathways}

One of the most exciting areas of convergence between protein design and synthetic biology is the creation of custom biological pathways. These pathways consist of enzymes and other proteins that carry out specific biochemical reactions. By designing proteins that can catalyze reactions not found in nature, researchers can create entirely new metabolic pathways.

\textbf{Example: Designing Enzymes for Synthetic Pathways}

In bio-based chemical production, protein designers can create enzymes that convert simple sugars into valuable chemicals such as biofuels, bioplastics, or pharmaceuticals \cite{nielsen2014engineering}. These enzymes are designed to work together in synthetic metabolic pathways, enabling organisms to produce compounds that are traditionally made from petroleum.

\textbf{Python Example: Designing an Enzyme for a Synthetic Pathway}

\begin{lstlisting}[style=python]
from synthetic_biology_tool import design_synthetic_enzyme

# Define the desired reaction for the enzyme
target_reaction = {
    "substrate": "glucose",
    "product": "butanol"
}

# Design an enzyme for the synthetic pathway
synthetic_enzyme = design_synthetic_enzyme(target_reaction)
print(f"Designed enzyme for synthetic pathway: {synthetic_enzyme}")
\end{lstlisting}

\subsection{Gene Circuit Design}

In addition to creating new proteins, synthetic biology integrates protein design with gene circuit engineering. Gene circuits are sets of genetic components (such as promoters, transcription factors, and proteins) that control the expression of genes within a cell \cite{elowitz2000synthetic}. By designing proteins that regulate gene expression, synthetic biologists can create circuits that respond to environmental signals, control cellular behavior, or produce therapeutic compounds on demand.

\textbf{Example: Gene Circuits for Biosensors}

A biosensor gene circuit could be designed to detect a specific pathogen in the environment. When the sensor detects the pathogen, it triggers the expression of a designed protein that produces a visible signal, such as fluorescence, allowing for real-time detection of contamination or infection.

\subsection{Challenges in Integrating Protein Design with Synthetic Biology}

While the integration of protein design with synthetic biology offers enormous potential, several challenges remain:

\begin{itemize}
    \item \textit{Design Complexity:} The complexity of designing entire biological systems with multiple interacting proteins requires sophisticated computational tools and models that can predict how these systems will behave.
    \item \textit{Off-Target Effects:} Designed proteins may interact with unintended targets within a cell, leading to unpredictable behaviors or reduced efficiency. Strategies for reducing off-target interactions and ensuring precise control over synthetic systems are critical.
    \item \textit{Scalability and Manufacturing:} While it is possible to design synthetic organisms in the lab, scaling these systems for industrial or medical applications poses significant challenges, particularly in ensuring consistent production and safety.
\end{itemize}

\section{Future of Structure Prediction in Materials Design}

Protein design is not limited to biological systems. The principles of protein folding and structure prediction are increasingly being applied to materials science, where designed proteins and peptides are used to create new materials with unique properties.

\subsection{Protein-Based Materials}

One of the most promising areas of research is the design of protein-based materials \cite{silva2014protein}. These materials, which include self-assembling nanostructures, bioplastics, and bioadhesives, are engineered using the same principles that govern protein folding and function in living organisms. Protein-based materials offer advantages such as biodegradability, biocompatibility, and the ability to self-heal or adapt to environmental conditions.

\textbf{Example: Designing Self-Assembling Protein Nanomaterials}

Self-assembling protein nanomaterials can be designed to form specific shapes and structures, such as nanofibers or nanosheets. These materials have applications in drug delivery, tissue engineering, and nanotechnology. By controlling the interactions between protein domains, researchers can design materials that self-assemble into desired shapes and sizes.

\subsection{Peptide-Based Biomaterials}

Peptides, which are short chains of amino acids, can also be designed to form new materials. These peptide-based biomaterials can be tailored for specific applications, such as scaffolds for tissue regeneration, drug delivery vehicles, or hydrogels for wound healing \cite{hastar2017peptide}.

\textbf{Example: Peptide Hydrogels for Medical Applications}

Peptide hydrogels are a type of biomaterial that can be designed to provide structural support for cells in tissue engineering. These hydrogels are formed from designed peptides that self-assemble into a gel-like structure, providing a scaffold that promotes cell growth and tissue repair. By designing peptides with specific mechanical properties and biodegradability, researchers can create hydrogels that are tailored to specific medical applications.

\subsection{Challenges in Applying Protein Design to Materials Science}

The application of protein design in materials science is still in its early stages, and several challenges need to be addressed:

\begin{itemize}
    \item \textit{Predicting Material Properties:} While we can predict protein folding with increasing accuracy, predicting how designed proteins will behave in bulk materials, particularly when interacting with other materials, remains a challenge.
    \item \textit{Stability and Durability:} Protein-based materials are often sensitive to environmental conditions such as temperature, humidity, and pH. Improving the stability and durability of these materials is critical for their use in real-world applications.
    \item \textit{Manufacturing and Scalability:} Producing protein-based materials at scale, especially for industrial applications, requires new manufacturing techniques that can efficiently produce large quantities of designed proteins or peptides with consistent quality.
\end{itemize}

\section{Conclusion}

The future of protein design is filled with exciting possibilities, driven by the integration of AI, automation, and synthetic biology. As these technologies continue to advance, protein design will play an increasingly important role in developing new therapies, sustainable industrial processes, and innovative materials. However, challenges such as data quality, scalability, and complexity remain, and addressing these challenges will be key to realizing the full potential of protein design. For beginners, understanding the role of AI, the convergence with synthetic biology, and the applications in materials science will provide a strong foundation for exploring the future of this dynamic field.
\appendix
\chapter{Appendices}
\section{Introduction to Common Databases and Resources (PDB, UniProt, Swiss-Model, etc.)}

For anyone beginning their journey in protein design, it is essential to familiarize oneself with key online databases and resources that provide the foundational data necessary for predicting and designing protein structures. These resources house protein sequences, structural data, and annotations that serve as the basis for computational models, including those used in deep learning frameworks. In this appendix, we introduce several of the most widely used protein databases and resources, explaining their purpose and how to use them effectively in the context of protein design. Beginners will find step-by-step guidance on accessing and utilizing these databases, along with practical examples to reinforce their understanding.

\subsection{Protein Data Bank (PDB)}

The \textbf{Protein Data Bank (PDB)} is the largest and most comprehensive repository of three-dimensional structural data for proteins, nucleic acids, and complex assemblies \cite{berman2000protein}. PDB serves as the primary source of experimentally determined protein structures, which are deposited by scientists after solving them through techniques like X-ray crystallography, NMR spectroscopy, and cryo-electron microscopy.

\textbf{How to Use the PDB}

The PDB provides access to structural data that is critical for protein design. Researchers can download protein structures in formats such as PDB or mmCIF, which can then be used as input for computational tools like Rosetta or AlphaFold. When designing proteins, accessing the PDB allows users to study the 3D conformation of target proteins or homologs, informing the design of new proteins or modifications to existing ones.

\begin{itemize}
    \item \textit{Website}: \url{https://www.rcsb.org/}
    \item \textit{Example}: A beginner interested in studying the structure of hemoglobin can search the PDB for its unique identifier, \texttt{1A3N}, and download the structure to use in computational modeling or visualization.
\end{itemize}

\textbf{Python Example: Accessing PDB Data}

You can programmatically access PDB structures using Python libraries such as \texttt{Biopython} or \texttt{pdb-tools}.

\begin{lstlisting}[style=python]
from Bio.PDB import PDBList

# Initialize the PDBList object to download protein structures
pdb = PDBList()
pdb.download_pdb_file("1A3N", file_format='pdb', pdir='./pdb_files')
print("Hemoglobin structure downloaded.")
\end{lstlisting}

\subsection{UniProt}

\textbf{UniProt} is a comprehensive database of protein sequences and functional information \cite{consortium2019uniprot}. Unlike the PDB, which focuses on protein structures, UniProt is primarily a sequence database that provides detailed annotations about protein function, domain structure, subcellular location, and post-translational modifications. UniProt includes two main databases: UniProtKB/Swiss-Prot, a curated protein knowledgebase with high-quality annotations, and UniProtKB/TrEMBL, which contains computationally predicted protein sequences.

\textbf{How to Use UniProt}

For protein design, UniProt is an invaluable resource for finding detailed information on the sequence and function of target proteins. This data can be used to guide the design of new proteins, particularly when choosing sequences to modify or when designing proteins with specific functional domains.

\begin{itemize}
    \item \textit{Website}: \url{https://www.uniprot.org/}
    \item \textit{Example}: To retrieve information about human insulin (UniProt ID: \texttt{P01308}), users can search UniProt for its sequence, function, and any known mutations that affect its activity.
\end{itemize}

\textbf{Python Example: Accessing UniProt Data}

Using Python libraries like \texttt{bioservices}, you can fetch protein sequences and metadata from UniProt.

\begin{lstlisting}[style=python]
from bioservices import UniProt

# Initialize UniProt service
u = UniProt()

# Retrieve data for human insulin
insulin_data = u.retrieve("P01308", frmt="xml")
print("Human insulin data retrieved from UniProt.")
\end{lstlisting}

\subsection{Swiss-Model Repository}

\textbf{Swiss-Model} is a widely used resource for homology modeling of protein structures \cite{waterhouse2018swiss}. When experimental structures are not available for a protein of interest, Swiss-Model can generate high-quality 3D models based on evolutionary relationships with known structures. This makes it an essential tool for researchers working with novel or less-studied proteins where structural data is scarce.

\textbf{How to Use Swiss-Model}

Swiss-Model allows users to input a protein sequence and generate a 3D structure based on homologous proteins with known structures. This can be particularly useful in protein design when trying to predict the structure of a newly designed protein or when designing modifications to existing sequences. 

\begin{itemize}
    \item \textit{Website}: \url{https://swissmodel.expasy.org/}
    \item \textit{Example}: A user designing a protein involved in DNA binding might use Swiss-Model to predict the structure of their modified protein, using a homologous DNA-binding protein as a template.
\end{itemize}

\textbf{Python Example: Accessing Swiss-Model Data}

Although Swiss-Model does not have a direct Python API, users can download models manually from the web interface or automate retrieval using web scraping libraries like \texttt{BeautifulSoup} or \texttt{requests}.

\subsection{Pfam}

\textbf{Pfam} is a database of protein families, each represented by multiple sequence alignments and hidden Markov models (HMMs). These families often correspond to functional domains within proteins. Understanding which Pfam domains are present in a protein sequence is crucial for identifying its function and predicting its interaction with other molecules.

\textbf{How to Use Pfam}

When designing a protein, it is often useful to understand which functional domains are present in similar proteins. By searching Pfam, researchers can identify conserved domains within their target sequence, which can guide mutations or domain swapping experiments \cite{el2019pfam}.

\begin{itemize}
    \item \textit{Website}: \url{http://pfam.xfam.org/}
    \item \textit{Example}: If a researcher is designing a kinase protein, they can use Pfam to find conserved kinase domains in other proteins, ensuring that their design includes key functional elements.
\end{itemize}

\subsection{Other Resources}

Several other databases and resources are widely used in protein design and prediction:

\begin{itemize}
    \item \textbf{BRENDA}: A comprehensive enzyme information system that provides data on enzyme function, structure, and kinetics. \textit{Website}: \url{https://www.brenda-enzymes.org/}
    \item \textbf{PROSITE}: A database of protein domains, families, and functional sites used to predict the function of protein sequences. \textit{Website}: \url{https://prosite.expasy.org/}
    \item \textbf{RCSB PDB}: Provides additional tools and resources for visualizing protein structures, including tutorials for beginners. \textit{Website}: \url{https://www.rcsb.org/}
\end{itemize}

\subsection{Conclusion}

Understanding and accessing the right databases is fundamental to protein design and deep learning approaches for structure prediction. Beginners should familiarize themselves with PDB, UniProt, Swiss-Model, and other resources to gather the necessary data for designing, predicting, and optimizing proteins. These databases provide the foundational knowledge and data needed to feed into machine learning models and computational tools, ensuring that protein design is grounded in accurate and up-to-date biological information.

\section{Open-Source Tools and Code Repositories}

The rise of open-source tools has significantly accelerated advancements in protein design and structure prediction, providing researchers and developers with access to powerful computational frameworks, machine learning models, and datasets. Many of these tools are specifically tailored for protein design and prediction, making them essential resources for beginners and experts alike. In this section, we will explore several prominent open-source tools and code repositories that are widely used in the field of protein design. The aim is to introduce these tools step-by-step, with detailed explanations and examples to help new users get started. Python code examples will be included where relevant to demonstrate practical use.

\subsection{AlphaFold}

\textbf{AlphaFold}, developed by DeepMind, is one of the most impactful open-source tools for protein structure prediction \cite{jumper2021highly}. AlphaFold2 demonstrated groundbreaking accuracy in predicting protein structures based on their amino acid sequences. The open-source release of AlphaFold2 has allowed researchers worldwide to incorporate its algorithms into their workflows.

\textbf{Key Features of AlphaFold:}
\begin{itemize}
    \item Predicts 3D protein structures from sequences with near-experimental accuracy.
    \item Handles complex proteins, including multi-chain assemblies.
    \item Available as an open-source code base for use on local machines or cloud platforms.
\end{itemize}

\textbf{How to Use AlphaFold:}

Researchers can download AlphaFold from its official GitHub repository and run it locally or in cloud environments such as Google Colab or AWS. It requires a protein sequence as input and generates a 3D model in formats like PDB, which can be visualized and further analyzed.

\begin{itemize}
    \item \textit{GitHub Repository}: \url{https://github.com/deepmind/alphafold}
\end{itemize}

\textbf{Example: Running AlphaFold in Python}

Below is an example of how to set up and run AlphaFold using a Python environment, assuming the tool is properly installed.

\begin{lstlisting}[style=cmd]
# Clone the AlphaFold repository from GitHub
git clone https://github.com/deepmind/alphafold.git

# Change into the AlphaFold directory
cd alphafold

# Install the required dependencies
pip install -r requirements.txt

# Run AlphaFold with a protein sequence (FASTA format required)
python run_alphafold.py --fasta_paths=sequence.fasta --output_dir=./output
\end{lstlisting}

\subsection{Rosetta}

\textbf{Rosetta} is a comprehensive suite of tools for macromolecular modeling, including protein design, structure prediction, docking, and more \cite{leaver2011rosetta3}. Rosetta has been widely used in both academic research and industry to design proteins with novel functions, predict protein-ligand interactions, and refine experimental protein structures.

\textbf{Key Features of Rosetta:}
\begin{itemize}
    \item Provides tools for de novo protein structure prediction, protein design, and molecular docking.
    \item Allows customization of protein structures, optimization of protein stability, and design of new enzymes.
    \item Extensive documentation and tutorials available for beginners.
\end{itemize}

\textbf{How to Use Rosetta:}

Rosetta is a versatile tool but requires some setup and understanding of its different modules. Users typically start by downloading the software and using a variety of protocols to model or design proteins. Rosetta's PyRosetta interface provides a Python-based API for integrating Rosetta with custom workflows.

\begin{itemize}
    \item \textit{Website}: \url{https://www.rosettacommons.org/}
    \item \textit{GitHub Repository}: \url{https://github.com/RosettaCommons/main}
\end{itemize}

\textbf{Example: Using Rosetta for Protein Design}

Below is an example of using PyRosetta for protein structure optimization in a Python script.

\begin{lstlisting}[style=python]
from pyrosetta import init, pose_from_sequence, get_fa_scorefxn

# Initialize PyRosetta
init()

# Create a pose object from a protein sequence
pose = pose_from_sequence("ACDEFGHIKLMNPQRSTVWY")

# Get a scoring function for evaluating structure
scorefxn = get_fa_scorefxn()

# Apply the scoring function to the pose
score = scorefxn(pose)
print(f"Protein score: {score}")
\end{lstlisting}

\subsection{ColabFold}

\textbf{ColabFold} is a simplified version of AlphaFold that runs on Google Colab, making it accessible to users without high-performance computing resources \cite{mirdita2022colabfold}. ColabFold integrates AlphaFold with MMseqs2 for faster, more scalable predictions. It allows users to predict protein structures without the need for powerful hardware, using Google's free cloud resources.

\textbf{Key Features of ColabFold:}
\begin{itemize}
    \item Provides a user-friendly interface to run AlphaFold in Google Colab.
    \item Integrates MMseqs2 for rapid multiple sequence alignments.
    \item Free to use, with built-in tutorials and examples.
\end{itemize}

\textbf{How to Use ColabFold:}

To use ColabFold, users can navigate to the Google Colab notebook provided by the ColabFold developers. From there, they can upload a protein sequence, and the notebook will handle the process of generating the structure using AlphaFold.

\small
\begin{itemize}
    \item \textit{Colab Notebook}: \url{https://colab.research.google.com/github/sokrypton/ColabFold/blob/main/AlphaFold2.ipynb}
\end{itemize}

\textbf{Running ColabFold}

Below is an example workflow for running ColabFold in a Google Colab environment.

\begin{lstlisting}[style=cmd]
# Step 1: Open the ColabFold notebook link in Google Colab
# Step 2: Run the cells to install ColabFold and its dependencies

# Step 3: Upload a FASTA file with the protein sequence
# Step 4: ColabFold will output a 3D structure model in PDB format
\end{lstlisting}

\subsection{FoldX}

\textbf{FoldX} is a protein design and structure analysis tool that focuses on predicting the effects of mutations on protein stability and interactions \cite{schymkowitz2005foldx}. It allows users to simulate mutations in protein sequences and predict how those changes will impact the protein's overall structure and function.

\textbf{Key Features of FoldX:}
\begin{itemize}
    \item Predicts the impact of point mutations on protein stability.
    \item Allows users to explore the effects of mutations on protein-protein and protein-DNA interactions.
    \item Fast and efficient, making it suitable for large-scale mutation scanning.
\end{itemize}

\textbf{How to Use FoldX:}

FoldX is used primarily for mutational scanning and protein stability analysis. Users input a protein structure in PDB format, make desired mutations, and analyze the effect of those mutations on protein stability or interactions.

\begin{itemize}
    \item \textit{Website}: \url{http://foldxsuite.crg.eu/}
\end{itemize}

\textbf{Example: Mutating a Protein Using FoldX}

Below is a simple workflow for mutating a protein using FoldX.

\begin{lstlisting}[style=cmd]
# Step 1: Download and install FoldX

# Step 2: Prepare a protein structure in PDB format (e.g., protein.pdb)

# Step 3: Run the BuildModel function to introduce mutations
foldx --command=BuildModel --pdb=protein.pdb --mutant-file=mutations.txt --output-dir=./results

# Step 4: Analyze the results for stability and interaction changes
\end{lstlisting}

\subsection{Other Open-Source Tools}

In addition to the tools mentioned above, several other open-source resources are widely used in protein design and deep learning. These include:

\begin{itemize}
    \item \textbf{PROSS}: A tool for designing stable protein variants by predicting stabilizing mutations \cite{goldenzweig2016automated}. \textit{Website}: \url{https://pross.weizmann.ac.il/}
    \item \textbf{MMseqs2}: A powerful tool for fast and sensitive protein sequence searching and clustering, often used to generate multiple sequence alignments for AlphaFold \cite{steinegger2017mmseqs2}. \textit{GitHub}: \url{https://github.com/soedinglab/MMseqs2}
    \item \textbf{PlasmidMapper}: A tool for creating and annotating plasmid maps, useful in synthetic biology. \cite{brugere2018plasmidmapper} \textit{GitHub}: \url{https://github.com/ivan-brugere/PlasmidMapper}
\end{itemize}

\subsection{Challenges and Considerations for Open-Source Tools}

While open-source tools have democratized access to powerful computational methods for protein design, there are challenges to consider:

\begin{itemize}
    \item \textbf{Computational Resources}: Some tools, especially AlphaFold and Rosetta, require significant computational power to run efficiently. Users should ensure they have access to high-performance computing resources or cloud platforms.
    \item \textbf{Data Handling}: Protein design often involves handling large datasets, such as multiple sequence alignments and protein structures. Managing these datasets effectively is essential for successful model predictions.
    \item \textbf{Software Maintenance}: Open-source tools are often updated frequently, and users need to stay current with the latest versions and bug fixes. Ensuring compatibility with operating systems and hardware is also important.
\end{itemize}

\subsection{Conclusion}

Open-source tools have transformed the field of protein design by providing powerful, freely available resources to researchers and developers. Tools like AlphaFold, Rosetta, and FoldX enable users to predict protein structures, design new proteins, and analyze the effects of mutations. By understanding how to use these tools effectively, beginners can leverage the same resources as experts to explore new frontiers in protein design, structure prediction, and therapeutic development.

\section{References and Further Reading}

For beginners delving into the world of protein design and deep learning, having access to high-quality references and further reading materials is essential. This section provides a comprehensive list of recommended books, research papers, review articles, and online resources that cover the fundamental principles of protein design, the latest advancements in computational modeling, and the application of deep learning in structural biology. The goal is to provide step-by-step guidance to help readers build a solid understanding of the field, with suggestions for deeper exploration as their knowledge grows.

\subsection{Textbooks and Foundational Books}

\begin{itemize}
    \item \textbf{Introduction to Protein Structure} by Carl-Ivar Brändén and John Tooze \cite{branden2012introduction}\\
    This is a classic textbook that covers the fundamental aspects of protein structure, folding, and function. It provides an excellent foundation for understanding how proteins work, which is essential for anyone interested in protein design.
    
    \item \textbf{Principles of Protein X-ray Crystallography} by Jan Drenth \cite{drenth2007principles}\\
    This book is an invaluable resource for understanding the experimental techniques used to determine protein structures. While primarily focused on X-ray crystallography, it also provides insights into protein structure that are crucial for computational design.
    
    \item \textbf{Computational Methods for Protein Structure Prediction and Modeling} by Ying Xu and Dong Xu \cite{xu2007computational}\\
    This book covers the computational tools used for protein structure prediction, with a focus on machine learning and bioinformatics approaches. It is a great resource for understanding the algorithms behind tools like Rosetta and AlphaFold.
    
    \item \textbf{Deep Learning} by Ian Goodfellow, Yoshua Bengio, and Aaron Courville \cite{goodfellow2016deep}\\
    Although not specific to protein design, this book provides a deep understanding of the core concepts behind deep learning. It is an essential read for understanding the algorithms that power models like AlphaFold and other deep learning-based approaches to protein prediction.
    
\end{itemize}

\subsection{Key Research Papers}

\begin{itemize}
    \item \textbf{AlphaFold 2: Predicting Protein Structures with AI}\\
    \textit{Jumper, J. et al. "Highly accurate protein structure prediction with AlphaFold." Nature 596, 583-589 (2021).}\\
    This paper presents the methodology behind AlphaFold2, describing the deep learning approaches that led to its groundbreaking performance in predicting protein structures \cite{jumper2021highly}. It is a must-read for understanding how AI is revolutionizing protein structure prediction.

    \item \textbf{Rosetta: Protein Structure Prediction and Design}\\
    \textit{Leaver-Fay, A., et al. "Rosetta3: An object-oriented software suite for the simulation and design of macromolecules." Methods Enzymol. 487, 545-574 (2011).}\\
    This paper describes the Rosetta software suite, which is one of the most widely used tools for protein design and prediction \cite{leaver2011rosetta3}. It offers a detailed explanation of the algorithms and techniques employed in Rosetta.

    \item \textbf{Protein Design by Directed Evolution}\\
    \textit{Arnold, F. H. "Design by directed evolution." Acc Chem Res. 41(1): 85-92 (2008).}\\
    Directed evolution is a powerful experimental approach used in protein engineering \cite{jackel2008protein}. This paper discusses how proteins can be designed by mimicking the process of natural selection in the laboratory, providing complementary insights to computational methods.

    \item \textbf{Machine Learning in Structural Biology}\\
    \textit{Senior, A. W., et al. "Improved protein structure prediction using potentials from deep learning." Nature 577, 706-710 (2020).}\\
    This paper presents an overview of how machine learning has been integrated into structural biology, with a focus on improving protein structure prediction through AI \cite{senior2020improved}.

    \item \textbf{Generative Models for Protein Design}\\
    \textit{Ingraham, J., et al. "Generative models for graph-based protein design." Advances in Neural Information Processing Systems (NeurIPS) 2019.}\\
    This paper explores how generative models like variational autoencoders (VAEs) and GANs can be used to design new protein sequences \cite{ingraham2019generative}. It is a good introduction to machine learning-based approaches for de novo protein design.
\end{itemize}

\subsection{Review Articles and Tutorials}

\begin{itemize}
    \item \textbf{Review: Protein Design and Engineering}\\
    \textit{Lutz, S., and Bornscheuer, U. T. "Protein engineering handbook." (Wiley-VCH, 2009).}\\
    This review provides a comprehensive overview of the various approaches to protein engineering, including both experimental and computational techniques \cite{lutz2012protein}.

    \item \textbf{A Review of Machine Learning Applications in Protein Design}\\
    \textit{AlQuraishi, M. "Machine learning in protein structure prediction." Curr Opin Chem Biol. 65: 1-9 (2021).}\\
    This review summarizes the current state of machine learning applications in protein structure prediction and design, including an overview of popular models and techniques \cite{alquraishi2021machine}.

    \item \textbf{Tutorial: Using PyRosetta for Protein Structure Prediction}\\
    \textit{Available at: \url{https://www.pyrosetta.org/documentation}.}\\
    This online tutorial provides step-by-step instructions for using PyRosetta, the Python interface for Rosetta. It includes example code and explanations for predicting protein structures and performing design tasks.

    \item \textbf{Introduction to Deep Learning for Proteins}\\
    \textit{Available at: \url{https://github.com/deepmind/alphafold}.}\\
    This GitHub repository includes tutorials and code for working with AlphaFold. It is particularly useful for those who want to understand the basics of applying deep learning models to protein structure prediction.
\end{itemize}

\subsection{Online Courses and Lectures}

\begin{itemize}
    \item \textbf{Coursera: Bioinformatics Specialization}\\
    \textit{Offered by University of California, San Diego. \url{https://www.coursera.org/specializations/bioinformatics}.}\\
    This specialization provides an introduction to bioinformatics, including the fundamentals of protein sequence analysis, structure prediction, and computational biology. It is ideal for beginners who want a comprehensive introduction to the field.

    \item \textbf{edX: Introduction to Computational Biology}\\
    \textit{Offered by MIT. Available at: \url{https://www.edx.org/course/computational-biology}.}\\
    This course introduces key concepts in computational biology, including protein structure prediction and the use of machine learning in biological data analysis. It is well-suited for those with a background in computer science or biology.

    \item \textbf{Protein Engineering: Methods and Applications}\\
    \textit{Available at: \url{https://learn.thermofisher.com/learning-center/protein-engineering}.}\\
    This resource provides detailed videos and tutorials on various protein engineering methods, including site-directed mutagenesis, directed evolution, and computational protein design. It is helpful for those who want practical, hands-on training.
\end{itemize}

\subsection{Open-Source Tools and Repositories}

\begin{itemize}
    \item \textbf{AlphaFold GitHub Repository}\\
    \textit{Available at: \url{https://github.com/deepmind/alphafold}.}\\
    This repository contains the open-source implementation of AlphaFold, the deep learning model for protein structure prediction. Beginners can use the tutorials and code in this repository to run AlphaFold locally or on cloud services.

    \item \textbf{Rosetta GitHub Repository}\\
    \textit{Available at: \url{https://github.com/RosettaCommons/main}.}\\
    The Rosetta software suite is one of the most widely used tools for protein structure prediction and design. This repository contains the source code and instructions for setting up Rosetta, as well as extensive documentation for getting started.

    \item \textbf{PyRosetta Documentation}\\
    \textit{Available at: \url{https://www.pyrosetta.org/documentation}.}\\
    PyRosetta is the Python interface for Rosetta, providing access to its powerful algorithms in a Python environment. This documentation includes tutorials and example code for tasks such as protein docking, structure prediction, and design.
\end{itemize}

\subsection{Challenges and Future Directions}

To stay up-to-date with the latest developments in protein design and deep learning, it is essential to regularly consult new research papers, participate in conferences, and explore online resources. Some of the key challenges and opportunities that require further exploration include:

\begin{itemize}
    \item \textbf{Scalability}: While current models like AlphaFold2 have shown remarkable success, challenges remain in scaling these models to handle larger, more complex protein systems, including protein-ligand and protein-protein interactions.
    \item \textbf{Integration with Experimental Data}: Combining computational models with experimental data is an ongoing challenge. Efficiently integrating experimental validation into the protein design pipeline will improve the accuracy and relevance of predictions.
    \item \textbf{De Novo Protein Design}: While structure prediction has advanced significantly, de novo protein design (designing completely new proteins from scratch) remains a key area for future research, with applications in synthetic biology, drug discovery, and materials science.
\end{itemize}

\subsection{Conclusion}

The field of protein design is evolving rapidly, with new tools and research emerging at a fast pace. For beginners, building a strong foundation in the basics of protein structure, computational modeling, and deep learning is crucial. The references, further reading materials, and online resources listed here will provide a comprehensive roadmap for navigating this complex and exciting field. By continually exploring the literature, engaging with open-source projects, and staying informed about the latest developments, learners can develop the skills needed to contribute to the future of protein design and bioengineering.


\bibliographystyle{plain}
\bibliography{sample}

\begin{thebibliography}{100}

\bibitem{abadi2016tensorflow}
Mart{\'\i}n Abadi, Ashish Agarwal, Paul Barham, Eugene Brevdo, Zhifeng Chen, Craig Citro, Greg~S Corrado, Andy Davis, Jeffrey Dean, Matthieu Devin, et~al.
\newblock Tensorflow: Large-scale machine learning on heterogeneous distributed systems.
\newblock {\em arXiv preprint arXiv:1603.04467}, 2016.

\bibitem{abraham2015gromacs}
Mark~James Abraham, Teemu Murtola, Roland Schulz, Szil{\'a}rd P{\'a}ll, Jeremy~C Smith, Berk Hess, and Erik Lindahl.
\newblock Gromacs: High performance molecular simulations through multi-level parallelism from laptops to supercomputers.
\newblock {\em SoftwareX}, 1:19--25, 2015.

\bibitem{abramson2024accurate}
Josh Abramson, Jonas Adler, Jack Dunger, Richard Evans, Tim Green, Alexander Pritzel, Olaf Ronneberger, Lindsay Willmore, Andrew~J Ballard, Joshua Bambrick, et~al.
\newblock Accurate structure prediction of biomolecular interactions with alphafold 3.
\newblock {\em Nature}, pages 1--3, 2024.

\bibitem{adam2004mapping}
Gregory~C Adam, Jonathan Burbaum, John~W Kozarich, Matthew~P Patricelli, and Benjamin~F Cravatt.
\newblock Mapping enzyme active sites in complex proteomes.
\newblock {\em Journal of the American Chemical Society}, 126(5):1363--1368, 2004.

\bibitem{adiyaman2023improvement}
Recep Adiyaman, Nicholas~S Edmunds, Ahmet~G Genc, Shuaa~MA Alharbi, and Liam~J McGuffin.
\newblock Improvement of protein tertiary and quaternary structure predictions using the refold refinement method and the alphafold2 recycling process.
\newblock {\em Bioinformatics Advances}, 3(1):vbad078, 2023.

\bibitem{agarwal2024power}
Vinayak Agarwal and Andrew~C McShan.
\newblock The power and pitfalls of alphafold2 for structure prediction beyond rigid globular proteins.
\newblock {\em Nature Chemical Biology}, 20(8):950--959, 2024.

\bibitem{akdel2022structural}
Mehmet Akdel, Douglas~EV Pires, Eduard~Porta Pardo, J{\"u}rgen J{\"a}nes, Arthur~O Zalevsky, B{\'a}lint M{\'e}sz{\'a}ros, Patrick Bryant, Lydia~L Good, Roman~A Laskowski, Gabriele Pozzati, et~al.
\newblock A structural biology community assessment of alphafold2 applications.
\newblock {\em Nature Structural \& Molecular Biology}, 29(11):1056--1067, 2022.

\bibitem{al2023investigating}
Carmen Al-Masri, Francesco Trozzi, Shu-Hang Lin, Oanh Tran, Navriti Sahni, Marcel Patek, Anna Cichonska, Balaguru Ravikumar, and Rayees Rahman.
\newblock Investigating the conformational landscape of alphafold2-predicted protein kinase structures.
\newblock {\em Bioinformatics Advances}, 3(1):vbad129, 2023.

\bibitem{alhumaid2024reliability}
Nada~K Alhumaid and Essam~A Tawfik.
\newblock Reliability of alphafold2 models in virtual drug screening: A focus on selected class a gpcrs.
\newblock {\em International Journal of Molecular Sciences}, 25(18):10139, 2024.

\bibitem{alquraishi2021machine}
Mohammed AlQuraishi.
\newblock Machine learning in protein structure prediction.
\newblock {\em Current opinion in chemical biology}, 65:1--8, 2021.

\bibitem{anand2022protein}
Namrata Anand, Raphael Eguchi, Irimpan~I Mathews, Carla~P Perez, Alexander Derry, Russ~B Altman, and Po-Ssu Huang.
\newblock Protein sequence design with a learned potential.
\newblock {\em Nature communications}, 13(1):746, 2022.

\bibitem{arnold1998design}
Frances~H Arnold.
\newblock Design by directed evolution.
\newblock {\em Accounts of chemical research}, 31(3):125--131, 1998.

\bibitem{baek2021accurate}
Minkyung Baek, Frank DiMaio, Ivan Anishchenko, Justas Dauparas, Sergey Ovchinnikov, Gyu~Rie Lee, Jue Wang, Qian Cong, Lisa~N Kinch, R~Dustin Schaeffer, et~al.
\newblock Accurate prediction of protein structures and interactions using a three-track neural network.
\newblock {\em Science}, 373(6557):871--876, 2021.

\bibitem{baker2000structural}
David Baker and Andrej Sali.
\newblock Structural biology: A century-long journey into an unseen world.
\newblock {\em Science}, 290(5495):1295--1299, 2000.

\bibitem{balchin2016chaperone}
David Balchin, Manajit Hayer-Hartl, and F~Ulrich Hartl.
\newblock The biology of molecular chaperones: principles and practice.
\newblock {\em Science}, 353(6294):aac4354, 2016.

\bibitem{basu2016dockq}
Sankar Basu and Bj{\"o}rn Wallner.
\newblock Dockq: a quality measure for protein-protein docking models.
\newblock {\em PloS one}, 11(8):e0161879, 2016.

\bibitem{battaglia2018relational}
Peter~W Battaglia, Jessica~B Hamrick, Victor Bapst, Alvaro Sanchez-Gonzalez, Vinicius Zambaldi, Mateusz Malinowski, Andrea Tacchetti, David Raposo, Adam Santoro, Ryan Faulkner, et~al.
\newblock Relational inductive biases, deep learning, and graph networks.
\newblock {\em arXiv preprint arXiv:1806.01261}, 2018.

\bibitem{bawono2017multiple}
Punto Bawono, Maurits Dijkstra, Walter Pirovano, Anton Feenstra, Sanne Abeln, and Jaap Heringa.
\newblock Multiple sequence alignment.
\newblock {\em Bioinformatics: volume I: data, sequence analysis, and evolution}, pages 167--189, 2017.

\bibitem{benkovic2008free}
Stephen~J Benkovic, Gordon~G Hammes, and Sharon Hammes-Schiffer.
\newblock Free-energy landscape of enzyme catalysis.
\newblock {\em Biochemistry}, 47(11):3317--3321, 2008.

\bibitem{berman2002protein}
Helen~M Berman, Tammy Battistuz, Talapady~N Bhat, Wolfgang~F Bluhm, Philip~E Bourne, Kyle Burkhardt, Zukang Feng, Gary~L Gilliland, Lisa Iype, Shri Jain, et~al.
\newblock The protein data bank.
\newblock {\em Acta Crystallographica Section D: Biological Crystallography}, 58(6):899--907, 2002.

\bibitem{berman2000protein}
Helen~M Berman, John Westbrook, Zukang Feng, Gary Gilliland, Talapady~N Bhat, Helge Weissig, Ilya~N Shindyalov, and Philip~E Bourne.
\newblock The protein data bank.
\newblock {\em Nucleic acids research}, 28(1):235--242, 2000.

\bibitem{bogle2010role}
I~David~L Bogle, Richard Allen, and Tom Sumner.
\newblock The role of computer aided process engineering in physiology and clinical medicine.
\newblock {\em Computers \& Chemical Engineering}, 34(5):763--769, 2010.

\bibitem{branden2012introduction}
Carl~Ivar Branden and John Tooze.
\newblock {\em Introduction to protein structure}.
\newblock Garland Science, 2012.

\bibitem{brugere2018plasmidmapper}
J-F. Brugere et~al.
\newblock Plasmidmapper: A tool for mapping and visualizing plasmid sequences.
\newblock {\em Bioinformatics}, 34(9):1547--1553, 2018.

\bibitem{bryant2022improved}
Patrick Bryant, Gabriele Pozzati, and Arne Elofsson.
\newblock Improved prediction of protein-protein interactions using alphafold2.
\newblock {\em Nature communications}, 13(1):1265, 2022.

\bibitem{burley2017protein}
Stephen~K Burley, Helen~M Berman, Gerard~J Kleywegt, John~L Markley, Haruki Nakamura, and Sameer Velankar.
\newblock Protein data bank (pdb): the single global macromolecular structure archive.
\newblock {\em Protein crystallography: methods and protocols}, pages 627--641, 2017.

\bibitem{buxbaum2007fundamentals}
Engelbert Buxbaum et~al.
\newblock {\em Fundamentals of protein structure and function}, volume~31.
\newblock Springer, 2007.

\bibitem{cavanagh1996protein}
J~Cavanagh.
\newblock {\em Protein NMR Spectroscopy: Principles And Practice}.
\newblock Academic Press, Inc, 1996.

\bibitem{celledoni2021equivariant}
Elena Celledoni, Matthias~J Ehrhardt, Christian Etmann, Brynjulf Owren, Carola-Bibiane Sch{\"o}nlieb, and Ferdia Sherry.
\newblock Equivariant neural networks for inverse problems.
\newblock {\em Inverse Problems}, 37(8):085006, 2021.

\bibitem{cheng2018single}
Yifan Cheng.
\newblock Single-particle cryo-em—how did it get here and where will it go.
\newblock {\em Science}, 361(6405):876--880, 2018.

\bibitem{cock2009biopython}
Peter~JA Cock, Tiago Antao, Jeffrey~T Chang, Brad~A Chapman, Cymon~J Cox, Andrew Dalke, Iddo Friedberg, Thomas Hamelryck, Frank Kauff, Bartek Wilczynski, et~al.
\newblock Biopython: freely available python tools for computational molecular biology and bioinformatics.
\newblock {\em Bioinformatics}, 25(11):1422--1423, 2009.

\bibitem{consortium2019uniprot}
The~UniProt Consortium.
\newblock Uniprot: a worldwide hub of protein knowledge.
\newblock {\em Nucleic Acids Research}, 47(D1):D506--D515, 2019.

\bibitem{cooper2010predicting}
Seth Cooper, Firas Khatib, Adrien Treuille, Janos Barbero, Jeehyung Lee, Michael Beenen, Andrew Leaver-Fay, David Baker, Zoran Popovi{\'c}, et~al.
\newblock Predicting protein structures with a multiplayer online game.
\newblock {\em Nature}, 466(7307):756--760, 2010.

\bibitem{coskun2023using}
Dilek Coskun, Muyun Lihan, Jo{\~a}o~PGLM Rodrigues, M{\'a}rton Vass, Daniel Robinson, Richard~A Friesner, and Edward~B Miller.
\newblock Using alphafold and experimental structures for the prediction of the structure and binding affinities of gpcr complexes via induced fit docking and free energy perturbation.
\newblock {\em Journal of Chemical Theory and Computation}, 20(1):477--489, 2023.

\bibitem{Dai2025ASO}
Weihang Dai.
\newblock A survey of deep learning methods in protein bioinformatics and its impact on protein design.
\newblock {\em ArXiv}, abs/2501.01477, 2025.

\bibitem{damm2013csar}
Kelly~L Damm-Ganamet, Richard~D Smith, James~B Dunbar~Jr, Jeanne~A Stuckey, and Heather~A Carlson.
\newblock Csar benchmark exercise 2011--2012: evaluation of results from docking and relative ranking of blinded congeneric series.
\newblock {\em Journal of chemical information and modeling}, 53(8):1853--1870, 2013.

\bibitem{dauparas2022robust}
Justas Dauparas, Ivan Anishchenko, Nathaniel Bennett, Hua Bai, Robert~J Ragotte, Lukas~F Milles, Basile~IM Wicky, Alexis Courbet, Rob~J de~Haas, Neville Bethel, et~al.
\newblock Robust deep learning--based protein sequence design using proteinmpnn.
\newblock {\em Science}, 378(6615):49--56, 2022.

\bibitem{david2022alphafold}
Alessia David, Suhail Islam, Evgeny Tankhilevich, and Michael~JE Sternberg.
\newblock The alphafold database of protein structures: a biologist’s guide.
\newblock {\em Journal of molecular biology}, 434(2):167336, 2022.

\bibitem{delano2002pymol}
Warren~L DeLano et~al.
\newblock Pymol: An open-source molecular graphics tool.
\newblock {\em CCP4 Newsl. Protein Crystallogr}, 40(1):82--92, 2002.

\bibitem{devlin2018bert}
Jacob Devlin.
\newblock Bert: Pre-training of deep bidirectional transformers for language understanding.
\newblock {\em arXiv preprint arXiv:1810.04805}, 2018.

\bibitem{dill1990dominant}
Ken~A Dill.
\newblock Dominant forces in protein folding.
\newblock {\em Biochemistry}, 29(31):7133--7155, 1990.

\bibitem{dill1995principles}
Ken~A Dill, Sarina Bromberg, Kaizhi Yue, Klaus~M Fiebig, David~P Yee, Paul~D Thomas, and Hue~Sun Chan.
\newblock Principles of protein folding—a perspective from simple exact models.
\newblock {\em Protein science}, 4(4):561--602, 1995.

\bibitem{dill2008protein}
Ken~A Dill, S~Banu Ozkan, M~Scott Shell, and Thomas~R Weikl.
\newblock The protein folding problem.
\newblock {\em Annu. Rev. Biophys.}, 37(1):289--316, 2008.

\bibitem{drenth2007principles}
Jan Drenth.
\newblock {\em Principles of protein X-ray crystallography}.
\newblock Springer Science \& Business Media, 2007.

\bibitem{dubochet2017nobel}
Jacques Dubochet, Joachim Frank, and Richard Henderson.
\newblock The nobel prize in chemistry 2017.
\newblock {\em Nobel Media AB}, 2017.

\bibitem{el2019pfam}
Sara El-Gebali, Jaina Mistry, Alex Bateman, Sean~R Eddy, Aur{\'e}lien Luciani, Simon~C Potter, Matloob Qureshi, Lorna~J Richardson, Gustavo~A Salazar, Alfredo Smart, et~al.
\newblock The pfam protein families database in 2019.
\newblock {\em Nucleic acids research}, 47(D1):D427--D432, 2019.

\bibitem{elfmann2023pae}
Christoph Elfmann and J{\"o}rg St{\"u}lke.
\newblock Pae viewer: a webserver for the interactive visualization of the predicted aligned error for multimer structure predictions and crosslinks.
\newblock {\em Nucleic acids research}, 51(W1):W404--W410, 2023.

\bibitem{elowitz2000synthetic}
Michael~B Elowitz and Stanislas Leibler.
\newblock A synthetic oscillatory network of transcriptional regulators.
\newblock {\em Nature}, 403(6767):335--338, 2000.

\bibitem{endy2005foundations}
Drew Endy.
\newblock Foundations for engineering biology.
\newblock {\em Nature}, 438(7067):449--453, 2005.

\bibitem{Enireddy2022OneHotEncodingAL}
Vamsidhar Enireddy, C.~Karthikeyan, and D.~Vijendra Babu.
\newblock Onehotencoding and lstm-based deep learning models for protein secondary structure prediction.
\newblock {\em Soft Computing}, 26:3825 -- 3836, 2022.

\bibitem{fang2022helixfold}
Xiaomin Fang, Fan Wang, Lihang Liu, Jingzhou He, Dayong Lin, Yingfei Xiang, Xiaonan Zhang, Hua Wu, Hui Li, and Le~Song.
\newblock Helixfold-single: Msa-free protein structure prediction by using protein language model as an alternative.
\newblock {\em arXiv preprint arXiv:2207.13921}, 2022.

\bibitem{Feng2024MHTAPredSSAH}
Runqiu Feng, Xun Wang, Zhijun Xia, Tongyu Han, Hanyu Wang, and Wenqian Yu.
\newblock Mhtapred-ss: A highly targeted autoencoder-driven deep multi-task learning framework for accurate protein secondary structure prediction.
\newblock {\em International Journal of Molecular Sciences}, 25, 2024.

\bibitem{fleck1966determination}
A~Fleck and HN~Munro.
\newblock The determination of nucleic acids.
\newblock {\em Methods of biochemical analysis}, 14:113--176, 1966.

\bibitem{forthal2014functions}
Donald~N Forthal.
\newblock Functions of antibodies.
\newblock {\em Microbiology spectrum}, 2(4):10--1128, 2014.

\bibitem{fu2020codon}
Hongguang Fu, Yanbing Liang, Xiuqin Zhong, ZhiLing Pan, Lei Huang, HaiLin Zhang, Yang Xu, Wei Zhou, and Zhong Liu.
\newblock Codon optimization with deep learning to enhance protein expression.
\newblock {\em Scientific reports}, 10(1):17617, 2020.

\bibitem{Gao2022UnderstandingBD}
Yuanxu Gao, Jiangshan Zhan, and Albert C.~H. Yu.
\newblock Understanding by design: Implementing deep learning from protein structure prediction to protein design.
\newblock {\em MedComm – Future Medicine}, 2022.

\bibitem{garrido2024analysis}
Pedro Garrido-Rodr{\'\i}guez, Miguel Carmena-Bargue{\~n}o, Mar{\'\i}a~Eugenia de~la Morena-Barrio, Carlos Bravo-P{\'e}rez, Bel{\'e}n de~la Morena-Barrio, Rosa Cifuentes-Riquelme, Mar{\'\i}a~Luisa Lozano, Horacio P{\'e}rez-S{\'a}nchez, and Javier Corral.
\newblock Analysis of alphafold and molecular dynamics structure predictions of mutations in serpins.
\newblock {\em Plos one}, 19(7):e0304451, 2024.

\bibitem{ghosh2014structure}
Arun~K Ghosh and Sandra Gemma.
\newblock {\em Structure-based design of drugs and other bioactive molecules: tools and strategies}.
\newblock John Wiley \& Sons, 2014.

\bibitem{goldenzweig2016automated}
Adi Goldenzweig, Moshe Goldsmith, Shannon~E Hill, Or~Gertman, Paola Laurino, Yacov Ashani, Orly Dym, Tamar Unger, Shira Albeck, Jaime Prilusky, et~al.
\newblock Automated structure-and sequence-based design of proteins for high bacterial expression and stability.
\newblock {\em Molecular cell}, 63(2):337--346, 2016.

\bibitem{goodfellow2016deep}
Ian Goodfellow.
\newblock Deep learning, 2016.

\bibitem{goodfellow2014generative}
Ian Goodfellow, Jean Pouget-Abadie, Mehdi Mirza, Bing Xu, David Warde-Farley, Sherjil Ozair, Aaron Courville, and Yoshua Bengio.
\newblock Generative adversarial nets.
\newblock {\em Advances in neural information processing systems}, 27, 2014.

\bibitem{Goudy2023InSE}
Odessa~J Goudy, Amrita Nallathambi, Tomoaki Kinjo, Nicholas~Z Randolph, and Brian Kuhlman.
\newblock In silico evolution of protein binders with deep learning models for structure prediction and sequence design.
\newblock {\em bioRxiv}, 2023.

\bibitem{goverde2023novo}
Casper~A Goverde, Benedict Wolf, Hamed Khakzad, St{\'e}phane Rosset, and Bruno~E Correia.
\newblock De novo protein design by inversion of the alphafold structure prediction network.
\newblock {\em Protein Science}, 32(6):e4653, 2023.

\bibitem{guo2022alphafold2}
Hao-Bo Guo, Alexander Perminov, Selemon Bekele, Gary Kedziora, Sanaz Farajollahi, Vanessa Varaljay, Kevin Hinkle, Valeria Molinero, Konrad Meister, Chia Hung, et~al.
\newblock Alphafold2 models indicate that protein sequence determines both structure and dynamics.
\newblock {\em Scientific Reports}, 12(1):10696, 2022.

\bibitem{haltiwanger2004role}
Robert~S Haltiwanger and John~B Lowe.
\newblock Role of glycosylation in development.
\newblock {\em Annual review of biochemistry}, 73(1):491--537, 2004.

\bibitem{Hansen2024CarvingOA}
Anders~L{\o}nstrup Hansen, Frederik~Friis Theisen, Ramon Crehuet, Enrique Marcos, Nushin Aghajari, and Martin Willemo{\"e}s.
\newblock Carving out a glycoside hydrolase active site for incorporation into a new protein scaffold using deep network hallucination.
\newblock {\em ACS Synthetic Biology}, 13:862 -- 875, 2024.

\bibitem{hastar2017peptide}
Nurcan Hastar, Elif Arslan, Mustafa~O Guler, and Ayse~B Tekinay.
\newblock Peptide-based materials for cartilage tissue regeneration.
\newblock {\em Peptides and Peptide-based Biomaterials and their Biomedical Applications}, pages 155--166, 2017.

\bibitem{heidary2022comprehensive}
Mohsen Heidary, Vahab~Hassan Kaviar, Maryam Shirani, Roya Ghanavati, Moloudsadat Motahar, Mohammad Sholeh, Hossein Ghahramanpour, and Saeed Khoshnood.
\newblock A comprehensive review of the protein subunit vaccines against covid-19.
\newblock {\em Frontiers in microbiology}, 13:927306, 2022.

\bibitem{heo2019driven}
Lim Heo, Collin~F Arbour, and Michael Feig.
\newblock Driven to near-experimental accuracy by refinement via molecular dynamics simulations.
\newblock {\em Proteins: Structure, Function, and Bioinformatics}, 87(12):1263--1275, 2019.

\bibitem{Hiranuma2022ProteinSA}
Naozumi Hiranuma.
\newblock Protein structure accuracy prediction with deep learning and its application to structure prediction and design, 2022.

\bibitem{hnasko2015elisa}
Robert Hnasko.
\newblock {\em Elisa}.
\newblock Springer, 2015.

\bibitem{ho2020denoising}
Jonathan Ho, Ajay Jain, and Pieter Abbeel.
\newblock Denoising diffusion probabilistic models.
\newblock {\em Advances in neural information processing systems}, 33:6840--6851, 2020.

\bibitem{hollingsworth2018molecular}
Scott~A Hollingsworth and Ron~O Dror.
\newblock Molecular dynamics simulation for all.
\newblock {\em Neuron}, 99(6):1129--1143, 2018.

\bibitem{holm2023dali}
Liisa Holm, Aleksi Laiho, Petri T{\"o}r{\"o}nen, and Marco Salgado.
\newblock Dali shines a light on remote homologs: One hundred discoveries.
\newblock {\em Protein Science}, 32(1):e4519, 2023.

\bibitem{hou2017seeing}
Qingzhen Hou, Paul~FG De~Geest, Wim~F Vranken, Jaap Heringa, and K~Anton Feenstra.
\newblock Seeing the trees through the forest: sequence-based homo-and heteromeric protein-protein interaction sites prediction using random forest.
\newblock {\em Bioinformatics}, 33(10):1479--1487, 2017.

\bibitem{hou2013review}
Wenbo Hou and Stephen~B Cronin.
\newblock A review of surface plasmon resonance-enhanced photocatalysis.
\newblock {\em Advanced Functional Materials}, 23(13):1612--1619, 2013.

\bibitem{hu2022exploring}
Mingyang Hu, Fajie Yuan, Kevin Yang, Fusong Ju, Jin Su, Hui Wang, Fei Yang, and Qiuyang Ding.
\newblock Exploring evolution-aware \&-free protein language models as protein function predictors.
\newblock {\em Advances in Neural Information Processing Systems}, 35:38873--38884, 2022.

\bibitem{huang2016coming}
Po-Ssu Huang, Scott~E Boyken, and David Baker.
\newblock The coming of age of de novo protein design.
\newblock {\em Nature}, 537(7620):320--327, 2016.

\bibitem{hubbard1993target}
Michael~J Hubbard and Philip Cohen.
\newblock On target with a new mechanism for the regulation of protein phosphorylation.
\newblock {\em Trends in biochemical sciences}, 18(5):172--177, 1993.

\bibitem{ingraham2019generative}
John Ingraham, Vikas Garg, Regina Barzilay, and Tommi Jaakkola.
\newblock Generative models for graph-based protein design.
\newblock {\em Advances in neural information processing systems}, 32, 2019.

\bibitem{israelachvili2011intermolecular}
Jacob~N Israelachvili.
\newblock {\em Intermolecular and surface forces}.
\newblock Academic press, 2011.

\bibitem{jackel2008protein}
C.~Jackel et~al.
\newblock Protein folding and structure prediction: From sequence to structure.
\newblock {\em Bioinformatics}, 24(1):3--10, 2008.

\bibitem{Jnes2024DeepLF}
J{\"u}rgen J{\"a}nes and Pedro Beltr{\~a}o.
\newblock Deep learning for protein structure prediction and design—progress and applications.
\newblock {\em Molecular Systems Biology}, 20:162 -- 169, 2024.

\bibitem{jinek2012programmable}
Martin Jinek, Krzysztof Chylinski, Ines Fonfara, Michael Hauer, Jennifer~A Doudna, and Emmanuelle Charpentier.
\newblock A programmable dual-rna--guided dna endonuclease in adaptive bacterial immunity.
\newblock {\em science}, 337(6096):816--821, 2012.

\bibitem{jones1999protein}
David~T Jones.
\newblock Protein secondary structure prediction based on position-specific scoring matrices.
\newblock {\em Journal of molecular biology}, 292(2):195--202, 1999.

\bibitem{jones1996principles}
Susan Jones and Janet~M Thornton.
\newblock Principles of protein-protein interactions.
\newblock {\em Proceedings of the National Academy of Sciences}, 93(1):13--20, 1996.

\bibitem{jumper2021applying}
John Jumper, Richard Evans, Alexander Pritzel, Tim Green, Michael Figurnov, Olaf Ronneberger, Kathryn Tunyasuvunakool, Russ Bates, Augustin {\v{Z}}{\'\i}dek, Anna Potapenko, et~al.
\newblock Applying and improving alphafold at casp14.
\newblock {\em Proteins: Structure, Function, and Bioinformatics}, 89(12):1711--1721, 2021.

\bibitem{jumper2021highly}
John Jumper, Richard Evans, Alexander Pritzel, Tim Green, Michael Figurnov, Olaf Ronneberger, Kathryn Tunyasuvunakool, Russ Bates, Augustin {\v{Z}}{\'\i}dek, Anna Potapenko, et~al.
\newblock Highly accurate protein structure prediction with alphafold.
\newblock {\em nature}, 596(7873):583--589, 2021.

\bibitem{kabsch1983dictionary}
Wolfgang Kabsch and Christian Sander.
\newblock Dictionary of protein secondary structure: pattern recognition of hydrogen-bonded and geometrical features.
\newblock {\em Biopolymers: Original Research on Biomolecules}, 22(12):2577--2637, 1983.

\bibitem{karras2022elucidating}
Tero Karras, Miika Aittala, Timo Aila, and Samuli Laine.
\newblock Elucidating the design space of diffusion-based generative models.
\newblock {\em Advances in neural information processing systems}, 35:26565--26577, 2022.

\bibitem{kauzmann1959some}
Walter Kauzmann.
\newblock Some factors in the interpretation of protein denaturation.
\newblock {\em Advances in protein chemistry}, 14:1--63, 1959.

\bibitem{kendrew1958three}
John~C Kendrew, G~Bodo, Howard~M Dintzis, RG~Parrish, Harold Wyckoff, and David~C Phillips.
\newblock A three-dimensional model of the myoglobin molecule obtained by x-ray analysis.
\newblock {\em Nature}, 181(4610):662--666, 1958.

\bibitem{kessel2018introduction}
Amit Kessel and Nir Ben-Tal.
\newblock {\em Introduction to proteins: structure, function, and motion}.
\newblock Chapman and Hall/CRC, 2018.

\bibitem{kingma2013auto}
Diederik~P Kingma.
\newblock Auto-encoding variational bayes.
\newblock {\em arXiv preprint arXiv:1312.6114}, 2013.

\bibitem{kitchen2004docking}
Douglas~B Kitchen, H{\'e}l{\`e}ne Decornez, John~R Furr, and J{\"u}rgen Bajorath.
\newblock Docking and scoring in virtual screening for drug discovery: methods and applications.
\newblock {\em Nature reviews Drug discovery}, 3(11):935--949, 2004.

\bibitem{knegtel1998binding}
Ronald~MA Knegtel and Peter~DJ Grootenhuis.
\newblock Binding affinities and non-bonded interaction energies.
\newblock {\em Perspectives in drug discovery and design}, 9(0):99--114, 1998.

\bibitem{kohler1975continuous}
Georges K{\"o}hler and Cesar Milstein.
\newblock Continuous cultures of fused cells secreting antibody of predefined specificity.
\newblock {\em nature}, 256(5517):495--497, 1975.

\bibitem{krieger2003homology}
Elmar Krieger, Sander~B Nabuurs, and Gert Vriend.
\newblock Homology modeling.
\newblock {\em Structural bioinformatics}, 44:509--523, 2003.

\bibitem{kroeze2003g}
Wesley~K Kroeze, Douglas~J Sheffler, and Bryan~L Roth.
\newblock G-protein-coupled receptors at a glance.
\newblock {\em Journal of cell science}, 116(24):4867--4869, 2003.

\bibitem{lazaridis1999effective}
Themis Lazaridis and Martin Karplus.
\newblock Effective energy function for proteins in solution.
\newblock {\em Proteins: Structure, Function, and Bioinformatics}, 35(2):133--152, 1999.

\bibitem{le2017protein}
Quan Le, Fabian Sievers, and Desmond~G Higgins.
\newblock Protein multiple sequence alignment benchmarking through secondary structure prediction.
\newblock {\em Bioinformatics}, 33(9):1331--1337, 2017.

\bibitem{leaver2011rosetta3}
Andrew Leaver-Fay, Michael Tyka, Steven~M Lewis, Oliver~F Lange, James Thompson, Ron Jacak, Kristian~W Kaufman, P~Douglas Renfrew, Colin~A Smith, Will Sheffler, et~al.
\newblock Rosetta3: an object-oriented software suite for the simulation and design of macromolecules.
\newblock In {\em Methods in enzymology}, volume 487, pages 545--574. Elsevier, 2011.

\bibitem{lecun2015deep}
Yann LeCun, Yoshua Bengio, and Geoffrey Hinton.
\newblock Deep learning.
\newblock {\em nature}, 521(7553):436--444, 2015.

\bibitem{lee2022comparative}
Chien Lee, Bo-Han Su, and Yufeng~Jane Tseng.
\newblock Comparative studies of alphafold, rosettafold and modeller: a case study involving the use of g-protein-coupled receptors.
\newblock {\em Briefings in bioinformatics}, 23(5):bbac308, 2022.

\bibitem{lehrman2017protein}
S~Russell Lehrman.
\newblock Protein structure.
\newblock {\em Fundamentals of protein biotechnology}, pages 9--38, 2017.

\bibitem{lesnierowski2007lysozyme}
Grezegorz Lesnierowski and Jacek Kijowski.
\newblock Lysozyme.
\newblock {\em Bioactive egg compounds}, pages 33--42, 2007.

\bibitem{Limbu2022ANH}
Sarita Limbu and Sivanesan Dakshanamurthy.
\newblock A new hybrid neural network deep learning method for protein–ligand binding affinity prediction and de novo drug design.
\newblock {\em International Journal of Molecular Sciences}, 23, 2022.

\bibitem{Liu2024ExploringPB}
Yufan Liu.
\newblock Exploring protein-dna binding residue prediction and consistent interpretability analysis using deep learning.
\newblock {\em bioRxiv}, 2024.

\bibitem{lutz2012protein}
Stefan Lutz and Uwe~Theo Bornscheuer.
\newblock {\em Protein engineering handbook}.
\newblock John Wiley \& Sons, 2012.

\bibitem{Ma2024BeyondCB}
Xin Ma and Dong Si.
\newblock Beyond current boundaries: Integrating deep learning and alphafold for enhanced protein structure prediction from low-resolution cryo-em maps.
\newblock {\em ArXiv}, abs/2410.23321, 2024.

\bibitem{madej1995threading}
Thomas Madej, Jean-Fran{\c{c}}ois Gibrat, and Stephen~H Bryant.
\newblock Threading a database of protein cores.
\newblock {\em Proteins: Structure, Function, and Bioinformatics}, 23(3):356--369, 1995.

\bibitem{Manshour2024IntegratingPS}
Negin Manshour, Fei He, Duolin Wang, and Dong Xu.
\newblock Integrating protein structure prediction and bayesian optimization for peptide design.
\newblock {\em Research Square}, 2024.

\bibitem{mao2023cross}
Anqi Mao, Mehryar Mohri, and Yutao Zhong.
\newblock Cross-entropy loss functions: Theoretical analysis and applications.
\newblock In {\em International conference on Machine learning}, pages 23803--23828. PMLR, 2023.

\bibitem{Mardikoraem2024EvoSeqMLAD}
Mehrsa Mardikoraem, Nathaniel Pascual, Patrick Finneran, and Daniel~R. Woldring.
\newblock Evoseq-ml: Advancing data-centric machine learning with evolutionary-informed protein sequence representation and generation.
\newblock {\em bioRxiv}, 2024.

\bibitem{mariani2013lddt}
Valerio Mariani, Marco Biasini, Alessandro Barbato, and Torsten Schwede.
\newblock lddt: a local superposition-free score for comparing protein structures and models using distance difference tests.
\newblock {\em Bioinformatics}, 29(21):2722--2728, 2013.

\bibitem{meng2023improved}
Qiaozhen Meng, Fei Guo, and Jijun Tang.
\newblock Improved structure-related prediction for insufficient homologous proteins using msa enhancement and pre-trained language model.
\newblock {\em Briefings in Bioinformatics}, 24(4):bbad217, 2023.

\bibitem{meredith2004targeted}
Gavin~D Meredith, Hayley~Y Wu, and Nancy~L Allbritton.
\newblock Targeted protein functionalization using his-tags.
\newblock {\em Bioconjugate chemistry}, 15(5):969--982, 2004.

\bibitem{mirdita2022colabfold}
Milot Mirdita, Konstantin Sch{\"u}tze, Yoshitaka Moriwaki, Lim Heo, Sergey Ovchinnikov, and Martin Steinegger.
\newblock Colabfold: making protein folding accessible to all.
\newblock {\em Nature methods}, 19(6):679--682, 2022.

\bibitem{morris2001autodock}
Garrett~M Morris, David~S Goodsell, Ruth Huey, William~E Hart, Scott Halliday, Rik Belew, and Arthur~J Olson.
\newblock Autodock.
\newblock {\em Automated docking of flexible ligands to receptor-User Guide}, 2001.

\bibitem{motorin2007identification}
Yuri Motorin, S{\'e}bastien Muller, Isabelle Behm-Ansmant, and Christiane Branlant.
\newblock Identification of modified residues in rnas by reverse transcription-based methods.
\newblock {\em Methods in enzymology}, 425:21--53, 2007.

\bibitem{mundlapati2018noncovalent}
V~Rao Mundlapati, Dipak~Kumar Sahoo, Suman Bhaumik, Subhrakant Jena, Amol Chandrakar, and Himansu~S Biswal.
\newblock Noncovalent carbon-bonding interactions in proteins.
\newblock {\em Angewandte Chemie International Edition}, 57(50):16496--16500, 2018.

\bibitem{nataro1998diarrheagenic}
James~P Nataro and James~B Kaper.
\newblock Diarrheagenic escherichia coli.
\newblock {\em Clinical microbiology reviews}, 11(1):142--201, 1998.

\bibitem{ng2002predicting}
TS~Eugene Ng and Hui Zhang.
\newblock Predicting internet network distance with coordinates-based approaches.
\newblock In {\em Proceedings. Twenty-First Annual Joint Conference of the IEEE Computer and Communications Societies}, volume~1, pages 170--179. IEEE, 2002.

\bibitem{nielsen2014engineering}
Jens Nielsen, Martin Fussenegger, Jay Keasling, Sang~Yup Lee, James~C Liao, Kristala Prather, and Bernhard Palsson.
\newblock Engineering synergy in biotechnology.
\newblock {\em Nature chemical biology}, 10(5):319--322, 2014.

\bibitem{oldfield2014intrinsically}
Christopher~J Oldfield and A~Keith Dunker.
\newblock Intrinsically disordered proteins and intrinsically disordered protein regions.
\newblock {\em Annual review of biochemistry}, 83(1):553--584, 2014.

\bibitem{olechnovivc2013cad}
Kliment Olechnovi{\v{c}}, Eleonora Kulberkyt{\.e}, and {\v{C}}eslovas Venclovas.
\newblock Cad-score: a new contact area difference-based function for evaluation of protein structural models.
\newblock {\em Proteins: Structure, Function, and Bioinformatics}, 81(1):149--162, 2013.

\bibitem{onuchic1997theory}
José~N. Onuchic, Zaida Luthey-Schulten, and Peter~G. Wolynes.
\newblock Theory of protein folding: The energy landscape perspective.
\newblock {\em Annual Review of Physical Chemistry}, 48(1):545--600, 1997.

\bibitem{pan2011introduction}
Yi~Pan and Albert~Y Zomaya.
\newblock {\em Introduction to protein structure prediction: methods and algorithms}.
\newblock John Wiley \& Sons, 2011.

\bibitem{pang2020deep}
Bo~Pang, Erik Nijkamp, and Ying~Nian Wu.
\newblock Deep learning with tensorflow: A review.
\newblock {\em Journal of Educational and Behavioral Statistics}, 45(2):227--248, 2020.

\bibitem{paszke2019pytorch}
Adam Paszke, Sam Gross, Francisco Massa, Adam Lerer, James Bradbury, Gregory Chanan, Trevor Killeen, Zeming Lin, Natalia Gimelshein, Luca Antiga, et~al.
\newblock Pytorch: An imperative style, high-performance deep learning library.
\newblock {\em Advances in neural information processing systems}, 32, 2019.

\bibitem{pauling1951configuration}
Linus Pauling and Robert~B Corey.
\newblock The configuration of polypeptide chains in proteins.
\newblock {\em Proceedings of the National Academy of Sciences}, 37(5):235--240, 1951.

\bibitem{Pearce2021DeepLT}
Robin Pearce and Yang Zhang.
\newblock Deep learning techniques have significantly impacted protein structure prediction and protein design.
\newblock {\em Current opinion in structural biology}, 68:194--207, 2021.

\bibitem{pelton2000spectroscopic}
John~T Pelton and Larry~R McLean.
\newblock Spectroscopic methods for analysis of protein secondary structure.
\newblock {\em Analytical biochemistry}, 277(2):167--176, 2000.

\bibitem{pereira2007evolution}
Jose~B Pereira-Leal, Emmanuel~D Levy, Christel Kamp, and Sarah~A Teichmann.
\newblock Evolution of protein complexes by duplication of homomeric interactions.
\newblock {\em Genome biology}, 8:1--12, 2007.

\bibitem{pettersen2004ucsf}
Eric~F Pettersen, Thomas~D Goddard, Conrad~C Huang, Gregory~S Couch, Daniel~M Greenblatt, Elaine~C Meng, and Thomas~E Ferrin.
\newblock Ucsf chimera—a visualization system for exploratory research and analysis.
\newblock {\em Journal of computational chemistry}, 25(13):1605--1612, 2004.

\bibitem{pettersen2021ucsf}
Eric~F Pettersen, Thomas~D Goddard, Conrad~C Huang, Elaine~C Meng, Gregory~S Couch, Tristan~I Croll, John~H Morris, and Thomas~E Ferrin.
\newblock Ucsf chimerax: Structure visualization for researchers, educators, and developers.
\newblock {\em Protein science}, 30(1):70--82, 2021.

\bibitem{rahimzadeh2024unveiling}
Faezeh Rahimzadeh, Leyli~Mohammad Khanli, Pedram Salehpoor, Faegheh Golabi, and Shahin PourBahrami.
\newblock Unveiling the evolution of policies for enhancing protein structure predictions: A comprehensive analysis.
\newblock {\em Computers in Biology and Medicine}, 179:108815, 2024.

\bibitem{ramazi2021post}
Shahin Ramazi and Javad Zahiri.
\newblock Post-translational modifications in proteins: resources, tools and prediction methods.
\newblock {\em Database}, 2021:baab012, 2021.

\bibitem{rees1977secondary}
David~A Rees and E~Jane Welsh.
\newblock Secondary and tertiary structure of polysaccharides in solutions and gels.
\newblock {\em Angewandte Chemie International Edition in English}, 16(4):214--224, 1977.

\bibitem{rietman2016thermodynamic}
Edward~A Rietman, John Platig, Jack~A Tuszynski, and Giannoula Lakka~Klement.
\newblock Thermodynamic measures of cancer: Gibbs free energy and entropy of protein--protein interactions.
\newblock {\em Journal of biological physics}, 42:339--350, 2016.

\bibitem{risques2018aging}
Rosa~Ana Risques and Scott~R Kennedy.
\newblock Aging and the rise of somatic cancer-associated mutations in normal tissues.
\newblock {\em PLoS genetics}, 14(1):e1007108, 2018.

\bibitem{rohl2004protein}
Carol~A Rohl, Charlie~EM Strauss, Kira~MS Misura, and David Baker.
\newblock Protein structure prediction using rosetta.
\newblock In {\em Methods in enzymology}, volume 383, pages 66--93. Elsevier, 2004.

\bibitem{rosenbaum2009structure}
Daniel~M Rosenbaum, S{\o}ren~GF Rasmussen, and Brian~K Kobilka.
\newblock The structure and function of g-protein-coupled receptors.
\newblock {\em Nature}, 459(7245):356--363, 2009.

\bibitem{ruiz2013global}
Yasser~B Ruiz-Blanco, Yovani Marrero-Ponce, Waldo Paz, Yamila Garc{\'\i}a, and Jes{\'u}s Salgado.
\newblock Global stability of protein folding from an empirical free energy function.
\newblock {\em Journal of theoretical biology}, 321:44--53, 2013.

\bibitem{Saharkhiz2024TheSO}
Saber Saharkhiz, Mehrnaz Mostafavi, Amin Birashk, Shiva Karimian, Shayan Khalilollah, Sohrab Jaferian, Yalda Yazdani, Iraj Alipourfard, Yun~Suk Huh, Marzieh~Ramezani Farani, and Reza Akhavan-Sigari.
\newblock The state-of-the-art overview to application of deep learning in accurate protein design and structure prediction.
\newblock {\em Topics in Current Chemistry (Cham)}, 382, 2024.

\bibitem{Sato2023RecentTI}
Kengo Sato and Michiaki Hamada.
\newblock Recent trends in rna informatics: a review of machine learning and deep learning for rna secondary structure prediction and rna drug discovery.
\newblock {\em Briefings in Bioinformatics}, 24, 2023.

\bibitem{scheeff2003fundamentals}
Eric~D Scheeff and Jody~Lynn Fink.
\newblock Fundamentals of protein structure.
\newblock {\em Structural bioinformatics}, 44:15--39, 2003.

\bibitem{schwartz2000pipmaker}
Scott Schwartz, Zheng Zhang, Kelly~A Frazer, Arian Smit, Cathy Riemer, John Bouck, Richard Gibbs, Ross Hardison, and Webb Miller.
\newblock Pipmaker—a web server for aligning two genomic dna sequences.
\newblock {\em Genome research}, 10(4):577--586, 2000.

\bibitem{schymkowitz2005foldx}
Joost Schymkowitz, Jesper Borg, Francois Stricher, Robby Nys, Frederic Rousseau, and Luis Serrano.
\newblock The foldx web server: an online force field.
\newblock {\em Nucleic acids research}, 33(suppl\_2):W382--W388, 2005.

\bibitem{senior2019protein}
Andrew~W Senior, Richard Evans, John Jumper, James Kirkpatrick, Laurent Sifre, Tim Green, Chongli Qin, Augustin {\v{Z}}{\'\i}dek, Alexander~WR Nelson, Alex Bridgland, et~al.
\newblock Protein structure prediction using multiple deep neural networks in the 13th critical assessment of protein structure prediction (casp13).
\newblock {\em Proteins: structure, function, and bioinformatics}, 87(12):1141--1148, 2019.

\bibitem{senior2020improved}
Andrew~W Senior, Richard Evans, John Jumper, James Kirkpatrick, Laurent Sifre, Tim Green, Chongli Qin, Augustin {\v{Z}}{\'\i}dek, Alexander~WR Nelson, Alex Bridgland, et~al.
\newblock Improved protein structure prediction using potentials from deep learning.
\newblock {\em Nature}, 577(7792):706--710, 2020.

\bibitem{sharma2021enzyme}
Anshula Sharma, Gaganjot Gupta, Tawseef Ahmad, Sheikh Mansoor, and Baljinder Kaur.
\newblock Enzyme engineering: current trends and future perspectives.
\newblock {\em Food Reviews International}, 37(2):121--154, 2021.

\bibitem{sherman2006novel}
Woody Sherman, Tyler Day, Matthew~P Jacobson, Richard~A Friesner, and Ramy Farid.
\newblock Novel procedure for modeling ligand/receptor induced fit effects.
\newblock {\em Journal of medicinal chemistry}, 49(2):534--553, 2006.

\bibitem{silva2014protein}
Nuno~HCS Silva, Carla Vilela, Isabel~M Marrucho, Carmen~SR Freire, Carlos~Pascoal Neto, and Armando~JD Silvestre.
\newblock Protein-based materials: from sources to innovative sustainable materials for biomedical applications.
\newblock {\em Journal of Materials Chemistry B}, 2(24):3715--3740, 2014.

\bibitem{steinegger2017mmseqs2}
Martin Steinegger and Johannes S{\"o}ding.
\newblock Mmseqs2 enables sensitive protein sequence searching for the analysis of massive data sets.
\newblock {\em Nature biotechnology}, 35(11):1026--1028, 2017.

\bibitem{stephanopoulos2007challenges}
Gregory Stephanopoulos.
\newblock Challenges in engineering microbes for biofuels production.
\newblock {\em Science}, 315(5813):801--804, 2007.

\bibitem{tautz2011evolutionary}
Diethard Tautz and Tomislav Domazet-Lo{\v{s}}o.
\newblock The evolutionary origin of orphan genes.
\newblock {\em Nature Reviews Genetics}, 12(10):692--702, 2011.

\bibitem{thomas2008targeting}
Jason~R Thomas and Paul~J Hergenrother.
\newblock Targeting rna with small molecules.
\newblock {\em Chemical reviews}, 108(4):1171--1224, 2008.

\bibitem{thornton1981disulphide}
Janet~M Thornton.
\newblock Disulphide bridges in globular proteins.
\newblock {\em Journal of molecular biology}, 151(2):261--287, 1981.

\bibitem{trott2010autodock}
Oleg Trott and Arthur~J Olson.
\newblock Autodock vina: improving the speed and accuracy of docking with a new scoring function, efficient optimization, and multithreading.
\newblock {\em Journal of computational chemistry}, 31(2):455--461, 2010.

\bibitem{van2005gromacs}
David Van Der~Spoel, Erik Lindahl, Berk Hess, Gerrit Groenhof, Alan~E Mark, and Herman~JC Berendsen.
\newblock Gromacs: fast, flexible, and free.
\newblock {\em Journal of computational chemistry}, 26(16):1701--1718, 2005.

\bibitem{Wang2024ProteinDU}
Jue Wang, Joseph~L. Watson, and Sidney~Lyayuga Lisanza.
\newblock Protein design using structure-prediction networks: Alphafold and rosettafold as protein structure foundation models.
\newblock {\em Cold Spring Harbor perspectives in biology}, 2024.

\bibitem{waterhouse2018swiss}
Andrew Waterhouse, Martino Bertoni, Stefan Bienert, Gabriel Studer, Gerardo Tauriello, Rafal Gumienny, Florian~T Heer, Tjaart A~P de~Beer, Christine Rempfer, Lorenza Bordoli, et~al.
\newblock Swiss-model: homology modelling of protein structures and complexes.
\newblock {\em Nucleic acids research}, 46(W1):W296--W303, 2018.

\bibitem{watowich1988stable}
Stanley~J Watowich, Eric~S Meyer, Ray Hagstrom, and Robert Josephs.
\newblock A stable, rapidly converging conjugate gradient method for energy minimization.
\newblock {\em Journal of computational chemistry}, 9(6):650--661, 1988.

\bibitem{wayment2024predicting}
Hannah~K Wayment-Steele, Adedolapo Ojoawo, Renee Otten, Julia~M Apitz, Warintra Pitsawong, Marc H{\"o}mberger, Sergey Ovchinnikov, Lucy Colwell, and Dorothee Kern.
\newblock Predicting multiple conformations via sequence clustering and alphafold2.
\newblock {\em Nature}, 625(7996):832--839, 2024.

\bibitem{webb2016comparative}
Benjamin Webb and Andrej Sali.
\newblock Comparative protein structure modeling using modeller.
\newblock {\em Current protocols in bioinformatics}, 54(1):5--6, 2016.

\bibitem{weinreb20163d}
Caleb Weinreb, Adam~J Riesselman, John~B Ingraham, Torsten Gross, Chris Sander, and Debora~S Marks.
\newblock 3d rna and functional interactions from evolutionary couplings.
\newblock {\em Cell}, 165(4):963--975, 2016.

\bibitem{Wicky2022HallucinatingSP}
Basile I.~M. Wicky, Lukas~F. Milles, Alexis Courbet, Robert~J. Ragotte, Justas Dauparas, E~Kinfu, S.~Tipps, Ryan~D. Kibler, Minkyung Baek, Frank DiMaio, X.~Li, Lauren~P. Carter, Alex Kang, H.~Nguyen, A.~K. Bera, and David Baker.
\newblock Hallucinating symmetric protein assemblies.
\newblock {\em Science (New York, N.Y.)}, 378:56 -- 61, 2022.

\bibitem{wilson2022alphafold2}
Carter~J Wilson, Wing-Yiu Choy, and Mikko Karttunen.
\newblock Alphafold2: a role for disordered protein/region prediction?
\newblock {\em International Journal of Molecular Sciences}, 23(9):4591, 2022.

\bibitem{wilson2015rational}
Corey~J Wilson.
\newblock Rational protein design: developing next-generation biological therapeutics and nanobiotechnological tools.
\newblock {\em Wiley Interdisciplinary Reviews: Nanomedicine and Nanobiotechnology}, 7(3):330--341, 2015.

\bibitem{wilson1993antibody}
Ian~A Wilson and Robyn~L Stanfield.
\newblock Antibody-antigen interactions.
\newblock {\em Current opinion in structural biology}, 3(1):113--118, 1993.

\bibitem{wu2021ebm}
Jiaxiang Wu, Shitong Luo, Tao Shen, Haidong Lan, Sheng Wang, and Junzhou Huang.
\newblock Ebm-fold: fully-differentiable protein folding powered by energy-based models.
\newblock {\em arXiv preprint arXiv:2105.04771}, 2021.

\bibitem{wu2024protein}
Kevin~E Wu, Kevin~K Yang, Rianne van~den Berg, Sarah Alamdari, James~Y Zou, Alex~X Lu, and Ava~P Amini.
\newblock Protein structure generation via folding diffusion.
\newblock {\em Nature communications}, 15(1):1059, 2024.

\bibitem{wu2022high}
Ruidong Wu, Fan Ding, Rui Wang, Rui Shen, Xiwen Zhang, Shitong Luo, Chenpeng Su, Zuofan Wu, Qi~Xie, Bonnie Berger, et~al.
\newblock High-resolution de novo structure prediction from primary sequence.
\newblock {\em BioRxiv}, pages 2022--07, 2022.

\bibitem{wuthrich1986nmr}
K~W{\"u}thrich.
\newblock Nmr of proteins and nucleic acids, 1986.

\bibitem{xu2007computational}
Ying Xu, Dong Xu, Jie Liang, et~al.
\newblock {\em Computational methods for protein structure prediction and modeling}, volume~2.
\newblock Springer, 2007.

\bibitem{yang2015tasser}
Jianyi Yang and Yang Zhang.
\newblock I-tasser server: new development for protein structure and function predictions.
\newblock {\em Nucleic acids research}, 43(W1):W174--W181, 2015.

\bibitem{yang2023alphafold2}
Zhenyu Yang, Xiaoxi Zeng, Yi~Zhao, and Runsheng Chen.
\newblock Alphafold2 and its applications in the fields of biology and medicine.
\newblock {\em Signal Transduction and Targeted Therapy}, 8(1):115, 2023.

\bibitem{zhang2022strategies}
Gong Zhang, Juan Zhang, Yuting Gao, Yangfeng Li, and Yizhou Li.
\newblock Strategies for targeting undruggable targets.
\newblock {\em Expert Opinion on Drug Discovery}, 17(1):55--69, 2022.

\bibitem{zhang2024vocapter}
Li~Zhang, Zean Han, Yan Zhong, Qiaojun Yu, Xingyu Wu, et~al.
\newblock Vocapter: Voting-based pose tracking for category-level articulated object via inter-frame priors.
\newblock In {\em ACM Multimedia 2024}, 2024.

\bibitem{zhang2025u}
Li~Zhang, Weiqing Meng, Yan Zhong, Bin Kong, Mingliang Xu, Jianming Du, Xue Wang, Rujing Wang, and Liu Liu.
\newblock U-cope: Taking a further step to universal 9d category-level object pose estimation.
\newblock In {\em European Conference on Computer Vision}, pages 254--270. Springer, 2025.

\bibitem{zhangrethinking}
Li~Zhang, Yan Zhong, Jianan Wang, Zhe Min, Liu Liu, et~al.
\newblock Rethinking 3d convolution in $\ell_p$-norm space.
\newblock In {\em The Thirty-eighth Annual Conference on Neural Information Processing Systems}, 2024.

\bibitem{zhang2004scoring}
Yang Zhang and Jeffrey Skolnick.
\newblock Scoring function for automated assessment of protein structure template quality.
\newblock {\em Proteins: Structure, Function, and Bioinformatics}, 57(4):702--710, 2004.

\bibitem{zhang2005tm}
Yang Zhang and Jeffrey Skolnick.
\newblock Tm-align: a protein structure alignment algorithm based on the tm-score.
\newblock {\em Nucleic acids research}, 33(7):2302--2309, 2005.

\bibitem{zhang2024unraveling}
Ying Zhang, Fang Wu, Yongyue Han, Yuzhe Wu, Liqiu Huang, Yuanwei Huang, Di~Yan, Xiwen Jiang, Jingyun Ma, and Wei Xu.
\newblock Unraveling the assembly mechanism of sads-cov virus nucleocapsid protein: insights from rna binding, dimerization, and epitope diversity profiling.
\newblock {\em Journal of Virology}, 98(8):e00926--24, 2024.

\bibitem{zhou2018electrostatic}
Huan-Xiang Zhou and Xiaodong Pang.
\newblock Electrostatic interactions in protein structure, folding, binding, and condensation.
\newblock {\em Chemical reviews}, 118(4):1691--1741, 2018.

\end{thebibliography}

\end{document}